\documentclass[
 groupedaddress,
frontmatterverbose,
 showpacs,
 nofootinbib,
 amsmath,
 amssymb,
 aps,
 tightenlines,
 twocolumn,
 rmp,
 floatfix,
]{revtex4-1}

\usepackage{txfonts}
\usepackage{bbm}
\usepackage{amsmath}
\usepackage{amsfonts}
\usepackage{graphicx}
\usepackage{dcolumn}
\usepackage{bm}
\usepackage{color}
\usepackage{epstopdf}

\newcommand{\tr}{\mathrm{tr}}
\newcommand{\CC}{\mathcal{CC}}
\newcommand{\iden}{\openone}

\begin{document}


\title{Quantum coherence and geometric quantum discord}

\author{Ming-Liang Hu}
\affiliation{Institute of Physics, Chinese Academy of Sciences, Beijing 100190, China}
\affiliation{School of Science, Xi'an University of Posts and Telecommunications, Xi'an 710121, China}

\author{Xueyuan Hu}
\affiliation{School of Information Science and Engineering, Shandong University, Jinan 250100, China}

\author{Jieci Wang}
\email{jcwang@hunnu.edu.cn}
\affiliation{Department of Physics, and Synergetic Innovation Center for Quantum Effects,
             Hunan Normal University, Changsha, Hunan 410081, China}
\affiliation{Institute of Physics, Chinese Academy of Sciences, Beijing 100190, China}

\author{Yi Peng}
\affiliation{Institute of Physics, Chinese Academy of Sciences, Beijing 100190, China}

\author{Yu-Ran Zhang}
\affiliation{Institute of Physics, Chinese Academy of Sciences, Beijing 100190, China}
\affiliation{Theoretical Quantum Physics Laboratory, RIKEN Cluster for Pioneering Research, Wako-shi, Saitama 351-0198, Japan}
\affiliation{Beijing Computational Science Research Center, Beijing 100094, China}

\author{Heng Fan}
\email{hfan@iphy.ac.cn}
\affiliation{Institute of Physics, Chinese Academy of Sciences, Beijing 100190, China}
\affiliation{CAS Center for Excellence in Topological Quantum Computation, University of Chinese Academy of Sciences, Beijing 100190, China}
\affiliation{Collaborative Innovation Center of Quantum Matter, Beijing 100190, China}

\date{\today}

\begin{abstract}
Quantum coherence and quantum correlations are of fundamental and
practical significance for the development of quantum mechanics.
They are also cornerstones of quantum computation and quantum
communication theory. Searching physically meaningful and
mathematically rigorous quantifiers of them are long-standing
concerns of the community of quantum information science, and
various faithful measures have been introduced so far. We review in
this paper the measures of discordlike quantum correlations for
bipartite and multipartite systems, the measures of quantum
coherence for any single quantum system, and their relationship in
different settings. Our aim is to provide a full review about the
resource theory of quantum coherence, including its application in
many-body systems, and the discordlike quantum correlations which
were defined based on the various distance measures of states. We
discuss the interrelations between quantum coherence and quantum
correlations established in an operational way, and the fundamental
characteristics of quantum coherence such as their complementarity
under different basis sets, their duality with path information of
an interference experiment, their distillation and dilution under
different operations, and some new viewpoints of the superiority of
the quantum algorithms from the perspective of quantum coherence.
Additionally, we review properties of geometric quantum correlations
and quantum coherence under noisy quantum channels. Finally, the
main progresses for the study of quantum correlations and quantum
coherence in the relativistic settings are reviewed. All these
results provide an overview for the conceptual implications and
basic connections of quantum coherence, quantum correlations, and
their potential applications in various related subjects of physics.
\end{abstract}


\maketitle
\tableofcontents

\section{Introduction} \label{sec:1}
Quantum correlations and quantum coherence are two fundamental
concepts in quantum theory \cite{Nielsen}. While quantum
correlations characterize the quantum features of a bipartite or
multipartite system, quantum coherence was defined for the integral
system. Despite this obvious difference and as the different
embodiments of the unique characteristics of a quantum system, they
are also intimately related to each other which can be scrutinized
from different perspectives. Moreover, from a practical point of
view, quantum correlations and quantum coherence are also invaluable
physical resources for quantum information and computation tasks
\cite{Nielsen}, hence remain research focuses since the early days
of quantum mechanics.

The quantum correlation in a system can be characterized and
quantified from different perspectives. Historically, there are
various categories of quantum correlation measures being proposed,
prominent examples include the widely-studied Bell-type nonlocal
correlation \cite{rev-Bell} and quantum entanglement \cite{RMPE}. At
the beginning of this century, a new framework for quantifying
quantum correlations was formulated by \citet{qd01}, as well as by
\citet{qd02} in the study of quantum discord (QD). Within this
framework, an abundance of discordlike quantum correlation measures
were proposed and studied from different aspects in the past years
\cite{RMP,Adesso-jpa}.

Quantum coherence, another embodiment of the superposition principle
of quantum states, is essential for many novel and intriguing
characteristics of quantum systems \cite{Ficek}. It is also of equal
importance as quantum correlations in the studies of both bipartite
and multipartite systems \cite{RMP}. Constructing a mathematically
rigorous and physically meaningful framework for its
characterization and quantification was a main pursue of researchers
in quantum community, as this is not only essential for quantum
foundations, but can also provides the basis for its potential
applications in a wide variety of promising subjects, such as
quantum thermodynamics \cite{ther1,ther2,ther3,ther4}, reference
frames \cite{frame}, and quantum biology \cite{biolo}.

Coherence is a main concern of quantum optics historically, and it
was usually studied from the perspective of phase space
distributions or multipoint correlations. Instead of reviewing these
approaches, we focus on the recent developments about quantitative
characterization of coherence \cite{coher}, the essence of which is
the adoption of the viewpoint of treating quantum coherence as a
physical resource, just like the resource theory of entanglement. A
big advantage of this approach is that the related resource theory
of entanglement can be adapted for introducing similar measures of
quantum coherence \cite{RMPE}.

The geometric characterization of quantum correlations in a
bipartite system have been studied extensively in the past five
years, with the corresponding measures being defined based on
various (pseudo) distance measures of two states. The central idea
for this kind of definition is that the distance between the
considered state and the state without the desired property is a
measure of that property \cite{reqd}. The similar idea can also be
used to characterize quantum coherence of a system, and along this
line, many quantum coherence measures have been proposed. We will
review these coherence measures and show their interrelations with
quantum correlations. Moreover, we will also discuss the other major
progresses achieved in studying fundamental problems of geometric
quantum correlations and quantum coherence, their dynamics under
noisy quantum channels, and their applications in the study of
many-body systems.

During the preparation of this review, we became aware of another
nice review work by \citet{Adesso}. Though seemingly discussing a
similar topic, our concerns are different. \citet{Adesso} reviewed
the resource theory of coherence, while our concerns are quantum
coherence, quantum correlations, and their intrinsic connections. In
our work, besides presenting a comprehensive view of the main
developments of quantum coherence and quantum correlations, we try
to summarize and reformulate some calculations scattering
in a large number of literature. In particular, we have also
reviewed in detail the main progresses for the study of quantum
correlations and quantum coherence in the relativistic settings. We
hope that this review may be useful for both beginners and seniors
in quantum information science.

This review is organized as follows. In Section \ref{sec:2}, we
first briefly recall the framework for defining QD and the seminal
measure of QD defined via the discrepancy between quantum mutual
information (QMI) and the classical aspect of correlations. Then, we
present in detail the recently introduced geometric quantum discords
(GQDs) and measurement-induced nonlocality (MIN) defined by virtue
of various distance measures of quantum states, e.g., the
Hilbert-Schmidt (HS) norm, the trace norm, the Bures distance, the
Hellinger distance, the relative entropy, and the Wigner-Yanase (WY)
skew information. We will also present a review concerning the
fundamental connections among these correlation measures and the
recent developments about their potential role in typical tasks of
quantum information processing.

In Section \ref{sec:3}, we first review the basic aspects related to
the resource theory of quantum coherence including structure of the
corresponding free states and free operations. We then present a
detailed review on those recent developments of the quantification
of quantum coherence. They can be classified roughly into the
following categories: the distance-based measure of coherence, the
entanglement-based measure of coherence, the convex roof measure of
coherence, the robustness of coherence, the Tsallis relative entropy
measure of coherence, and the skew-information-based measure of
coherence. For every category, there are also different definitions
of coherence measures. Moreover, some of their extensions valid for
infinite-dimensional systems and some other coherence measures
defined within slightly different frameworks, for example, the
different construction of free operations, will also be reviewed in
this section.

In Section \ref{sec:4}, we review recent developments on
interpretations of the aforementioned quantum coherence measures.
First, we show from an operational perspective that quantum
coherence is intimately related to quantum correlations in bipartite
and multipartite systems, highlighting the fundamental position of
coherence in quantum theory. Then, we review the complementarity
relations of quantum coherence under both mutually unbiased bases
(MUBs) and incompatible bases, as well as the complementarity
relation between coherence and mixedness and between coherence and
path distinguishability. Finally, the distillation and dilution of
quantum coherence with different forms of restricted operations and
the average coherence of randomly sampled states are discussed.

In Section \ref{sec:5}, we consider the role of quantum coherence in
some quantum information processing tasks, including quantum state
merging, Deutsch-Jozsa algorithm, Grover search algorithm, and
deterministic quantum computation with one qubit (DQC1). The role of
coherence in the quantum metrology tasks such as phase
discrimination and subchannel discrimination is also reviewed. Of
course, as quantum coherence underlies different forms of quantum
correlations which are essential for quantum information, we focus
here only on those of the closely related topics.

In Section \ref{sec:6}, we discuss dynamics of quantum coherence and
QD, mainly concentrating on their singular behaviors in open quantum
systems. In this section, we first review frozen phenomena of
quantum coherence and QD which are preferable for information
processing tasks. Then, we discuss potential ways for protecting and
enhancing quantum coherence and QD. Two closely related problems,
i.e., the resource creating and breaking powers of quantum channel
and the factorization relation for the evolution equation of QD and
quantum coherence are described in detail.

In Section \ref{sec:7}, we consider quantum coherence in explicit
physical systems. We employ the various spin-chain model to show
that quantum coherence can be used to study the long-range order,
valence-bond-solid states, localized and thermalized states, and
quantum phase transitions of the many-body systems. This shows that
the resource theory of quantum coherence is not only of fundamental
but is also of practical significance.

In Section \ref{sec:8}, we present a review on recent progresses of
quantum coherence and QD in relativistic settings, including their
behaviors for the free field modes, for curved spacetime and
expanding universe, and for noninertial cavity modes. We provide a
summary of the Unruh temperature, Hawking temperature, expansion
rate of the universe, accelerated motion of cavities and detectors,
and boundary conditions of the field on quantum correlations and
quantum coherence. Quantum correlations for particle detectors and
the dynamical Casimir effects on these correlations are also
provided here.

Finally, in Section \ref{sec:9}, we present a concluding remark on
the main results of this review. We hope the review be helpful for
further exploration in these fields. Several open questions are also
raised for possible future research.

\section{Geometric quantum correlation measures} \label{sec:2}
The widely used discordlike quantum correlation measures proposed in
the past ten years can be categorized roughly into two different
families, namely, those based on the entropy theory, and those based
on various distance measures of quantum states. A detailed overview
of the first category has already been given by \citet{RMP}, and
there are no new measures being proposed along this line in recent
years, so we will recall only the original definition of QD and
several of the related entropic measures for self consistency of
this review. Our main concern will be the second category of
discordlike quantum correlation measures. Most of them are proposed
after the year 2012, and have been proven to be well defined. The
related notions and approaches used in their definitions have also
been proven useful for introducing coherence measures. In
particular, this allows us to put the discordlike correlations and
quantum coherence on an equal footing, which facilitates one's
investigation of the interrelation between these two different
quantifiers of quantumness. Moreover, as the definitions of GQD and
quantum coherence have something in common, the well developed
methods for calculating GQD are also enlightening for deriving
analytical expressions of the related coherence measures.

To begin with, we recall the concept of QD that was framed
from the viewpoints of information theory. Within the seminal
framework formulated by \citet{qd01} as well as \citet{qd02}, it was
defined based on the partition of the total correlation in state
$\rho_{AB}$ into two different parts, that is, the classical part
and the quantum part. The QMI $I(\rho_{AB})$ was used as a measure
of total correlation, and it reads
\begin{equation}\label{new1-1}
 I(\rho_{AB})=S(\rho_A)+S(\rho_B)-S(\rho_{AB}),
\end{equation}
with $S(\rho_X)=-\tr(\rho_X\log_2 \rho_X)$ ($X=A$, $B$, or $AB$)
being the von Neumann entropy.

Similar to most correlation measures whose quantification implies
measurement, the classical correlation was also defined from a
measurement perspective. \citet{qd01} proposed four defining
conditions for a classical measure $J(\rho_{AB})$: (\romannumeral+1)
$J(\rho_{AB}) = 0$ for any $\rho_{AB}=\rho_A\otimes \rho_B$,
(\romannumeral+2) it should be locally unitary invariant,
(\romannumeral+3) It is nonincreasing under local operations, and
(\romannumeral+4) $J(\rho_{AB})=S(\rho_A)=S(\rho_B)$ for pure
states. Based on these conditions, they defined the classical
correlation as the maximum information about one party (say $B$) of
a bipartite system that can be extracted by performing the positive
operator valued measure (POVM) on the other party (say $A$). If the
POVM $\{E_k^A\}$ with elements $E_k^A=M_k^{A\dagger} M_k^A$ is
performed on party $A$, one can obtain the postmeasurement state of
the system and the conditional state of party $B$ as
\begin{equation}\label{new1-2}
 \rho'_{AB}=\sum_k M_k^A\rho_{AB}M_k^{A\dagger},~~
 \rho_{B|E_k^A}=\frac{\mathrm{tr}_{A}(E_k^A\rho_{AB})}{p_k},
\end{equation}
where $p_k=\mathrm{tr}(E_k^A\rho_{AB})$ is the probability for
obtaining the outcome $k$. The classical correlation is given by
\begin{equation}\label{new1-3}
 J_A(\rho_{AB})=S(\rho_B)-\min_{\{E_k^A\}}S(B|\{E_k^A\}),
\end{equation}
where
\begin{equation}\label{new1-4}
 S(B|\{E_k^A\})=\sum_{k}p_{k}S(\rho_{B|E_k^A}),
\end{equation}
is the averaged conditional entropy of the postmeasurement state
$\rho'_{AB}$.

The QD is then defined by the discrepancy between
$I(\rho_{AB})$ and $J_A(\rho_{AB})$ as
\begin{equation}\label{new1-5}
 D_A(\rho_{AB})=I(\rho_{AB})-J_A(\rho_{AB})=\min_{\{E_k^A\}}S(B|\{E_k^A\})-S(B|A),
\end{equation}
where $S(B|A)=S(\rho_{AB})-S(\rho_A)$ is the conditional entropy. Of
course, one can also define $D_B(\rho_{AB})$ by performing the
measurements on party $B$. In general, $D_A(\rho_{AB}) \neq
D_B(\rho_{AB})$, that is, the QD is an asymmetric quantity.

The QD defined above is based on POVM, but the corresponding
maximization is generally a notoriously challenging task. So
sometimes one can also consider the set of rank-one projectors
$\{\Pi_k^A\}$, for which the postmeasurement state turns out to
\begin{equation}\label{new1-6}
 \rho'_{AB}=\sum_k p_k \Pi_k^A \otimes \rho_{B|\Pi_k^A}.
\end{equation}
This yields $I(\rho'_{AB})= S(\rho_B)-\sum_k p_k
S(\rho_{B|\Pi_k^A})$, and the QD becomes
\begin{equation}\label{new1-7}
 D_A(\rho_{AB})=I(\rho_{AB})-\max_{\{\Pi_k^A\}}I(\rho'_{AB}).
\end{equation}
Hence the intuitive meaning of QD can be interpreted as the minimal
loss of correlations due to the local projective measurements
$\{\Pi_k^A\}$. Indeed, it suffices to consider only the
projective measurements $\{\Pi_k^A\}$ for two-qubit states in most
cases \cite{QD-ana1,QD-ana2}. Moreover, the optimal measurement
strategy (over four-element POVM) for obtaining classical
correlation $J_A(\rho_{AB})$ in many-body system has been studied by
\citet{Amico}.

Similarly, a symmetric version of QD was also proposed. It reads
\cite{qdsym1,qdsym2}
\begin{equation}\label{new1-8}
 D_s(\rho_{AB})=I(\rho_{AB})-\max_{\{\Pi_k^A\otimes \Pi_l^B\}}I(\rho''_{AB}),
\end{equation}
where $\rho''_{AB}=\sum_k p_{kl} \Pi_k^A \otimes \otimes \Pi_l^B$.

Apart from the above QD measure, \citet{thermalqd} presented a
slightly different discordlike correlation measure which was called
thermal QD, and is defined as
\begin{equation}\label{new1-9}
 \tilde{D}_A(\rho_{AB})=\min_{\{\Pi_k^A\}}[S(\rho'_A)+S(B|\{\Pi_k^A\})]-S(\rho_{AB}),
\end{equation}
where $\rho'_A$ is the reduced state of $\rho'_{AB}$ in Eq.
\eqref{new1-6}, and $S(\rho'_A)=H(\{p_k\})$, with $H(\{p_k\})$ being
the Shannon entropy function and the probability $p_k=\tr(\Pi_k^A
\rho_{AB}\Pi_k^A)$.

Apart from the above entropic measure and the other related entropic
measures summarized in detail by \citet{RMP}, QD can also be
measured from a geometric aspect. The motivation for this approach
is very similar to the geometric measure of entanglement first
introduced by \citet{GME1} and further extended by \citet{GME2}. For
pure state $|\psi\rangle$, they proposed to adopt the minimal
squared distance between $|\psi\rangle$ and the set of separable
pure states $|\phi\rangle$ to characterize its entanglement, that
is, by minimizing $\min_{|\phi\rangle}\||\psi\rangle-
|\phi\rangle\|^2$. Based on this, one can derive the following
geometric entanglement measure
\begin{equation}\label{new1-10}
 E_g(\psi)= \min_{|\phi\rangle} (1-|\langle\psi|\phi\rangle|^2)
                  = 1-\max_{|\phi\rangle} |\langle\psi|\phi\rangle|^2,
\end{equation}
where $\psi=|\psi\rangle\langle\psi|$. If $\psi$ is a bipartite
state, $E_g(\psi)=1- \lambda_{\max}^{1/2}$ \cite{GME1}, where
$\lambda_{\max}^{1/2}$ is the maximal Schmidt coefficient
corresponding to the Schmidt decomposition of $|\psi\rangle$ of the
following form
\begin{equation} \label{schmidt}
 |\psi\rangle=\sum_i \sqrt{\lambda_i}
 |\varphi_i^A\rangle\otimes|\varphi_i^B\rangle.
\end{equation}

For a mixed state described by density operator $\rho$, the
geometric entanglement measure can be defined in terms of the convex
roof construction
\begin{equation}\label{eq2b-5}
 E_g(\rho)= \min_{\{p_i,\psi_i \}}\sum_i p_i E_g(\psi_i),
\end{equation}
where $\psi_i=|\psi_i\rangle\langle\psi_i|$, and the minimization is
with respect to the possible decompositions of
\begin{equation}\label{decomp}
 \rho= \sum_i p_i \psi_i.
\end{equation}

Though the calculation of $E_g(\rho)$ for general mixed states is a
daunting task, for any two-qubit state $\rho$, it can be evaluated
analytically as
\begin{equation}\label{GME-qubit}
 E_g(\rho)= \frac{1-\sqrt{1-C^2(\rho)}}{2},
\end{equation}
where $C(\rho)$ is the concurrence of $\rho$ \cite{concur2}. We
refer to the work of \citet{GME3} for a comparison of different
geometric entanglement measures.

In the following, we review in detail the geometric measure of
discordlike correlations. The motivation for such kind of measures
may be fourfold. First, the definition of geometric correlations are
based on the idea that \emph{a distance from a given state to the
closest state without the desired property is a measure of that
property} \cite{reqd}, thereby one can quantify amount of
correlations by the distance of the considered state to the set of
states without the desired property. This endows the resulting
geometric measure a clear geometric interpretation. Second, the
geometric measures are preferable due to their analytically
computable for a wide regime of states. In particular, the theory
for geometric entanglement measure is historically well developed,
while the features of various distance measures of states are also
intensively investigated. The corresponding results can be borrowed
for studying geometric discordlike correlations. Thirdly, it is
hard to generalize the concept of the entropic discord to
multipartite scenario as it is based on QMI which is not defined for
multipartite systems. But the geometric approach enables one to
define discordlike correlations which are completely applicable for
multipartite states. Finally, the geometric discordlike
correlations have also been shown to be related to some quantum
information processing tasks, thereby endows them with an actual
meaning.

Once a distance measure of quantum states is chosen, the
corresponding GQD measure will be determined by the set of classical
states. In general, the definition of classical states is not unique
and different types are studied in different contexts, see, e.g.,
the work of \citet{QD-ana2} and references therein. We refer the
following two slightly different types of them which are within the
theory of discordlike correlations: partial classical states and
total classical states. In the case where there are only two
subsystems, they are usually called one-sided (classical-quantum or
quantum-classical) and two-sided (classical-classical) classical
states. The set contains mixtures of locally distinguishable states
and include the set of product states as its subset. They are
defined to be classical as the total correlation (measured by QMI)
contained in them is the same as the classical correlation
\cite{qd01,qd02}. In fact, for any partial (total) classical state,
there exist local (tensor product of local) POVM such that the
postmeasurement state is the same as the premeasurement one.
Contrary, if there dose not exist such a POVM, the considered state
is said to be discordant.

Apart from classical states, we will also consider the sets of
locally invariant states and locally invariant projective
measurements which are utilized in defining MIN. Here, by saying a
projective measurement to be locally invariant we mean that it does
not disturb the reduced state (say $\rho_A$) of a bipartite system
$AB$. It constitutes a subset of the full set of local projective
measurements. Moreover, the set of locally invariant states are also
different from that of the above-mentioned classical (i.e.,
zero-discord) states, as some of the classical states may have
nonvanishing MIN \cite{min}.

In the following discussion of discordlike correlations other than
that measured by relative entropy, we focus our attention mainly on
bipartite states. But most of them can be generalized directly to
multipartite scenario due to the definite structure of total
classical states \cite{reqd}.

\subsection{Geometric quantum discord} \label{sec:2A}
The starting point for the definition of GQD is the identification
of the set $\mathcal{CQ}$ of classical-quantum (i.e., zero-discord
with respect to subsystem $A$) states. For a bipartite state in the
Hilbert space $\mathcal {H}_{AB}$, the classical-quantum states can
be written as
\begin{equation}\label{eq1-1}
 \chi=\sum_i p_k \Pi_k^A\otimes \rho_k^B,
\end{equation}
which is a convex combination of the tensor products of the
orthogonal projector $\Pi_k^A$ in $\mathcal{H}_A$ and an arbitrary
density operator $\rho_k^B$ in $\mathcal{H}_B$, with $\{p_k\}$ being
any probability distribution. Intuitively, $\chi$ of Eq.
\eqref{eq1-1} is said to be classical-quantum as there exists at
least one measurement on subsystem $A$ for which $B$ is not
affected. or in other words, by measuring $A$ one extracts no
information about $B$ as the entropy $S(\rho_B)$ and the residual
entropy $\sum_k p_kS(\rho_k^B)$ for the conditional ensemble
$\{p_k,\rho_k^B\}$ after an optimal local POVM is performed on $A$
are the same. Indeed, within the framework of \citet{qd01} and
\citet{qd02}, one can also check directly that the classical
correlation contained in $\chi$ is zero.

\begin{figure}
\centering
\resizebox{0.46 \textwidth}{!}{%
\includegraphics{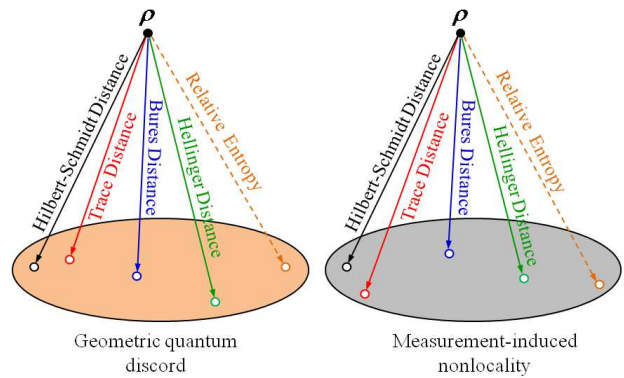}}
\caption{Geometric quantum correlations of a state $\rho$ can be
quantified by the closest (pseudo) distance between it and the set
of classical-quantum states (for GQD) and the locally invariant
states (for MIN).}\label{fig:hu1}
\end{figure}

With $\mathcal{CQ}$ in hand, the category of GQDs for a state $\rho$
can be characterized by its closest (pseudo) distance to the
zero-discord state in set $\mathcal{CQ}$ (see Fig. \ref{fig:hu1}).
More specifically, it can be formalized in the general form
\begin{equation}\label{eq1-2}
 D_\mathcal{D}(\rho)= \min_{\chi\in\mathcal{CQ}}\mathcal{D}(\rho,\chi),
\end{equation}
where $\mathcal{D}(\rho,\chi)$ is a suitable distance measure of
quantum states which should satisfy certain natural restrictions in
order for the GQD to be well defined, for example, it should be
nonnegative, and  should be nonincreasing under the action of
completely positive and trace preserving (CPTP) map. In certain
specific situations, some equivalent forms of $\mathcal{D}(\rho,
\chi)$ may be used as well. As the distance between two quantum
states can be measured from different aspects, the GQDs can be
defined accordingly, provided that they satisfy the conditions for a
faithful measure of quantum correlation \cite{qd01}. Moreover, while
the GQD defined in Eq. \eqref{eq1-2} can increase under local
operations on party $A$ by its definition, it should not be
increased by local operations on the unmeasured party $B$
\cite{Piani}.

Likewise, one could write directly the set $\mathcal{QC}$ of
quantum-classical states and define the GQD with respect to
subsystem $B$, or the set $\mathcal{CC}$ of classical-classical
states and define the GQD with respect to total system $AB$. The
classical-classical states can be written as $\chi'=\sum_i p_{kl}
\Pi_k^A\otimes \Pi_l^B$, and there exists at least one local
measurements for which it is not affected. In what follows we
consider the GQD defined with respect to $A$. Its definition with
respect to $B$ or $AB$ is similarly.

This definition of GQD is somewhat different from the initially
proposed entropic measure of QD \cite{qd01}. But it should also
satisfy the similar necessary conditions in order for it to be a
bona fide measure of quantum correlation, e.g., it is non-negative,
vanishes only for zero-discord states, keeps invariant under local
unitary transformations, and is nonincreasing under local
operations.

\subsubsection{Hilbert-Schmidt norm of discord}
By using the HS norm as a measure of the distance between two
states, \citet{GQD}defined the GQD of $\rho$ as
\begin{equation}\label{gqd-1}
 D_G(\rho)=\min_{\chi\in \mathcal{CQ}}\|\rho-\chi\|_2^2,
\end{equation}
with $\|X\|_2$ denoting the HS norm which is defined as
$\|X\|_2=\sqrt{\tr(X^\dag X)}$.

\citet{bound1} further proved that the above definition of GQD is
completely equivalent to
\begin{equation}\label{gqd-v2}
 D_G(\rho)=\min_{\Pi^A}\|\rho-\Pi^A(\rho)\|_2^2,
\end{equation}
where $\Pi^A=\{\Pi_k^A\}$ is the local von Neumann measurements on
party $A$ which sum to the identity (i.e., $\sum_k \Pi_k^A=
\iden_A$), and
\begin{equation}\label{poststate}
 \Pi^A(\rho)= \sum_k (\Pi_k^A\otimes\iden_B)\rho (\Pi_k^A\otimes\iden_B).
\end{equation}
As the set $\{\Pi^A(\rho)\}$ of postmeasurement states is generally
a subset of the full set $\mathcal{CQ}$ of classical states, the
equivalence between the above two definitions implies that one only
need to take the minimization over $\{\Pi^A(\rho)\}$, and this
greatly simplifies the estimation of $D_G(\rho)$. Moreover, Eq.
\eqref{gqd-v2} also reveals that $D_G(\rho)$ basically measures how
much a measurement on party $A$ does disturbs other parts of the
state.

\citet{qdiscord} put forward another discord measure which was
termed as $q$-discord. It reads
\begin{equation} \label{eq-qdiscord}
 D_q(\rho)=\min_{\Pi^A} S_q(\Pi^A[\rho])- S_q(\rho),
\end{equation}
where $S_q(\rho)$ is the Tsallis $q$-entropy defined as
\cite{qentropy}
\begin{equation} \label{eq-qentropy}
 S_q(\rho)=\frac{1-\tr\rho^q}{q-1}.
\end{equation}
It reduces to $-\tr(\rho\ln\rho)$ when $q\rightarrow 1$. Moreover,
one can obtain immediately from Eq. \eqref{eq-qdiscord} that
$D_2(\rho)=\tr\rho^2-\tr(\Pi^A[\rho])^2$, thus $D_G(\rho)$ can also
be retrieved from the $q$-discord by setting $q = 2$.

This GQD measure is favored for its ease of computation. In
particular, by noting that any two-qubit state $\rho$ can be
represented as
\begin{equation}\label{gqd-ad1}
 \rho=\frac{1}{4} \Bigg(\iden_4 +\vec{x}\cdot\vec{\sigma}\otimes\iden_2
      +\iden_2\otimes\vec{y}\cdot\vec{\sigma}
      +\sum_{i,j=1}^3r_{ij}\sigma_i\otimes\sigma_j\Bigg),
\end{equation}
\citet{GQD} derived the explicit formula for $D_G(\rho)$, which is
given by
\begin{equation}\label{gqd-2}
 D_G(\rho)=\frac{1}{4}\left(\|x\|_2^2+\|R\|_2^2-k_{\max} \right),
\end{equation}
where $\|\vec{x}\|_2^2=\sum_{i=1}^3 x_i^2$, $\|R\|_2^2=\tr(R^T R)$,
and $k_\mathrm{max}$ is the largest eigenvalue of the matrix
$K=\vec{x}\vec{x}^T+RR^T$, where $R=(r_{ij})$ is a $3\times 3$ real
matrix, and the superscript $T$ denotes transpose of vectors or
matrices.

Using the same method, the GQD for any qubit-qutrit state $\rho$ was
obtained as \cite{gqd-ana}
\begin{equation}\label{gqd-3}
 D_G(\rho)=\frac{1}{6}\|\vec{x}\|_2^2+\frac{1}{4}\|R\|_2^2-k_{\max},
\end{equation}
where $\|\vec{x}\|_2^2$, $\|R\|_2^2$, and $k_\mathrm{max}$ are
similar to those for the two-qubit case, with however $x_i=\tr
\rho(\sigma_i\otimes \iden_3)$, $R$ becomes a $3\times 8$ matrix
with the elements $r_{ij}=\tr\rho(\sigma_i\otimes \lambda_j)$,
$\lambda_j$ ($j=1,2,\ldots,8$) are the Gell-Mann matrices, and
$K=\vec{x}\vec{x}^T/6+RR^T/4$.

Moreover, for the $(d\times d)$-dimensional Werner state $\rho_W$
and isotropic state $\rho_I$ of the following form
\begin{equation}\label{mina-4}
 \begin{split}
  &\rho_W=\frac{d-x}{d^3-d}\iden_{d^2}+\frac{dx-1}{d^3-d}\sum_{ij}|ij\rangle\langle ji|, ~~x\in[-1,1],\\
  &\rho_I=\frac{1-x}{d^2-1}\iden_{d^2}+\frac{d^2 x-1}{d^3-d}\sum_{ij} |ii\rangle\langle jj|, ~~x\in[0,1],
  \end{split}
\end{equation}
the HS norm of GQD can be obtained analytically as \cite{bound1}
\begin{equation}\label{mina-v2}
 \begin{split}
  &D_G(\rho_W)=\frac{(dx-1)^2}{d(d-1)(d+1)^2},\\
  &D_G(\rho_I)=\frac{(d^2 x-1)^2}{d(d-1)(d+1)^2}.
  \end{split}
\end{equation}

For any $(m\times n)$-dimensional bipartite state, it can always be
decomposed as
\begin{eqnarray}\label{mina-1}
\rho=\sum_{ij}r_{ij}X_i\otimes Y_j,
\end{eqnarray}
where $\{X_i\!:i=0,1, \ldots,m^2-1\}$ ($X_0=\iden_m/ \sqrt{m}$) is
the orthonormal operator basis for subsystem $A$ that satisfy ${\rm
tr} (X_i^\dag X_{i'})=\delta_{ii'}$ (likewise for $Y_j$), the HS
norm of GQD is showed to be lower bounded by \cite{bound1}
\begin{equation}\label{mina-v3}
 D_G(\rho)\geq \tr(CC^T) -\sum_{j=1}^m \lambda_j=\sum_{j=m+1}^{m^2} \lambda_j,
\end{equation}
with $\lambda_j$ representing eigenvalues of the matrix $CC^T$
arranged in nonincreasing order (counting multiplicity), and
$C=(r_{ij})$ is a $m^2\times n^2$ matrix.

\citet{bound2} also obtained a different tight lower bound of
$D_G(\rho)$, which is given by
\begin{equation}\label{mina-v4}
 D_G(\rho)\geq  \|x\|^2+ \|R\|^2-\sum_{j=1}^{m-1}\eta_j,
\end{equation}
where $\eta_j$ are eigenvalues of the matrix $m^2 n(xx^T+RR^T)/2$
arranged in nonincreasing order (counting multiplicity). Here, we
have denoted by $x=(r_{10},r_{20},\cdots,r_{m^2-1,0})^T$, and
$R=(r_{kl})$ with $k=1,2,\cdots, m^2-1$, and $l=1,2,\cdots, n^2-1$.
Moreover, note that our decomposed form of $\rho$ in Eq.
\eqref{mina-1} is slightly different from that given by
\citet{bound2}, thus induces the seemingly different but essentially
the same expressions of the lower bound of $D_G(\rho)$.

Different from the GQD of Eq. \eqref{gqd-v2}, \citet{Guo2015ijtp}
proposed another measure of quantumness by using the average
distance between the reduced state $\rho_B= \tr_A \rho$ and the
$i$th output reduced state of subsystem $B$ after the local von
Neumann measurements were performed on $A$. Let $\mathcal{H}_A
\otimes \mathcal{H}_B$ with $\dim \mathcal{H}_A=m$ and $\dim
\mathcal{H}_B =n\geq m$ be the state space of a bipartite system.
The measure is then defined by
\begin{eqnarray} \label{avhs}
 D_G^{av}(\rho)= \sup_{\Pi^A} \sum_k p_k \|\rho_B-\rho^B_k\|_2^2,
\end{eqnarray}
where the supremum is taken over the full set of local von Neumann
measurements $\Pi^A=\{\Pi_k^A\}$, and $\rho_k^B=\tr_A(\Pi_k^A\otimes
\iden_B)\rho(\Pi_k^A\otimes \iden_B)$. It was showed that only the
product states do not contain this kind of quantumness, that is,
$D_G^{av}(\rho)=0$ only for $\rho=\rho_A\otimes\rho_B$. So it
captures quantumness of a state which is different from that
captured by the QD defined within the framework of \citet{qd02}.

While the GQD given in Eq. \eqref{gqd-1} is analytical computable
for any two-qubit state, it is noncontractive, i.e., its value may
be changed even by local reversible operations on the unmeasured
party $B$, so it was thought to be not well defined \cite{Piani}.
But it does play a role in some quantum information tasks, see Sec.
\ref{sec:2C}. Due to this reason, it is desirable to find ways of
characterizing and quantifying GQD using other distance measures of
states.

\subsubsection{Trace norm of discord}
In stead of using the HS norm, \citet{TDD} considered the
possibility of using the general Schatten $p$-norm to measure
quantum correlations. The Schatten $p$-norm for a matrix $M$ is
defined as
\begin{equation} \label{Schatten}
 \|M\|_p=\{\tr[M^\dag M]^{p/2}\}^{1/p},
\end{equation}
which reduces to the HS norm if $p=2$, and the trace norm if $p=1$.
By using multiplicative property of the Schatten $p$-norm under
tensor products, \citet{TDD} showed that the corresponding GQD is
well defined only for $p=1$. Based on this fact, they introduced the
trace norm of discord as
\begin{equation}\label{eq1a-1}
 D_T(\rho)=\min_{\chi\in \mathcal{CQ}}\|\rho-\chi\|_1,
\end{equation}
and for $2\times n$ dimensional state $\rho$ (i.e., $A$ is a qubit),
the optimal $\chi$ can also be obtained from the subset
$\Pi^A(\rho)$ \cite{TDD1}, with $\Pi^A=\{\Pi_k^A\}$ being the set of
local projective measurements, i.e.,
\begin{equation}\label{eq-tdd}
 D_T(\rho)=\min_{\Pi^A} \|\rho-\Pi^A(\rho)\|_1.
\end{equation}

The calculation of $D_T(\rho)$ is a hard task, and there is no
analytical solution for it in general cases. For the two-qubit
Bell-diagonal states
\begin{equation}\label{eq-bell}
 \rho^{\rm Bell}=\frac{1}{4}\left(\iden_4+\sum_{i=1}^3 c_i
                 \sigma_i\otimes\sigma_i\right),
\end{equation}
it can be derived as
\begin{equation} \label{eq1a-bell2}
 D_T(\rho^{\rm Bell})= {\rm int} \{|c_1|,|c_2|, |c_3|\},
\end{equation}
with ${\rm int}\{\cdot\}$ denoting the intermediate value. The
closest $\chi_\rho$ is still a Bell-diagonal state with the only
nonzero parameter $c_k$ corresponding to
$|c_k|=\max\{|c_1|,|c_2|,|c_3|\}$.

Moreover, for two-qubit \emph{X} state $\rho^X$ which contains
nonzero elements only along the main diagonal and anti-diagonal in
the computational basis $\{|00\rangle,|01\rangle,
|10\rangle,|11\rangle\}$, the trace norm of discord is given by
\cite{TDD-ana}
\begin{eqnarray} \label{eq1a-2}
 D_{T}(\rho^X)=\sqrt{\frac{\xi_1^2\xi_{\rm max}-\xi_2^2\xi_{\rm min}}
                {\xi_{\rm max}-\xi_{\rm min}+\xi_1^2-\xi_2^2}},
\end{eqnarray}
where
\begin{eqnarray} \label{eq1a-v2}
 \begin{aligned}
 & \xi_{1,2}=2(|\rho_{23}|\pm |\rho_{14}|),~
   \xi_3=1-2(\rho_{22}+\rho_{33}),\\
 & \xi_{\rm max}=\max\{\xi_3^2,\xi_2^2+x_{A3}^2\},~
   x_{A3}=2(\rho_{11}+\rho_{22})-1.
 \end{aligned}
\end{eqnarray}

For higher-dimensional states, \citet{TDD-ana1} considered a
simplified version of $D_T(\rho)$ defined also by Eq.
\eqref{eq-tdd}, and obtained its analytical solution for certain
very special kinds of qutrit-qutrit states, e.g., the maximally
entangled states and the Werner states.

The trace norm of discord could also be connected to quantum
correlations such as entanglement witness. We refer to the work of
\citet{witness} for a comprehensive review of entanglement
witnesses. In general, an entanglement witness $W$ is an Hermitian
operator for which $\tr(W\rho)$ takes negative value for at least
one entangled state and non-negative values for all separable
states. By minimizing over the compact subset $\mathcal{M}$ of the
set of entanglement witnesses $\mathcal {W}$, one can obtain the
optimal entanglement witness, and define the quantifier
\begin{eqnarray}\label{eq1a-3}
 E_w(\rho)=\max\{0,-\min_{W\in\mathcal {M}}\tr(W\rho)\},
\end{eqnarray}
as an entanglement measure \cite{OEW}.

\citet{TDD-en} proved that $D_T(\rho)$ is lower bounded by
\begin{eqnarray} \label{eq1a-v3}
 D_T(\rho)\geq \max\{0, -\min_{\{W\in\mathcal{W}| -\iden\leq W\leq\iden\}}\tr(W\rho)\}.
\end{eqnarray}
As $E_w(\rho)$ is in fact the negativity $\mathcal{N}(\rho)$
\cite{negat} for $\mathcal{M}= \{W^{T_A}\in\mathcal{W}| 0\leq
W^{T_A}\leq \iden\}$ \cite{OEW}, and the robustness of entanglement
$R_r(\rho)/d$ for $\mathcal{M}= \{W\in\mathcal{W}|\tr W=1\}$
\cite{roe,TDD-en1}, both of which are obviously equal to or smaller
than the optimal entanglement witness showed on the right-hand side
of Eq. \eqref{eq1a-v3}, we also have
\begin{eqnarray} \label{eq1a-vv4}
 D_T(\rho)\geq \mathcal {N}(\rho),~~ D_T(\rho)\geq R_r(\rho)/d.
\end{eqnarray}

\begin{figure}
\centering
\resizebox{0.33 \textwidth}{!}{%
\includegraphics{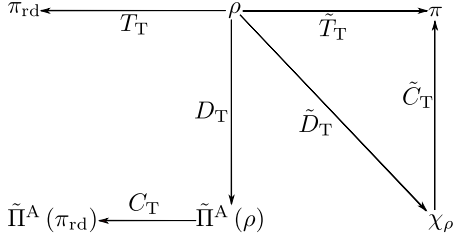}}
\caption{Geometric correlations in the state $\rho$, where
$\tilde{\Pi}^A$ is the optimal measurement operator for obtaining
$D_T(\rho)$, $\pi_{\rm rd}=\rho^A\otimes\rho^B$, $\chi_\rho$ is the
state with closest trace distance to $\rho$, and $\pi$ is the local
product state.}\label{fig:hu2}
\end{figure}

While Eq. \eqref{eq1a-1} gives a proper quantum correlation measure,
\citet{togd} further defined the corresponding geometric classical
and total correlations using the trace norm. By fixing $\chi\in
\Pi^A(\rho)$ and denoting $\tilde{\Pi}^A$ the corresponding optimal
measurement operator for obtaining $D_T(\rho)$ (the minimization
over different $\Pi^A(\rho)$ is equivalent to the minimization over
$\mathcal{CQ}$ for qubit states), they defined the geometric
classical correlation $C_T(\rho)$ and total correlation $T_T(\rho)$
as (see Fig. \ref{fig:hu2})
\begin{eqnarray}\label{eq1a-4}
 \begin{aligned}
 & C_T(\rho)=\|\tilde{\Pi}^A (\rho)-\tilde{\Pi}^A(\pi_{\rm rd})\|_1, \\
 & T_T(\rho)=\|\rho-\pi_{\rm rd}\|_1,
 \end{aligned}
\end{eqnarray}
with $\pi_{\rm rd}=\rho_A\otimes\rho_B$ being product of the reduced
density matrices of $\rho$.

For the Bell-diagonal states $\rho^{\rm Bell}$ of Eq.
\eqref{eq-bell}, \citet{togd} further obtained
\begin{eqnarray} \label{eq1a-v4}
 \begin{aligned}
  & C_T(\rho^{\rm Bell})=c_+, \\
  & T_T(\rho^{\rm Bell})=\frac{1}{2}[c_{+} +\max\{c_+, c_0+c_-\}],
 \end{aligned}
\end{eqnarray}
with $c_+$, $c_-$, and $c_0$ being the maximum, minimum, and
intermediate values of $\{|c_1|,|c_2|, |c_3|\}$, respectively. This
yields the superadditivity relation: $T_T\leq C_T+D_T$.

In fact, $\tilde{\Pi}^A(\pi_{\rm rd})$ in Eq. \eqref{eq1a-4} may be
not the closest state to $\tilde{\Pi}^A (\rho)$, and $\pi_{\rm rd}$
composed of the reduced density matrices may also be not the closest
product state to $\rho$. This stimulates more general definitions of
geometric classical correlation and total correlation. Without loss
of generality, one can denote by $\chi_\rho$ for the state with
closest trace distance to $\rho$ [note that $\tilde{\Pi}^A (\rho)$
is optimal only for $A$ being a qubit], and $\mathcal {P}$ the set
of local product states of the subsystems. Based on these,
\citet{traceqd2} defined (see Fig. \ref{fig:hu2})
\begin{eqnarray}\label{eq1a-5}
 \begin{aligned}
  & \tilde{C}_T(\rho)=\min_{\pi\in \mathcal{P}}\|\chi_\rho-\pi\|_1, \\
  & \tilde{T}_T(\rho)=\min_{\pi\in \mathcal{P}}\|\rho-\pi\|_1,
 \end{aligned}
\end{eqnarray}
and derived analytically
\begin{equation} \label{eq1a-v5}
 \tilde{C}_T(\rho^{\rm Bell})=\sqrt{1+c_+}-1,
\end{equation}
where the closest state $\pi_{\chi_\rho}$ to $\chi_\rho$ is given by
$\pi_{\chi_\rho}=\tilde{\rho}_A\otimes\tilde{\rho}_B$ with
$\tilde{\rho}_A=(\iden_2+a_k\sigma_k)/2$ and $\tilde{\rho}_B=
(\iden_2+b_k\sigma_k)/2$. The index $k$ corresponds to the maximum
of $|c_k|=c_+$, and $a_k=b_k |c_k|/c_k$. Clearly, $\pi_{\chi_\rho}$
is not the product of its marginals, and is not even a Bell-diagonal
state in general.

For the family of classical-quantum or quantum-classical bipartite
states, \citet{noncommutativity} introduced another quantum
correlation measure based on the non-commutativity of quantum
observables, where the trace norm of the commutators of the ensemble
state of one subsystem is used. To be explicit, for
classical-quantum state $\chi$ of Eq. \eqref{eq1-1} described by the
ensemble $\{X_i\}$ with $X_i=p_i\rho_i$, they found the quantity
\begin{eqnarray} \label{eq-nc}
 D(\chi)=\sum_{i>j} \|[X_i,X_j]\|_1,
\end{eqnarray}
satisfy the following properties of correlations: $D(\chi)\geq 0$,
and the equality holds when subsystem $B$ is also classical.
Moreover, it is local unitary invariant, and is nonincreasing when
an ancillary system is introduced.

The above result was further extended by \citet{Guo2016srep}, who
proposed to define GQD for any $\rho$ in a similar manner. Let
$\{|i^A\rangle\}$ be an orthonormal basis of $\mathcal{H}_A$. Then
any state $\rho$ acting on $\mathcal{H}_A \otimes \mathcal{H}_B$ can
be represented by
\begin{eqnarray} \label{eq-guo}
\rho=\sum_{i,j}|i^A\rangle\langle j^A| \otimes B_{ij}.
\end{eqnarray}
with $B_{ij}$ being operators in $\mathcal{H}_B$. The
non-commutativity measure of GQD for $\rho$ is defined by
\begin{eqnarray} \label{eq-guo2}
 D_N(\rho)\coloneqq \frac{1}{2}\sum_{(ij)\neq(kl)}\|[B_{ij},B_{kl}]\|_1,
\end{eqnarray}
under the trace norm, and
\begin{eqnarray} \label{eq-guo3}
 D'_N(\rho)\coloneqq \frac{1}{2}\sum\limits_{(ij)\neq(kl)}\|[B_{ij},B_{kl}]\|_2,
\end{eqnarray}
under the HS norm, where the commutator $[X,Y]=XY-YX$, and the
summation is over all different pairs of $\{B_{ij}\}$.

These two measures of GQD can be calculated easily for $\rho$ of
arbitrary dimension. In particular, for pure state $\psi=
|\psi\rangle\langle\psi|$ with Schmidt decomposition of Eq.
\eqref{schmidt}, analytical solutions of them are given by
\begin{eqnarray} \label{eq-guo4}
 \begin{aligned}
  & D_N(\psi)=2\sum_{i,j}\lambda_i\lambda_j\left(\sum_{(k,l)\in \Omega} \lambda_k\lambda_l\right), \\
  & D'_N(\psi)=2\sum_{i,j}\lambda_i\lambda_j\left(\sum_{(k,l)\in \Omega'} \lambda_k\lambda_l\right)+\sqrt{2},
 \end{aligned}
\end{eqnarray}
where $\Omega=\{(k,l)\}$ with $i<k\leq j\leq l$ or $k=i$, $l=j$ if
$i<j$, and $i\leq k<l$ if $i=j$, while $\Omega'=\{(k,l)\}$ with
$i<k\leq j\leq l$ if $i<j$, and $i\leq k<l$ if $i=j$.

It was showed via several examples that they can reflect the amount
of the original QD. In particular, these two measures disappear if
and only if the corresponding state is zero discordant. Here, we
would like to further point out that this is in fact a direct
consequence of the result of \citet{chenlin}, in which a necessary
and sufficient condition for vanishing QD has been proven. It says
that $\rho$ has zero QD if and only if all the operators
$\rho_{B|ij}$ commute with each other for any orthonormal basis
$\{|i^A\rangle\}$ in $\mathcal {H}_A$, where
\begin{equation} \label{chenlin}
 \rho_{B|ij}\coloneqq \langle i^A|\rho|j^A\rangle.
\end{equation}
It is obvious that $B_{ij}$ in Eq. \eqref{eq-guo} is the same as
$\rho_{B|ij}$ of the above equation.

\subsubsection{Bures distance of discord}
The distance between two states $\rho$ and $\sigma$ can also be
quantified by the Bures distance
\begin{eqnarray} \label{Bures}
 \mathcal{D}_{B}(\rho,\sigma)= 2[1- \sqrt{F(\rho,\sigma)}],
\end{eqnarray}
where
\begin{eqnarray} \label{fidelity}
 F(\rho,\sigma)=\big[\tr(\sqrt{\rho} \sigma\sqrt{\rho})^{1/2}\big]^2,
\end{eqnarray}
is the Uhlmann fidelity \cite{Nielsen}. The Bures distance satisfy
the preferable properties of joint convexity, i.e.,
\begin{equation} \label{fidelity2}
 \begin{aligned}
  \mathcal{D}_B(p_1\rho_1+p_2\rho_2,p_1\sigma_1+p_2\sigma_2)
       \leq &  p_1 \mathcal{D}_B(\rho_1,\sigma_1) \\
            &+ p_2 \mathcal{D}_B(\rho_2,\sigma_2),
 \end{aligned}
\end{equation}
and it is also monotonous under CPTP maps. It has been used to
quantify entanglement \cite{bures-e1, bures-e2}, and there are some
equivalent definitions of Bures distance discord. \citet{BDD}
proposed to define it as
\begin{eqnarray}\label{eq1b-1}
 D_B (\rho)=(2+\sqrt{2})\big[1-\sqrt{F_{\rm max}(\rho)}\big],
\end{eqnarray}
where $F_{\rm max}(\rho)=\max_{\chi\in\mathcal{CQ}}F(\rho,\chi)$
represents the maximum of the Uhlmann fidelity, and the constant $2+
\sqrt{2}$ is introduced for the normalization of it for two-qubit
maximally discordant states. Moreover, the square root of
$D_B(\rho)$ in Eq. \eqref{eq1b-1} equals to that defined by
\citet{bures2}.

There are several cases that the evaluation of $F_{\rm max}(\rho)$,
and thus $D_B (\rho)$ can be simplified:

(1) For arbitrary pure state $|\psi\rangle$, the maximum Uhlmann
fidelity can be obtained as $F_{\rm max} (|\psi\rangle
\langle\psi|)=\mu_{\rm max}$, with $\mu_{\rm max}$ being the largest
Schmidt coefficient of $|\psi\rangle$ \cite{BDD}.

(2) For any Bell-diagonal state $\rho^{\rm Bell}$ of Eq.
\eqref{eq-bell}, we have \cite{bures2,bures3}
\begin{eqnarray}\label{eq1b-2}
 F_{\rm max}(\rho^{\rm Bell})&=&\frac{1}{2}+\frac{1}{4}\max_{\langle ijk\rangle}
 \bigg[\sqrt{(1+c_i)^2-(c_j-c_k)^2} \nonumber\\
 &&+\sqrt{(1-c_i)^2-(c_j+c_k)^2}\bigg],
\end{eqnarray}
where the maximum is taken over all the cyclic permutations of
$\{1,2,3\}$.

(3) For general $(2\times n)$-dimensional state, although there is
no analytical solution, the maximum of the Uhlmann fidelity can be
calculated as \cite{bures3}
\begin{eqnarray}\label{eq1b-3}
 F_{\rm max}(\rho)=\frac{1}{2}\max_{||\vec{u}=1||}
                   \left(1-\tr \Lambda(\vec{u})
                   +2\sum_{k=1}^{n_B}\lambda_k(\vec{u})\right),
\end{eqnarray}
where $\lambda_k(\vec{u})$ are eigenvalues of
\begin{eqnarray} \label{eq1b-4}
 \Lambda(\vec{u})= \sqrt{\rho}(\sigma_{\vec{u}}\otimes \iden_B)\sqrt{\rho}
\end{eqnarray}
arranged in nonincreasing order, and $\sigma_{\vec{u}}=\vec{u}\cdot
\vec{\sigma}$, with $\vec{u}=(\sin\theta\cos\phi,\sin\theta\sin\phi,
\cos\theta)$ being a unit vector in $\mathbb{R}^3$, and $n_B$ the
dimension of $\mathcal {H}_B$.

\subsubsection{Relative entropy of discord}
The relative entropy of a state $\rho$ to another state $\sigma$ is
defined as
\begin{eqnarray}\label{new2-1}
 S(\rho\|\sigma)=\tr(\rho\log_2\rho)- \tr(\rho\log_2\sigma),
\end{eqnarray}
which is non-negative, and can sometimes be infinite. Though
technically the relative entropy does not has a geometric
interpretation as $S(\rho\|\sigma)\neq S(\sigma\|\rho)$ in general,
it can be recognized as a (pseudo) distance measure of quantum
states.

The relative entropy has been used to define quantum entanglement,
\begin{eqnarray}\label{new2-2}
 E_R=\min_{\sigma\in \mathcal{S}} S(\rho\|\sigma),
\end{eqnarray}
which is indeed the minimal relative entropy of $\rho$ to the set
$\mathcal{S}$ of separable states \cite{re-en1,bures-e1}. In the
same spirit, one can use it to define the discordlike correlation
measures. \citet{reqd} made the first attempt in this direction by
introducing the relative entropy of discord $D_R$ and the relative
entropy of dissonance $Q_R$, They are defined, respectively, to be
the minimal relative entropy of $\rho$ and $\sigma$ (the closest
separable state to $\rho$) to the set of classical states
$\mathcal{C}$ (here, by saying a state to be classical, we mean that
it is classical with respect to all of its subsystems, which is
similar to the set of $\mathcal{CC}$ states for bipartite systems),
and can be written explicitly as
\begin{eqnarray}\label{new2-3}
  D_R =\min_{\chi\in \mathcal{C}} S(\rho\|\chi), ~~
  Q_R =\min_{\chi\in \mathcal{C}} S(\sigma\|\chi),
\end{eqnarray}
which are applicable for the general bipartite and multipartite
states. In particular, $Q_R(\sigma)$ reveals a kind of quantum
correlation excluding quantum entanglement.

\citet{reqd} also showed that $D_R$ and $Q_R$ are equivalent to
\begin{eqnarray}\label{new2-4}
  D_R=S(\chi_\rho)-S(\rho), ~~
  Q_R=S(\chi_\sigma)-S(\rho),
\end{eqnarray}
where $S(\chi_\rho)=\min_{|\vec{k}\rangle}S(\sum_{\vec{k}}
|\vec{k}\rangle\langle \vec{k}|\rho|\vec{k}\rangle\langle \vec{k}|)$
with $\{|\vec{k}\rangle\}$ forming the eigenbasis of $\chi_\rho$,
and likewise for $S(\chi_\sigma)$. So the optimization in Eq.
\eqref{new2-2} is reduced to the optimization of the von Neumann
entropy $S(\chi_\rho)$ and $S(\chi_\sigma)$.

In a similar manner to Eqs. \eqref{new2-3} and \eqref{new2-4},
\citet{reqd} defined the total correlation and classical correlation
as
\begin{eqnarray}\label{new2-5}
 \begin{aligned}
  & T_\rho=S(\rho\|\pi_\rho)=S(\pi_\rho)-S(\rho), \\
  & T_\sigma=S(\sigma\|\pi_\sigma)=S(\pi_\sigma)-S(\sigma), \\
  & C_\rho=S(\chi_\rho\|\pi_{\chi_\rho})=S(\pi_{\chi_\rho})-S(\chi_\rho), \\
  & C_\sigma=S(\chi_\sigma\|\pi_{\chi_\sigma})=S(\pi_{\chi_\sigma})-S(\chi_\sigma),
 \end{aligned}
\end{eqnarray}
where $\pi_\rho=\pi_1\otimes \ldots\otimes\pi_N$ ($\pi_k$ is the
reduced density operator of the $k$th subsystem of $\rho$), and
likewise for $\pi_\sigma$, $\pi_{\chi_\rho}$, and
$\pi_{\chi_\sigma}$. From these definitions, one can obtain the
following additivity relations
\begin{eqnarray}\label{new2-6}
 T_\rho+L_\rho=D_R+C_\rho,~~
 T_\sigma+L_\sigma=Q_R+C_\sigma,
\end{eqnarray}
where $L_\rho= S(\pi_{\chi_\rho})-S(\pi_\rho)$ and $L_\sigma=
S(\pi_{\chi_\sigma})-S(\pi_\sigma)$.

For two-qubit Bell-diagonal states of Eq. \eqref{eq-bell}, if we
rewrite it as $\rho^{\mathrm{Bell}}=\sum_i \lambda_i
|\Psi_i\rangle\langle\Psi_i|$, where $\lambda_i$ are arranged in
nonincreasing order and $|\Psi_i\rangle$ ($i=1,2,3,4$) are the four
Bell states, then the closest separable state to it is given by
\cite{bures-e1}
\begin{eqnarray}\label{new2-7}
 \sigma= \sum_{i=1}^4 p_i |\Psi_i\rangle\langle\Psi_i,
\end{eqnarray}
where $p_1=1/2$ and $p_i=\lambda_i/[2(1-\lambda_1)]$ for $i\neq 1$.
Similarly, the closest classical state to $\rho^{\mathrm{Bell}}$ is
given by \cite{reqd}
\begin{eqnarray}\label{new2-8}
 \begin{aligned}
 \chi_\rho =&\frac{q_\rho}{2}[|\Psi_1\rangle\langle\Psi_1|+|\Psi_2\rangle\langle\Psi_2|] \\
         &+\frac{1-q_\rho}{2}[|\Psi_3\rangle\langle\Psi_3|+|\Psi_4\rangle\langle\Psi_4|],
 \end{aligned}
\end{eqnarray}
with $q_\rho=\lambda_1+\lambda_2$, and the closest classical state
to $\sigma$ can be obtained directly by substituting $q_\rho$ with
$q_\sigma=p_1+p_2$.

Moreover, it is worthwhile to note that for any bipartite state
$\rho_{AB}$, the relative entropy of discord $D_R(\rho_{AB})$ equals
to the zero-way quantum deficit
\begin{eqnarray}\label{new2-9}
\Delta^{\O}(\rho_{AB}) = \min_{\Pi^A \otimes \Pi^B}
S(\rho_{AB}\|\Pi^A \otimes \Pi^B [\rho_{AB}]),
\end{eqnarray}
which is also a discordlike quantum correlation measure and was
defined originally from the perspective of work extraction from the
quantum system coupled to a heat bath \cite{deficit}. The above
equation thereby endows $\Delta^{\O}(\rho_{AB})$ a geometric
interpretation, that is, it corresponds to the minimal relative
entropy between $\rho_{AB}$ and the full set of postmeasurement
states $\Pi^A \otimes \Pi^B [\rho_{AB}]$.

The one-way quantum deficit can also be expressed by using the
quantum relative entropy as \cite{deficit}
\begin{eqnarray}\label{new2-10}
  \Delta^{\rightarrow}(\rho_{AB}) = \min_{\Pi^A}S(\rho_{AB}\|\Pi^A[\rho_{AB}]),
\end{eqnarray}
and it also equals to the minimal relative entropy between
$\rho_{AB}$ and the set $\mathcal{CQ}$ of classical-quantum states.
Furthermore, $\Delta^{\rightarrow}(\rho_{AB})$ also equivalents to
the thermal QD $\tilde{D}_A(\rho_{AB})$ introduced by
\citet{thermalqd}.

\subsubsection{Hellinger distance of discord}
Although in most cases the quantum correlation measure is defined as
a direct function of the density operator $\rho$ itself, its other
forms may also be very useful. For example, with roots in the
well-know notion of WY skew information \cite{WYD}, the square root
$\sqrt{\rho}$ has been used to study the local quantum uncertainty
(LQU) of a single system \cite{LQU}.

By using the square root form of a density operator, \citet{HDD}
introduced a new quantifier of the GQD, for which we call it the
Hellinger distance discord. It can be recognized as a modified
version of the GQD proposed by \citet{GQD}, and reads
\begin{equation}\label{eq1c-1}
 D_H(\rho)=2\min_{\Pi^A}\parallel\sqrt{\rho}-\Pi^A(\sqrt{\rho})\parallel_2^2,
\end{equation}
where the minimum is taken over $\Pi^A= \{\Pi_k^A\}$, with
\begin{equation}\label{eq1c-2}
 \Pi^A(\sqrt{\rho})= \sum_k (\Pi_k^A\otimes \iden_B)\sqrt{\rho}(\Pi_k^A\otimes \iden_B),
\end{equation}
and $\iden_B$ is the identity operator in $\mathcal {H}_B$.

The Hellinger distance discord is well defined. It is locally
unitary invariant, and vanishes if and only if $\rho$ is a
classical-quantum state. It also keeps invariant when adding a local
ancilla to the unmeasured party, i.e., $D_H(\rho_{A:BC})=
D_H(\rho_{A:B})$ for $\rho_{A:BC}=\rho_{AB} \otimes\rho_C$. This
property averts the fault encountered when measuring GQD via the HS
norm \cite{Piani}. Moreover, it is similar to the squared form of
the Hellinger distance defined as
\begin{equation} \label{eq1c-v2}
 d_H^2(\rho,\chi)=\frac{1}{2}\tr\{(\sqrt{\rho}-\sqrt{\chi})^2\},
\end{equation}
and this is the reason for it to be called the Hellinger distance
discord.

For pure state $\psi=|\psi\rangle\langle\psi|$ with the Schmidt
decomposition of Eq. \eqref{schmidt}, the Hellinger distance discord
can be obtained as $D_H(\psi)=1-\sum_i \lambda_i^2$, which is the
same as that of the GQD based on the HS norm \cite{GQD}. Moreover,
for the Bell-diagonal state $\rho^{\rm Bell}$, it is given by
\begin{eqnarray}\label{eq1c-3}
 D_H(\rho^{\rm Bell})=1-\frac{1}{4}(h^2+\max_i\{d_i^2\}),
\end{eqnarray}
where
$h=\sum_i \sqrt{\lambda_i}$, $d_i=h-2\sqrt{\lambda_4}
-2\sqrt{\lambda_i}$ ($i=1,2,3$), and $\lambda_i$ are eigenvalues of
$\rho^{\rm Bell}$ given by
\begin{eqnarray} \label{eq1c-v3}
 \begin{aligned}
  & \lambda_1=\frac{1}{4}(1-c_1+c_2+c_3),~ \lambda_2=\frac{1}{4}(1+c_1-c_2+c_3), \\
  & \lambda_3=\frac{1}{4}(1+c_1+c_2-c_3),~ \lambda_4=\frac{1}{4}(1-c_1-c_2-c_3).
 \end{aligned}
\end{eqnarray}

For $(2\times n)$-dimensional state $\rho$ with the decomposed form
of
\begin{eqnarray} \label{eq-sqrt}
 \sqrt{\rho}=\sum_{ij} \gamma_{ij}X_i\otimes Y_j,
\end{eqnarray}
where $\{X_i\!:i=0,1,2,3\}$ and $\{Y_j\!:j=0,1,\ldots,n^2-1\}$
constitutes the orthonormal operator bases for the Hilbert spaces
$\mathcal {H}_A$ and $\mathcal {H}_B$, the Hellinger distance
discord can be calculated as \cite{HDD}
\begin{eqnarray}\label{eq1c-4}
 D_H(\rho)=2(1-||\bold{r}||_2^2-\mu_{\max}),
\end{eqnarray}
where $||\bold{r}||_2^2=\sum_j \gamma_{0j}^2$, and $\mu_{\max}$
represents the largest eigenvalue of the matrix $\Gamma\Gamma^\dag$,
with $\Gamma=(\gamma_{ij})_{ i=1,2,3;j=0,1,\cdots,n^2-1}$.

\subsubsection{Local quantum uncertainty}
The WY skew information was defined as follows \cite{WYD}
\begin{eqnarray}\label{eq2f-1}
 \mathsf{I}^p(\rho,K)= -\frac{1}{2}\tr\{[\rho^p,K][\rho^{1-p},K]\},
\end{eqnarray}
with $p\in(0,1)$, and when $p=1/2$ (we omit the superscript in
$\mathsf{I}^p(\rho,K)$ for brevity),
\begin{eqnarray} \label{eq2f-2}
 \mathsf{I}(\rho,K)= -\frac{1}{2}\tr\{[\sqrt{\rho},K]^2\}=\frac{1}{2}\|[K,\sqrt{\rho}]\|_2^2,
\end{eqnarray}
was also termed the WY skew information, where $K$ denotes the
observable to be measured (a self-adjoint operator).
$\mathsf{I}(\rho,K)$ measures the information content embodied in a
state that is skewed to the chosen observable $K$, and is bounded
above by the variance of $K$, i.e.,
\begin{eqnarray} \label{eq2f-v2}
 \mathsf{I}(\rho,K)\leq \langle K^2\rangle_\rho-\langle K\rangle^2_\rho,
\end{eqnarray}
where the equality holds for pure states. This equation shows that
$\mathsf{I}(\rho,K)$ is indeed a lower bound of the weighted
statistical uncertainty about $K$ (measured by the variance of $K$)
for any possible state preparation. For $\rho=\sum_i \lambda_i
|\psi_i\rangle\langle\psi_i|$, one can further obtain
\begin{equation}\label{eq1d-2}
 \mathsf{I}(\rho,K)=\frac{1}{2}\sum_{ij}(\sqrt{\lambda_i}-\sqrt{\lambda_j})^2 K_{ij}^2,
\end{equation}
where the overlap $K_{ij}=|\langle\psi_i|K|\psi_j\rangle|$.

Compared to the variance of $K$, the skew information has many
advantages. In particular, it possesses preferable properties which
are useful for defining quantum correlations, e.g., it is
nonnegative and vanishes if and only if $[\rho,K]=0$, it is convex,
that is, $\mathsf{I}(\sum_i p_i \rho_i,K)\leq \sum_i p_i
\mathsf{I}(\rho_i,K)$. Indeed, the skew information can used to
characterize uncertainty relation \cite{luoprl}, while a correlation
measure based on it has also been introduced \cite{skew}.

\citet{LQU} proposed to use the WY skew information to quantify LQU
of a bipartite state $\rho$. They chose the local observable
$K^\Lambda=K_A^\Lambda\otimes \iden_B$ ($\Lambda$ denotes spectrum
of $K_A^\Lambda$ that are nondegenerate as this corresponds to
maximally informative observables on $A$) and defined the LQU as
\begin{equation}\label{eq1d-3}
 \mathcal {U}_A^\Lambda(\rho)=\min_{K^\Lambda} \mathsf{I}(\rho,K^\Lambda),
\end{equation}
which is not only a measure of uncertainty, but also a well-defined
quantum correlation measure. In particular, for $(2\times
n)$-dimensional state $\rho$, by dropping the superscript $\Lambda$
for brevity and choosing the nondegenerate observables as
$K_A=\vec{n}\cdot\vec{\sigma}^A$, with $\vec{\sigma}=(\sigma_1,
\sigma_2,\sigma_3)$ being the vector of Pauli operators, the LQU can
be derived as \cite{LQU}
\begin{eqnarray}\label{eq1d-4}
 \mathcal{U}_A(\rho)=1-\lambda_{\max}(W_{AB}),
\end{eqnarray}
where $\lambda_{\max}(W_{AB})$ denotes the maximum eigenvalue of the
matrix $W_{AB}$ whose elements are given by
\begin{eqnarray}\label{eq1d-5}
 (W_{AB})_{ij}=\tr\{\sqrt{\rho}(\sigma_i^A\otimes \iden_B)
               \sqrt{\rho}(\sigma_j^B\otimes \iden_B)\},
\end{eqnarray}
from which one can obtain that for pure state $|\psi\rangle$, the
LQU reduces to the linear entropy of entanglement
\begin{eqnarray} \label{eq1d-v5}
 \mathcal{U}_A(|\psi\rangle\langle\psi|)= 2[1-\tr(\rho_A)^2],
\end{eqnarray}
where $\rho_A=\tr_B (|\psi\rangle\langle\psi|)$.

In fact, for arbitrary $(2\times n)$-dimensional state $\rho$, as
$K_A=\vec{n}\cdot\vec{\sigma}^A$ is a root-of-unity local unitary
operation, which implies $K_Af(\rho)K_A=f(K_A\rho K_A)$ for
arbitrary function $f(\cdot)$, thus
\begin{equation}\label{eq1d-6}
 \begin{aligned}
  \mathcal{U}_A(\rho)=& 1-\tr\{\sqrt{\rho}K_A\sqrt{\rho}K_A\},\\
                     =& 1-\tr\{\sqrt{\rho}\sqrt{K_A\rho K_A}\},\\
                     =& d_H^2(\rho,K_A\rho K_A),
 \end{aligned}
\end{equation}
while $\{\Pi_k^A\}$ of Eq. \eqref{eq1c-2} can be written as
$\Pi_{1,2}^A= (\iden_2\pm K_A)/2$, which gives
\begin{equation}\label{eq1d-7}
 \begin{aligned}
  \big[\sqrt{\rho}-\Pi^A(\sqrt{\rho})\big]^2
    = & \frac{1}{4}(\rho+K_A\rho K_A-\sqrt{\rho}K_A\sqrt{\rho}K_A \\
      & -K_A\sqrt{\rho}K_A\sqrt{\rho}),
 \end{aligned}
\end{equation}
then by combining this with Eqs. \eqref{eq1c-1}, \eqref{eq1d-3}, and
\eqref{eq1d-6}, one can obtain
\begin{equation} \label{eq1d-v7}
 \mathcal{U}_A(\rho)= 2D_H(\rho).
\end{equation}
This equation establishes a direct connection between LQU and the
Hellinger distance discord, thereby gives LQU a geometric
interpretation, although it applies only for $(2\times
n)$-dimensional states.

By restricting $K_A$ to rank-one projectors, \citet{yucslqu} further
defined a measure of quantum correlation for arbitrary bipartite
state as follows
\begin{equation}\label{eq1d-8}
  Q_A(\rho)= \min_{K_A}\sum_{i=1}^{m}
             \mathsf{I}(\rho,K_A^i\otimes \iden_B),
\end{equation}
where the minimization is taken over the set of $K_A=
\{|i^A\rangle\langle i^A|\}$, and $m$ is the dimension of
$\mathcal{H}_A$. The measure $Q_A(\rho)$ vanishes if and only if
$\rho$ is classical-quantum correlated (i.e., $\rho\in
\mathcal{CQ}$). Moreover, it is locally unitary invariant, and is
contractive under CPTP map on the unmeasured party $B$.

\citet{yucslqu} also gave a numerical method for calculating
$Q_A(\rho)$ which uses the technique of approximate joint
diagonalization. Moreover, for any pure state $|\psi\rangle$ and
$(2\times n)$-dimensional state $\rho$, analytical solutions of
$Q_A(\rho)$ can be obtained, which equal half of
$\mathcal{U}_A(\rho)$. For the $(d\times d)$-dimensional Werner
state $\rho_W$ and isotropic state $\rho_I$ of the form of Eq.
\eqref{mina-4}, one has
\begin{equation}\label{eq1d-9}
 \begin{aligned}
  & Q_A(\rho_W)=\frac{d-x-\sqrt{(d^2-1)(1-x^2)}}{2(d+1)},\\
  & Q_A(\rho_I)=\frac{1-2\sqrt{(d^2-1)(1-x)x}+(d^2-2)x}{d(d+1))}.
 \end{aligned}
\end{equation}

\subsubsection{Negativity of quantumness}
The quantumness in a bipartite or multipartite state can also be
quantified by virtue of the amount of entanglement created between
the considered system and the measurement apparatus in a local
measurement. \citet{entqd} made such an attempt along this line. For
a bipartite state $\rho_{AB}$ and a measurement apparatus $M$
prepared in an initial state $|0^M\rangle$, they proved that the
created minimum distillable entanglement between $M$ and $AB$ equals
to the one-way deficit $\Delta^{\rightarrow}(\rho^{AB})$ [see Eq.
\eqref{new2-10}], that is,
\begin{equation}\label{eq1h-0}
 \Delta^{\rightarrow}(\rho^{AB})= \min_{U_{MAB}}E_D^{M:AB}(U_{MAB}\rho_{MAB}U_{MAB}^\dagger),
\end{equation}
where $\rho_{MAB}=|0^M\rangle\langle 0^M|\otimes \rho_{AB}$,
$U_{MAB}=U_{MA}\otimes\iden_B$ and $U_{MA}$ denotes only those
unitary operators which give $\sum_k \Pi_k^A\rho_{AB}\Pi_k^A=\tr_M
(U\rho_{MAB}U^\dagger)$. Therefore, the above equation establishes a
quantitative connection between discordlike quantum correlation and
entanglement.

\citet{TDD1} further proposed several discordlike measures of
quantumness by using this approach. First, they introduced the
measurement interaction $V_{A\mapsto AA'}$ described by a linear
isometry from $A$ to a bipartite system $AA'$, i.e.,
\begin{equation}\label{eq1h-1}
 V_{A\mapsto AA'}|a_i\rangle= |a_i\rangle |i'\rangle, ~ \forall i,
\end{equation}
with $\{|a_i\rangle\}$ being the basis of system $A$, and
$\{|i'\rangle\}$ is the computational basis of system $A'$. Then,
for any $N$-partite system described by density operator
$\rho_{\bm{A}}$ (we denote $\bm{A}={A_1 A_2 \ldots A_N}$ for short)
and the chosen subsystems $\Sigma\subseteq\{A_1 A_2 \ldots A_N\}$
for which the measurements are performed, the corresponding
premeasurement state reads
\begin{equation}\label{eq1h-2}
 \tilde{\rho}_{\Xi}=\Bigg(\bigotimes_{i\in\Sigma}V_{i\mapsto ii'}\Bigg)
                    \rho_{\bm{A}} \Bigg(\bigotimes_{i\in\Sigma}V_{i\mapsto
                    ii'}\Bigg)^\dagger,
\end{equation}
where $\Xi=\bm{A} \cup \Sigma'$.

By using negativity as a measure of entanglement \cite{negat},
\citet{TDD1} defined negativity of quantumness as the minimum
entanglement created between the system and the apparatus, that is,
\begin{equation}\label{eq1h-3}
 Q_\mathcal {N}^\Sigma(\rho_{\bm{A}})\coloneqq \min_{\tilde{\rho}^{\Xi}}
                    \mathcal{N}_{\bm{A}:\Sigma'}(\tilde{\rho}_{\Xi}),
\end{equation}
where the minimization is taken over all possible
$\tilde{\rho}_{\Xi}$ obtained with different choice of basis for the
system $\bm{A}$. This measure is showed to be nonnegative for any
$\rho_{\bm{A}}$, and vanishes if and only if $\rho_{\bm{A}}$ is
classical on the subsystems $\Sigma$ to be measured. When $\Sigma=
\{k_1,k_2,\ldots, k_n\}$ ($n<N$), $Q_\mathcal {N}^\Sigma
(\rho_{\bm{A}})$ is said to be the partial negativity of quantumness
and is equivalent to \cite{TDD1}
\begin{equation}\label{eq1h-4}
 Q_\mathcal {N}^\Sigma(\rho_{\bm{A}})=\min_{\bigotimes_{k\in\Sigma}\mathcal{B}_k}
                       \frac{1}{2} \left( \sum_{i_{k_1},i_{k_2},\ldots,i_{k_n}}
                      \| \rho_{i_{k_1}, i_{k_2},\ldots, i_{k_n}}\|_1 -1 \right),
\end{equation}
where $\mathcal{B}_k= \{|a^{(k)}_{i_k}\rangle\}$ denotes basis of
the $k$th subsystem, and
\begin{equation}\label{eq1h-5}
 \rho_{i_{k_1}, i_{k_2},\ldots, i_{k_n}}=\langle a_{i_{k_1}}^{(k_1)}
       a_{i_{k_2}}^{(k_2)}\dots a_{i_{k_n}}^{(k_n)}| \rho_{\bm{A}} |
       a_{i_{k_1}}^{(k_1)} a_{i_{k_2}}^{(k_2)}\dots
       a_{i_{k_n}}^{(k_n)}\rangle.
\end{equation}
When $\Sigma=\bm{A}$, one obtains the total negativity of
quantumness, and it is equivalent to \cite{TDD1}
\begin{equation}\label{eq1h-6}
 Q_\mathcal {N}^{\bm{A}}(\rho^{\bm{A}})= \min_{\bigotimes_{k=1}^N \mathcal{B}_k}
              \frac{1}{2} \Big( \| \rho^{\bm{A}}\|_{l_1} -1 \Big),
\end{equation}
where the minimization is taken over different choices of factorized
basis $\bigotimes_{k=1}^N \mathcal{B}_k$ and the $l_1$ norm is also
calculated in the same basis.

For the case of bipartite state $\rho_{AB}$ with $\dim A=2$ (i.e.,
$A$ is a qubit), \citet{TDD1} further showed that
\begin{equation}\label{eq1h-7}
 \begin{aligned}
 Q_\mathcal {N}^A(\rho_{AB})& = \frac{1}{2}\min_{\Pi^A}\| \rho_{AB}-\Pi^A(\rho_{AB})\|_1, \\
                            & = \frac{1}{2}\min_{\sigma\in \mathcal {C}\mathcal {Q}}\|
                            \rho_{AB}-\sigma\|_1,
 \end{aligned}
\end{equation}
where $\Pi^A=\{\Pi_i^A\}$ with $\Pi_i^A= |a_i\rangle\langle a_i|$
being the local projective measurements on $A$. It implies that the
minimization over the full set of classical-quantum states can be
simplified to the minimization only over the full set of
postmeasurement states. Similarly, the total negativity of
quantumness for the above-mentioned $\rho_{AB}$ is given by
\begin{equation}\label{eq1h-8}
 \begin{aligned}
 Q_\mathcal{N}^{AB}(\rho_{AB})& = \frac{1}{2}\min_{\Pi^A\otimes\Pi^B}\|\rho_{AB}-\Pi^A\otimes\Pi^B(\rho_{AB})\|_{l_1}, \\
                              & = \frac{1}{2}\min_{\sigma\in \mathcal {CC}}\| \rho_{AB}-\sigma\|_{l_1}
 \end{aligned}
\end{equation}
where the local projective measurements $\Pi^A\otimes \Pi^B$ are
defined with respect to the factorized basis $\mathcal{B}_A\otimes
\mathcal{B}_B$, and the $l_1$ norm in the first line is also
calculated with the same basis, while that in the second line is
calculated with respect to the eigenbasis of $\sigma$ (if the
eigenbasis are degenerate then it is chosen optimally to minimize
the distance by default).

For certain special $\rho_{AB}$, analytical solutions of $Q_\mathcal
{N}^A(\rho_{AB})$ and $Q_\mathcal {N}^{AB}(\rho_{AB})$ can be
obtained. For example, for the two-qubit $\rho_{AB}$ with
$\rho_A=\iden_2/2$, one has $Q_\mathcal {N}^A(\rho_{AB})=
\mathrm{int} \{s_1, s_2, s_3\}/2$, where $s_i$ is the singular value
of $R=(r_{ij})$ with $r_{ij}=\tr (\rho_{AB}\sigma_i \otimes
\sigma_j)$. If $\rho_{AB}$ belongs to the Bell-diagonal states, one
can further obtain $Q_\mathcal {N}^{AB}(\rho_{AB})=\mathrm{int}
\{s_1, s_2, s_3\}/2$. Moreover, for Werner states and isotropic
states given in Eq. \eqref{mina-4}, one has \cite{TDD1}
\begin{equation}\label{eq1h-9}
 \begin{aligned}
  & Q_\mathcal {N}^A(\rho_W)= Q_\mathcal {N}^{AB}(\rho_W) = \frac{|dx-1|}{2(d+1)}, \\
  & Q_\mathcal {N}^A(\rho_I)= Q_\mathcal {N}^{AB}(\rho_I) = \frac{|d^2 x-1|}{d+1}.
 \end{aligned}
\end{equation}

\subsection{Measurement-induced nonlocality} \label{sec:2B}
Apart from the various Bell-type nonlocality widely studied in the
literature \cite{rev-Bell}, the nonlocality of a system can also be
studied from other aspects. One typical research direction in recent
years is initialized by \citet{min}, who proposed the notion of MIN.
In this subsection, we will review in detail various geometric
measures of them. They were all defined from the measurement
perspective, and were motivated by those of the discordlike
correlation measures \cite{RMP}. We shall focus mainly on the
bipartite systems described by the density operator $\rho$ in the
Hilbert space $\mathcal {H}_A\otimes \mathcal {H}_B$. But the
related concepts and ideas can in fact be generalized to
multipartite systems straightforwardly.

Different from the definitions of GQDs in the above section, and
motivated by the consideration that the state of a bipartite system
may be disturbed by a measurement on one party (say $A$) of the
considered system, one can define the MIN as the maximal distance
that a state $\rho$ to the set $\mathcal {L}$ of locally invariant
quantum states, namely
\begin{equation}\label{min}
 N(\rho)=\max_{\delta\in\mathcal{L}}\mathcal{D}(\rho,\delta),
\end{equation}
where the locally invariant of $\delta$ means that $\delta=\sum_k
\Pi_k^A\rho\Pi_k^A$ for all $\Pi^A=\{\Pi_k^A\}$ satisfying $\sum_k
\Pi_k^A \rho_A \Pi_k^A=\rho_A$.

By adopting different distance measures, one can define different
measures of MIN which possess distinct novel characteristics.
Moreover, for bipartite state $\rho$ with nondegenerate reduced
state $\rho_A$, the MIN measures can readily be obtained as the
optimal measurements $\tilde{\Pi}^A=\{\tilde{\Pi}^A_i\}$ are induced
by the spectral resolutions of $\rho_A=\sum_i p_i^A
\tilde{\Pi}_i^A$. But when $\rho_A$ is degenerate, an optimization
procedure should be performed. In fact, seeking the optimal
measurements in order to extract various measurement-based
correlations (including MIN) is an important task for characterizing
quantumness of a state \cite{QD-ana2, Amico}.

\subsubsection{Hilbert-Schmidt norm of MIN}
The notion of MIN was introduced by \citet{min}. They used the HS
norm as a measure of distance, and defined the MIN as
\begin{equation}\label{smin}
 N_G(\rho)=\max_{\Pi^A} \|\rho-\Pi^A(\rho)\|_2^2,
\end{equation}
with $\Pi^A$ being the locally invariant projective measurements.
$N_G(\rho)$ characterizes the maximal global disturbance caused by
the locally invariant measurements, in the sense that it corresponds
to the maximal square HS distance between the postmeasurement state
$\Pi^A(\rho)$ and the premeasurement state $\rho$. The way for
revealing nonlocal feature of a state $\rho$ by doing local
measurements on one of its subsystem is somewhat similar to the
notion of localizable entanglement which was also defined based on
local measurements on fixed subsystems of $\rho$ \cite{localent1,
localent2}.

This MIN measure is showed to have the basic properties:
(\romannumeral+1) $N_G(\rho)\geqslant 0$, and the inequality holds
for any product state. (\romannumeral+2) it is locally unitary
invariant, namely, $N_G(U_{AB}\rho U_{AB}^\dag)=N_G(\rho)$,
$\forall$ $U_{AB}=U_A\otimes U_B$. (\romannumeral+3) For the case of
nondegenerate reduced state $\rho_A=\sum_k \lambda_k
|k\rangle\langle k|$, the optimal $\tilde{\Pi}^A$ is given by
$\tilde{\Pi}^A(\rho)=\sum_k |k\rangle\langle k|\rho |k\rangle\langle
k|$.

For $(m\times n)$-dimensional states of Eq. \eqref{mina-1},
$N_G(\rho)$ is upper bounded by $\sum_{i=1}^{m^2-m}\lambda_i$, where
$\lambda_i$ ($i=1,2,\ldots, m^2-1$) denote the eigenvalues of $RR^T$
in nonincreasing order, $R=(r_{ij})$ with $i,j\geq1$ is a real
matrix.

This MIN measure can be derived analytically for a wide range of
quantum states, which include the pure states, the bipartite states
$\rho_{AB}$ with $A$ being a qubit, certain higher dimensional
states with symmetry, as well as certain bound entangled states
\cite{minbd} and other special states with degenerate $\rho_A$
\cite{minar}. Some of the results are summarized as follows:

(1) For pure state $|\psi\rangle$ with the Schmidt decomposition of
Eq. \eqref{schmidt}, one has
\begin{equation}\label{mina-2}
 N_G(|\psi\rangle\langle\psi|)= 1-\sum_k \lambda_k^2.
\end{equation}

(2) For bipartite state $\rho$ of Eq. \eqref{mina-1} with
$\mathrm{dim} \mathcal{H}_A=2$, one has
\begin{equation}\label{mina-3}
  N_G(\rho)=\left\{
    \begin{aligned}
     &||R||_2^2-\frac{1}{||\vec{x}||_2^2}\vec{x}^T R R^T \vec{x}  &\text{if}~\vec{x}\neq 0,\\
     &||R||_2^2-\lambda_{\text{min}}(RR^T) &\text{if}~\vec{x}=0.
    \end{aligned} \right.
\end{equation}
where $\lambda_{\text{min}}(RR^T)$ denotes the smallest eigenvalue
of $R R^T$, and $\vec{x}=(r_{10}, r_{20}, r_{30})^T$.

(3) For $\rho_W$ and  $\rho_I$ of Eq. \eqref{mina-4}, one has
\cite{min2}
\begin{equation}\label{mina-5}
 \begin{aligned}
  & N_G(\rho_W)=\frac{(dx-1)^2}{d(d+1)(d^2-1)},\\
  & N_G(\rho_I)=\frac{(d^2 x-1)^2}{d(d+1)(d^2-1)}.
 \end{aligned}
\end{equation}

\citet{Guo2013jpa-min} given a necessary and sufficient condition
for nullity of the HS norm of MIN. Let $\rho$ be a bipartite state
acting on $\mathcal{H}_A\otimes \mathcal{H}_B$, and write
$\rho=\sum_{i,j}A_{ij}\otimes |i^B\rangle\langle j^B|$ [similar to
Eq. \eqref{eq-guo}], they showed that $N_G(\rho)=0$ if and only if
the $A_{ij}$s are mutually commuting normal operators, and each
eigenspace of $\rho_A=\sum_i A_{ii}$ is contained in some eigenspace
of $A_{ij}$, $\forall~ i,j$.

Furthermore, it is showed that for a zero-MIN state $\rho$ with
$\dim \mathcal{H}_A \geqslant 3$, any local channel acting on party
$A$ cannot create MIN if and only if either it is a completely
contractive channel or it is a nontrivial isotropic channel
\cite{Guo2013jpa-min}. For the qubit case this property is an
additional characteristic of the completely contractive channel or
the commutativity-preserving unital channel. That is, MIN can also
be created under local operations and classical communication
(LOCC).

\subsubsection{Trace norm of MIN}
Similar to the GQD measured with the HS norm, the flaw of
$N_G(\rho)$ is that it is also noncontractive under CPTP maps.
Explicitly, it can be increased or decreased by trivial local
reversible operations on the unmeasured party $B$. For example, a
map $\mathcal{E}_B(\rho)= \rho\otimes \rho_C$ leads to
$N_G(\rho_{A:BC})= N_G(\rho) \tr(\rho_C)^2$. As the purity $\tr
\rho_C^2\leq 1$, this equality means that the MIN is decreased by
simply introducing an uncorrelated local ancillary system. As a
matter of fact, the flaw of the HS norm of MIN, being noncontractive
under CPTP maps, is the same flaw for every HS norm measure of
correlation. Despite this flaw, the MIN defined originally using the
HS norm, in particular its motivation, inspires one to introduce
most of the subsequent MIN measures.

Motivated by using the trace norm to measure GQD \cite{TDD},
\citet{trmin} proposed that this norm can also be used to measure
MIN, with the explicit expression
\begin{equation}\label{minb-1}
 N_T(\rho)=\max_{\Pi^A}\|\rho-\Pi^A(\rho)\|_1.
\end{equation}
This definition, although amends slightly the definition of Eq.
\eqref{min}, avoids successfully its non-contractivity problem. One
can show that $N_T(\rho)$ is nonincreasing under any CPTP map
$\mathcal {E}_B$ \cite{trmin}, i.e., $N_T(\rho)\geq N_T(\mathcal
{E}_B[\rho])$. The proof is as follows: Let $\bar{\Pi}^A$ be the
optimal measurement for obtaining $N_T(\rho)$, and $\tilde{\Pi}^A$
be the optimal measurement for obtaining $N_T(\mathcal
{E}_B[\rho])$, then as $\mathcal{E}_B$ and $\tilde{\Pi}^A$ commute,
we obtain $\tilde{\Pi}^A (\mathcal{E}_B [\rho])=\mathcal{E}_B
(\tilde{\Pi}^A [\rho])$, and therefore
\begin{equation}\label{minb-2}
 \begin{aligned}
 N_T(\rho)&= \|\rho-\bar{\Pi}^A (\rho)\|_1 \\
          &\geq  \|\rho-\tilde{\Pi}^A (\rho)\|_1 \\
          &\geq  \|\mathcal{E}_B(\rho)-\mathcal{E}_B (\tilde{\Pi}^A [\rho])\|_1 \\
          &= N_T(\mathcal {E}_B[\rho]),
 \end{aligned}
\end{equation}
where the first inequality comes from the fact that $\tilde{\Pi}^A
\neq \bar{\Pi}^A$ in general, and the second inequality is due to
the contractivity of the trace norm under CPTP map. Therefore,
$N_T(\rho)$ circumvents successfully the problem incurred for
$N_G(\rho)$.

In purse of the analytical solutions of $N_T(\rho)$, some main
results are as follows:

(1) For $(2\times n)$-dimensional state $|\psi\rangle$ with the
Schmidt decomposition of Eq. \eqref{schmidt}, one has
$N_T(|\psi\rangle\langle\psi|)=2\sqrt{\lambda_1 \lambda_2}$.

(2) For two-qubit state $\rho$ decomposed as Eq. \eqref{mina-1},
with the addition $r_{ij}=0$ for $i\neq j$, we have
\begin{equation}\label{minb-4}
  N_T(\rho)=\left\{
  \begin{aligned}
   &\frac{\sqrt{\chi_{+}}+\sqrt{\chi_{-}}}{||\vec{x}||_1}
                              &\text{if}~\vec{x}\neq 0,\\
   &2\max\{|r_{11}|,|r_{22}|,|r_{33}|\} &\text{if}~\vec{x}=0,
  \end{aligned} \right.
\end{equation}
where the corresponding parameter are
\begin{equation}\label{minb-v4}
   \begin{aligned}
   &\chi_\pm=\alpha\pm 4\sqrt{\beta} |\vec{x}|,~
    \alpha=|\vec{r}|^2 |\vec{x}|^2-|\vec{r}\cdot \vec{x}|^2,\\
   &\vec{r}=(r_{11},r_{22},r_{33}),~
    \beta=\sum_{\langle ijk\rangle}x_i^2 r_{jj}^2 r_{kk}^2,
  \end{aligned}
\end{equation}
and the summation in the second line of the above equation runs over
all the cyclic permutations of $\{1,2,3\}$.

(3) For $\rho_W$ and $\rho_I$ of Eq. \eqref{mina-4}, solutions of
the the trace norm MIN are given, respectively, by
\begin{eqnarray}\label{minb-5}
 N_T(\rho_{W})=\frac{|dx-1|}{d+1}, ~~N_T(\rho_I)=\frac{2|d^2 x-1|}{d(d+1)},
\end{eqnarray}
and by comparing them with Eq. \eqref{mina-5}, one can see that for
the present cases, the two MIN measures $N_G$ and $N_T$ give
qualitatively the same descriptions of nonlocality.

\subsubsection{Bures distance of MIN}
By changing the maximization of Eq. \eqref{eq1b-1}, one can define
the Bures distance of MIN as follows \cite{trmin}
\begin{eqnarray}\label{minc-1}
 N_B(\rho)= \max_{\Pi^A}\{1 -\sqrt{F(\rho,\Pi^A(\rho)}\},
\end{eqnarray}
where $\Pi^A$ is still the locally invariant measurements on party
$A$, and $F(\rho,\sigma)$ is the Uhlmann fidelity defined in Eq.
\eqref{fidelity}.

Compared with the former two measures of MIN, the calculation of the
present MIN is more complicated. But when $A$ is a qubit, the
minimum Uhlmann fidelity $F_{\min}(\rho,\Pi^A[\rho])= \min_{\Pi^A}
F(\rho,\Pi^A[\rho])$ can be calculated via Eq. \eqref{eq1b-3}, with
however, the maximization being replaced by the minimization.

For Bell-diagonal state $\rho^{\rm Bell}$ of Eq. \eqref{eq-bell},
its square root can be derived explicitly, from which
$F_{\min}(\rho^{\rm Bell}, \Pi^A[\rho^{\rm Bell}])$ can be
calculated as
\begin{equation}\label{minc-3}
 F_{\min} =\frac{1}{2} \left(1+\min_{\{\theta,\phi\}}
         \sqrt{b_3^2+(b_{13}^2+b_{21}^2 \sin^2\phi)\sin^2\theta}\right),
\end{equation}
where $b_{ij}^2=b_i^2-b_j^2$, $b_i=8(t_0^2+t_i^2)-1$ ($i=1,2,3$),
and by writing $c_{\rm sum}=c_1+c_2+c_3$, we have
\begin{eqnarray}\label{minc-4}
 \begin{aligned}
    t_0=&\frac{1}{8}\sqrt{1-c_{\rm sum}}+\frac{1}{8}\sum_{k=1}^3 \sqrt{1+c_{\rm sum}-2c_k},\\
    t_i=&-\frac{1}{8}\sqrt{1-c_{\rm sum}}+\frac{1}{8}\sum_{k=1}^3\sqrt{1+c_{\rm sum}-2c_k}\\
        &-\frac{1}{4}\sqrt{1+c_{\rm sum}-2c_i}.
 \end{aligned}
\end{eqnarray}

From Eq. \eqref{minc-4} one can see that $F_{\min}$ equals to
$(1+|b_1|)/2$ if $|b_1|\leqslant\min\{|b_2|,|b_3|\}$, $(1+|b_2|)/2$
if $|b_2|\leqslant \min\{|b_1|, |b_3|\}$, and $(1+|b_3|)/2$
otherwise.

\subsubsection{Relative entropy of MIN}
The relative entropy can also be recognized as a (pseudo) distance
measure of quantum states, though technically it does not has a
geometric interpretation as it is not symmetric, i.e.,
$S(\rho\|\sigma) \neq S(\sigma\|\rho)$ in general. It has been used
to define the relative entropy of discord and quantum dissonance
\cite{reqd}.

\citet{remin} introduced the relative entropy of MIN as
\begin{equation}\label{mind-1}
 N_R(\rho)=\max_{\Pi^A}S(\rho||\Pi^A[\rho]),
\end{equation}
where $\Pi^A(\rho)=\sum_i \Pi_i^A \rho \Pi_i^A$, and $\{\Pi_i^A\}$
is the set of locally invariant projective measurements.

This MIN measure has been showed to be well defined. It possesses
the same basic properties (\romannumeral+1), (\romannumeral+2), and
(\romannumeral+3) as that of the HS norm of MIN. Furthermore,
$N_R(\mathcal {E}_B[\rho])\leq N_R(\rho)$ for any CPTP map $\mathcal
{E}_B$ on the unmeasured party $B$ \cite{enmin}. It is also
intimately related to the HS norm of MIN, $N_{E}(\rho)\geq
N_G^2(\rho)/(2\ln 2)$ \cite{remin}.

It vanishes for the classical-quantum state $\chi$ with
nondegenerate reduced density operator $\chi_A=\tr_B\chi$, or for
$\chi$ with degenerate $\chi_A$ and $\rho^B_k=\rho^B_l$ ($\forall$
$k,l$), see Eq. \eqref{eq1-1}. Moreover, it is lower bounded by
$-S(A|B)$ and upper bounded by $\min\{I(\rho), S(\rho_A)\}$, with
$S(A|B)=S(\rho)-S(\rho_B)$ the conditional entropy \cite{enmin}. For
$\rho^{\rm Bell}$ of Eq. \eqref{eq-bell}, analytical solution of it
is given by
\begin{equation}\label{mind-2}
 \begin{split}
 N_R(\rho^{\rm Bell})=& 1+H\left(\frac{1+c_{-}}{2}\right)+\frac{1-c_{\rm sum}}{4}\log_2\frac{1-c_{\rm sum}}{4} \\
                      &+\sum_{k=1}^3\frac{1+c_{\rm sum}-2c_k}{4}\log_2 \frac{1+c_{\rm sum}-2c_k}{4},
 \end{split}
\end{equation}
with $H(\cdot)$ being the binary Shannon entropy function.

The measure $N_R(\rho)$ is equivalent to that of the entropic MIN
defined as the maximal discrepancy between QMI of the pre- and
post-measurement states as \cite{enmin}
\begin{equation}\label{mind-3}
 N_{E}(\rho)=I(\rho)-\min_{\Pi^A}I[\Pi^A(\rho)],
\end{equation}
where $I(\rho)$ is the QMI given by Eq. \eqref{new1-1}.

This MIN quantifies in fact, the maximal loss of total correlations
under locally non-disturbing measurements $\Pi^A$. Moreover, as
$\rho$ and $\Pi^A(\rho)$ have the same reduced states, $N_E(\rho)$
defined above is equivalent to
\begin{equation}\label{mind-4}
 N_E(\rho)=\max_{\Pi^A} S(\Pi^A[\rho])-S(\rho).
\end{equation}
Thus, this measure of MIN quantifies also the maximal increment of
von Neumann entropy induced by $\Pi^A$. Moreover, as the entropy of
a state measures how much uncertainty there is in it, $N_E(\rho)$
can also be interpreted as the maximal increment of our uncertainty
about the considered system induced by the locally invariant
measurements.

\subsubsection{Skew information measure of MIN}
Apart from measuring uncertainty in a state, the WY skew information
has also been proposed to measure MIN. Its definition is as follows
\cite{simin}
\begin{equation}\label{mine-1}
 N_{SI}(\rho)=\max_{\tilde{K}^A}\sum_{i=1}^m
    \mathsf{I}(\rho,\tilde{K}_i^A\otimes \iden_B),
\end{equation}
which is in some sense dual to the correlation measure given in Eq.
\eqref{eq1d-8}, with however the rank-one projectors
$\tilde{K}^A=\{\tilde{K}_i^A\}$ are restricted to those which do not
disturb $\rho_A=\tr_B \rho$.

This MIN measure is invariant under locally unitary operations,
contractive under CPTP map $\mathcal {E}_B$ on party $B$, and
vanishes for all the product states and the classical-quantum states
with nondegenerate reduced state $\rho_A$. For general state the
calculation of $N_{SI}(\rho)$ is difficult. But if we decompose
$\sqrt{\rho}$ as Eq. \eqref{eq-sqrt}, an upper bound can be obtained
as follows \cite{simin}
\begin{equation}\label{mine-2}
N_{SI}(\rho)\leq1- \sum_{i=1}^{m-1}\mu_i,
\end{equation}
with $\mu_i$ ($i=1,2,\ldots,m^2$) being the eigenvalues of
$\Gamma\Gamma^T$ listed in decreasing order (counting multiplicity),
and $\Gamma=(\gamma_{ij})$ is the $(m^2\times n^2)$-dimensional
correlation matrix.

For the pure states $\psi=|\psi\rangle\langle\psi|$,
$N_{SI}(\psi)=N_G(\psi)$, while for the bipartite states $\rho$ with
$A$ being a qubit, one has $N_{SI}(\rho)=1-\mu_1$ if $\vec{u}= 0$,
and
\begin{equation}\label{mine-3}
  N_{SI}(\rho)=1-\frac{1}{2}{\rm tr}\left(\left(\begin{array}{cc}
              1    &   \vec{u}_0 \\
              1    &   -\vec{u}_0
              \end{array}\right)
              \Gamma\Gamma^T\left(\begin{array}{cc}
              1    &   \vec{u}_0 \\
              1    &   -\vec{u}_0
              \end{array}\right)^T\right),
\end{equation}
if $~\vec{u}\neq 0$. Here, $\vec{u}=(u_1,u_2,u_3)$ with
$u_i=\tr(\rho_A\sigma_i) /\sqrt{2}$, and $\vec{u}_0=\vec{u}
/|\vec{u}|$. Moreover, for $\rho_W$ and $\rho_I$ of Eq.
\eqref{eq-bell}, one has
\begin{eqnarray}\label{mine-4}
 \begin{split}
 & N_{SI}(\rho_{W})=\frac{1}{2}\left(\frac{d-x}{d+1}-\sqrt{\frac{d-1}{d+1}(1-x^2)}\right),\\
 & N_{SI}(\rho_I)=\frac{1}{d}\left(\sqrt{(d-1)x}-\sqrt{\frac{1-x}{d+1}}\right)^2.
 \end{split}
\end{eqnarray}

Similar to the above measure, \citet{uin} introduced another
MIN-like nonlocality measure which was termed as uncertainty-induced
nonlocality. It takes the form
\begin{eqnarray}\label{mine-5}
 \mathcal {U}_{SI}(\rho)=\max_{K^A} \mathsf{I}(\rho,K^A\otimes \iden_B),
\end{eqnarray}
where $K^A$ is a Hermitian observable with nondegenerate spectrum,
and $[K^A,\rho_A]=0$. This measure is locally unitary invariant,
nonincreasing under any CPTP map on the unmeasured party $B$.
Moreover, it can also be interpreted by the Hellinger distance via
the equality
\begin{eqnarray}\label{mine-6}
 \mathcal {U}_{SI}(\rho)= \max_{K^A}d_H^2(\rho,K^A\rho K^A).
\end{eqnarray}

For $(2\times n)$-dimensional state of Eq. \eqref{mina-1}, the
uncertainty-induced nonlocality can be obtained explicitly as
\begin{equation}\label{mine-7}
 \mathcal {U}_{SI}(\rho)=\left\{
    \begin{aligned}
         & 1-\lambda_{\min}(W_{AB})                      &\text{if}~\vec{x}= 0, \\
         & 1-\frac{1}{|\vec{x}|^2}\vec{x}^T W_{AB} \vec{x} &\text{if}~\vec{x}\neq
         0,
    \end{aligned} \right.
\end{equation}
where $\vec{x}=(r_{10},r_{20},r_{30})^T$, and $\lambda_{\min} (W)$
is the smallest eigenvalue of the $3\times 3$ matrix $W_{AB}$, the
elements of which is given by Eq. \eqref{eq1d-5}.

\subsubsection{Generalization of the MIN measures}
The MIN measures we reviewed in the above sections reveal in fact
only partial information about nonlocal features of a state, as they
are defined based on the one-sided locally invariant measurements,
thus those measures are all asymmetric. But a local state with
respect to one party may be nonlocal with respect to another party.
From this respect of view, it is significant to extend their
definitions to more general case of two-sided locally invariant
measurements. This gives the symmetric measure of MIN which can be
written as
\begin{eqnarray}\label{minf-1}
 \tilde{N}(\rho)= \max_{\tilde{\delta}\in\mathcal{L}}D(\rho,\tilde{\delta}),
\end{eqnarray}
with $\tilde{\delta}$ being the two-sided locally invariant states
in the sense that $\Pi^{AB}\tilde{\delta} \Pi^{AB}= \tilde{\delta}$
(with $\Pi^{AB}=\Pi^A\otimes\Pi^B$) should be satisfied, and $\sum_k
\Pi_k^A \rho_A \Pi_k^A=\rho_A$ and $\sum_k \Pi_k^B \rho_B
\Pi_k^B=\rho_B$ for any bipartite state $\rho$.

As an explicit example, we list the symmetric MIN measure defined
based on the HS norm, i.e., $\tilde{N}_G(\rho)= \max_{\Pi^{AB}}
\|\rho-\Pi^{AB}(\rho)\|_2^2$. This measure is locally unitary
invariant, and vanishes for the product states. For the pure state
$\psi=|\psi\rangle \langle\psi|$, we have $\tilde{N}_G(\psi)=
N_G(\psi)$ \cite{tsmin}. In fact, $\tilde{N}_G(\rho)$ can also be
extended to $N$-partite quantum states. The definition can be
written in the same form of Eq. \eqref{minf-1}, with however the
locally invariant measurements $\Pi^{A_1} \otimes\Pi^{A_2}\otimes
\ldots\otimes \Pi^{A_N}$, with $\sum_k \Pi_k^{A_i} \rho_{A_i}
\Pi_k^{A_i}= \rho_{A_i}$ for $i=\{1,2,\ldots,N\}$, and $\rho_{A_i}$
the reduced state of the subsystem $A_i$. But now the evaluation of
their analytical expression becomes a hard work.

\subsection{Applications of geometric quantum discord} \label{sec:2C}
Up to now, we have presented an overview of the formal definitions
and related formulae of the discordlike correlations defined via
different distances. In general, these measures are conceptually
different, and it is natural to wonder in what context one is more
or less useful than the other. As a matter of fact, these measures
capture different characteristic features of a state, and may have
different physical implications and potential applications, e.g.,
the LQU (equivalent to the Hellinger distance of discord for any
two-qubit state) guarantees a minimum precision of phase estimation
\cite{LQU}, the GQD defined via the relative entropy enables a
direct comparison of it with the relative entropy of entanglement,
and the negativity of quantumness can be connected to the negativity
of entanglement. The above correlations defined with different
distances may play role in different quantum information protocols,
e.g., the HS norm of discord bounds from above fidelity of quantum
teleportation and remote state preparation. Moreover, these
discordlike correlations may reveal different aspects of the
physical properties of a many-body system, and this will be
discussed in Sec. \ref{sec:7} of this review.

\subsubsection{Quantum teleportation}
To teleport a state from one party to another spatially separated
party, the sender Alice and the receiver Bob should share a quantum
channel $\rho$, and one can achieve a perfect teleportation if
$\rho$ is maximally entangled \cite{qtp0}. However, entanglement of
$\rho$ is the prerequisite but not the only key elements for
accomplishing the teleportation protocol. This is because for the
non-maximally entangled channel, the fidelity of teleportation is
not proportional to the amount of entanglement in $\rho$, e.g., it
has been showed that the purity of $\rho$ is also a crucial element
in determining the quality of the teleportation protocol
\cite{Hupla}.

When the channel is composed of a general two-qubit state $\rho$ as
given by Eq. \eqref{gqd-ad1}, the average teleportation fidelity,
based on the assumption that Bob can perform all kinds of recovery
operations to his qubit, can be derived as $\bar{F}=1/2+\tr
\sqrt{R^\dagger R}/6$ \cite{fidelity1}. By considering a normalized
version of the HS norm of discord
\begin{equation}\label{new4-1}
 \tilde{D}_{G}(\rho)=\frac{d}{d-1}D_{G}(\rho),
\end{equation}
\citet{int-gqd} identified a connection between an upper bound of
$\tilde{D}_{G}(\rho)$ and $\bar{F}(\rho)$. The bound of
$\tilde{D}_{G}(\rho)$ was derived by using the Weyl's theorem, and
is given by
\begin{equation}\label{new4-2}
  \tilde{D}_{G}^{\max}(\rho)=\frac{1}{3}\left[\|R^2\|-k_{\max}(RR^T)\right],
\end{equation}
where $k_{\max}(RR^T)$ represents the largest eigenvalue of $RR^T$.
As $\tr\sqrt{R^\dagger R}\geqslant \|R^2\|$, one can show that
\begin{equation}\label{new4-3}
 \bar{F}(\rho)\geqslant \frac{1+\tilde{D}_{G}^{\max}(\rho)}{2}.
\end{equation}

On the other hand, by using the relations $3\bar{F}(\rho)-2\leq
\mathcal {N}(\rho)$ \cite{nega1} and $\mathcal{N}^2(\rho) \leq
\tilde{D}_G(\rho)$ \cite{nega2} for all two-qubit states, with
$\mathcal{N}(\rho)$ being an entanglement measure called negativity
\cite{negat}, one can obtain
\begin{equation}\label{new4-4}
 \bar{F}(\rho)\leqslant \frac{2+\sqrt{\tilde{D}_G(\rho)}}{3}.
\end{equation}
These two equations show that the GQD bounds the average
teleportation fidelity. But a direct quantitative connection between
the various discordlike quantum correlation measures and $\bar{F}$
does not exist.

\subsubsection{Remote state preparation}
Remote state preparation (RSP) is a quantum protocol for remotely
preparing a quantum state by LOCC \cite{rsp0}. To accomplish this
task, the two participants, Alice and Bob, also need to share a
correlated channel. But different from the protocol of quantum
teleportation \cite{qtp0}, Alice knows what state to be transmitted
in advance, so the amount of required classical information can be
reduced.

If the shared state is maximally entangled, one can accomplish a
perfect state preparation. Otherwise, the fidelity of the protocol
may be reduced. \citet{rsp} considered such a problem. They
considered the channel to be a general two-qubit state of the form
of Eq. \eqref{gqd-ad1}, and Alice wants to prepare a qubit state
$\rho(\vec{s})=(I+\vec{s}\cdot \sigma)/2$ with the Bloch vector
$\vec{s}$ in the plane orthogonal to the direction $\vec{\beta}$. To
this purpose, she performs the local measurements
$\Pi_\alpha^A=[I+\alpha\vec{\alpha}\cdot \vec{\sigma}]/2$ along the
direction $\vec{\alpha}$ and informs Bob of her outcome $\alpha=\pm
1$. The Bloch vector of Bob's state can then be obtained as
\begin{equation}\label{new4-5}
 \vec{y}_\alpha= \frac{\vec{y}+\alpha R^T\vec{\alpha}}
                 {1+\alpha\vec{x}\cdot \vec{\alpha}}.
\end{equation}
If Alice's outcome is $\alpha=-1$, Bob applies a $\pi$ rotation
about $\vec{\beta}$ to his system, whereas no operation is required
for $\alpha=1$. After these conditional operations, the Bloch vector
of Bob's resulting state becomes the following mixture
\begin{equation}\label{new4-6}
 \vec{r}=p_{+} \vec{y}_{+} + p_{-}R(\pi)\vec{y}_{-}.
\end{equation}
where $p_\alpha=(1+\alpha\vec{\alpha}\cdot \vec{x})/2$ is the
probability for Alice's measurement outcome $\alpha$.

To evaluate the efficiency of the RSP protocol, \citet{rsp} defined
the payoff-function $\mathcal{P}=(\vec{r}\cdot \vec{s})^2$ which is
proportional to the fidelity $F=\tr[\rho(\vec{r})\rho(\vec{s})]=(1+
\vec{r} \cdot \vec{s})/2$. For the present case, $\mathcal{P}$ can
be derived explicitly as
\begin{equation}\label{new4-7}
 \mathcal {P}= (\vec{\alpha}^T R \vec{s})^2
             = \sum_{j=1}^3[\alpha_j (r_{j1}s_1+r_{j2}s_2+r_{j3}s_3)]^2,
\end{equation}
and by optimizing over Alice's choice of $\vec{\alpha}$, one can
obtain
\begin{equation}\label{new4-8}
 \mathcal {P}_{opt}=\sum_{j=1}^3 (r_{j1}s_1+r_{j2}s_2)^2.
\end{equation}

Finally, the expected payoff is averaged over the distribution
$\vec{s}$ and minimized over all possible choices of $\vec{\beta}$.
The corresponding RSP-fidelity is given by
\begin{equation}\label{new4-9}
 \mathcal{F}=\frac{1}{2}(E_2+E_3),
\end{equation}
where $E_1\geqslant E_2\geqslant E_3$ are the eigenvalues of $R^T R$
arranged in nonincreasing order. Clearly, $\mathcal{F}$ vanishes if
and only if $E_2=E_3=0$, which corresponds to a zero-discord state.

Moreover, if the local Bloch vector $\vec{x}$ of $\rho_{AB}$ is
parallel to the eigenvector corresponding to largest eigenvalue of
$R^T R$, the HS norm of discord is given by $D_G=\mathcal{F}/2$
\cite{rsp}, which endows the GQD an operational interpretation.

Note that the nonvanishing GQD in the channel state is a necessary
but not sufficient condition for RSP, as it has been found that
there are discordant states which yields zero RSP-fidelity, e.g.,
the family of two-qubit states described by the real density matrix
with $\rho_{11}-\rho_{22} = \rho_{44}- \rho_{33}$, $\rho_{14}=
\rho_{23}=0$, and $\rho_{12}=\rho_{13}=\rho_{24}=\rho_{34}$
\cite{rsp1}.

\subsubsection{Phase estimation}
\citet{LQU} considered a phase estimation task in which a bipartite
state $\rho$ is utilized as a probe. In this task, a local unitary
operation $U_\phi$ is performed on subsystem $A$ of this system,
therefore an unknown phase $\phi$ is encoded to it and $\rho$ is
transformed to $\rho_\phi=(U_\phi\otimes \iden) \rho (U_\phi^\dagger
\otimes\iden)$. One's goal is to estimate as precisely as possible
the parameter $\phi$. For a given probe state $\rho$, one can
optimize the measurements performed on $\rho_\phi$ to achieve the
Cram\'{e}r-Rao bound \cite{PD}
\begin{equation}\label{pd01}
 \mathrm{Var}(\tilde{\phi}_{\mathrm{best}})= \frac{1}{N\mathcal{F} (\rho_\phi)},
\end{equation}
where $\mathrm{Var} (\tilde{\phi}_{\mathrm{best}})$ is the variance
of the best unbiased estimator $\tilde{\phi}_{\mathrm{best}}$, $N$
is the times of independent measurements, and
\begin{equation}\label{QFI}
 \mathcal{F}(\rho_\phi)=\tr(\rho_\phi L_\phi^2),
\end{equation}
is the quantum fisher information, with $L_\phi$ being the symmetric
logarithmic derivative determined by
\begin{equation}\label{SLD}
 \frac{\partial{\rho_\phi}}{\partial{\phi}}= \frac{1}{2}(L_\phi\rho_\phi+\rho_\phi
 L_\phi).
\end{equation}

For the above phase estimation task, \citet{LQU} proved that
\begin{equation}\label{pd02}
 \mathrm{Var} (\tilde{\phi}_{\mathrm{best}}) \leq \frac{1}{4N\mathcal{U}_A^\Lambda(\rho)},
\end{equation}
hence the inverse of the LQU limits the achievable precision of the
estimated phase parameter $\phi$. This gives an operational
interpretation of the LQU.

\section{Quantum coherence measures}\label{sec:3}
Different from quantum correlations which are defined in the
framework of bipartite and multipartite scenarios, quantum coherence
is related to the characteristics of the whole system. In general,
the starting point for the resource theoretic characterization of a
quantum character, e.g., quantum entanglement \cite{RMPE} and QD
\cite{RMP}, is the identification of free states which can be
created at no cost and free operations which transform any free
state into free state.

In a manner similar to the resource framework of entanglement where
the free states are identified as those of the separable one and the
free operations are specified by the LOCC, the set $\mathcal {I}$ of
free states for quantum coherence encompasses those of the
incoherent states which are diagonal in the prefixed reference basis
$\{|i\rangle\}_{i=1}^d$, and take the form \cite{coher}
\begin{equation}\label{eq2-1}
 \delta=\sum_{i=1}^d \delta_i |i\rangle\langle i|,
\end{equation}
for a $d$-dimensional Hilbert space.

Within the framework of \citet{coher}, the set of free operations
are those of the incoherent operations (IO) which can be specified
by the Kraus operators $\{K_i\}$ satisfying $\sum_i K_i^\dag
K_i=\iden$. Based on the measurements with and without subselection,
\citet{coher} further identified two different classes of IO:

(A) The incoherent completely positive and trace preserving (ICPTP)
operations which act as $\Lambda(\rho)= \sum_i K_i \rho K_i^\dag$.
Here, all $K_i$ are of the same dimension, and should obey the
property $K_i \delta K_i^\dag/p_i\in\mathcal{I}$ for arbitrary
$\delta\in\mathcal{I}$, with $p_i=\tr (K_i \rho K_i^\dag)$ being the
probability for obtaining the result $i$.

(B) The incoherent operations with subselection for which the output
measurement results are retained. They also require $K_i \delta
K_i^\dag/p_i\in \mathcal{I}$ to be satisfied for any
$\delta\in\mathcal{I}$. But the dimension of $K_i$ may be different,
that is, different $K_i$ may corresponds to different output spaces.

In general, a Kraus operator for an IO can be represented as
$K_i=\sum_i c_i |f(i)\rangle\langle i|$, with the coefficient
$c_i\in [0,1]$ and $f(i)$ a function on the index set \cite{qcd1}.

As showed through explicit examples by \citet{meas0} and proved
strictly by \citet{Sun}, the Kraus operators related to incoherent
operations $\Lambda$ are very limited. There is at most one nonzero
entry in every column of $K_i$, and the number of possible structure
of $K_i$ (a legal structure stands for a possible arrangement of
nonzero entries in $K_i$) is $m^n$ for $K_i$ being the $m\times n$
matrices. \citet{structure} further discussed the problem relevant
to the number of Kraus operators in a general quantum operation. For
a system of dimension $d$, it has been found that any IO admits a
decomposition with at most $d^4+1$ Kraus operators. For $d=2$ and 3,
this number can be improved to 5 and 39, respectively.

Equipped with the sets of incoherent states and IO, \citet{coher}
presented the defining conditions for a faithful coherence measure
$C(\rho)$ which is a function that maps state $\rho$ to a
nonnegative real value:

(C1) Nonnegativity, i.e., $C(\rho)\geq 0$, and $C(\delta)=0$ iff
$\delta\in\mathcal {I}$.

(C2a) Monotonicity under ICPTP map, $C(\rho)\geq C(\Lambda[\rho])$.

(C2b) Monotonicity under selective IO on average, that is,
$C(\rho)\geq \sum_i p_i C(\rho_i)$.

(C3) Convexity under mixing of states, i.e., $\sum_i p_i C(\rho_i)
\geq C(\sum_i p_i \rho_i)$, with $\{p_i\}$ being the probability
distribution.

Note that condition (C2b) is stronger than (C2a), as its combination
with (C3) automatically imply (C2a). In general, a real-valued
function $C(\rho)$ is called a coherence measure if it satisfies the
above four conditions. If only the conditions (C1), (C2a), and (C2b)
are satisfied, $C(\rho)$ is usually called a coherence monotone.

A dual notion to incoherent states is the maximally coherent state,
which can serve as a unit for defining coherence measure
\cite{coher}. It takes the form
\begin{equation}\label{eq2-2}
 |\Psi_d\rangle=\frac{1}{\sqrt{d}}\sum_{i=1}^d |i\rangle,
\end{equation}
for which any other $\rho$ in the same Hilbert space can be
generated with certainty by merely IO on it.

\citet{cotr} considered problem of general pure states
transformation under IO by using the majorization theory
\cite{major}. For states $|\psi\rangle= \sum_{i=1}^d \psi_i
|i\rangle$ and $|\phi\rangle=\sum_{i=1}^d \phi_i |i\rangle$, with
the parameters $\{|\psi_i|\}$ and $\{|\phi_i|\}$ being arranged in
nonincreasing order, they proved that $|\psi\rangle$ can be
transformed to $|\phi\rangle$ via IO if and only if $(|\psi_1|^2,
|\psi_2|^2, \cdots, |\psi_d|^2)^T$ is majorized by
$(|\phi_1|^2,|\phi_2|^2, \cdots, |\phi_d|^2)^T$, i.e.,
\begin{equation} \label{majorization}
 (|\psi_1|^2,|\psi_2|^2,\cdots,|\psi_d|^2)^T \prec (|\phi_1|^2,|\phi_2|^2,\cdots,|\phi_d|^2)^T.
\end{equation}

Moreover, by applying the general unitary incoherent operations
\begin{equation} \label{UIO}
 U_I=\sum_j e^{i\theta_j}|\alpha_j\rangle\langle j|
\end{equation}
on $|\Psi_d\rangle$, with $\{\alpha_j\}$ being a relabeling of
$\{j\}$, \citet{mcvs} found that the complete set $\mathcal {M}$ of
maximally coherent states is composed of $\rho^{\rm mcs}
=|\Psi_d^{\rm mcs}\rangle\langle\Psi_d^{\rm mcs}|$, with
\begin{equation}\label{Peng}
 |\Psi_d^{\rm mcs}\rangle= \frac{1}{\sqrt{d}} \sum_je^{i\theta_j} |j\rangle.
\end{equation}
Building upon this, they proposed that $U_I$ are the unique quantum
operations that preserve the coherence of a state, and suggested an
additional condition for a valid coherence measure, i.e.,

(C4) $C(\rho)$ should assign a maximal value only to $\rho^{\rm
mcs}$.

\citet{Tong} proposed an alternative framework for defining
coherence, in which their first two conditions are the same as (C1)
and (C2a), while (C2b) and (C3) are replaced by one condition, that
is, the additivity requirement of coherence for subspace-independent
states. To be precise, for $\rho_1$ and $\rho_2$ in two different
subspaces, the amount of coherence contained in $\rho=p_1\rho_1
\oplus p_2\rho_2$ (with $p_1$ and $p_2$ being probabilities) should
be neither more nor less than the average coherence of $\rho_1$ and
$\rho_2$ due to the block-diagonal structure of $\rho$. Hence, a
reasonable measure of coherence should satisfy the condition
\begin{equation}\label{eq2-3}
 C(p_1\rho_1 \oplus p_2\rho_2)=p_1 C(\rho_1)+p_2 C(\rho_2),
\end{equation}
the above condition, together with (C1) and (C2a), have been showed
to be equivalent to the four conditions introduced by \citet{coher}.

\begin{figure}
\centering
\resizebox{0.46 \textwidth}{!}{%
\includegraphics{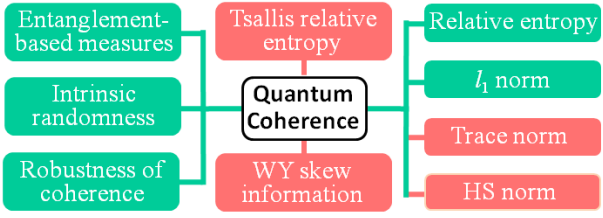}}
\caption{Different quantum coherence measures. Here, the measures
based on relative entropy, $l_1$ norm, entanglement, intrinsic
randomness, and the robustness of coherence satisfy all the required
conditions, while the others satisfy only partial of
them.}\label{fig:coh}
\end{figure}

While the the set of free or incoherent states is widely accepted,
there is no general consensus on the set of free operations in the
resource theory of coherence. Apart from the above mentioned IO,
there are other forms of free operations being introduced based on
different physical or mathematical motivations up to date. Three
typical ones are as follows:

(1) Maximally incoherent operations (MIO). It refers to the set of
physically realizable quantum operations $\Phi$ which maps
incoherent states into incoherent states, i.e., $\Phi(\mathcal{I})
\in \mathcal {I}$ \cite{cof}. Obviously, this is the most general
class of operations which do not create coherence from incoherent
states.

(2) Dephasing-covariant incoherent operations (DIO). The relevant
set of it is a subset of MIO with the additional property
$[\Delta,\Phi]=0$ \cite{DIO1,DIO2,Marvianpra}. That is, it admits
$\Lambda[\Delta(\rho)]=\Delta[\Lambda(\rho)]$.

(3) Strictly incoherent operations (SIO). This type of operations
also admit an incoherent Kraus decomposition $\{K_i\}$ for which not
only $K_i$ but also $K_i^\dagger$ ($\forall~ i$) is incoherent
\cite{{qcd1}}. That is, $\Delta(K_i\rho K_i^\dagger)= K_i
\Delta(\rho) K_i^\dagger$, where
\begin{equation}\label{eq2d-v2}
 \Delta(\rho)=\sum_i \langle i|\rho|i\rangle |i\rangle\langle i|,
\end{equation}
denotes full dephasing of $\rho$ in the basis $\{|i\rangle
\}_{i=1}^d$. It is the most general class of operations which do not
use coherence, and admits a decomposition with at most $\min\{d^4+1,
\sum_{k=1}^d d!/(k-1)!\}$ Kraus operators \cite{structure} .

The inclusion relation of the above free operations are given by
$\mathrm{SIO}\subset \mathrm{IO} \subset \mathrm{MIO}$ and
$\mathrm{SIO}\subset \mathrm{DIO} \subset \mathrm{MIO}$.

The definitions of incoherent state $\delta$ and maximally coherent
state $|\Psi_d\rangle$ imply that the related coherence measure will
be a basis dependent quantity. This is because any density operator
can be diagonalized in the reference basis spanned by its
eigenvectors, hence casting a doubt on the rationality of this
framework. But the recent progresses, particularly those studied
from an operational perspective, still yields physically meaningful
results. Moreover, in practice the reference basis are usually
chosen according to the physical problem under consideration. All
these indicate that the study of coherence measure has its own
irreplaceable role.

Up to now, there are a number of quantum coherence measures being
proposed in the literature (see Fig. \ref{fig:coh}), where some of
them satisfy the defining conditions, while the others satisfy only
partial of these conditions. We review them in detail in the
following.

\subsection{Distance-based measures of coherence} \label{sec:3A}
With the advent of quantum information science, geometric approaches
are used to treat a huge class of problems such as the
characterization and quantification of various quantum features.
Analogously to the resource theory of entanglement for which the
free operations are described by LOCC, the free states correspond to
the separable states, and the entanglement can be defined by a
distance between the considered state and the set of separable
states, it is natural to quantify coherence of a state by utilizing
a distance measure because coherence is also placed in a resource
theoretic framework. To be explicit, one can quantify the amount of
coherence contained in a state $\rho$ by using the minimal distance
between $\rho$ and the set $\mathcal {I}$ of incoherent states,
i.e.,
\begin{equation}\label{eq2a-1}
 C_\mathcal{D}(\rho)= \min_{\delta\in \mathcal{I}}\mathcal{D}(\rho,\delta),
\end{equation}
where $\mathcal{D}(\rho,\delta)$ denotes certain distance measures
of quantum states.

By its definition of Eq. \eqref{eq2a-1}, the condition (C1) is
fulfilled for the distance measure which gives
$\mathcal{D}(\rho,\delta) =0$ if and only if $\rho=\delta$, while
(C2a) can be fulfilled when $\mathcal{D}$ is monotonous under the
action of CPTP maps, i.e., $\mathcal{D} (\rho,\delta)\geq
\mathcal{D}(\Lambda[\rho],\Lambda[\delta])$. Moreover, (C3) is also
fulfilled if $\mathcal{D}$ is jointly convex, i.e., $\mathcal{D}
(\sum_i p_i \rho_i,\sum_i p_i\sigma_i)\leq \sum_i p_i
\mathcal{D}(\rho_i,\sigma_i)$.

\subsubsection{Relative entropy of coherence}
The relative entropy has been adopted to quantify entanglement, QD,
and MIN. \citet{coher} showed that it can also serve as a valid tool
for quantifying coherence. To be explicit, they defined
\begin{equation}\label{eq2a-2}
 C_r(\rho)= \min_{\delta\in \mathcal{I}}S(\rho\|\delta)
          = S(\rho_{\rm diag})-S(\rho),
\end{equation}
where $\rho_{\rm diag}$ denotes the diagonal part of $\rho$.

This is an entropic measure of coherence which has a clear physical
interpretation, as $C_r(\rho)$ equals to the optimal rate of the
distilled maximally coherent states by IO in the asymptotic limit of
many copies of $\rho$ \cite{qcd1}.

For the one-qubit state $\rho= (\iden_2+\vec{r}\cdot
\vec{\sigma})/2$, with $\vec{r}\in \mathbb{R}^3$ and $|\vec{r}|\leq
1$, and $\vec{\sigma}=(\sigma_1,\sigma_2,\sigma_3)$ is the vector of
Pauli matrices, if one chooses the reference basis as the eigenbasis
of $\hat{n}\cdot \vec{\sigma}$ ($\hat{n}$ is a unit vector in $
\mathbb{R}^3$), then
\begin{equation} \label{eq2a-v2}
 C_r(\rho)= H\left(\frac{1+\vec{r}\cdot\hat{n}}{2}\right)
            -H\left(\frac{1+\vec{r}}{2}\right),
\end{equation}
and $H(\cdot)$ is the binary Shannon entropy function.

For the two-qubit Bell-diagonal states of Eq. \eqref{eq-bell},
\citet{froz1} found that the closest incoherent state $\delta$ with
respect to the bona fide distance measure of coherence (e.g., the
relative entropy of coherence) is still a Bell-diagonal state with
vanishing local Bloch vectors along $x$ and $y$ directions, and for
the particular case of $c_2=-c_1 c_3$, $\delta$ reduces to the
diagonal part of $\rho^{\rm Bell}$.

\citet{RQC} and \citet{maxco} studied maximal coherence of a state
$\rho$ under generic reference basis. When the dimension of $\rho$
is $d$, they showed that the maximal coherence is given by
\begin{equation} \label{eq-maxrc}
 C_r^{\max}(\rho)= \log_2 d -S(\rho).
\end{equation}
\citet{maxco} also derived the corresponding unitary operator which
transforms the computational basis to the optimal basis such that
the maximal $C_r^{\max}(\rho)$ is obtained. It is given by
$U=VH^\dag$, with the column vectors of $V$ being the eigenvectors
of $\rho$, and $H$ is the rescaled complex Hadamard matrices. In
fact, the Fourier matrix (a subset of the complex Hadamard matrices)
$\bold{F}_d$ with elements $[\bold{F}_d]_{\mu\nu}= e^{i2\pi
\mu\nu/d}/\sqrt{d}$ is suffice for this purpose.

\subsubsection{$l_1$ norm of coherence}
Intuitively, the superposition corresponds to the nonvanishing
off-diagonal elements of the density operator description of a
quantum state with respect to the selected reference basis. Starting
from this consideration, \citet{coher} showed that the $l_1$ norm
can also serve as a bona fide measure of coherence. To be explicit,
they defined it as
\begin{equation}\label{eq2a-3}
 C_{l_1}(\rho)= \min_{\delta\in \mathcal {I}}\|\rho-\delta\|_{l_1}
              = \sum_{i\neq j}|\langle i|\rho|j\rangle|,
\end{equation}
which equals to sum of the absolute values of the off-diagonal
elements of $\rho$, and is favored for its ease of evaluation. Apart
from convexity, it also satisfy the inequality $C_{l_1}(p_1
\rho_1+p_2\rho_2) \geq |p_1 C_{l_1}(\rho_1)-p_2 C_{l_1}(\rho_2)|$
\cite{triangle}. Moreover, for any bipartite pure state
$|\psi\rangle_{AB}$, the $l_1$ norm of coherence equals to twice of
its negativity which is a measure of quantum entanglement
\cite{negat}.

For pure state $\psi=|\psi\rangle\langle\psi|$, a relation between
$C_{l_1}(\psi)$ and $C_r(\psi)$ has also been established
\cite{meas1}, which is given by
\begin{equation}\label{eq2a-4}
 C_{l_1}(\psi)\geq \max\{C_r(\psi),2^{C_r(\psi)}-1\},
\end{equation}
where $C_{l_1}(\psi)$ equals to $C_r(\psi)$ if and only if the
diagonal elements of $\psi$ are (up to permutation) either
$\{1,0,\ldots,0\}$ or $\{1/2,1/2,0, \ldots ,0\}$. It has also been
proven that \cite{Rana}
\begin{equation} \label{eq2a-n1}
 C_{l_1}^2(\psi)\leq \frac{d(d-1)C_r(\psi)}{\sqrt{2}},
\end{equation}
where $d=\mathrm{rank}(\psi)$ is the rank of $\psi$. If $d>2$, one
can further obtain $C_{l_1}(\psi)-C_r(\psi)\leq d-1-\log_2 d$.
\citet{Rana} also proved a sharpest bound of $C_r(\psi)$ in terms of
$C_{l_1}(\psi)$ as follows:
\begin{equation} \label{eq2a-n2}
 \begin{aligned}
 H(\alpha)+(1-\alpha)\log_2 (d-1) & \leq C_r(\psi) \\
                                  & \leq H(\beta)+(1-\beta)\log_2 (n-1),
 \end{aligned}
\end{equation}
where $n$ equals $C_{l_1}+1$ if $C_{l_1}$ is an integer, and
$[C_{l_1}]+2$ ($[C_{l_1}]$ is the integer part of $C_{l_1}$)
otherwise. By denoting $x=C_{l_1}+1$, the other two parameters are
given by
\begin{equation} \label{eq2a-n3}
 \begin{aligned}
  & \alpha=\frac{2+(d-2)(d-C_{l_1})+2\sqrt{(d-1)(d-x)x}}{d^2},\\
  & \beta=\frac{2+(n-2)(n-C_{l_1}) -2\sqrt{(n-1)(n-x)x}}{n^2}.
 \end{aligned}
\end{equation}

For general states $\rho$, one has
\begin{equation}\label{eq2a-5}
 C_{l_1}(\rho)\geq C_r(\rho)/ \log_2 d.
\end{equation}
\citet{meas1} have also conjectured $C_{l_1}(\rho)\geq C_r(\rho)$,
but it was proved only for pure states, single-qubit states, and
pseudopure states $\rho=p\psi+(1-p)\delta$ ($\forall \delta\in
\mathcal {I}$ and $p\in[0,1]$), while for general case, one can only
prove $C_{l_1}(\rho)\geq 2^{C_r(\rho)}-1$ \cite{Rana}, which is
somewhat similar to the case for pure states showed in Eq.
\eqref{eq2a-4}.

Moreover, for any single-qubit state $\rho$, by using convexity of
$C_r$ and the inequality $2\min\{x,1-x\}\leq H(x)\leq
2\sqrt{x(1-x)}$, $\forall x\in[0,1]$, \citet{Rana} further proved
the following relation
\begin{equation} \label{eq2a-n4}
 \begin{aligned}
 1-H\left(\frac{1-C_{l_1}(\rho)}{2} \right)& \leq C_r(\rho)\leq H\left(\frac{1-\sqrt{1-C_{l_1}^2(\rho)}}{2} \right) \\
                                           & \leq C_{l_1}(\rho),
 \end{aligned}
\end{equation}
where the equality holds when $\rho$ is either incoherent or
maximally coherent.

In fact, any state $\rho$ can be decomposed as
\begin{equation}\label{eq2d-4}
 \rho = \frac{1}{d}\iden_d+\frac{1}{2}\sum_{i=1}^{d^2 -1} x_i X_i,
\end{equation}
where $x_i=\tr(\rho X_i)$, and $\{X_i/\sqrt{2}\}$ ($X_0=\iden_d
/\sqrt{d}$) is the orthonormal operator bases for $\mathcal {H}$
(e.g., $X_i$ is the Pauli matrices when $d=2$, and the Gell-Mann
matrices when $d=3$). Then if one arranges elements of $\vec{X}$ as
\begin{equation} \label{eq2a-n5}
 \vec{X}=\{u_{12},v_{12}, \ldots, u_{d-1,d},v_{d-1,d}, w_1,\ldots, w_{d-1}\},
\end{equation}
with the elements
\begin{equation}\label{eq2d-5}
 \begin{split}
  & u_{jk}= |j\rangle\langle k| + |k\rangle\langle j|, ~
    v_{jk}= -i(|j\rangle\langle k| - |k\rangle\langle j|), \\
  &  w_{l}= \sqrt{\frac{2}{l(l+1)}}\sum_{j=1}^l (|j\rangle\langle j|
            - l |l+1\rangle\langle l+1|),
 \end{split}
\end{equation}
where $j,k\in \{1,2,\ldots, d\}$ with $j<k$, $l\in \{1,2,\ldots,
d-1\}$ (the symbol $i$ in $v_{jk}$ is the imaginary unit), one has
\cite{comp2}
\begin{equation}\label{eqa-01}
 C_{l_1}(\rho)=\sum_{r=1}^{(d^2-d)/2}\sqrt{x_{2r-1}^2+ x_{2r}^2}.
\end{equation}
By using Eq. \eqref{eqa-01}, \citet{RQC} showed that the maximal
$C_{l_1}(\rho)$ under generic basis is upper bounded by
\begin{equation} \label{eq2a-n6}
 C_{l_1}^{\max}(\rho)\leq \sqrt{\frac{d^2-d}{2}}|\vec{x}|,
\end{equation}
where $|\vec{x}|$ is length of the vector $(x_1,x_2,\cdots,
x_{d^2-1})$.

\subsubsection{Trace norm of coherence}
Apart from the $l_1$ norm, one may wonder whether the general $l_p$
and Schatten-$p$ matrix norm can be adopted for defining coherence
measures. In general, the answer to this question is negative. For
example, \citet{coher} have considered the HS norm (i.e., $p=2$, for
which it is also known as the Frobenius norm) measure of coherence
defined as
\begin{equation}\label{eq2a-nn3}
 C_{l_2}(\rho)=\min_{\delta\in \mathcal {I}}\|\rho-\delta\|^2_{2}
              =\sum_{i\neq j}|\langle i|\rho|j\rangle|^2,
\end{equation}
and showed through an counterexample that it does not satisfy
condition (C2b). \citet{meas1} further showed that any coherence
measure defined via the $l_p$ norm or the Schatten-$p$ norm with
$p\geq 2$ violates (C2b).

For the case of $p=1$ which corresponds to the trace norm (i.e., the
Schatten-$1$ norm), if one defines
\begin{equation}\label{eq2a-6}
 C_{tr}(\rho)=\min_{\delta\in \mathcal {I}}\|\rho-\delta\|_1,
\end{equation}
with $\|M\|_1=\tr\sqrt{M^\dag M}$ denoting the trace norm of the
matrix $M$, the conditions (C1), (C2a), and (C3) for $C_{tr}(\rho)$
to be a proper coherence measure have been proven, but the work of
\citet{Tong} showed that the condition of Eq. \eqref{eq2-3} may be
violated, thus proved it not to be a proper coherence measure.

For certain special classes of states, e.g., $\rho$ of one qubit or
having possible nonzero elements along only the main diagonal and
anti-diagonal (i.e., the \emph{X} states), $C_{tr}(\rho)$ has
already been proven to be a coherence monotone, and the
corresponding optimal incoherent state is given by $\rho_{\rm diag}$
\cite{froz1}. Moreover, for the state $\rho$ with all of its
non-diagonal elements equal to each other, i.e., $\rho_{ij}=a$
($\forall i\neq j$), the trace norm of coherence can be derived
analytically as
\begin{equation} \label{eq2a-v6}
 C_{tr}(\rho)= 2(d-1)|a|,
\end{equation}
where $d=\dim\rho$, and the closest incoherent state is
$\delta^\star= \rho_\mathrm{diag}$ \cite{traceana}. For a restricted
family of SIO \cite{qcd1}, i.e., those of the SIO whose Kraus
operators are $(2\times d)$-dimensional, the trace norm of coherence
for this family of $\rho$ was also showed to satisfy the four
conditions for a reliable quantum coherence measure \cite{traceana}.
But its monotonicity under general IO may does not hold.

When restricted to pure states $|\psi\rangle$, it is possible to
identify structure of the optimal incoherent state under the trace
norm of coherence. As for any pure state $|\psi\rangle$, one can
find a diagonal unitary matrix $U$ and a permutation matrix $P$
which gives $PU|\psi\rangle=|x\rangle$, with entries $x_1\geq
\cdots\geq x_d\geq 0$, the calculation can be performed to
$|x\rangle$ only. By using the approximation theory, \citet{pure}
found that $|\Psi_d\rangle$ of Eq. \eqref{eq2-2} is the unique state
that maximizing the trace norm of coherence, for which
$C_{tr}^{\max} =2(1-1/d)$. The optimal incoherent state to
$|x\rangle$ and the corresponding trace norm of coherence are given,
respectively, by
\begin{equation}\label{eq2a-7}
 \begin{aligned}
  & \delta_{\rm opt}={\rm diag} \{\alpha_1,\cdots,\alpha_k,0,\cdots,0\},\\
  & C_{tr}(|x\rangle\langle x|)=2(q_k s_k+m_k),
  \end{aligned}
\end{equation}
where
\begin{equation}\label{eq2a-8}
 \alpha_i=\frac{x_i-q_k}{s_k-k q_k},
\end{equation}
and $k$ is the maximum integer satisfying
\begin{equation}\label{eq2a-9}
 x_k>q_k\coloneqq \frac{1}{2ks_k}\left( p_k+\sqrt{p_k^2+4km_k s_k^2}\right),
\end{equation}
with the parameters
\begin{equation} \label{eq2a-v9}
 s_l=\sum_{i=1}^l x_i,~m_l=\sum_{i=l+1}^d x_i^2,
  ~ p_l=s_l^2-l m_l-1
\end{equation}
for all $l\in\{1,2,\cdots,d\}$.

To avoid the perplexity for $C_{tr}(\rho)$ of Eq. \eqref{eq2a-6},
i.e., the non-monotonicity of the trace norm of coherence under
general incoherent operations, \citet{Tong} further proposed a
modified version of trace norm of coherence by introducing a control
parameter $\lambda$, and defined it as
\begin{equation}\label{eq-mtc}
 C'_{tr}(\rho)= \min_{\lambda\geq 0, \delta\in \mathcal {I}}
                                   \|\rho-\lambda\delta\|_1,
\end{equation}
and proved that it satisfies the conditions (C1), (C2a) and Eq.
\eqref{eq2-3}, that is to say, it satisfies all the conditions for a
reliable measure of quantum coherence. As its relation with other
coherence measures, we have
\begin{equation} \label{eq-mtc2}
 C'_{tr}(\rho)\leq  C_{tr}(\rho)\leq C_{l_1}(\rho),
\end{equation}
where the first inequality is obvious from their definitions in Eq.
\eqref{eq2a-6} and \eqref{eq-mtc}, and the second one is due to
$\|\cdot\|_1\leq\|\cdot\|_{l_1}$ for any Hermitian operator.

For one-qubit state, $ C'_{tr}(\rho)= C_{tr}(\rho)=C_{l_1}(\rho)$,
and the optimal parameter $\lambda^\star=1$ and the optimal
$\delta^\star = \rho_\mathrm{diag}$ \cite{froz1,feism}. For general
state $\rho$, determination of the analytical solution of
$C'_{tr}(\rho)$ is possible only for certain special family of
states. For example, the class of maximally coherent mixed states
(MCMS) with respect to the $l_1$ norm of coherence, up to incoherent
unitaries, is given by \cite{comp2}
\begin{equation}\label{eq4b-4}
 \rho_{\rm mcms}=\frac{1-p}{d}\iden+p|\Psi_d\rangle\langle\Psi_d|,
\end{equation}
for which the modified trace norm of coherence can be obtained
analytically as $ C'_{tr}(\rho_{\rm mcms})=p$, with the optimal
$\lambda^\star=1-p$ and $\delta^\star=\iden_d/d$.

\subsection{Entanglement-based measure of coherence} \label{sec:3B}
In a way analogous to the entanglement activation via local von
Neumann measurements \cite{{entqd}}, one can also introduce the
operational coherence measure with the help of IO.

Given a system $S$ in the state $\rho_S$ and an ancilla $A$
initialized in the pure state $|0^A\rangle$, \citet{meas2}
considered incoherent operations $\Lambda^{SA}$ on the combined
system $SA$. By denoting $E_\mathcal{D}= \min_{\chi\in\mathcal {S}}
\mathcal{D}(\rho,\chi)$ a distance-based entanglement monotone and
$C_\mathcal{D}$ the corresponding coherence monotone as given in Eq.
\eqref{eq2a-1}, with $\mathcal{D}$ any contractive distance measure
of quantum states and $\mathcal{S}$ the set of separable states,
they found that the generated entanglement $E_\mathcal{D}^{S:A}$ is
bounded from above by
\begin{equation}\label{eq2b-1}
 E_\mathcal{D}^{S:A}(\Lambda^{SA}[\rho^S\otimes |0^A\rangle\langle 0^A|])\leq C_\mathcal{D}(\rho^S),
\end{equation}
which means that when $\rho^S$ is incoherent, the IO cannot generate
entanglement between $S$ and $A$.

Particularly, when $\mathcal{D}$ is the relative entropy and
$d_A\geq d_S$ with $d_{A,S}=\dim\mathcal {H}_{A,S}$, then there
always exists an incoherent operation (i.e., the generalized
\textsc{cnot} operation)
\begin{equation}\label{eq-cnot}
 U_{\textsc{cnot}}= \sum_{i=0}^{d_S-1}\sum_{j=0}^{d_S-1}|i,i\oplus j\rangle^{SA}\langle ij|
            + \sum_{i=0}^{d_S-1}\sum_{j=d_S}^{d_A-1}|ij \rangle^{SA}\langle ij|,
\end{equation}
where $\oplus$ represents addition modulo $d_S$. This unitary
operation maps the state $\rho^S\otimes |0^A\rangle\langle 0^A|$ to
\begin{equation}\label{eq2b-2}
 \Lambda^{SA}[\rho^S\otimes |0^A\rangle\langle 0^A|] = \sum_{ij}\rho^S_{ij}
                           |ij\rangle^{SA}\langle ij|,
\end{equation}
and henceforth Eq. \eqref{eq2b-1} is saturated:
\begin{equation}\label{eq2b-3}
 E_{r}^{S:A}(\Lambda^{SA}[\rho^S\otimes |0^A\rangle\langle 0^A|]) = C_{r}(\rho^S),
\end{equation}
which can be proved immediately by Eq. \eqref{eq2b-1} and the lower
bound $-S(A|S)$ of $ E_{r}^{S:A}(\rho^{SA})$, where $S(A|S)$ is the
conditional entropy. Then, \citet{meas2} proposed to define
coherence of $\rho^S$ as the maximal entanglement of $SA$ generated
by IO, that is,
\begin{equation}\label{eq2b-4}
 C_E(\rho^S)=\lim_{d_A\rightarrow \infty}\{\sup_{\Lambda^{SA}}E^{S:A}
             (\Lambda^{SA}[\rho^S\otimes|0^A\rangle\langle 0^A|]) \},
\end{equation}
with $E$ being an arbitrary entanglement measure, and $C_E$ will
satisfy the four conditions of \citet{coher} if $E$ is convex as
well.

For the geometric measure of entanglement $E_g(\rho)=
1-\max_{\sigma\in \mathcal{S}}F(\rho,\sigma)$ [see also Eq.
\eqref{eq2b-5}] \cite{fident}, the associated coherence measure can
be evaluated as
\begin{equation}\label{eq2b-6}
 C_g(\rho)=1-\max_{\delta\in\mathcal{I}}F(\rho,\delta),
\end{equation}
and for $\rho$ of single-qubit state,
\begin{equation}\label{eq2b-7}
 C_g(\rho)=\frac{1}{2}\left(1-\sqrt{1-4|\rho_{12}|^2}\right),
\end{equation}
which is an increasing function of $C_{l_1}(\rho)= 2|\rho_{12}|$.

Moreover, for pure state $\psi=|\psi\rangle\langle\psi|$, as
$F(\psi,\sigma)= |\langle\psi|\sigma|\psi\rangle|$, we have
\begin{equation} \label{eq-cg}
 C_g(\psi)=1-\max_i\{\psi_{ii}\},
\end{equation}
with $\psi_{ii}$ being the diagonal elements of $\psi$. For general
mixed states, the calculation of $C_g(\rho)$ is formidably
difficult, hence derives some bounds of it is significant. By using
the relations among fidelity $F(\rho,\sigma)$, sub-fidelity
$E(\rho,\sigma)$, and super-fidelity $G(\rho,\sigma)$,
\citet{longgl} obtained lower and upper bounds of $C_g(\rho)$. The
sub-fidelity and super-fidelity were defined as \cite{subsuper}
\begin{equation} \label{eq-ss}
 \begin{aligned}
  & E(\rho,\sigma)=\tr(\rho\sigma)+\sqrt{2[(\tr(\rho\sigma))^2-\tr(\rho\sigma\rho\sigma)]}, \\
  & G(\rho,\sigma)=\tr(\rho\sigma)+\sqrt{(1-\tr\rho^2)(1-\tr\sigma^2)},
 \end{aligned}
\end{equation}
and based on the relation $E(\rho,\sigma)\leq F(\rho,\sigma) \leq
G(\rho,\sigma)$ (the equality holds for one-qubit state or at least
one of $\rho$ and $\sigma$ is pure), they found
\begin{equation} \label{eq-longgl}
 \begin{aligned}
  & 1-\frac{1}{d}-\frac{d-1}{d}\sqrt{1-\frac{d}{d-1}\left(\tr\rho^2-\sum\nolimits_i \rho_{ii}^2\right)} \\
  & \leq C_g(\rho) \leq \min\left\{1-\max_i\{\rho_{ii}\},1-\sum\nolimits_i b_{ii}^2\right\},
  \end{aligned}
\end{equation}
where $b_{ii}$ is related to the square root of $\sqrt{\rho}=
\sum_{ij}b_{ij}|i\rangle\langle j|$.

\subsection{Convex roof measure of coherence} \label{sec:3C}

\subsubsection{Intrinsic randomness of coherence}
In the framework of quantum theory, measurement of quantum states
induce intrinsically random outputs in general, and this randomness
indicates genuine quantumness of a system. Based on this
consideration, \citet{meas4} proposed a convex roof measure for
coherence, which has been proved to satisfy the four conditions of
\citet{coher}. We call it intrinsic randomness of coherence.

For pure states $\psi=|\psi\rangle\langle\psi|$, the intrinsic
randomness can be quantified by Shannon entropy of the probability
distribution $\{p_i\}$ of the measurement outcomes which reads
\begin{equation}\label{eq2c-1}
 R_I(\psi)=H(\{p_i\})=-\sum_i p_i\log_2 p_i,
\end{equation}
with $H(\{p_i\})$ denotes the Shannon entropy of the probability
distribution $\{p_i\}$, with $p_i=\tr(E_i \psi)$ and $\{E_i\}$ is
the set of measurement operators. $R_I$ characterizes also the
degree of uncertainty related to the measurement outcomes, namely,
the outcomes that cannot be predicted by blindly guessing. When
restricted to projective measurements for which
$E_i=|i\rangle\langle i|$ with $\{|i\rangle\}$ the reference basis,
the right-hand side of Eq. \eqref{eq2c-1} is $S(\psi_{\rm diag})$,
henceforth, the intrinsic randomness $R_I(\psi)$ equals the relative
entropy of coherence $C_r(\psi)$.

For general case of mixed states $\rho$, \citet{meas4} defined the
intrinsic randomness $R_I(\rho)$ by utilizing convex roof
construction, that is,
\begin{equation}\label{eq2c-2}
 R_I(\rho)=\min_{\{p_i,\psi_i\}}\sum_i p_i R_I(\psi_i),
\end{equation}
where $\sum_i p_i=1$, $\psi_i=|\psi_i\rangle\langle\psi_i|$, and the
minimum is taken over all possible pure state decompositions of
$\rho$ given in Eq. \eqref{decomp}.

Eq. \eqref{eq2c-2} establishes a convex roof definition of quantum
coherence, which bears some resemblance with the convex roof
measures of entanglement such as entanglement of formation
$E_f(\rho)=\min_{\{p_i, \psi_i\rangle}\sum_i p_i S(\tr_B \psi_i)$
and the geometric entanglement $E_g(\rho)=\min_{\{p_i,
\psi_i\}}\sum_i p_i E_g (\psi_i)$. It was also termed as
superposition of formation by \citet{cof}, and coherence of
formation by \citet{qcd1}.

Apart from pure states, $R_I(\rho)$ is analytically computable for
one-qubit states, i.e.,
\begin{equation} \label{eq-ri}
 R_I(\rho)=H\left(\frac{1+\sqrt{1-C_z^2(\rho)}}{2} \right),
\end{equation}
where $C_z(\rho)=2|\rho_{12}|$ is termed as coherence concurrence,
as it is given by
\begin{equation}\label{eq-rand}
 C_z(\rho)= |\lambda_1 -\lambda_2|,
\end{equation}
with $\lambda_i$ being square roots of the eigenvalues of the
product matrix
\begin{equation}\label{eq-cc}
 R=\rho\sigma_x\rho^*\sigma_x,
\end{equation}
where $\rho^*$ is the conjugation of $\rho$, and $\sigma_x$ is the
first Pauli matrix.

\subsubsection{Coherence concurrence}
Using the fact that any $\rho$ can be decomposed as Eq.
\eqref{eq2d-4}, \citet{gaoyan} found that the $l_1$ norm of
coherence for $\rho$ is equivalent to
\begin{equation} \label{gaoyan}
 C_{l_1}(\rho)=\sum_{j<k}\bigg |\sqrt{\eta_1^{jk}}-\sqrt{\eta_2^{jk}}\bigg |,
\end{equation}
where $\eta_1^{jk}$ and $\eta_2^{jk}$ are nonzero eigenvalues of the
matrix
\begin{equation} \label{gaoyan2}
 R^{jk}=\rho u_{jk}\rho^* u_{jk},
\end{equation}
and the generalized Gell-Mann matrices $\{u_{jk}\}$ as given in Eq.
\eqref{eq2d-5}. When $d=2$, $R^{jk}$ is just that of $R$ given in
Eq. \eqref{eq-cc}. For pure state $\psi=|\psi\rangle\langle\psi|$,
the above equation further gives
\begin{equation} \label{gaoyan3}
 C_{l_1}(\psi)=\sum_{j<k} |\langle\psi|u_{jk}|\psi^*\rangle |,
\end{equation}
which is direct consequence of $R^{jk}=|\psi\rangle\langle\psi|
u_{jk}|\psi^*\rangle\langle\psi^*|u_{jk}$. Motivated by this fact,
\citet{gaoyan} proposed a convex roof measure of coherence
\begin{equation}\label{eq-gaoyan}
 C_{\rm con}(\rho)=\min_{\{p_i,\psi_i\}}\sum_i p_i C_{l_1}(\psi_i),
\end{equation}
where the minimization is with respect to the possible pure state
decompositions of $\rho$ given in Eq. \eqref{decomp}. $C_{\rm
con}(\rho)$ is termed as coherence concurrence, as it is very
similar to the entanglement concurrence given by
\begin{equation}\label{gaoyan4}
 C_E(\rho)=\min_{\{p_i,\psi_i\}}\sum_i p_i C_E(\psi_i),
\end{equation}
with $C_E(\psi_i)=[2(1-\tr\rho_A^2)]^{1/2}$, and $\rho_A=\tr_B
\psi_i$ is the reduced density matrix of $\psi_i$. For two-qubit
state, $C_E(\rho)$ is analytically solved as $C_E(\rho)=\max\{0,
2\lambda_{\max}-\sum_{j=1}^4 \lambda_j\}$ (with $\lambda_j$s being
eigenvalues of $\rho(\sigma_y\otimes\sigma_y)\rho^*(\sigma_y\otimes
\sigma_y)$), and it is linked to the entanglement of formation as
$E_f=H([1+(1-C_E^2)^{1/2}]/2)$, with $H(\cdot)$ being the binary
Shannon entropy function \cite{concur1,concur2}.

It can be proved that this measure satisfy all the conditions (C1),
(C2a), (C2b), and (C3) for a reliable measure of quantum coherence.
Moreover, it is bounded from below by the $l_1$ norm of coherence,
i.e., $C_{\rm con}(\rho) \geq C_{l_1}(\rho)$, which can be derived
directly by combining the definition \eqref{eq-gaoyan} and the
convexity of it.

\subsubsection{Fidelity-based measure of coherence}
\citet{liuqip} proposed a convex roof measure of coherence based on
fidelity and showed that it fulfills the four conditions introduced
by \citet{coher}. For any pure state $\psi=|\psi\rangle
\langle\psi|$, it is defined as
\begin{equation} \label{eq-cf1}
 C_F(\psi)=\min_{\delta\in \mathcal {I}} \sqrt{1-F(\psi,\delta)},
\end{equation}
which is very similar to $C_g(\rho)$ of Eq. \eqref{eq2b-6}, and can
be derived analytically as $C_F(\psi)=C_g^{1/2}(\psi)$ [see Eq.
\eqref{eq-cg}]. For general mixed state $\rho$, it is defined via a
convex roof construction, that is
\begin{equation} \label{eq-cf2}
 C_F(\rho)=\min_{\{p_i,\psi_i\}}\sum_i p_i C_F(\psi_i),
\end{equation}
and the minimization is taken over all the possible pure state
decompositions of $\rho$, see Eq. \eqref{decomp}. Moreover, when
$\rho$ is of a single-qubit state, $C_F(\rho)= C_g^{1/2}(\rho)$, see
Eq. \eqref{eq2b-7}.

\subsubsection{The rank-measure of coherence}
The Schmidt rank for a pure state and the Schmidt number which is an
extension of the Schmidt rank to mixed states have been shown to be
useful for defining entanglement measures \cite{snumber}. The
Schmidt rank $r(\psi)$ for a $(d\times d')$-dimensional pure state
$\psi=|\psi\rangle\langle\psi|$ is the number of nonzero
coefficients in its Schmidt decomposition of Eq. \eqref{schmidt},
and the Schmidt number for a general $(d\times d')$-dimensional
mixed state $\rho_{AB}$ is defined as \cite{srank}
\begin{equation} \label{rank1}
 r(\rho_{AB})=\min_{\{p_i,\psi_i\}}\max_i r(\psi_i),
\end{equation}
where $\psi_i=|\psi_i\rangle\langle\psi_i|$, and the minimization is
taken over all the pure state decompositions of $\rho$, see Eq.
\eqref{decomp}.

In analogous to Schmidt rank and Schmidt number, the coherence rank
$r_C(\psi)=\mathrm{rank}(\psi)$ is the number of nonzero
coefficients $\alpha_i$ for a pure state $|\psi\rangle=\sum_i
\alpha_i |i\rangle$. It can serve as a coherence measure of $\psi$.
For a general mixed state $\rho$, \citet{cnumber} introduced a
convex roof measure of coherence termed as coherence number. It
reads
\begin{equation}\label{eq-cnumber}
 r_C(\rho)=\min_{\{p_i,\psi_i\}}\max_i r_C(\psi_i),
\end{equation}
where the minimization is taken over the pure state decompositions
of $\rho$ showed in Eq. \eqref{decomp}. This is a coherence monotone
under IO, as it satisfies the conditions (C1), (C2a), and (C2b) of
\citet{coher}.

It is obvious that $r_C(|\Psi_d\rangle)=d$ for the maximally
incoherent state $|\Psi_d\rangle$, and $r_C(\delta)=1$ for any
incoherent state $\delta$. Of course, one can also take logarithm to
$r_C(\rho)$ and define the coherence monotone as \cite{oneshot}
\begin{equation}\label{eq-oneshot}
 C_0(\rho)= \log_2 r_C(\rho),
\end{equation}
and now $C_0(\delta)$ equals zero when $\delta$ is incoherent.

\subsection {Robustness of coherence} \label{sec:3D}
Given an arbitrary state $\rho$ on the Hilbert space $\mathcal {H}$,
its convex mixture with another state $\tau$ on the same space may
be coherent or incoherent. In another word, a proper choice of
$\tau$ and weight factor $s$ of mixing may destroy the coherence in
$\rho$. Based on this fact and stimulated by similar definitions for
various quantum correlation monotones, \citet{meas6} introduced a
new coherence measure which was called robustness of coherence
(RoC). It is defined as
\begin{equation}\label{eq2d-1}
 C_R(\rho)= \min_{\tau\in\mathcal{D}(\mathbb{C}^d)}\left\{ s\geq 0 \bigg|
             \frac{\rho+s\tau}{1+s}\coloneqq \delta\in\mathcal{I}\right\},
\end{equation}
where $\mathcal{D}(\mathbb{C}^d)$ is the convex set of density
operators on $\mathcal{H}$.

Alternatively, the RoC can also be defined as \cite{asymm}
\begin{equation}\label{eq2d-new1}
 C_R(\rho)=\min_{\delta\in \mathcal {I}}\{s\geq 0| \rho\leq
 (1+s)\delta\}.
\end{equation}

The RoC is proved to be a full coherence monotone \cite{meas6}. That
is to say, it satisfies the conditions required by the framework for
a resource theory of quantum coherence \cite{coher}. It is also
analytically computable for $\rho$ being arbitrary one-qubit and
pure states, as well as for those $\rho$ with possible nonzero
elements along only the main diagonal and anti-diagonal (i.e., the
so-called $X$ states). For those $\rho$, $C_R(\rho)= \sum_{i\neq
j}|\rho_{ij}|$ in the reference basis $\{|i\rangle\}$, hence equals
to the related $l_1$ norm of coherence $C_{l_1}(\rho)$.

For general $\rho$, \citet{meas6} constructed a semidefinite program
for calculating $C_R(\rho)$ numerically, and obtained tight lower
bounds of RoC of the following form \cite{asymm}
\begin{equation}\label{eq2d-2}
 C_R(\rho)\geq \frac{\|\rho-\Delta(\rho)\|_2^2}{\|\Delta(\rho)\|_\infty}
          \geq \frac{\|\rho-\Delta(\rho)\|_2^2}{\|\Delta(\rho)\|_2}
          \geq \|\rho-\Delta(\rho)\|_2^2,
\end{equation}
where the operator norm $\|M\|_\infty=\lambda_{\max}$, with
$\lambda_{\max}$ being the largest singular value of $M$, and
$\|M\|_2$ is the HS norm.

\citet{meas6} also obtained bounds of $C_R(\rho)$ via the $l_1$ norm
of coherence which is analytically computable, i.e.,
\begin{equation}\label{eq2d-3}
 (d-1)^{-1}C_{l_1}(\rho)\leq C_R(\rho)\leq C_{l_1}(\rho),
\end{equation}
where the upper bound is tight obviously, and the lower bound is
saturated by the family of
$\rho=(1+p)\iden/d-p|\Psi_d\rangle\langle\Psi_d|$, with $0\leq p\leq
1/(d-1)$. It has also been proven that \cite{Rana}
\begin{equation} \label{eq2d-v3}
 C_r(\rho)\leq \log_2 [1+C_R(\rho)].
\end{equation}

Moreover, the RoC is also showed to be upper bounded by \cite{asymm}
\begin{equation} \label{roc2}
 C_R(\rho)\leq d\|\rho\|_\infty -1,
\end{equation}
where $\|\rho\|_\infty$ denotes the largest singular value of $\rho$
(also known as the operator norm). This inequality implies that
$C_R(\rho)$ takes its maximum value $d-1$ only if $\rho$ is a
maximally coherent pure state.

From Eq. \eqref{eq2d-5} one can verify directly that the
optimization in Eq. \eqref{eq2d-1} can be restricted to the subset
of $\tau\in\mathcal{D}(\mathbb{C}^d)$ given by
\begin{equation}\label{eq2d-6}
 \tau_{\rm sub} = \frac{1}{d}\iden_d-\frac{1}{2}\sum_{i=1}^{d_0} x_i X_i
                 +\frac{1}{2}\sum_{i=d_0+1}^{d^2-1} y_i X_i,
\end{equation}
where $d_0=(d^2-d)/2$. That is to say, the optimization can be
performed over all possible $\{y_i\}$ such that $\tau_{\rm sub}$ is
a physically allowed state.

Experimentally, \citet{roc-exp} have explored the RoC for one-qubit
states. They developed two different methods to measure directly the
quantum coherence, i.e., the interference-fringe method and the
witness-observable method. For the former one, they showed
experimentally that the sweeping on ancilla state is necessary only
along the equatorial pure states, while for the latter one, the
optimal witness operator is
\begin{equation} \label{eq2d-v6}
W^*=\cos\varphi_\rho \sigma_1+ \sin\varphi_\rho\sigma_2.
\end{equation}
They have also compared the experimental results with those
calculated via state tomography and found a high coincidence of
them.

Stimulated by the resource weight-based quantification of quantum
features such as the best separable approximation of entangled
states \cite{weight1}, the steering weight \cite{weight2}, and the
measurement incompatibility weight \cite{weight2},
\citet{coh-weight} proposed a similar measure of coherence which
they termed as coherence weight. It is defined as
\begin{equation} \label{eq2d-cw}
 C_w(\rho)= \min_{\delta\in \mathcal {I}, \tau\in\mathcal{D}(\mathbb{C}^d)}
            \{s\geq 0| \rho= (1-s)\delta+s\tau\},
\end{equation}
which can be interpreted operationally as the minimal number of
coherent states (combination with the maximal number of incoherent
states) needed to prepare the considered state $\rho$ on average.
$C_w(\rho)$ was showed to satisfy all the four conditions proposed
by \citet{coher}, thus constitutes a bona fide measure of coherence.
For certain special states, the coherence weight can be obtained
analytically, e.g., for the pure states we always have
$C_w(\rho)=1$, while for Werner state $\rho_W$ of Eq.
\eqref{mina-4}, we have
\begin{equation} \label{eq2d-cw2}
 C_w(\rho_W)=C_R(\rho_W)=C_{l_1}(\rho_W)=\frac{1-dx}{d+1}.
\end{equation}

Similar to the definition of RoC in Eq. \eqref{eq2d-new1}, the
coherence weight can also be defined as
\begin{equation} \label{eq2d-cw3}
 C_w(\rho)=\min_{\delta\in \mathcal {I}}\{s\geq 0| \rho\geq
 (1-s)\delta\}.
\end{equation}

\citet{coh-weight} also obtained lower bounds of $C_w(\rho)$ as
\begin{equation}\label{eq2d-cw4}
 C_w(\rho)\geq \frac{\|\rho-\Delta(\rho)\|_2^2}{\|\rho\|_\infty}
          \geq \|\rho-\Delta(\rho)\|_2^2,
\end{equation}
and proved its relation with the other quantum coherence measures,
i.e.,
\begin{equation}\label{eq2d-cw5}
 \begin{aligned}
  & C_w(\rho)\geq \frac{1}{d-1}C_{l_1}(\rho)\geq \frac{1}{d-1}C_R(\rho),\\
  & C_w(\rho)\geq \frac{1}{\ln d}C_r(\rho).
 \end{aligned}
\end{equation}

Moreover, the coherence weight $C_w$ is showed to satisfy the
following relation
\begin{equation}\label{eq2d-cw6}
  C_w(\rho_1\otimes \rho_2)\leq C_w(\rho_1)+C_w(\rho_2)-C_w(\rho_1)C_w(\rho_2),
\end{equation}
for any quantum states $\rho_1$ and $\rho_2$, while for RoC, we have
\begin{equation} \label{eq2d-cw7}
  C_R(\rho_1\otimes \rho_2)\leq C_R(\rho_1)+C_R(\rho_2)+C_R(\rho_1)C_R(\rho_2).
\end{equation}

\subsection{Tsallis relative entropy measure of coherence} \label{sec:3E}
\citet{meas5} studied the problem for quantifying coherence via the
Tsallis $\alpha$ relative entropies, which are functionals of powers
of density matrices, and formulated family of coherence measures
defined by quantum divergences of the Tsallis type.

As an extension of the standard quantum relative entropy, the
Tsallis $\alpha$ divergence is given by
\begin{equation}\label{eq2e-1}
\begin{small}
 D_{\alpha}(\rho\|\sigma)=\left\{ \begin{aligned}
    &\frac{\tr(\rho^\alpha\sigma^{1-\alpha})-1}{\alpha-1},
       &&{\rm if~} {\rm ran}(\rho)\subseteq {\rm ran}(\sigma), \\
    & +\infty, &&{\rm otherwise},
    \end{aligned} \right.
\end{small}
\end{equation}
with $\alpha$ being a positive number. The trace is taken over ${\rm
ran}(\sigma)$ for $\alpha>1$, while for $\alpha\in(0,1)$ this
condition is not necessary. Here, ${\rm ran}(\rho)$ denotes the
range of $\rho$, and likewise for ${\rm ran}(\sigma)$.

Then, motivated by Eq. \eqref{eq2a-1}, \citet{meas5} proposed to
quantify coherence of $\rho$ as
\begin{equation} \label{eq2e-v1}
 C_\alpha(\rho)= \min_{\delta\in \mathcal{I}}
                 D_\alpha(\rho\|\delta),
\end{equation}
and further proved that it can be evaluated analytically as
\begin{equation}\label{eq2e-2}
 C_\alpha(\rho)= \frac{1}{\alpha-1}\left\{ \left(\sum_i \langle i|
                 \rho^\alpha|i\rangle^\alpha\right)^{1/\alpha}-1\right\},
\end{equation}
and for $\alpha=1$, it reduces to the relative entropy of coherence
in Eq. \eqref{eq2a-2}, while for the specific case of $\alpha=2$, we
have
\begin{equation}\label{eq2e-3}
 C_2(\rho)=\left( \sum_j \sqrt{\sum_i |\langle i|\rho|j\rangle|^2} \right)^2-1,
\end{equation}

The coherence measure based on the quantum $\alpha$ divergence is
bounded above by
\begin{equation}\label{eq2e-4}
  C_{\alpha}(\rho) \leq\left\{ \begin{aligned}
    & -\ln_\alpha \frac{1}{d\tr\rho^2}, &&{\rm if~} \alpha\in(0,2], \\
    & \frac{d\tr \rho^2 \varsigma^{\alpha-2}-1}{\alpha-1},
      &&{\rm if~} \alpha\in(2,+\infty),
    \end{aligned} \right.
\end{equation}
where $\varsigma=\{(d-1)(d\tr\rho^2-1)\}^{1/2}+1$, which is
intimately related to the mixedness of $\rho$, hence is
experimentally accessible.

To be a reliable quantifier of coherence, the proposed functional
should obey the four conditions derived via the resource theoretic
framework of coherence \cite{coher}. For $C_\alpha(\rho)$, it
vanishes if and only if $\rho\in \mathcal{I}$, namely, it satisfies
the condition (C1). Furthermore, (C2a) is satisfied for
$\alpha\in(0,2]$, which is a direct result of $D_\alpha
(\Lambda[\rho],\Lambda[\sigma])\leq D_\alpha(\rho,\sigma)$ for the
Tsallis $\alpha$ divergence. Thirdly, by denoting $\delta^*$ the
closest incoherent state to $\rho$, and $q_i=\tr(K_i \delta^*
K_i^\dag)$, $p_i=\tr(K_i \rho K_i^\dag)$, and $\rho_i=E_i\rho
E_i^\dag /p_i$, \citet{meas5} derived a generalized form of the
monotonicity formulation applicable for $C_{\alpha}(\rho)$. It is
given by
\begin{equation}\label{eq2e-5}
  C_{\alpha}(\rho)\geq \sum_i p_i^\alpha q_i^{1-\alpha}C_\alpha
  (\rho_i),
\end{equation}
which reduces to the usual monotonicity formula for standard
relative entropy when $\alpha=1$. Note that the coherence measure of
Eq. \eqref{eq2e-1} may violate the monotonicity condition (C2b) for
$\alpha\neq 1$. Finally, the Tsallis $\alpha$ relative entropy
measure of coherence is also convex for $\alpha\in (0,2]$, this is
because the Tsallis divergence $D_\alpha$ is a convex function of
density matrices in the same region of $\alpha$.

If one takes logarithm (with base 2) to the first term of Eq.
\eqref{eq2e-1}, then in the limit of $\alpha\rightarrow \infty$, we
can recover the maximum relative entropy defined as
\begin{equation} \label{eq2e-6}
  D_{\max}(\rho\|\sigma)= \min\{\lambda| \rho\leq 2^\lambda \sigma\},
\end{equation}
where $\lambda\geq 0$. It is also an important concept in quantum
information science \cite{MRE1,MRE2,MRE3,MRE4}. In a recent work,
\citet{Buprl} and \citet{oneshot} proposed to use it as a basis for
defining a coherence measure which was termed as the maximum
relative entropy of coherence. It reads
\begin{equation} \label{eq-mre1}
 C_\mathrm{max}(\rho)\coloneqq \min_{\delta\in \mathcal{I}}D_{\max}(\rho\|\delta),
\end{equation}
and has been proven to obey the conditions (C1), (C2a), and C(2b)
introduced by \citet{coher}, so it is a coherence monotone under MIO
\cite{oneshot}. Nevertheless, it does not satisfy the convexity
condition (C3), and it is only quasiconvex, that is, $C_\mathrm{max}
(\sum_i p_i \rho_i)\leq \max_i C_\mathrm{max}(\rho_i)$. Moreover, by
comparing the above equation with Eq. \eqref{eq2d-new1}, one can
also found that $C_\mathrm{max}(\rho)$ can be linked quantitatively
to the RoC as
\begin{equation} \label{eq-mre2}
 C_\mathrm{max}(\rho)=\log_2 [1+C_R(\rho)],
\end{equation}
thus similar to RoC, if $\exists U$ such that $(U\rho
U^\dagger)_{kl}=|\rho_{kl}|$, then a closed formula for
$C_\mathrm{max}(\rho)$ can also be obtained. A class of $\rho$ where
such a requirement is satisfied consists all the one-qubit states,
the pure states, and the $X$ states.

Besides $C_\mathrm{max}(\rho)$, \citet{Buprl} further proposed the
$\varepsilon$-smoothed maximum relative entropy of coherence, which
was defined as
\begin{equation} \label{eq-mre3}
 C_\mathrm{max}^\varepsilon(\rho)\coloneqq \min_{\rho'\in B_\varepsilon
 (\rho)}C_\mathrm{max}(\rho'),
\end{equation}
where $B_\varepsilon (\rho) \coloneqq \{\rho'\geq 0 \!: \|\rho'-
\rho\|_1 \leq \varepsilon, \tr\rho'\leq \tr\rho\}$. It has been
shown that $\lim_{\varepsilon\rightarrow 0, n\rightarrow\infty}
C_\mathrm{max}^\varepsilon(\rho^{\otimes n})=n C_r(\rho)$. That is,
$C_\mathrm{max}^\varepsilon(\rho)$ is equivalent to the relative
entropy of coherence in the asymptotic limit.

\citet{DIO2} put forward another similar coherence measure as
\begin{equation} \label{eq-mre4}
 C_{\Delta,\mathrm{max}}(\rho)\coloneqq \min\{\lambda| \rho\leq 2^\lambda \Delta(\rho)\},
\end{equation}
which has also been proven to be a coherence monotone under DIO
\cite{oneshot,DIO2} and it is also quasiconvex.

\subsection{Skew-information-based measure of coherence} \label{sec:3F}
Soon after the work of \citet{coher}, \citet{meas8} proposed a new
method to quantify the amount of coherence in a state. It is defined
based on the WY skew information, and has later been proven to
violate one of the reliability criteria for a bona fide coherence
monotone \cite{meas9}. But due to the experimentally accessibility,
it may still be of interest to the quantum community.

\subsubsection{Definition and properties}
For a system to be measured, the uncertainty of the outputs comes
from both ignorance of the mixture of the state and the truly
quantum part related to state collapse induced by measurements.
\citet{meas8} proposed that the latter feature (i.e., the truly
quantum uncertainty of a measurement) is an embodiment of quantum
coherence, and can be reliably quantified by the WY skew
information.

The skew information $\mathsf{I}(\rho,K)$ is nonnegative and
vanishes if and only if $[\rho,K]=0$, namely, if and only if $\rho$
is diagonal in the basis defined by $K$, hence it fulfills condition
(C1). Furthermore, $\mathsf{I}(\rho,K)$ is a convex function of
density matrices, thus (C3) is also fulfilled. But $\mathsf{I}
(\rho,K)$ does not fulfill the other axiomatic postulates for a
faithful coherence measure \cite{coher}, e.g., \citet{meas9} have
constructed a series of phase sensitive IO $\Lambda$ for which
$\mathsf{I}(\rho,K)\leq \mathsf{I}(\Lambda[\rho],K)$.

But the above fact does not affect the status of
$\mathsf{I}(\rho,K)$ as a well-accepted measure of asymmetry, which
is defined with respect to a given symmetry group $\mathsf{G}$, and
includes as a special case the group $\mathsf{U}(1)$ used for
defining quantum coherence. In fact, previously \citet{Marvian} have
proposed such an asymmetry measure, where they used the more general
$\mathsf{I}^p(\rho,L)$ of Eq. \eqref{eq2f-1} with
$p\in(0,1)\cup(1,2]$, i.e.,
\begin{equation} \label{wysk1}
 S_{L,p}(\rho)=\tr(\rho L^2)-\tr(\rho^p L \rho^{1-p}L),
\end{equation}
and $L$ represents an arbitrary generator of the group. If $\rho$ is
symmetric relative to group $\mathsf{G}$ [$U_g(\rho)=\rho$ for all
group elements $g\in\mathsf{G}$], then $S_{L,p}(\rho)=0$, see
\citet{Marvian} and \citet{Marvian-phd}. In the same work, the
authors also proposed several other measures of asymmetry, for
example, those based on the Holevo quantity, the trace norm, and the
relative R\'{e}nyi entropy [see Eq. \eqref{eq2e-1}]. Moreover,
\citet{asymm} also proposed a measure of asymmetry which they called
robustness of asymmetry. It is defined in a manner very similar to
RoC in Eq. \eqref{eq2d-1}, with only $\delta$ being replaced by the
symmetric state relative to a group $\mathsf{G}$.

In a more general sense, all the coherence measures defined in the
framework of \citet{coher} constitute a proper subset of measures of
asymmetry. Asymmetry measures the extent to which the symmetry
relative to a group of translations such as time translations or
phase shifts is broken, wherein the translationally invariant
operations $\Lambda_{\rm TI}$ play a central role. To be explicit,
an asymmetry measure $f$ from states to reals should satisfy the
inequality \cite{Marvianpra}
\begin{equation} \label{wysk2}
  f(\Lambda_\mathrm{TI}[\rho])\leq f(\rho).
\end{equation}
This answers why the WY skew information which measures asymmetry
relative to the group of translations generated by an observable $H$
cannot serve as a measure of coherence, as there are IO which are
not translationally invariant. For more detailed explanations about
the relations between coherence and asymmetry, and a comparison of
different notions of coherence, see the recent works of
\citet{Marvianpra} and \citet{Marvianpra2}.

\subsubsection{Tight lower bounds}
By adopting the inequality
\begin{equation}  \label{wysk3}
 \tr\{[\rho,K]^2\}\geq 2\tr \{[\rho^{1/2},K]^2\},
\end{equation}
\citet{meas8} derived the following lower bound of
$\mathsf{I}(\rho,K)$,
\begin{equation}\label{eq2f-3}
 \mathsf{I}(\rho,K)\geq \mathsf{I}^L (\rho,K)=-\frac{1}{4}\tr\{[\rho,K]^2\},
\end{equation}
and demonstrated that it can be experimentally evaluated efficiently
without tomographic state reconstruction of the full density matrix.

\citet{bound} also established a lower bound for the skew
information. By discussing a dynamical process with the evolved
state $\rho_\phi=U_\phi \rho_0 U_\phi^\dag$ and the observable
$K_\phi$ generating its evolution, they derived
\begin{equation}\label{eq2f-4}
 \mathsf{I}(\rho_\phi,K_\phi)\geq \frac{\hbar^2}{2}\bigg|\frac{\partial}
               {\partial\phi}\cos[\mathcal{L}(\rho_0,\rho_\phi)]\bigg|^2,
\end{equation}
where $\mathcal{L}(\rho_0,\rho_\phi)=\arccos[\tr(\rho_0^{1/2}
\rho_\phi^{1/2})]$ is the Hellinger angle, and $K_\phi$ is connected
to the unitary operator $U_\phi$ via
\begin{equation}\label{eq2f-5}
 K_\phi=-i\hbar U_\phi\frac{\partial U_\phi^\dag}{\partial\phi},
\end{equation}
with $\phi$ being an arbitrary parameter encoded in $U_\phi$. Eq.
\eqref{eq2f-4} indicates that the lower bound of the evolved WY skew
information $\mathsf{I}(\rho_\phi,K_\phi)$ is determined by change
rate of the distinguishability between the initial and the evolved
states.

\subsubsection{Modified version of coherence measure}
To avoid the problem occurred for $\mathsf{I}(\rho,K)$,
\citet{Yucsskw} further proposed a similar definition of coherence
measure still by using the WY skew information, which is very
similar to the definition of quantum correlation measure $Q_A(\rho)$
given in Eq. \eqref{eq1d-8}. To be explicit, by denoting
$\{|k\rangle\}$ the reference basis,  the new coherence measure is
defined as
\begin{equation}\label{eq2f-6}
 C_{sk}(\rho)=\sum_k \mathsf{I}(\rho,|k\rangle\langle k|).
\end{equation}
which has been proven to satisfy all the conditions for a bona fide
measure of quantum coherence \cite{coher}. It can be linked to the
task of quantum phase estimation. For the special case of
single-qubit state, $C_{sk}(\rho)$ is also qualitatively equivalent
to the asymmetry measure $\mathsf{I}(\rho,K)$ given by
\citet{meas8}.

To calculate $C_{sk}(\rho)$ for a given state $\rho$, one can also
use its equivalent form
\begin{equation} \label{yucs1}
 C_{sk}(\rho)=1-\sum_k \langle k|\sqrt{\rho} |k\rangle^2,
\end{equation}
which can be further written in a distance-based form
$C_{sk}(\rho)=1- [\max_{\delta\in\mathcal {I}}f(\rho,\delta)]^2$,
where $f(\rho,\delta)= \tr(\sqrt{\rho}\sqrt{\delta})$. The optimal
$\delta^\star$ can be derived as
\begin{equation} \label{yucs2}
 \delta^\star=\sum_k \frac{\langle k|\sqrt{\rho} |k\rangle^2}
              {\sum_{k'}\langle k'|\sqrt{\rho} |k'\rangle^2}.
\end{equation}

Moreover, by using Eq. \eqref{eq2f-3} and the inequality $\langle
k|\sqrt{\rho}|k\rangle \geq \langle k|\rho |k\rangle$, a connection
between $C_{sk}(\rho)$ and the HS norm of coherence measure
$C_{l_2}(\rho)$ given in Eq. \eqref{eq2a-nn3} can be established as
follows
\begin{equation}\label{eq2f-7}
 \frac{1}{2}C_{l_2}(\rho)\leq C_{sk}(\rho) \leq 1-\tr\rho^2 + C_{l_2}(\rho).
\end{equation}
Since $C_{l_2}(\rho)$ is experimentally measurable, the above
relation provides a way for estimating bounds of $C_{sk}(\rho)$.

\subsection{Coherence of Gaussian states} \label{sec:3G}
In real experiments, there exists very relevant physical situations
for which the systems under scrutiny are of infinite-dimensional
(e.g., quantum optics states of light and Gaussian states). Hence,
the characterization and quantification of coherence in these
systems are also required.

\subsubsection{Coherence in the Fock space}
A typical class of infinite-dimensional system is the bosonic system
in the Fock space, which is describable using the Fock basis
$\{|n\rangle\}_{n=0}^\infty$. Here, $|n\rangle$ is the eigenstate of
the number operator $\hat{a}^\dag \hat{a}$, and $\hat{a}^\dag$ and
$\hat{a}$ are the bosonic creation and annihilation operators.

By generalizing the set $\mathcal{I}$ of incoherent states as those
with $\delta=\sum_{n=0}^\infty \delta_n |n\rangle\langle n|$, and
incoherent operations described by the Kraus operators $\{K_n\}$
satisfying $\sum_{n=0}^\infty K_n^\dag K_n=\iden$ and
$K_n\mathcal{I}K_n^\dag \subset \mathcal{I}$, \citet{gaus1} studied
the problem of quantification of coherence in this system. For this
purpose, they first proposed a new criterion that $C(\rho)$ should
satisfy in order to circumvent the divergence problem of $C(\rho)$,
which may be termed as the mean energy constraints,

(C5) If the first-order moment, the average particle number
$\bar{n}=\langle \hat{a}^\dag \hat{a} \rangle$ is finite, it should
fulfill $C(\rho)<\infty$.

Based on this new criterion, \citet{gaus1} proved that the relative
entropy of coherence in Eq. \eqref{eq2a-2} is also a proper
coherence measure for the infinite-dimensional systems. But the
$l_1$ norm of coherence of Eq. \eqref{eq2a-3}, despite its simple
structure and intuitive meaning, does not satisfy the condition
(C5), hence fails to serve as a proper measure of coherence for the
infinite-dimensional systems.

Referring to the coherence measure in infinite-dimensional systems,
one may also concern about the counterparts of Eqs. \eqref{eq2-1}
and \eqref{eq2-2}, i.e., the incoherent state and the maximally
coherent state. For the $d$-mode Fock space $\mathcal {H}=\otimes_
{i=1}^d \mathcal {H}_i$ with the basis $|\boldsymbol{n} \rangle=
\otimes_{i=1}^d |n_i\rangle$ and probability distributions
$\{p_{\boldsymbol{n}}\}$, the incoherent state is given by
\cite{gaus2}
\begin{equation}\label{eq2g-1}
 \delta_{\rm th}(\boldsymbol{n})=\otimes_{i=1}^d \rho_{\rm
th}(\bar{n}_i),
\end{equation}
where  $\bar{n}_i=\langle \hat{a}_i^\dag\hat{a}_i\rangle$, and
$\rho_{\rm th}(\bar{n}_i)$ is just the thermal state for the
$i$th-mode Fock space,
\begin{equation}\label{eq2g-2}
\rho_{\rm th}(\bar{n}_i)=\sum_{n=0}^\infty \frac{\bar{n}_i^n}
                       {(\bar{n}_i+1)^{n+1}}|n\rangle\langle n|.
\end{equation}
Moreover, the maximally coherent state is given by \cite{gaus1}
\begin{equation}\label{eq2g-3}
 |\Psi_m^d\rangle=\sum_{\boldsymbol{n}} \frac{\bar{n}_t^{|\boldsymbol{n}|_1/2}}
                  {[(\bar{n}_t+1)^{|\boldsymbol{n}|_1+1}\mathbb{C}_{|\boldsymbol{n}|_1
                  +d-1}^{d-1}]^{1/2}}|\boldsymbol{n}\rangle,
\end{equation}
where $|\boldsymbol{n}|_1=\sum_{i=1}^d n_i$, and $\bar{n}_t=\sum_
{\boldsymbol{n}}p_{\boldsymbol{n}}|\boldsymbol{n}|_1$ denotes the
average total particle number which is finite. The corresponding
maximal coherence is given by
\begin{equation}\label{eq2g-4}
 C_r^{\max}=C_{r,d=1}^{\max}+\sum_{n=0}^{\infty}\frac{\bar{n}_t^n}
               {(\bar{n}_t+1)^{n+1}}\log_2\mathbb{C}_{n+d-1}^{d-1},
\end{equation}
with
\begin{equation} \label{eq2g-v4}
 C_{r,d=1}^{\max}=(\bar{n}+1)\log_2(\bar{n}+1)-\bar{n}\log_2\bar{n},
\end{equation}
being the maximal coherence for the single-mode case
($\bar{n}_t=\bar{n}$).

\subsubsection{Analytic formulas}
A state is said to be Gaussian if its characteristic function
$\chi(\rho,\lambda)=\tr[\rho D(\lambda)]$ is Gaussian, where
$D(\lambda)$ is the displacement operator. A Gaussian state is fully
describable using the covariance matrix $\gamma$ (with entries
$\gamma_{kl}$) and displacement vector $\vec{\upsilon}=(\upsilon_1,
\upsilon_2)^T$, $\rho=\rho(\gamma,\vec{\upsilon})$. The incoherent
thermal state $\rho_{\rm th}(\bar{n})$ corresponds to $\gamma=
(2\bar{n}+1)\iden$ and $\vec{\upsilon}=(0,0)^T$, where the
superscript $T$ denotes transpose.

For the case of $d$-mode Gaussian states $\rho(\gamma,
\vec{\upsilon})$, by denoting $x_i=[\det \gamma^{(i)}]^{1/2}$ square
of the determinant of the covariance matrix $\gamma^{(i)}$ for the
$i$th mode, \citet{gaus2} obtained analytical formula for the
relative entropy of coherence, which is given by
\begin{equation}\label{eq2g-5}
 \begin{aligned}
 C_r(\rho)=&\sum_{i=1}^d \left(\frac{x_i-1}{2}\log_2\frac{x_i-1}{2}-\frac{x_i+1}{2}\log_2\frac{x_i+1}{2}\right)\\
           &+\sum_{i=1}^d[(\bar{n}_i+1)\log_2(\bar{n}_i+1)-\bar{n}_i\log_2\bar{n}_i],
\end{aligned}
\end{equation}
where $\bar{n}_i$ can be written in terms of the covariance matrix
$\gamma^{(i)}$ and displacement vector $\vec{\upsilon}$ as
\begin{equation}\label{eq2g-6}
 \bar{n}_i=\frac{1}{4}\{\gamma_{11}^{(i)}+\gamma_{22}^{(i)}+[\upsilon^{(i)}_1]^2+[\upsilon^{(i)}_2]^2-2\},
\end{equation}
from which one can also see that the maximally coherent state should
be pure, i.e., $x_i=1$, $\forall~i\in\{1,\cdots,d\}$.

\subsubsection{Coherence of coherent states}
For the set $\{|\alpha\rangle\}$ of coherent states which spans an
infinite-dimensional Hilbert space, a direct application of the
resource theory of quantum coherence formulated by \citet{coher} is
not applicable. This is because states of the considered set are not
only overcomplete but also may be not linearly independent. To
circumvent this perplexity, \citet{continuous} developed an
orthogonalization procedure which allows for defining coherence
measure for arbitrary superposition of coherent states, whether they
are orthogonal or not.

For a given density operator $\rho_A$, let $\rho_{AB}^{(0)}=\rho_A
\otimes |0^B\rangle\langle 0^B|$ and $\rho_{AB}^{(i)}=
U_{\alpha^{(i)}} \rho_{AB}^{(i-1)} U_{\alpha^{(i)}}^\dagger$ ($i=1,
\ldots,N$). Then if one denotes $|\alpha^{(i)}\rangle$ for the
coherent state admitting $\tr(|\alpha^{(i)}\rangle \langle
\alpha^{(i)}|\otimes |0^B\rangle\langle 0^B|\rho_{AB}^{(i-1)})
=\max_{\{|\alpha\rangle\}} \tr(|\alpha \rangle \langle \alpha
|\otimes |0^B\rangle\langle 0^B|\rho_{AB}^{(i-1)})$, the $N$th
Gram-Schmidt unitary $U_{\mathrm{GS}}^{(N)}=U_{\alpha^{(N)}} \ldots
U_{\alpha^{(1)}}$, where the \textsc{cnot} type unitary is given by
\begin{equation}\label{eq2g-7}
 \begin{aligned}
 U_{\alpha^{(i)}}=& \mathbb{I}\otimes \mathbb{I}+|\alpha^{(i)}\rangle\langle \alpha^{(i)}|\otimes
                  (|\beta^{(i)}\rangle\langle 0^B|+|0^B\rangle\langle \beta^{(i)}| \\
                  &-|0^B\rangle\langle 0^B|-|\beta^{(i)}\rangle\langle \beta^{(i)}|).
\end{aligned}
\end{equation}

Using the Gram-Schmidt unitary, \citet{continuous} defined the $N$
coherence for a general state $\rho_A$ as follows
\begin{equation}\label{eq2g-8}
 C_{\alpha}(\rho_A, N)=\inf_{\rho_{AE}\in\mathcal {E}\atop U_{\mathrm{GS}}\in \mathcal{S}^{(N)}}
 C(\Phi_{\mathrm{GS}}^{(N)}[\rho_{AE}]),
\end{equation}
where $C$ denotes any faithful coherence measure,
$\mathcal{S}^{(N)}$ denotes the full set of $N$th Gram-Schmidt
unitaries, $\mathcal {E}=\{\rho_{AE}|\tr_E \rho_{AE}=\rho_A\}$ is
the set of extensions of $\rho_A$, and
\begin{equation}\label{eq2g-9}
 \Phi_{\mathrm{GS}}^{(N)}[\rho_A]=\frac{\Pi_{\mathrm{GS}}^{(N)}
       [U_{\mathrm{GS}}^{(N)}(\rho_A\otimes |0^B\rangle\langle 0^B|)
        U_{\mathrm{GS}}^{(N)\dagger}]\Pi_{\mathrm{GS}}^{(N)}}{\tr\{\Pi_{\mathrm{GS}}^{(N)}
       [U_{\mathrm{GS}}^{(N)}(\rho_A\otimes |0^B\rangle\langle 0^B|)
        U_{\mathrm{GS}}^{(N)\dagger}]\Pi_{\mathrm{GS}}^{(N)}\}},
\end{equation}
with the projector
\begin{equation}\label{eq2g-10}
 \Pi_{\mathrm{GS}}^{(N)}= \sum_{i=1}^N |\alpha^{(i)}\rangle\langle\alpha^{(i)}|
                          \otimes|\beta^{(i)}\rangle\langle\beta^{(i)}|.
\end{equation}
Then, an $\varepsilon$-smoothed version of $N$ coherence can be
written as
\begin{equation}\label{eq2g-11}
 C_{\alpha}^\varepsilon(\rho_A, N,)=\inf_{\rho'_A\in \mathcal{B}_\varepsilon(\rho_A)}
 C_{\alpha}(\rho'_A, N),
\end{equation}
where $\mathcal{B}_\varepsilon(\rho_A)=\{\rho'_A|\frac{1}{2}
\|\rho'_A-\rho_A\|_1\leq\varepsilon \}$. Finally, the $\alpha$
coherence is defined as the smoothed $N$ coherence in the asymptotic
limit, that is,
\begin{equation}\label{eq2g-12}
 C_{\alpha}(\rho_A)= \lim_{\varepsilon\rightarrow 0}\lim_{N\rightarrow\infty}
                     C_{\alpha}^\varepsilon(\rho_A, N,).
\end{equation}

\citet{continuous} proved that $C_{\alpha}(\rho_A)=0$ if and only if
$\rho_A$ is a classical state, and it is also a nonclassicality
measure.

\subsection{Generalized coherence measures} \label{sec:3H}
When the constraints imposed by the axiomatic-like postulates of
\citet{coher} are somewhat released, one may introduce other
measures of quantum coherence that are physically relevant. These
measures may also have potential applications under specific
contexts.

\subsubsection{Basis-independent coherence measure}
For a state $\rho$ in the $d$-dimensional Hilbert space,
\citet{frobe} formulated the following basis-independent measure of
quantum coherence
\begin{equation}\label{eq2h-1}
 C_{BI}(\rho)=\sqrt{\frac{d}{d-1}}\|\rho-\rho_{\rm mm}\|_2,
\end{equation}
which is proportional to the HS distance between $\rho$ and the
maximally mixed state $\rho_{\rm mm}=\iden_d/d$, and the parameter
before $\|\cdot\|_2$ is introduced for normalizing $C_{\rm
free}(\rho)$.

The $C_{BI}(\rho)$ can be calculated analytically as
\begin{equation}\label{eq2h-2}
 C_{BI}(\rho)= \sqrt{\frac{dP-1}{d-1}}
             = \sqrt{\frac{d\mathcal{I}_{BZ}}{d-1}},
\end{equation}
where $P=\tr\rho^2$ is the purity of $\rho$, and $P-1/d$ equals to
the Brukner-Zeilinger invariant information
\begin{equation}\label{eq2h-3}
\mathcal{I}_{BZ}(\rho)=\sum_{i=1}^m \sum_{j=1}^d \left[
                 \tr(\Pi_{ij}\rho)-\frac{1}{d}\right]^2,
\end{equation}
thereby endows $C_{BI}(\rho)$ a clear physical meaning. Here,
$\{\Pi_{ij}\}$ denote eigenvectors of the mutually complementary
observables, for example, for the case of $d=2$, they are those of
the Pauli operators $\sigma_x$, $\sigma_y$, and $\sigma_z$.

The coherence measure $C_{BI}(\rho)$ is unitary invariant (a trait
that distinguishes it from other coherence measures), takes the
maximum 1 for all the pure states, and is nonincreasing under the
action of any unital channel $\Lambda_u$, i.e., $C_{BI}(\rho) \geq
C_{BI}(\Lambda_u[\rho])$. Moreover, it also provides loose lower
bounds for the $l_1$ norm of coherence and trace norm of coherence,
i.e.,
\begin{equation}\label{eq2h-4}
 \begin{split}
 & C_{l_1}(\rho)\leq \sqrt{d(d-1)}C_{BI}(\rho),\\
 & C_{tr}(\rho)\leq \sqrt{d-1}C_{BI}(\rho).
 \end{split}
\end{equation}

As the measures of quantum coherence is basis dependent, it is of
particular interest to consider the maximum amount of coherence
attainable by varying the reference basis, and define
\begin{equation}\label{maxcoh}
 C^{\max}(\rho)=\max_{U} C(U\rho U^\dag),
\end{equation}
with $C$ being any valid coherence measure and $\{U\}$ the set of
unitary operations.

\citet{Yucs} investigated problems of such kind. For a
$d$-dimensional state $\rho$, they proved
\begin{equation}\label{maxcoh2}
  C_{r}^{\max}(\rho)=\log_2 d-S(\rho),~~
  C_{l_2}^{\max}(\rho)=\tr\rho^2-\frac{1}{d},
\end{equation}
where $C_r(\rho)$ and $C_{l_2}(\rho)$ denote, respectively, the
relative entropy and the HS norm measure of coherence. In
particular, $C_{r}^{\max}(\rho)$ equals to the relative entropy
between $\rho$ and the corresponding maximally mixed state
$\rho_{\rm mm}$, and $C_{2}^{\max} (\rho)$ equals to the squared HS
norm between $\rho$ and $\rho_{\rm mm}$. Both $C_{r}^{\max}(\rho)$
and $C_{2}^{\max}(\rho)$ take the maximum value for pure states and
zero for incoherent states. They also possess preferable features of
coherence measures such as: (\romannumeral+1) invariant under
unitary operations; (\romannumeral+2) convexity under mixing of
states; (\romannumeral+3) monotonicity under the unitary operations
$\{U_i| \sum p_i U_i^\dag U_i=\iden, U_i^\dag U_i=\iden\}$ and SIO.

For some other widely used coherence measures including the RoC, the
coherence weight, and the modified skew information measure of
coherence, $C^{\max}(\rho)$ defined in Eq. \eqref{maxcoh} can also
be obtained analytically, which reads \cite{Husf}
\begin{equation} \label{maxcoh3}
 \begin{aligned}
  & C_R^{\max}(\rho)=d\lambda_{\max}-1,~~
    C_w^{\max}(\rho)=1-d\lambda_{\min}, \\
  & C_{sk}^{\max}(\rho)= 1-\frac{1}{d}\left( \sum_i \sqrt{\lambda_i}\right)^2.
 \end{aligned}
\end{equation}
where $\{\lambda_i\}$ denote the eigenvalues of $\rho$, with
$\lambda_{\max}=\max\{\lambda_i\}$ and $\lambda_{\min}= \min
\{\lambda_i\}$.

\citet{coh-pur} also considered the maximal coherence achievable by
performing unitary operations on a state $\rho$, with however a
different set of free operations from that of \citet{coher} was used
when defining $C(\rho)$. To be explicit, they used the set of MIO
$\Lambda_{\mathrm{MIO}} [\delta]\in \mathcal{I}$ ($\forall \delta\in
\mathcal{I}$) instead of the traditional IO as the free operations,
which was first suggested by \citet{cof}. Clearly, the set of IO is
a subset of MIO, and thus any coherence monotone with respect to MIO
is also a coherence monotone with respect to IO, but the inverse may
not always be true, e.g., the $l_1$ norm of coherence and the
coherence of formation is an IO monotone but not a MIO monotone
\cite{ncgc}.

Based on the above setting, \citet{coh-pur} showed that the
following state
\begin{equation}\label{maxcohmio}
 \rho_{\max}=\sum_{n=1}^d p_n |n_+\rangle\langle n_+|,
\end{equation}
is a MCMS with respect to any MIO monotone, where $\{p_n\}$ is the
probability distribution, and $\{|n_+\rangle\}$ denotes a MUB with
respect to the incoherent basis $\{|i\rangle\}$, i.e., $|\langle
i|n_+\rangle|=1/d$, $\forall i, n_+$ \cite{mubs1, mubs2}. The Eq.
\eqref{maxcohmio} can be proved straightforwardly by noting that
$\Lambda_{\mathrm{MIO}}[\rho_{\max}]=U\rho_{\max}U^\dag$ if one
chooses Kraus operators of $\Lambda_{\mathrm{MIO}}$ as
$K_n=U|n_+\rangle\langle n_+|$. This is because
$C(U\rho_{\max}U^\dag)= C(\Lambda_{\mathrm{MIO}}[\rho_{\max}]) \leq
C(\rho_{\max})$, where the inequality is due to the monotonicity of
a coherence measure under MIO. By the way, one can also show that
$\Lambda_{\mathrm{MIO}}$ of this type yields $\Lambda_{\mathrm{MIO}}
[\delta]=\iden/d$ and $\Lambda_{\mathrm{MIO}}[\rho]\in \rho_{\max}$.

Furthermore, when the coherence is measured by the shortest distance
between $\rho$ and the set $\mathcal{I}$ of incoherent states [see
Eq. \eqref{eq2a-1}, with only the IO being replaced by MIO], we have
\begin{equation} \label{maxc1}
 C^{\max}(\rho)=C(\tilde{U}\rho \tilde{U}^\dag)\leq
\mathcal{D}(\tilde{U}\rho\tilde{U}^\dag,\iden/d)=
\mathcal{D}(\rho,\iden/d),
\end{equation}
where $\tilde{U}$ is the optimal unitary for obtaining
$C^{\max}(\rho)$, and the inequality comes from the fact that
$\iden/d$ is not necessary the closest incoherent state to
$\tilde{U}\rho \tilde{U}^\dag$, while the last equality is due to
the unitary invariance of $\mathcal{D}$. Moreover, by denoting
$\Delta_+$ the full dephasing of $\rho$ in the maximally coherent
basis $\{n_+\rangle\}$ [see the definition in Eq. \eqref{eq2d-v2}],
one can obtain $\Delta_+[\rho_{\max}]=\rho_{\max}$ and
$\Delta_+[\delta] =\iden/d$. Thus by denoting $\tilde{\delta}$ the
closest incoherent state to $\rho_{\max}$, we have
\begin{equation}\label{maxc2}
 \begin{split}
  C^{\max}(\rho)&\geq C(\rho_{\max})=\mathcal{D}(\rho_{\max},\tilde{\delta})\\
                &\geq \mathcal{D}(\Delta_+[\rho_{\max}],\Delta_+[\tilde{\delta}])\\
                &= \mathcal{D}(\rho_{\max},\iden/d)= \mathcal{D}(\rho,\iden/d),
 \end{split}
\end{equation}
then by combining the above two equations, one can obtain
\begin{equation} \label{maxcohana}
 C^{\max}(\rho)=C(\rho_{\max})=\mathcal{D}(\rho,\iden/d),
\end{equation}
for any contractive distance measure $\mathcal{D}$ of two states.

When $\rho$ is incoherent, i.e., $\rho=\delta$, Eq. \eqref{maxcoh}
corresponds to the maximum amount of coherence (quantified by any
faithful measures) generated from a given incoherent state $\delta$.
By focusing on the two-qubit states only, and arranging $\delta$'s
diagonal elements as $\delta_1\leq \delta_2\leq \delta_3\leq
\delta_4$ and taking the relative entropy as a measure of coherence,
\citet{Sun} obtained solutions of Eq. \eqref{maxcoh} for specified
types of $U$. They are
\begin{equation}\label{eq5b-2}
 \begin{aligned}
  & C^{\rm max}_{r,1}=1+H(\delta_1+\delta_3)-\sum_i \delta_i\log_2\delta_i,\\
  & C^{\rm max}_{r,2}=2-\sum_i \delta_i\log_2\delta_i,
 \end{aligned}
\end{equation}
where $C^{\rm max}_{r,1}$ ($C^{\rm max}_{r,2}$) is the optimal
coherence created under local unitaries $U_A\otimes \iden$
($U_A\otimes U_B$), with the corresponding
\begin{eqnarray}\label{eq5b-v2}
 U_A=U_B=|+\rangle\langle 1|+|-\rangle\langle 0|,
\end{eqnarray}
where $|\pm\rangle=(|0\rangle\pm |1\rangle)/\sqrt{2}$, and
$H(\cdot)$ is the binary Shannon entropy. For the kernel $U_d$ of
nonlocal $U$ in the Cartan decomposed form, \citet{Sun} found that
they cannot outperform the local unitaries on creating quantum
coherence, but it remains open if this is also true for more general
nonlocal $U$.

\subsubsection{Genuine quantum coherence}
The core for the resource theory of quantum coherence is incoherent
states and incoherent operations. \citet{gqc} introduced a slightly
different types of incoherent operations which he called genuinely
incoherent operations (GIO). These operations preserve all
incoherent states, i.e.,
\begin{equation} \label{GIO}
 \Lambda_{\rm gi}(\delta)=\delta,
\end{equation}
for all $\delta\in\mathcal{I}$. This definition of GIO ensures their
independency on the explicit forms of the Kraus decompositions.

It has been proved that the Kraus operators of $\Lambda_{\rm gi}$
are diagonal in the prefixed reference basis, and
\begin{equation} \label{GIO2}
 \Lambda_{\rm gi}(\cdot)=\sum_k p_k U_k (\cdot) U_k^\dag,
\end{equation}
for the single-qubit case, with $\{p_k\}$ being the probability
distribution and $U_k=\sum_l e^{i\phi_{lk}}|k\rangle\langle k|$ the
unitary. Moreover, the GIO includes the nondegenerate thermal
operations $\Lambda_{\rm th}$ as a special case, while itself
belongs to the set of translationally invariant operations
$\Lambda_{\rm ti}$ \cite{gqc}. Given a system state $\rho^S$ with
the Hamiltonian $\hat{H}_S$, and the thermal state $\rho_{\rm th}^E=
e^{-\beta\hat{H}_E}/\tr e^{-\beta\hat{H}_E}$ with $\hat{H}_E$ the
environmental Hamiltonian and $\beta=1/T$ the inverse temperature,
$\Lambda_{\rm th}$ and $\Lambda_{\rm ti}$ are defined by
\begin{equation}\label{eq2h-5}
\begin{aligned}
 & \Lambda_{\rm th}(\rho^S)=\tr_E (U\rho^S\otimes \rho_{\rm th}^E U^\dag),\\
 & \Lambda_{\rm ti}(e^{-i\hat{H}_S t}\rho e^{i\hat{H}_S t})=e^{-i\hat{H}_S t}
                                       \Lambda_{\rm ti}(\rho)e^{i\hat{H}_S t},
\end{aligned}
\end{equation}
where $U$ is the unitary which preserves the total energy of the
considered system plus its environment.

Under the set of GIO, \citet{gqc} proposed the prerequisites for a
function to be a genuine coherence measure. They are analogous to
those labeled as (C1), (C2a), (C2b), and (C3), with only the IO
being replaced by the GIO. \citet{gqc} called a measure respecting
the first three prerequisites a genuine coherence monotone. As the
GIO is a strict subset of the general IO, the $l_1$ norm, relative
entropy, and intrinsic randomness measures of coherence, as well as
the RoC are all genuine coherence monotones.

Apart from the above measures, the WY skew information
$\mathsf{I}(\rho,K)$ obeys (C1), (C2a), and (C3). The distance-based
measure of genuine coherence
\begin{equation}\label{eq2h-6}
G_p(\rho)= \min_{\delta\in\mathcal{I}} \|\rho- \delta\|_p,
\end{equation}
also obeys these three conditions, and for the special case of
$p=2$, $G_2(\rho)=\|\rho- \Delta(\rho)\|_2$, where the closest
incoherent state is $\Delta(\rho)=\sum_i \langle i|\rho|i\rangle
|i\rangle\langle i|$. For other cases, $\Delta(\rho)$ is not the
closest state for obtaining $G_p(\rho)$, but \citet{gqc} showed that
\begin{equation}\label{eq2h-7}
 \tilde{G}_p(\rho)= \|\rho- \Delta(\rho)\|_p,
\end{equation}
is also a valid genuine coherence measure, as it obeys the
conditions (C1), (C2a), and (C3).

\subsubsection{Quantification of superposition}
As a generalization of the resource theories of coherence,
\citet{superposition} introduced a similar framework for quantifying
superposition. In their framework, the set $\mathcal{F}$ of free
states is comprised of the states that can be represented as
statistical mixtures of linear independent (not necessarily
orthogonal) basis states $\{|c_i\rangle\}_{i=1}^d$. To be explicit,
these superposition-free states are given by
\begin{equation} \label{eq2h-8}
 \varsigma=\sum_{i=1}^d \varsigma_i |c_i\rangle\langle c_i|,
\end{equation}
where $\varsigma_i\geq 0$ and $\sum_i \varsigma_i=1$. Those states
that are not free are called superposition states. Similarly, the
quantum operation $\Phi(\rho)=\sum_i K_i \rho K_i^\dagger$ is said
to be superposition-free if the Kraus operator gives the map
$K_i\varsigma K_i^\dagger/\tr(K_i\varsigma K_i^\dagger) \in \mathcal
{F}$ ($\forall K_i$), that is, every $K_i$ (hence $\Phi$) maps the
superposition-free state to another superposition state.
\citet{superposition} showed that such a $K_i$ is of the following
general form:
\begin{equation} \label{eq2h-9}
 K_i=\sum_k c_i(k)|c_{f_i(k)}\rangle\langle c_k^\bot|,
\end{equation}
where $|c_k^\bot\rangle$ are called reciprocal states which satisfy
$\langle c_k^\bot|c_l\rangle=\delta_{kl}$, $c_i(k)$ are
coefficients, and $f_i(k)$ are index functions.

\citet{superposition} presented the defining conditions for a
faithful superposition measure $M(\rho)$. These conditions are very
similar to those for a faithful coherence measure \cite{coher}. The
difference is that $\Lambda(\rho)$ and $\delta$ in (C1), (C2a),
(C2b), and C(3) were replaced by $\Phi(\rho)$ and $\varsigma$,
respectively. Then, in a similar manner to the definition of
$C_{\mathcal{D}} (\rho)$ in Eq. \eqref{eq2a-1}, one can define
\begin{equation} \label{eq2h-10}
 M_{\mathcal{D}}(\rho)= \min_{\varsigma\in\mathcal{F}}D(\rho,\varsigma).
\end{equation}

For explicit distance measures, \citet{superposition} proved the
superposition measures including the relative entropy of
superposition, the $l_1$ norm of superposition, and the robustness
of superposition. They are similar to the coherence measures
$C_r(\rho)$, $C_{l_1}(\rho)$, and $C_R(\rho)$ defined respectively,
in Eqs. \eqref{eq2a-2}, \eqref{eq2a-3}, and \eqref{eq2d-1}. Apart
from these, \citet{superposition} also proved the rank-measure of
superposition
\begin{equation} \label{eq2h-11}
 M_\mathrm{rank}(\rho)=\min_{\{p_i,\psi_i\}}\sum_i p_i \log[r_S(\psi_i)],
\end{equation}
where the superposition rank $r_S(\psi_i)$ is the number of nonzero
$\alpha_n^{(i)}$ for $|\psi_i\rangle=\sum_n \alpha_n^{(i)}
|c_n\rangle$, and the minimization is taken over all pure state
decompositions of $\rho$ showed in Eq. \eqref{decomp}.

Before ending this section, we remark here that while various
coherence measures have been introduced, \citet{tongdm} proposed a
proposal for estimating their values with limited experimental data
available. Their approach is based on the optimization of a
Lagrangian function and the limited expectation value of certain
Hermitian operators, and can be applied to any coherence measure
$C(\rho)$ that is continuous and convex.

\section{Interpretation of quantum coherence} \label{sec:4}
Coherence is not only a basic feature signifying quantumness in an
integral system, but also a common prerequisite for different forms
of quantum correlations when composite systems are considered. Apart
from its characterization and quantification, it has also been shown
to be intimately related to many other quantities manifesting
quantumness of states, and fundamental problems of quantum mechanics
such as complementarity and uncertainty relations. All these have
triggered the community's interest in investigating it from
different perspectives, which endows quantum coherence clear
operational interpretations and physical meanings.

\subsection{Coherence and quantum correlations} \label{sec:4A}
\begin{figure}
\centering
\resizebox{0.45 \textwidth}{!}{%
\includegraphics{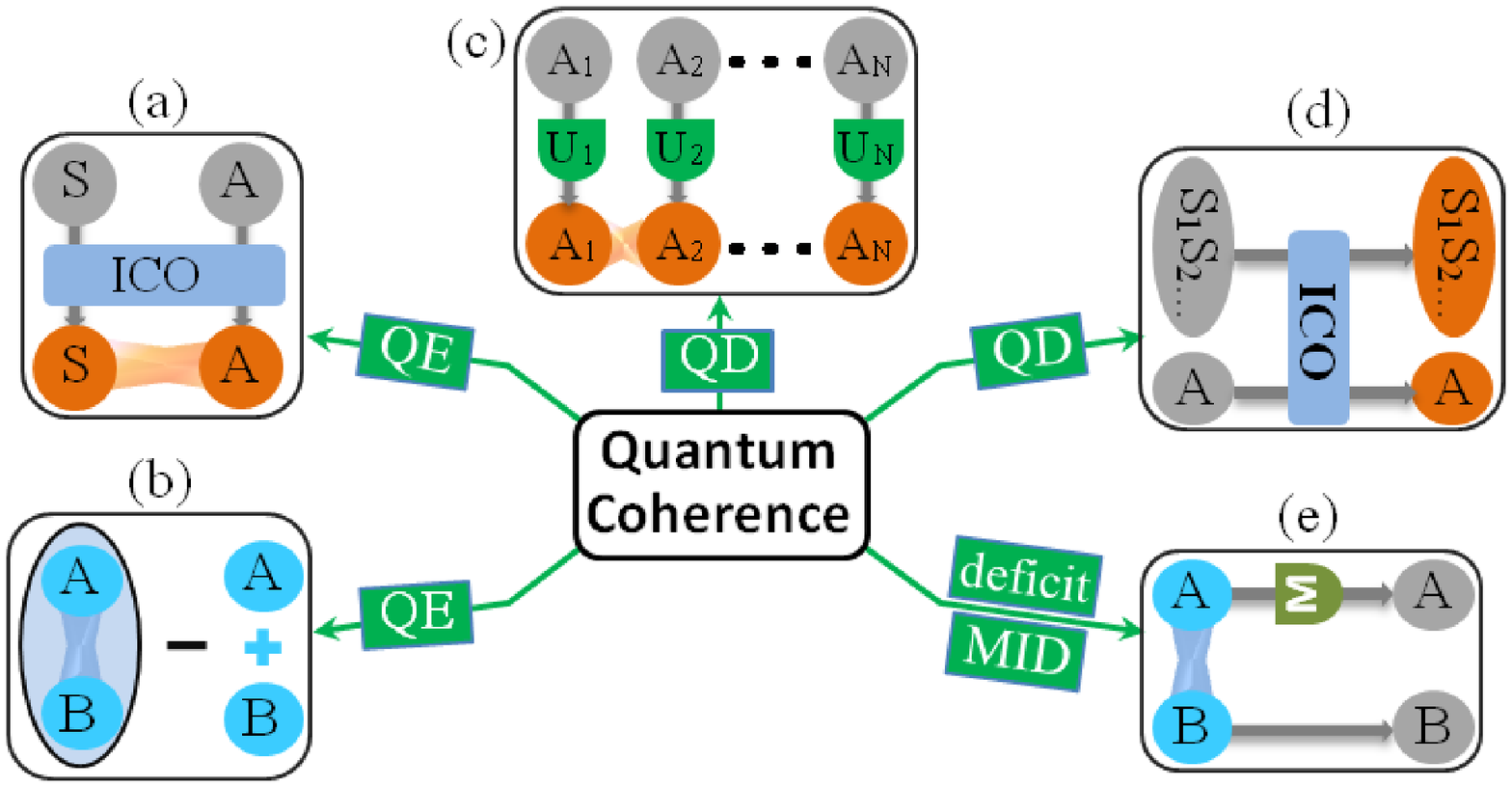}}
\caption{Schematic picture for the connections between quantum
coherence and quantum correlations. (a) Quantum entanglement (QE)
based measure of quantum coherence \cite{meas2}, (b) correlated
coherence and QE \cite{coqd}, (c) basis-free coherence measure and
relative entropy of QD \cite{Sun}, (d) quantum coherence consumption
and discord generation \cite{Mile}, (e) difference of quantum
coherence for the premeasurement and postmeasurement states and
one-way quantum deficit \cite{ulrc1}, as well as steering-induced
coherence and MID (in the basis spanned by local eigenvectors of
$\rho^B$) \cite{Huxy}.}\label{fig:coqd}
\end{figure}

In the seminal work of \citet{coher} and the subsequent stream of
works, the quantum coherence measures are defined for single
systems. Contrary to it, the traditional manifestation of
quantumness for a system, e.g., quantum correlations, are defined in
a scenario which involves at least two parties. In fact, both
quantum coherence and quantum correlations arise from the
superposition principle of quantum mechanics, hence it is essential
to study the interrelation between them. The main progresses up to
now are summarized in Fig. \ref{fig:coqd}, and we review them in
detail in the following.

\subsubsection{Coherence and entanglement}
\citet{meas2} made a first step toward the above problem. By
considering the setting where a coherent state $\rho^S$ is coupled
to an incoherent ancilla initially in the vacuum state
$|0^A\rangle$, they showed that the generated entanglement between
$S$ and $A$ is upper bounded by the coherence of $S$. The bound can
be saturated for certain contractive distance measures, hence yields
a natural family of entanglement-based coherence measures, see Sec.
\ref{sec:2B} for more detail.

\citet{gaoyan} considered a very similar problem to that of
\citet{meas2}. They used the coherence concurrence and entanglement
concurrence, and found that the generated entanglement concurrence
from the initial state $\rho^S\otimes |0^A\rangle\langle 0^A|$ is
upper bounded by
\begin{equation} \label{qi01}
 C_E(\Lambda^{SA}[\rho^S\otimes|0^A\rangle\langle 0^A|])\leq
 C_{con}(\rho^S),
\end{equation}
and when $\rho^S$ is a two-qubit state while the ancilla $A$ is also
a qubit, the above equality is saturated. Moreover, by applying the
generalized \textsc{cnot} gate of Eq. \eqref{eq-cnot}, they also
found a lower bound of the created entanglement
\begin{equation} \label{qi02}
 C_E(\Lambda^{SA}[\rho^S\otimes|0^A\rangle\langle 0^A|])\geq \sqrt{\frac{2}{d(d-1)}}
 C_{\rm con}(\rho^S).
\end{equation}

Apart from the link to entanglement, \citet{coqd} introduced the
concept of correlated coherence, and argued that it can be connected
to QD and entanglement. Their key idea is by distinguishing the
coherence in $\rho^{AB}$ as local and nonlocal, i.e., by dividing
the total coherence into two different portions: those stored
locally in the subsystems, and those stored only in the correlated
states. Based on this starting point, they defined the correlated
coherence as
\begin{equation}\label{eq3a-1}
 C_{cc}(\rho^{AB})=C(\rho^{AB})-C(\rho^A)-C(\rho^B),
\end{equation}
which is a nonnegative quantity.

By choosing tensor products of the eigenvectors $\{|i_A\rangle\}$
(for $\rho^A$) and $\{|j_B\rangle\}$ (for $\rho^B$) as reference
basis (for degenerate case, they will chosen to be those minimize
$C_{cc}$), $C_{cc}(\rho^{AB})=0$ if and only if $\rho^{AB}\in
\mathcal{CC}$. Similarly, $C_{cc}(\rho^{AB})= C_{cc}
(\Delta^A[\rho^{AB}])$ if and only if $\rho^{AB}\in\mathcal{CQ}$.
Based on these observations, \citet{coqd} defined
\begin{equation}\label{eq3a-2}
 E_{cc}(\rho^{AB})\coloneqq \min C_{cc}(\rho^{AA'BB'}),
\end{equation}
and showed that it possesses the preferable properties for an
entanglement monotone. Here, the minimization is taken over the full
set of unitarily symmetric extensions of $\rho^{AB}$ satisfying
\begin{equation}\label{eq3a-3}
 U_{AA'}\otimes U_{BB'} (U_{\rm SWAP}\rho^{AA'BB'}U_{\rm SWAP}^\dag)
 U_{AA'}^\dagger\otimes U_{BB'}^\dagger= \rho^{AA'BB'},
\end{equation}
where $\rho^{AB}=\tr_{A'B'} \rho^{AA'BB'}$ for all the local
unitaries $U_{AA'}$ and $U_{BB'}$, and $U_{\rm SWAP}$ is the swap
operator.

\citet{NAQC} examined the steered coherence from another
perspective. In their framework, Alice and Bob hold respectively,
qubit $A$ and $B$ of $\rho^{AB}$, and agree on the observables
$\{\sigma_1, \sigma_2,\sigma_3\}$ in advance. Alice then measures
$\sigma_i$ on her qubit and informs Bob of her choice $\sigma_i$ and
outcome $a\in\{0,1\}$. Bob computes the coherence of his conditional
states $\{p(a|\sigma_i), \rho_{B|\sigma_i^a}\}$ in the eigenbasis of
either $\sigma_j$ or $\sigma_k$ ($j,k\neq i$) randomly, which can be
written as $\sum_a p(a|\sigma_i) C^{\sigma_j}(\rho_{B|\sigma_i^a})$.
Here, $p(a|\sigma_i)$ is the probability for Alice's outcome $a$
when she measures $\sigma_i$, and $\rho_{B|\sigma_i^a}$ is the
corresponding postmeasurement state of $B$. By averaging over
Alice's possible measurements and Bob's allowable eigenbasis sets,
one can obtain
\begin{equation}\label{eq-hu3}
 \begin{aligned}
 & C_{l_1}^{na}(\rho^{AB})= \frac{1}{2}\sum_{i,j,a\atop i\neq j} p(a|\Pi_i) C^{\sigma_j}_{l_1}(\rho_{B|\sigma_i^a}),\\
 & C_{r}^{na}(\rho^{AB})=  \frac{1}{2}\sum_{i,j,a\atop i\neq j} p(a|\Pi_i) C^{\sigma_j}_{re}(\rho_{B|\sigma_i^a}).
 \end{aligned}
\end{equation}
As for any single-partite state $\rho$, we have
\begin{equation}\label{eq-hu4}
 \begin{aligned}
 & \sum_{j=1}^3 C^{\sigma_j}_{l_1}(\rho)\leq \sqrt{6},\\
 & \sum_{j=1}^3 C^{\sigma_j}_{r}(\rho)\leq C_2^m
 =3H(1/2+\sqrt{3}/6),
 \end{aligned}
\end{equation}
it is said that a nonlocal advantage of quantum coherence is
achieved on $B$ when $C_{l_1}^{na}(\rho^{AB})>\sqrt{6}$ or
$C_r^{na}(\rho^{AB}) >C_2^m$. \citet{NAQC} showed that any two-qubit
state that can achieve a nonlocal advantage of quantum coherence is
quantum entangled. Moreover, the interplay between nonlocal
advantage of quantum coherence and Bell nonlocality was also
established for two-qubit states \cite{NAQC1}.

The above framework was extended to $(d\times d)$-dimensional state
$\rho^{AB}$, in which the Pauli observables are replaced by the set
of mutually unbiased observables $\{A_i\}$ \cite{NAQC2}. Now, the
average coherence for Bob's conditional states is
\begin{equation} \label{eq-hu5}
  C^{na}(\rho_{AB})= \frac{1}{d}\sum_{i,j,a \atop i\neq j} p(a|A_i)
                     C^{A_j}(\rho_{B|A_i^a}),
\end{equation}
and $C^{na}(\rho_{AB})>C^m$ captures the existence of nonlocal
advantage of quantum coherence in state $\rho_{AB}$. When one adopts
the $l_1$ norm of coherence and relative entropy of coherence, the
state-independent bound $C^m$ is given by $(d-1)\sqrt{d(d+1)}$ and
$(d+1)\log_2{d}- {(d-1)^2 \log_2(d-1)}/ {d(d-2)}$, respectively
\cite{NAQC2}.

Similarly, one can also formulate other framework for capturing the
nonlocal advantage of quantum coherence in a state, e.g., after
Alice executing one round of measurements and announced her choice
$A_i$ and outcomes $a$, Bob can measure the coherence of his
conditional states only in the preagreed basis spanned by the
eigenvectors of $A_{\alpha_i}$, with $\{\alpha_i\}$ being any one of
the possible permutations of the elements of $\{i\}$. This can give
some new insights on the interrelation between coherence and quantum
correlations \cite{NAQC2}.

\subsubsection{Coherence and quantum discord}
Since the coherence measures reviewed in Sec. \ref{sec:3} are basis
dependent, they can be changed by unitary operations. Based on this
consideration, \citet{Sun} introduced a basis-free coherence measure
of the following form
\begin{equation}\label{eq3b-1}
 C^{\rm free}(\rho)=\min_{\vec{U}} C(\bm{U}\rho\bm{U}),
\end{equation}
where $\bm{U}=U_1\otimes U_2\otimes\cdots \otimes U_N$ for a
$N$-partite state $\rho$. It is in fact the minimum coherence
created by local unitary operations.

By putting the measures of coherence and QD on an equal footing,
that is, to quantify the both via relative entropy, \citet{Sun}
found that $C^{\rm free}(\rho)$ defined above equals to the QD
$D_r(\rho)= \min_{\chi\in\mathcal{C}} S(\rho\|\chi)$ ($\mathcal {C}$
is the set of classical states) introduced by \citet{reqd}, i.e.,
\begin{equation} \label{eq3b-v1}
 C^{\rm free}(\rho) =D_r(\rho),
\end{equation}
thereby establishes a direct connection between coherence of a
$N$-partite state in the product basis $\{|\bm{i} \rangle \coloneqq
\otimes_{k=1}^N |i_k\rangle\}$ (with $\{|i_k\rangle\}_{i=1}
^{d_k}\}$ and $d_k=\dim\mathcal{H}_k$) and QD of this state with the
same multipartite divisions.

For bipartite states $\rho$ and local von Neumann measurements
$\{\Pi_k^A\}$, \citet{ulrc1} found that the difference between
relative entropy of coherence for $\rho$ and the postmeasurement
state $\rho'=\sum_k (\Pi_k^A \otimes \iden_B)\rho(\Pi_k^A\otimes
\iden_B)$ equals to the one-way quantum deficit \cite{def1,def2},
i.e.,
\begin{equation} \label{eq3b-v2}
 C_r(\rho)- C_r(\rho')=\Delta^{\rightarrow}(\rho).
\end{equation}

For a system with $N$-partite division $S=\{S_1,S_2, \cdots,S_N\}$
and product reference basis $\{|\bm{i}\rangle\}$, one can define the
sets of $S_k$-incoherent states $\mathcal{I}_{S_k}$ with elements
$\delta^{S_k}=\sum_i p_i |i_k\rangle\langle i_k|\otimes
\rho_{\tilde{S}|k}$ ($\tilde{S}=S-S_k$), incoherent operations
$\{\Lambda^{S_k}\}$ which map $\mathcal{I}_{S_k}$ to itself, and
coherence measure $C_{\mathcal{D}}^{S_k}(\rho)= \min_{\delta\in
\mathcal{I}_{S_k}} \mathcal {D}(\rho,\delta)$, all with respect to
the local basis $\{|i_k\rangle\}$. When $\mathcal{D}$ is the
relative entropy, we have
\begin{equation}\label{eq3b-2}
 C_r^{S_k}(\rho)= S(\rho\|\Delta^{S_k}[\rho]),
\end{equation}
where $\Delta^{S_k}[\rho]$ is the full dephasing of $\rho=\rho^{S_1
S_2\cdots S_N}$ in the basis $\{|i_k\rangle\}$ of party $S_k$. The
GQD $D(\rho)\coloneqq \min_{\chi\in \mathcal{C}} \mathcal{D}
(\rho,\chi)$, and for $\mathcal{D}$ to be the relative entropy,
$D(\rho)= \min_{\Delta} S(\rho\|\Delta[\rho])$, with
$\Delta=\otimes_{k=1}^N \Delta^{S_k}$ \cite{RMP}. Moreover, the
global discord
\begin{equation} \label{eq3b-n2}
 \bar{D}(\rho)=\min_{\Delta} [S(\rho\|\Delta[\rho])- \sum_k
 S(\rho^{S_k}\|\Delta^{S_k}[\rho^{S_k}]),
\end{equation}
and the usual asymmetric discord
\begin{equation} \label{eq3b-n3}
 D_{\tilde{S}|S_k}(\rho)=\min_{\Delta^{S_k}}[S(\rho\|
\Delta^{S_k}[\rho])-S(\rho^{S_k}\|\Delta^{S_k}[\rho^{S_k}])].
\end{equation}

By considering an analogous setting to that constructed by
\citet{meas2}, i.e., the system $S$ and an incoherent ancilla is
prepared initially in the product state $\rho^S\otimes
|0^A\rangle\langle 0^A|$, \citet{Mile} studied, from the perspective
of coherence consumption and discord generation, the interplay
between quantum coherence and QD. First, they found that
\begin{equation}\label{eq3b-3}
 D(\Lambda^{SA}[\rho^S\otimes\rho^A])\leq C_{\mathcal {D}}(\rho^S),
\end{equation}
for any contractive measure of $\mathcal{D}$, i.e., the generated
discord is upper bounded by the initial coherence in $\rho^S$. In
particular, if $d_A\geq d_S$, the equality holds for $\mathcal{D}$
to be the relative entropy or the Bures distance. Second, for
$\rho^{S_1 S_2\cdots S_N}=\otimes_{k=1}^N \rho^{S_k}$, the sum of
coherence consumed for all subsystems bounding the amount of global
discord that can be generated by IO, i.e.,
\begin{equation}\label{eq3b-4}
 \bar{D}(\Lambda[\rho^{S_1 S_2\cdots S_N}])\leq \sum_k \delta C_r(\rho^{S_k}),
\end{equation}
where $\delta C_r =C_r^i-C_r^t$, with $C_r^i$ ($C_r^t$) being
coherence of the state prior to the measurement (after the
measurement). Similarly, for the asymmetric discord,
\begin{equation}\label{eq3b-as4}
 D_{S_2|S_1} (\Lambda^{S_1}[\rho^{S_1 S_2}])\leq \delta
 C(\rho^{S_1}),
\end{equation}
for $\rho^{S_1S_2}=\rho^{S_1}\otimes\rho^{S_2}$.

In a recent paper, \citet{RQC} introduced the concept of relative
quantum coherence (RQC), which is the coherence of one state in the
reference basis spanned by the eigenvectors of another one. To be
explicit, for $\rho$ and $\sigma$ in the same Hilbert space
$\mathcal {H}$, and the eigenvectors of $\sigma$ being given by
$\{|\psi_i \rangle\}$, with the corresponding eigenvalues
$\{\epsilon_i\}$, the RQC is given by
\begin{eqnarray} \label{eq-hu1}
 C(\rho,\sigma)= C^{\Xi}(\rho),
\end{eqnarray}
where $C^{\Xi}(\rho)$ denotes any \emph{bona fide} measure of
quantum coherence defined in the reference basis $\Xi$.

When the quantum coherence is measured by the $l_1$ norm, they
showed that the QD $D_A(\rho_{AB})$ \cite{{qd02}} is bounded from
above by the discrepancy between the RQC for the total system and
that for the subsystem to be measured in the definition of QD, that
is
\begin{equation} \label{eq-hu2}
  D_A(\rho_{AB}) \leqslant  C_{\rm re}(\rho_{AB},\tilde{\rho}_{\mathbb{PQ}})
          -C_{\rm re}(\rho_A,\tilde{\rho}_\mathbb{P}),
\end{equation}
where $\tilde{\rho}_{\mathbb{PQ}}$ denotes the optimal
postmeasurement state for obtaining the QD, and
$\tilde{\rho}_\mathbb{P}$ is the reduced state of
$\tilde{\rho}_{\mathbb{PQ}}$. This upper bound can also be saturated
when the state $\rho_{AB}$ is quantum-classical correlated.

Similarly, for the symmetric QD $D_{s}(\rho_{AB})= I(\rho_{AB})-
I(\tilde{\rho}_\mathbb{PQ})$ defined via two-sided optimal
measurements $\{\tilde{\Pi}_k^A \otimes \tilde{\Pi}_l^B\}$
\cite{qdsym1,qdsym2}, it was further showed that \cite{RQC}
\begin{equation}\label{eq3-9}
 \begin{split}
 D_{s}(\rho_{AB}) = C_{\rm re}(\rho_{AB},\tilde{\rho}_\mathbb{PQ})
                     -C_{\rm re}(\rho_A,\tilde{\rho}_\mathbb{P})
                   -C_{\rm re}(\rho_B,\tilde{\rho}_\mathbb{Q}),
 \end{split}
\end{equation}
which implies that $D_{s}(\rho_{AB})$ is nonzero if and only if
there exists RQC not localized in the subsystems. This establishes a
direct connection between the RQC discrepancy and the symmetric
discord.

For a bipartite system $AB$ described by the density operator
$\rho^{AB}$, \citet{steer} considered the maximum amount of
coherence created at party $B$ by the procedure of steering on $A$,
and defined the steered coherence
\begin{equation}\label{eq5b-4}
 C_{\rm str}(\rho^{AB})\coloneqq \inf_{e^B}[\max_{E^A} C_{l_1}(e^B,\rho^B_i)],
\end{equation}
where $E^A=\{E_k^A\}$ represents the set of POVM operators, and
$\rho^B_i=\tr_A (E_i^A\otimes\iden_B\rho^{AB})/p_i$, $p_i=\tr
(E_i^A\otimes\iden_B\rho^{AB})$. The infimum over the eigenbasis
$e^B=\{e_k^B\}$ of $\rho^B$ is necessary only when it is degenerate.
The motivation for the definition of $C_{\rm str}(\rho^{AB})$ is
very similar to the concept of localizable entanglement which is
indeed the maximum entanglement that can be localized, on average,
between two parties of a multipartite system, by performing local
measurements on the other parties \cite{localent1,localent2}.

The steered coherence is showed to have several preferable
properties of $C_{\rm str}(\rho^{AB})$, e.g., it vanishes when
$\rho^{AB}$ is quantum-classical correlated, takes the maximum for
all pure entangled states with full Schmidt rank $d_B$, and is
locally unitary invariant. Moreover, $C_{\rm str}(\rho^{AB})$ may be
increased by the local operations $\Lambda_B$ on $B$ prior to the
steering on $A$. For two-qubit states, this is achievable if and
only if $\Lambda_B$ is neither unital nor semi-classical. All these
properties are very similar to those of the various discordlike
quantum correlation measures \cite{RMP}.

For two-qubit states $\rho^{AB}$, the maximum steered coherence is
given by \cite{steer}
\begin{equation}\label{eq5b-5}
 C_{\rm str}(\rho^{AB})=\inf_{\vec{n}_B\in \mathbb{R}^3,|\vec{n}_B|=1}
     \left\{\max_{\vec{m}\in \mathbb{R}^3,|\vec{m}|=1}
     \bigg|\frac{R^T\vec{m}\times\vec{n}_B}{1+\vec{a}\cdot \vec{m}}\bigg|\right\},
\end{equation}
where $R$ is a $3\times 3$ matrix with elements $R_{ij}=\tr
\rho^{AB}(\sigma_i\otimes\sigma_j)$, $\vec{m}$ is a vector related
to the POVM $E^A=(\iden+\vec{m}\cdot \vec{\sigma})/2$,
$\vec{a}=\tr(\rho^A\vec{\sigma})$ is the local Bloch vector of
$\rho^A$  (similarly for $\vec{b}$), and $\vec{n}_B=\vec{b}/
|\vec{b}|$.

\subsubsection{Coherence and measurement-induced disturbance}
Based on quantum steering, \citet{Huxy} introduced the
steering-induced coherence, and explored its connection with a
quantum correlation measure known as measurement-induced disturbance
\cite{MID}. For a bipartite system described by density operator
$\rho^{AB}$ and shared by two players Alice and Bob, if Alice
performs a local quantum measurement $\Xi^A=\{|\xi^A_i\rangle\langle
\xi_i^A|\}$ on her subsystem, she will obtain the outcome $i$ with
probability $p_i=\tr (\Xi_i^A\otimes\iden_B\rho)$, and Bob's
subsystem is steered to $\rho_i^B=\tr_A(\Xi_i^A\otimes\iden_B
\rho)/p_i$. After multi-rounds of measurements, Bob will have the
ensemble $\{p_i,\rho_i^B\}$. see Fig. \ref{fig:coqd}(e) for an
illustration of the scheme

From the above scheme, \citet{Huxy} defined the steering-induced
coherence as
\begin{equation}\label{eq3c-1}
 \bar{C}(\rho^{AB})=\inf_{e^B}[\max_{\Xi^A}\sum_i p_i C(e^B,\rho^B_i)],
\end{equation}
where Bob's reference basis is chosen to be eigenbasis
$e^B=\{|e_k^B\rangle\}$ of $\rho^B$, and the infimum in Eq.
\eqref{eq3c-1} is necessary only when $\rho^B$ is degenerate.
$\bar{C}(\rho)$ characterizes Alice's ability to steer coherence on
Bob's side, and has also been proven satisfying the necessary
requirements of a faithful coherence measure.

Based on the observation that the symmetric measurement-induced
disturbance equals to coherence of $\rho^{AB}$ in the tensor-product
eigenbasis of $\rho^A$ and $\rho^B$, \citet{Huxy} further considered
the asymmetric measurement-induced disturbance
\begin{equation} \label{eq3c-a1}
 Q_B(\rho^{AB})=\inf_{E^B}\mathcal{D}(\rho^{AB}, E^B[\rho^{AB}]),
\end{equation}
with $E^B=\{|e_k^B\rangle\langle e_k^B|\}$ the locally invariant
projective measurements on $B$. Their results show that when being
quantified by the same distance measure, $\bar{C}(\rho^{AB})$ is
bound from above by $Q_B(\rho^{AB})$, i.e.,
\begin{equation}\label{eq3c-2}
 \bar{C}(\rho^{AB})\leq Q_B(\rho^{AB}),
\end{equation}
and the equality holds for the maximally correlated state
\begin{equation}\label{eq-mcs}
 \rho_{\rm mc}= \sum_{i,j}\rho_{ij}|ii\rangle\langle jj|,
\end{equation}
when the relative entropy quantifiers of them are adopted, for which
they both equal to $S(\rho^B_{\rm mc})-S(\rho^{AB}_{\rm mc})$.
Moreover, for the two-qubit state and the $l_1$ norm quantifiers,
the upper bound is also saturated.

\citet{MIAC} considered a very similar coherence steering scheme to
that of \citet{Huxy}. The difference is that they used the
computation basis $\{|i\rangle\}_{i=1}^d$, and discussed the amount
of coherence gain with classical correlation of the premeasurement
state $\rho^{AB}$. They defined the measurement-induced average
coherence $\bar{C}_r^P$ and measurement-induced average total
coherence $\bar{C}^T_r$ as
\begin{equation}\label{eq-mia1}
\begin{aligned}
 & \bar{C}^P_r(\rho^B)=\sum_i p_i C_r(\{|i\rangle\},\rho^B_i),\\
 & \bar{C}^T_r(\rho^B)=\sum_i p_i C^{\max}_r(\rho^B_i),
\end{aligned}
\end{equation}
where $C^{\max}_r(\rho^B_i)$ is the maximal attainable relative
entropy of coherence under generic basis, see Eq. \eqref{eq-maxrc}.
The corresponding coherence gain are given by
\begin{equation}\label{eq-mia2}
 \begin{aligned}
 & \Delta C_r^P= \bar{C}^P_r(\rho^B)-C_r(\rho^B),\\
 & \Delta C_r^T=\bar{C}^T_r(\rho^B)-C_r^T(\rho^B).
\end{aligned}
\end{equation}
Based on these definitions, they found that
\begin{equation} \label{eq-yucs}
 \Delta C_r^P\leq \Delta C_r^T,
\end{equation}
which can be seen from $\Delta C_r^P-\Delta C_r^T=\sum_i p_i
S((\rho_i^B)_\mathrm{diag})- S(\rho^B_{\mathrm{diag}})\geq 0$.
Moreover, the two coherence gains are proved to be upper bounded by
the classical correlation (with respect to subsystem $A$) present in
the premeasurement state $\rho^{AB}$, i.e.,
\begin{equation} \label{eq-yucs2}
\max\{\Delta C_r^P,\Delta C_r^T\}\leq J_A(\rho^{AB}),
\end{equation}
where $J_A(\rho^{AB})\coloneqq S(\rho^B)-\min_{\{E_k^A\}}\sum_k q_k
S(\rho_{B|E_k^A})$, and $\{E_k^A\}$ represents local positive
operator valued measurements for defining classical correlation and
QD \cite{qd02}. Here, we provide a slightly different proof of the
above equation from \citet{MIAC}. First, $\Delta C_r^T \leq
J_A(\rho^{AB})$ can in fact be obtained directly from $\Delta
C_r^T=S(\rho^B)- \sum_i p_i S(\rho_i^B)$ and the ensemble
$\{p_i,S(\rho_i^B)\}$ obtained with the measurement operators
$\Xi^A$ may not be optimal for attaining $J_A(\rho^{AB})$. Second,
$\Delta C_r^P \leq J_A(\rho^{AB})$ is due to Eq. \eqref{eq-yucs}.

Moreover, when a maximization process is performed over all possible
$\Xi^A$, just like that of Eq. \eqref{eq3c-1}, the statements in the
above two equations still holds. The optimized coherence gain
$\Delta C_r^P$ equals zero if and only if $\rho^{AB}=\sum_i
A_{ii}\otimes|i^A\rangle \langle i^A|$ [cf. Eq. \eqref{eq-guo} for
the meaning of $A_{ij}$] or a product state, while $\Delta C_r^T$
equals zero if and only if $\rho^{AB}=\rho^A\otimes \rho^B$.

\subsubsection{Distribution of quantum coherence}
The distribution of quantum coherence among the subsystems of a
multipartite system is also an interesting research direction. In
fact, for quantum correlation measures such as entanglement and QD,
similar problems have been studied via various monogamy
inequalities, see, e.g., the works of \citet{mono1,mono2,mono3}. For
different coherence measures, one can also derive monogamy
inequalities that impose limits on their shareability among
multipartite systems. For example, for the $l_1$ norm of coherence,
it is direct to show that
\begin{equation} \label{monoga}
  C_{l_1}(\rho^{A_1 A_2 \cdots A_N})\geq \sum_{i} C_{l_1} (\rho^{A_i}).
\end{equation}
for any multipartite system described by the density operator
$\rho^{A_1 A_2 \cdots A_N}$, with $\rho^{A_i}$ ($i=1,2,\ldots,N$)
being the reduced density operators.

For bipartite state $\rho^{AB}$ with the reduced states $\rho^A=
\tr_B\rho^{AB}$ and $\rho^B=\tr_A\rho^{AB}$, \citet{ulrc1} proved
that the relative entropy of coherence respects the monogamy
relation
\begin{equation}\label{eq3d-1}
 C_r(\rho^{AB})\geq C_r(\rho^A)+C_r(\rho^B).
\end{equation}

In fact, for the multipartite system described by the density
operator $\rho^{A_1 A_2 \cdots A_N}$, an application of Eq.
\eqref{eq3d-1} immediately yields
\begin{equation} \label{eq3d-v1}
 C_r(\rho^{A_1 A_2 \cdots A_N})\geq \sum_{i} C_r (\rho^{A_i}).
\end{equation}
But a similar monogamy relation does not hold for the general
bipartite division of tripartite states. Even for the pure state
$\rho^{ABC}=|\psi\rangle^{ABC} \langle\psi|$ with $\rho^{AB}=
\tr_C\rho^{ABC}$ and $\rho^{AC}=\tr_B\rho^{ABC}$, the relation
\begin{equation}\label{eq3d-2}
 C_r(\rho^{ABC})\geq C_r(\rho^{AB})+C_r(\rho^{AC}),
\end{equation}
holds with a very strong constraint, that is, there should exist
some real-valued parameters $\lambda\in[0,1]$ such that \cite{ulrc2}
\begin{equation} \label{eq3d-v2}
 \lambda S(\rho^{AB}_{\rm diag})\leq S(\rho^{AB}),~~
 (1-\lambda)S(\rho^{AC}_{\rm diag})\leq S(\rho^{AC}).
\end{equation}

\citet{dist} also explored the distribution of quantum coherence
among constituents of a $N$-partite system. By introducing a quantum
version of the Jensen-Shannon divergence (QJSD):
\begin{equation}\label{eq3d-3}
 \mathcal {J}(\rho,\sigma)=\frac{1}{2}[S(\rho\| (\rho+\sigma)/2)+
S(\sigma\|(\rho+\sigma)/2)],
\end{equation}
or equivalently,
\begin{equation}\label{eq3d-v3}
 \mathcal{J}
(\rho,\sigma)=S\left(\frac{\rho+\sigma}{2}\right)-\frac{1}{2}S(\rho)-\frac{1}{2}S(\sigma),
\end{equation}
and using its square root as the distance measure of two states,
i.e., $\mathcal {D}=\mathcal {J}^{1/2}$, they defined $C(\rho)$ of
Eq. \eqref{eq2a-1} as the total coherence, and
\begin{equation}\label{eq3d-v4}
 C_I (\rho)=\min_{\delta_S \in\mathcal{I}_S} \mathcal{D}(\rho,\delta_S),
\end{equation}
as the intrinsic coherence which excludes the contribution of the
subsystems, and
\begin{equation} \label{eq3d-v5}
 C_L (\rho)= \mathcal{D} (\delta_S^\star, \delta^\star),
\end{equation}
as the local coherence. Here, $\mathcal {I}_S$ is comprised of the
states $\delta_S=\sum_k p_k \tau_{k,1}^b \otimes \cdots \otimes
\tau_{k,N}^b$ obtained by choosing all the possible basis set
$\{|b_{l,n}\rangle \langle b_{l,n}|\}$ (with $\tau_{l,n}^b=\sum_l
p_{l,n} |b_{l,n}\rangle \langle b_{l,n}|$), while $\delta^\star$ and
$\delta_S^\star$ are the closest states for obtaining $C(\rho)$ and
$C_I(\rho)$, respectively.

Building upon the above definitions, one has
\begin{equation} \label{eq3d-v6}
 C\leq C_L+C_I\leq \sum_n C_{L,n}+C_I,
\end{equation}
where the first inequality is a direct result of the metric
properties of $\mathcal{D}$ (i.e., the triangle inequality), and the
second one is due to the subadditivity of $C_L$ for product
$\delta_S^\star$. In particular, for the $N$-partite system, one can
divide it into different subsystems, and calculate the corresponding
coherence. For example, we denote by $C_{1:2}$ and $C_{1:23}$ the
intrinsic coherence between subsystems 1 and 2, and between
subsystem 1 and the combined subsystem 23, and similarly for other
cases. In this way, \citet{dist} defined the multipartite monogamy
of coherence as
\begin{equation} \label{eq3d-v7}
 M=\sum_{n=2}^N C_{1:n}- C_{1:2\cdots N},
\end{equation}
which is monogamous for $M\leq 0$ and polygamous otherwise. They
have calculated $M$ for the three-qubit \emph{W} and GHZ states, and
showed the validity of it on analyzing coherence distribution in
spin systems of the Heisenberg model.

Considering the fact that the coherence of a state $\rho$ cannot be
larger than the average coherence of its ensemble states
$\{p_i,\rho_i\}$ \cite{coher}, \citet{acc-coh} studied the
distribution of quantum coherence from another perspective. For
$\rho=\sum_ip_i\rho_i$, they introduced a quantity which they called
it accessible coherence,
\begin{equation} \label{eq3d-v8}
 C^{\mathrm{acc}}(\rho)= \sum_i p_i C(\rho_i)-C(\rho),
\end{equation}
which characterizes the coherence one gains when knowing the
information of the ensemble $\{p_i,\rho_i\}$. Moreover, if a
maximization is taken over all state decompositions of
$\rho=\sum_i p_i\rho_i$, one can obtain the maximal accessible
coherence. In fact, the maximization is only necessary to be taken
over the pure state decompositions of $\rho$ due to compact
convexity of the density matrix set.

For a bipartite state $\rho^{AB}$, they further defined the
remaining coherence as
\begin{equation}\label{eq3d-v9}
 \begin{aligned}
 C^{\mathrm{rem}}(\rho^{AB})=&C(\rho^{AB})-C(\rho^A)-C(\rho^B)\\
                             &-C^{\mathrm{acc}}(\rho^A)-C^{\mathrm{acc}}(\rho^B).
 \end{aligned}
\end{equation}
That is to say, the coherence in $\rho^{AB}$ are divided into the
local coherence and the local accessible coherence in its subsystems
plus the remaining coherence. Through explicit examples, they showed
that there are states for which the local coherence and local
accessible coherence vanishes, while the remaining coherence
survives. Moreover, the remaining coherence can also be
qualitatively different when being measured by the relative entropy
and the $l_1$ norm, e.g., there are cases for which
$C^{\mathrm{rem}}_{l_1}(\rho^{AB})=0$ and $C^{\mathrm{rem}}_r
(\rho^{AB})>0$.

For the skew-information-based coherence measure of Eq.
\eqref{eq2f-6}, \citet{Yucsskw} showed that for bipartite pure state
$|\psi\rangle_{AB}$ with the reduced states $\rho^A$ and $\rho^B$,
the following polygamy relation holds
\begin{equation} \label{eq3d-v10}
 [1-C_{sk}(\rho^A)][1-C_{sk}(\rho^B)]\geq 1-C_{sk}(|\psi\rangle_{AB}),
\end{equation}
while for the mixed state $\rho^{AB}$, one has \cite{Yucsskw}
\begin{equation}\label{eq3d-5}
 [1-C_{sk}(\rho^A)][1-C_{sk}(\rho^B)]\geq \sum_{kk'}\langle kk'|\rho^{AB}|kk'\rangle^2,
\end{equation}
and
\begin{equation}\label{eq3d-new5}
 [1-C_{sk}(\rho^A)][1-C_{sk}(\rho^B)]\geq \frac{1}{c_s}[1-C_{sk}(\rho^{AB})]^2,
\end{equation}
where the right-hand side of Eq. \eqref{eq3d-5} equals to
$\tr(\rho^{AB})^2 -C_{l_2}(\rho^{AB})$, and $c_s=[r-\sum_i
C_{sk}(\rho^{Ai})][r-\sum_i C_{sk}(\rho^{Bi})]$, with
$r=\mathrm{rank}(\rho^{AB})$, and $\rho^{Ai}$ and $\rho^{Bi}$ are
the reduced states of the $i$th eigenstate of $\rho^{AB}$.

\subsubsection{State ordering under different coherence measures}
As various measures of quantum coherence have been put forward up to
now, one may wonder whether they impose the same state ordering or
not, just as the similar problem encountered when comparing various
entanglement \cite{ordere} and discordlike correlation measures
\cite{orderd1,orderd2}.

When considering two measures of quantumness of a state denoted by
$Q_1$ and $Q_2$, if
\begin{equation}\label{eq3e-1}
 Q_1(\rho_1)\leq Q_1(\rho_2) \Longleftrightarrow
                     Q_2(\rho_1)\leq Q_2(\rho_2),
\end{equation}
for arbitrary two states $\rho_1$ and $\rho_2$, then they are said
to give the same state ordering. Otherwise, they give inconsistent
descriptions of quantumness.

By concentrating on the coherence measures, \citet{orderc} examined
state ordering problem imposed by the $l_1$ norm of coherence,
relative entropy of coherence, and coherence of formation. Through
explicit examples, they found that these measures also impose
different orderings of states. In particular, for all measures of
coherence that are equivalent for pure states, they must impose
different orderings for general mixed states.

\subsection{Complementarity of quantum coherence} \label{sec:4B}
As the measures of quantum coherence are basis dependent, a natural
question that arises is how they behave when different bases are
involved?

\subsubsection{Mutually unbiased bases}
For the $l_1$ norm and relative entropy measures of coherence,
\citet{comp1} studied tradeoffs between coherence of the MUBs. Here,
two observables and the resulting basis sets are said to be mutually
unbiased if the measurement outcomes of either one with respect to
any eigenstate of the other is uniformly distributed, i.e., the
probability distribution is $\{1/d,\cdots, 1/d\}$ for a
$d$-dimensional Hilbert space $\mathcal {H}$. For example, for
qubits the three Pauli observables $\sigma_x$, $\sigma_y$, and
$\sigma_z$ are mutually unbiased. In fact, when $d=\dim\mathcal{H}$
is a prime power, there always exists a complete set of $d +1$ MUBs
\cite{mubs1,mubs2}.

If one uses the $l_1$ norm of coherence $C_{l_1}(A_j,\rho)$ as a
quantifier, with $\{A_j\}_{j=1}^{d+1}$ being the MUBs and $A_j=
\{|a_i^{(j)} \rangle\}_{i=1}^d$, \citet{comp1} obtained
\begin{equation}\label{eq4a-1}
 C_{l_1}(A_j,\rho)\leq \sqrt{d(d-1)[P(\rho)-P(A_j|\rho)]},
\end{equation}
where $P(\rho)=\tr\rho^2$ and $P(A_j|\rho)=\sum_i \langle
a_i^{(j)}|\rho|a_i^{(j)}\rangle^2$ are called the quantum and
classical purities, respectively. On the other hand, from the
equality $\rho=\sum_j \rho(A_j)- \iden$ \cite{mubs3}, with
$\rho(A_j)=\Delta(A_j,\rho)$ denoting full dephasing of $\rho$ in
the basis $A_j$ [see Eq. \eqref{eq2d-v2}], one can prove $\sum_j
P(A_j|\rho)=1+P(\rho)$, hence
\begin{equation}\label{eq4a-2}
 \sum_{j=1}^{d+1} C_{l_1}^2 (A_j,\rho)\leq d(d-1)[dP(\rho)-1].
\end{equation}
This is the complementarity relation for $l_1$ norm of coherence
under the complete set of MUBs. It establishes connection between
coherence and purity of a state, and bounds from above distribution
of coherence as well. This bound is tight as it is saturated by the
following states
\begin{equation}\label{eq4a-3}
 \rho_\epsilon=\frac{\epsilon}{d-1}\iden+\frac{d(1-\epsilon)-1}{d-1}
               |a_i^{(j)}\rangle\langle a_i^{(j)}|,
\end{equation}
with $0\leq\epsilon\leq 1$.

Similarly, \citet{comp1} derived a complementarity relation for the
relative entropy of coherence
\begin{equation}\label{eq4a-4}
 \begin{aligned}
  \sum_{j=1}^{d+1}C_r(A_j,\rho)\leq & (d+1)[\log_2 d-S(\rho)]\\
                   & -\frac{(d-1)[dP(\rho)-1]}{d(d-2)}\log_2(d-1),
  \end{aligned}
\end{equation}
this bound is saturated for the maximally coherent state
$|\Psi_d\rangle$, and the second term on the right-hand side reduces
to $[P(\rho)-0.5]\log_2 e$ for $d=2$.

\citet{comp1} also defined the mean coherence $\bar{C}(\rho)$ and
the root mean square coherence ${\rm rms}[C(\rho)]$ as
\begin{equation}\label{eq4a-5}
 \begin{aligned}
  & \bar{C}(\rho)=\int dU C(U\rho U^\dag,\rho),\\
  & {\rm rms}[C(\rho)]=\left[\int dU C^2(U\rho U^\dag,\rho)\right]^{1/2},
  \end{aligned}
\end{equation}
where $\{U\}$ denote the unitaries which transform one basis set to
another one, and $dU$ is the normalized invariant Haar measure over
$\{U\}$. Based on this, one can obtain
\begin{equation}\label{eq4a-6}
 \begin{aligned}
  & \bar{C}_{l_1}(\rho)\leq {\rm rms}[C_{l_1}(\rho)]\leq \sqrt{\frac{d(d-1)
                            [dP(\rho)-1]}{d+1}},\\
  & \bar{C}_r(\rho)=\sum_{n=2}^d \frac{1}{n}\log_2 e-[S(\rho)-Q(\rho)],
 \end{aligned}
\end{equation}
where $\{\lambda_i\}_{i=1}^d$ are eigenvalues of $\rho$, and
\begin{equation}\label{eq4a-7}
Q(\rho)=-\sum_{i=1}^d \frac{\lambda_i^d\log_2\lambda_i}
        {\prod_{j\neq i} (\lambda_i-\lambda_j)},
\end{equation}
is the quantum subentropy \cite{sube1}.

\subsubsection{Incompatible bases}
Following the established notions for various entropic uncertainty
relations (EURs) [see the review paper of \citet{eur0}], \citet{incbs}
further discussed tradeoff relations between quantum coherence of
the MUBs.

First, for single-partite quantum state $\rho$, it follows
immediately from the EUR $H(P)+H(Q)\geq \log_2 (1/c)+S(\rho)$ that
\begin{equation} \label{eq4a-new6}
 C_r(Q,\rho)+C_r(R,\rho)\geq \log_2(1/c)-S(\rho),
\end{equation}
where
\begin{equation} \label{EUR1}
 c=\max_{k,l}|\langle\psi_k^Q|\psi_l^R\rangle|^2,
\end{equation}
with $\{|\psi_k ^Q \rangle\}$ and $\{|\psi_l^R\rangle\}$ denoting
respectively, the eigenstates of the two incompatible observables
$Q$ and $R$. Similarly, by using the EUR derived by \citet{ur1} and
\citet{ur2}, one can obtain two new lower bounds for the sum of
coherence, which are as follows
\begin{equation} \label{EUR2}
 \begin{aligned}
  & C_r(Q,\rho)+C_r(R,\rho)\geq \log_2(1/c)[1-S(\rho)], \\
  & C_r(Q,\rho)+C_r(R,\rho)\geq H\left(\frac{1+\sqrt{2c-1}}{2}\right)-2S(\rho),
 \end{aligned}
\end{equation}
and they may be more or less optimal than the bound of Eq.
\eqref{eq4a-new6} due to the different values of $c$ and the form of
$\rho$.

For single-qubit state $\rho$, \citet{Maxf} obtained new lower
bounds for the sum of coherence measures under two incompatible
bases. By denoting $P'=2\tr \rho^2-1$, these bounds can be written
explicitly as
\begin{equation} \label{EUR3}
 \begin{aligned}
  & C_r(Q,\rho)+C_r(R,\rho)\geq H\left(1+\frac{\sqrt{P'}(2\sqrt{c}-1)}{2} \right)-S(\rho),\\
  & C_{l_1}(Q,\rho)+C_{l_1}(R,\rho)\geq 2\sqrt{P'c(1-c)}, \\
  & R_I(Q,\rho)+R_I(R,\rho)\geq H\left( \frac{1+\sqrt{1-4P'(\sqrt{c}-c)}}{2}\right).
 \end{aligned}
\end{equation}

Second, for the bipartite state $\rho^{AB}$, by using the
quantum-memory-assisted EUR \cite{eur1}
\begin{equation} \label{EUR4}
 S(Q|B)+S(R|B)\geq \log_2(1/c)+S(A|B),
\end{equation}
and taking the eigenstates $\Xi=\{|\psi_k^X\rangle
|\varphi_j^B\rangle\}$ ($X=\{Q,R\}$, and $|\varphi_j^B\rangle$ is
the eigenstate of $\rho^B=\tr_A \rho^{AB}$) as the basis, we have
\begin{equation}\label{eq4a-8}
 C_r(\Xi,\rho^{AB})+C_r(\Xi,\rho^{AB})\geq \log_2(1/c)-S(A|B),
\end{equation}
where the bound on the right-hand side can be further tightened by
using the concept of QD \cite{eur2}.

Similarly, if one considers the mutually incompatible observables
$\{Q_1,Q_2,\cdots,Q_n\}$, with the corresponding eigenstate bases
$\{|\psi_{k_1}^{Q_1}\rangle |\psi_{k_2}^{Q_2}\rangle\cdots
|\psi_{k_n}^{Q_n}\rangle\}$ and $\Xi_i=\{|\psi_{k_i}^{Q_i}\rangle
|\varphi_j^B\rangle\}$, then by using the formulas established by
\citet{eur3}, one can obtain
\begin{equation}\label{eq4a-9}
 \begin{aligned}
  & \sum_{i=1}^n C_r(Q_i,\rho)\geq \log_2 (1/b)-S(\rho),\\
  & \sum_{i=1}^n C_r(\Xi_i,\rho^{AB})\geq \log_2 (1/b)-S(A|B),
 \end{aligned}
\end{equation}
and this can be considered as an extension of the results for two
mutually unbiased observables.

\subsubsection{Complementarity between coherence and mixedness}
For the class of states with fixed mixedness, the amount of quantum
coherence contained in them may be different. Using the linear
entropy measure of mixedness
\begin{equation}\label{eq4b-1}
M_l(\rho)=\frac{d}{d-1}(1-\tr\rho^2),
\end{equation}
and the $l_1$ norm of coherence given in Eq. \eqref{eqa-01},
\citet{comp2} derived a tradeoff relation between the two
quantities,
\begin{equation}\label{eq4b-2}
 \frac{C_{l_1}^2(\rho)}{(d-1)^2}+M_l(\rho)\leq 1,
\end{equation}
where the first term on the left-hand side can be seen as the square
of the normalized coherence, $\tilde{C}_{l_1} (\rho) \coloneqq
C_{l_1} (\rho)/(d-1)$. It sets a fundamental limit to the amount of
coherence that can be extracted from the class of states with equal
mixedness, and vice versa.

Moreover, in the same vein with the definition of maximally
entangled mixed states \cite{mems1,mems2,mems3}, \citet{comp2}
considered the class of MCMS given in Eq. \eqref{eq4b-4}, and found
that the upper bound in Eq. \eqref{eq4b-3} is saturated, as it gives
\begin{equation} \label{eq4b-3}
 C_{l_1}(\rho_{\rm mcms})= p(d-1),~
 M_l(\rho_{\rm mcms})= 1-p^2,
\end{equation}
for $1\leq p\leq 1$. In fact, $\rho_{\rm mcms}$ constitutes also the
class of states with maximal mixedness for fixed coherence.

Although \citet{comp2} pointed out that similar tradeoffs apply to
the relative entropy of coherence [i.e., $C_r(\rho)+S(\rho) \leq 1$,
which is incorrect as $C_r(\rho)+ S(\rho)= S(\rho_{\rm diag})\leq
\log_2 d$], and the fidelity-based measure of coherence for single
qubit state [see Eq. \eqref{eq2b-6}], it remains as a challenge to
generalize the complementarity relation \eqref{eq4b-3} to other
coherence measures which are on equal footing with the $l_1$ norm of
coherence. Some other progress have been made, e.g., \citet{longgl}
have showed that if the mixedness of $\rho$ is defined via the
fidelity as
\begin{equation} \label{eq4b-v3}
 M_g(\rho)=F(\rho,\iden/d)=\frac{1}{d}(\tr\sqrt{\rho})^2,
\end{equation}
then by combing this with Eq. \eqref{eq-longgl} and further using
the mean inequality $\sum_i x_i\leq (d\sum\nolimits_i x_i^2)^{1/2}$
($\forall x_i\in\mathbb{R}$), it is easy to see that
\begin{equation} \label{eq4b-v4}
 C_g(\rho)+M_g(\rho)\leq 1-\sum_i b_{ii}^2+\frac{1}{d}\left(\sum_i b_{ii}\right)^2\leq 1,
\end{equation}
with equality holding for $\rho_\mathrm{mcms}$ of Eq.
\eqref{eq4b-4}.

Stimulated by the work of \citet{purity} and \citet{ther4}, a
resource theory of purity was established by \citet{coh-pur}. In
this framework, the only free state is the maximally mixed state
$\rho_{\mathrm{mm}}=\iden/d$, and the set of free operations is the
unital operations $\Lambda_{\mathrm{U}}$. A functional $P(\rho)$ is
said to be a purity monotone if it is nonnegative and
$P(\Lambda_{\mathrm{U}}[\rho]) \leq P(\rho)$. $P(\rho)$ is a purity
measure if it further satisfies the additivity property
$P(\rho\otimes\sigma)=P(\rho)+ P(\sigma)$ and normalization
condition $P(|\psi\rangle)=\log_2 d$. Moreover, it is convex if
$\sum_i p_i P(\rho_i)\geq P(\sum_i p_i\rho_i)$.

Based on the above framework, \citet{coh-pur} introduced a
coherence-based purity monotone
\begin{equation} \label{purity}
 P_C(\rho)\coloneqq \max{\Lambda_{\mathrm{U}}} C(\Lambda_{\mathrm{U}}[\rho])
          =C(\rho_{\max}),
\end{equation}
with $C$ being any MIO monotone. When $C$ is defined by the
contractive distance $\mathcal{D}$, the combination of the above
equation with Eq. \eqref{maxcohana} further gives
$P_C(\rho)=\mathcal{D}(\rho,\iden/d)$. This shows another connection
between purity of a state and the maximum amount of quantum
coherence achievable by suitable unitary operation.

Moreover, \citet{coh-pur} also introduced a R\'{e}nyi
$\alpha$-entropy purity measure
\begin{equation} \label{purity2}
 P_\alpha(\rho)= \log_2 d -\frac{1}{1-\alpha}\log_2\{\tr(\rho^\alpha)\},
\end{equation}
for $\alpha\geq 0$, which is also convex when $\alpha\in[0,1]$. In
particular, the R\'{e}nyi $2$-entropy purity measure $P_2(\rho)=
\log_2\{d\tr(\rho^2)\}$ is quantitatively related to the linear
entropy of purity $\tr\rho^2$, and when $\alpha\rightarrow 1$, we
have the traditional relative entropy of purity
\begin{equation} \label{purity3}
 P_r(\rho)=\log_2 d-S(\rho).
\end{equation}

\subsection{Duality of coherence and path distinguishability} \label{sec:4C}
With roots in quantum optics, quantum coherence lies at the heart of
interference phenomenon. The presence of coherence in a quantum
system can be seen as a manifestation of the wave nature of it
\cite{wapa1,wapa2}, while the path distinguishability or which-path
information signifies its complementarity aspect, i.e., the particle
nature of it. The quantitative connections between them can be
investigated in the context of unambiguous quantum state
discrimination (UQSD) or ambiguous quantum state discrimination
(AQSD), which are implementable in interference experiments.

\subsubsection{Unambiguous quantum state discrimination}
\citet{wapa1} proposed to use quantum coherence to signify the wave
nature of a particle, and the upper bound of the success probability
of UQSD to signify its particle aspect. Let $\{|\xi_i\rangle\}$ be a
collection of states which may be nonorthogonal, then the task of
UQSD is to find with certainty which of them is the given one, see
\citet{uqsd1,uqsd2} and references therein.

In the $N$-path interference experiment, we denote by
$\{|\psi_i\rangle\}$ the orthogonal basis state of the path. Then if
the initial state of the particle entering the interferometer is
\begin{equation}\label{eq4c-0}
 |\psi\rangle_s=\sum_{i=1}^N c_i |\psi_i\rangle,
\end{equation}
with $\sum_i |c_i|^2=1$, and the related detector state is
$|0_i\rangle$. Their combined state after the interaction operation
is
\begin{equation}\label{par-dec}
 |\psi\rangle_{sd}=\sum_{i=1}^N c_i |\psi_i\rangle\otimes |\xi_i\rangle.
\end{equation}
To discriminate the which-path information, the experimenter can
perform measurements on the detector states. The probability for
successfully discriminating them is proved to be bounded from above
by \cite{uqsd1,uqsd2}
\begin{equation}\label{eq4c-1}
 p_{\rm uqsd}\leq \mathcal{D}_Q \coloneqq 1-\frac{1}{N-1}\sum_{i\neq j}|c_i c_j|
                                          |\langle\xi_i|\xi_j\rangle|,
\end{equation}
where $\mathcal{D}_Q$ sets a limit to the ability of the
experimenter to distinguish between the states $\{|\xi_i\rangle\}$
(and hence $\{|\psi_i\rangle\}$), although it may not be achievable
in real experiments.

On the other hand, the postmeasurement state of the particle is
\begin{equation}\label{eq4c-2}
 \rho'_s= \sum_{i,j}c_i c_j^* \langle \xi_j|\xi_i\rangle |\psi_i
\rangle\langle\psi_j|,
\end{equation}
hence
\begin{equation}\label{eq4c-3}
 C_{l_1}(\rho'_s)=\sum_{i\neq j}|c_i c_j|  |\langle\xi_j|\xi_i\rangle|,
\end{equation}
in the path basis $\{|\psi_i\rangle\}$. Based on these,
\citet{wapa1} derived the following relation
\begin{equation}\label{eq4c-4}
 \frac{C_{l_1}(\rho'_s)}{N-1}+\mathcal{D}_Q=1,
\end{equation}
It characterizes in a quantitative way the wave-particle duality. In
particular, for two- and three-path situations with uniform $|c_i|$,
the normalized coherence $\tilde{C}_{l_1}= C_{l_1}/(N-1)$ is also
quantitatively related to the interference fringe visibility
$\mathcal{V}$, namely, $\tilde{C}_{l_1}=\mathcal {V}$ and
$\tilde{C}_{l_1}=2\mathcal {V}/(3-\mathcal {V})$, respectively
\cite{wapa1}.

Moreover, for the case of initially mixed particle state
$\rho_s=\sum_i \rho_{ij}|\psi_i\rangle\langle\psi_j|$ and pure
detector state, or the more general case of both initially mixed
particle and detector states, \citet{wapa1} showed that the equality
of Eq. \eqref{eq4c-3} becomes inequality
\begin{equation}\label{eq4c-5}
 \frac{C_{l_1}(\rho'_s)}{N-1}+\mathcal{D}_Q\leq 1.
\end{equation}

By using a slightly different path distinguishability
$\mathcal{D}=[\mathcal{D}_Q(2-\mathcal{D}_Q)]^{1/2}$,
\citet{waveparticle} further gave an equivalent form of Eq.
\eqref{eq4c-5}, i.e.,
\begin{equation} \label{eq4c-v5}
 \frac{C_{l_1}^2(\rho'_s)}{(N-1)^2} + \mathcal{D}^2 \leq 1,
\end{equation}
which is similar to the complementarity relation $\mathcal
{D}^2+\mathcal {V}^2\leq 1$ given by \citet{mierd}.

In we put a screen behind the slits, the interference pattern of the
particle is described by the probability density of particle hitting
the screen at particular position. \citet{multipath} considered one
such kind of problem. For the particle-detector state $
|\psi\rangle_{sd}$ of Eq. \eqref{par-dec}, the expression for the
pattern on the screen is given by $|\langle x|\psi\rangle_{sd}|^2$.
For an $N$-slit experiment with the width of the slits being
$\epsilon$, and the distance between any two neighboring slits
(between the slits and the screen) is $\ell$ ($D$), then if we
assume that the state that emerges from the $j$th slit is a Gaussian
along the $x$ axis and centered ad $x_j=j\ell$, the state of the
particle hitting the screen at a position $x$ will be
\cite{multipath}
\begin{equation} \label{eq4c-n5}
 \langle x|\psi(t)\rangle_{sd}= A_t \sum_{j=1}^N c_j \exp{\left[-\frac{(x-j\ell)^2}
                                {\epsilon^2+i\lambda D/\pi}\right]}|\xi_j\rangle,
\end{equation}
where $A_t=\{2/[\pi(\epsilon+i\lambda D/\pi\epsilon)]\}^{1/4}$ ($i$
is the imaginary unit), and $|\psi(t)\rangle_{sd}$ is the evolved
state with $|\psi(0)\rangle_{sd}$ being given by Eq.
\eqref{par-dec}.

Then by using the facts that $\epsilon^2 \ll (\lambda D/\pi)^2$ and
the distance between the primary maxima $\lambda D/\ell\gg j\ell$,
one can obtain that the intensity of the fringe $I(x)=|\langle
x|\psi(t)\rangle_{sd}|^2$ at position $x$ is given by
\citet{multipath}
\begin{equation}\label{intensity}
 \begin{split}
  I(x) =& |A_t|^2 \exp{\left[-\frac{2\epsilon^2 x^2}{(\lambda D/\pi)^2}\right]}
          \left\{1+\sum_{j\neq k} |c_j c_k|\right.  \\
        & \left.\times|\langle\xi_j|\xi_k\rangle|\cos\left[\frac{2\pi x\ell(k-j)}
          {\lambda D}+\theta_k-\theta_j\right]\right\},
 \end{split}
\end{equation}
where we have defined $c_k|\xi_k\rangle=|c_k| |\xi'_k\rangle
e^{i\theta_k}$, with $ |\xi'_k\rangle$ being real.

By choosing $\theta_k=\theta_j$ ($\forall k, j$), one can obtain
from the above equation that at positions $x_m=m\lambda D/\ell$
$(m\in\mathbb{Z})$ of the primary maxima, the intensity of the
fringe is given by
\begin{equation}\label{inten1}
 I_{\max} = |A_t|^2 \exp{\left[-\frac{2\epsilon^2 x_m^2}{(\lambda D/\pi)^2}\right]}
           \left(1+\sum_{j\neq k} |c_j c_k| |\langle\xi_j|\xi_k\rangle|\right).
\end{equation}

Moreover, when a phase randomizer is applied to the setup such that
the phases of the incoming state at different slits are randomized
(i.e., the incoming state becomes incoherent), the cosine term of
Eq. \eqref{intensity} will disappear, thus we have
\begin{equation}\label{inten2}
 I_{\mathrm{inc}} = |A_t|^2 \exp{\left[-\frac{2\epsilon^2 x_m^2}{(\lambda
 D /\pi)^2}\right]}.
\end{equation}

Finally, by combining the above two results with Eq. \eqref{eq4c-3},
one can obtain directly that
\begin{equation} \label{inten3}
 \frac{I_{\max}-I_{\mathrm{inc}}}{I_{\mathrm{inc}}}=C_{l_1}(\rho'_s),
\end{equation}

In fact, from Eq. \eqref{intensity} one can see directly that when
the which-path information is completely indistinguishable (i.e.,
$\langle\xi_j| \xi_k\rangle=1$, $\forall k,j$), the intensity of the
interference fringe at the primary maximum  $x_m=m\lambda D/\ell$ is
given by
\begin{equation} \label{inten4}
 I_{\max}^\parallel = |A_t|^2 \exp{\left[-\frac{2\epsilon^2 x_m^2}{(\lambda D/\pi)^2}\right]}
           \left(1+\sum_{j\neq k} |c_j c_k|\right).
\end{equation}
Similarly, when the which-path information is completely
distinguishable (i.e., $\langle\xi_j| \xi_k\rangle=0$, $\forall
k,j$), the intensity of the interference fringe at the primary
maximum $x_m=m\lambda D/\ell$ turns out to be
\begin{equation} \label{inten5}
 I_{\max}^\perp = |A_t|^2 \exp{\left[-\frac{2\epsilon^2 x_m^2}{(\lambda
 D/\pi)^2}\right]}.
\end{equation}
Then it is obvious that
\begin{equation} \label{inten6}
 \frac{I_{\max}^\parallel-I_{\max}^\perp}{I_{\max}^\perp}=C_{l_1}(\rho'_s).
\end{equation}

The implementation of the above scheme for measuring quantum
coherence depends essentially on whether there exits such a path
detector which is (at least) capable of making the which-path
information completely indistinguishable and distinguishable.

\subsubsection{Ambiguous quantum state discrimination}
Different from UQSD, one always has a result in the AQSD
experiments, but it may be right or wrong, and the task of the
experimenter is to minimize the probability of being wrong to its
theoretical limit \cite{mierd}, hence it is also known as
minimum-error state discrimination .

By using the $l_1$ norm of coherence to characterize the wave
nature, and an upper bound on the average probability $p_{\rm aqsd}$
of successfully discriminating the path information to characterize
the particle nature, \citet{wapa2} derived the following duality
relation between them
\begin{equation}\label{eq4c-6}
 \left(\frac{C_{l_1}(\rho'_s)}{N-1}\right)^2+
 \left(\frac{Np_{\rm aqsd}-1}{N-1}\right)^2 \leq 1,
\end{equation}
where $\rho'_s$ is the same as Eq. \eqref{eq4c-2}. To discriminate
the which-path information, the experimenter can perform POVM on the
reduced detector state $\rho_{d}=\sum_i |c_i|^2 \rho^\xi_i$ (with
$\rho^\xi_i= |\xi_i\rangle \langle \xi_i|$), and the average
probability $p_{\rm aqsd}=\sum_i |c_i|^2 \tr(\Pi_i|\xi_i \rangle
\langle\xi_i|)$ is showed to be upper bounded by
\begin{equation}\label{eq4c-8}
 p_{\rm aqsd}\leq \frac{1}{N}+\frac{1}{2N}\sum_{ij}\|\Lambda_{ij}\|_1,
\end{equation}
where $\{\Pi_i\}$ is the set of POVM, while $\Lambda_{ij}=|c_i|^2
\rho^\xi_i-|c_j|^2 \rho^\xi_j$ is the Helstron matrix of the state
pair $(\rho^\xi_i, \rho^\xi_j)$, and
\begin{equation} \label{eq4c-v8}
 \|\Lambda_{ij}\|_1=2\sqrt{\left(\frac{|c_i|^2+|c_j|^2}{2}\right)^2-|c_i
 c_j|^2|\langle\xi_i|\xi_j\rangle|^2}.
\end{equation}

\citet{wapa2} also derived a duality relation between the relative
entropy of coherence and path distinguishability, i.e.,
\begin{equation}\label{eq4c-9}
 C_r (\rho'_s)+H(M\!:\!D)\leq H(\{p_i\}),
\end{equation}
where $H(M\!:\!D)=H(\{p_j\})+H(\{q_i\})-H(\{p_{ij}\})$ is the mutual
information of $D=\{p_j\}$ and $M=\{q_i\}$, and $H(\cdot)$ is the
Shannon entropy, with the probabilities $p_{ij}= \tr(\Pi_i
\rho^\xi_j)p_j$, $p_j=\sum_i p_{ij}=|c_j|^2$, and $q_i=\sum_j
p_{ij}=\tr(\Pi_i\rho_d)$.

Eq. \eqref{eq4c-9} holds as well even if its second term on the
left-hand side is replaced by the accessible information $I_{acc}$,
which is defined as the maximum value of $H(M\!:\!D)$ over all
possible POVMs, and characterizes how well an experimenter can do at
inferring the detector states.

Recently, it has been pointed out by \citet{waveparticle} that the
upper bound of Eq. \eqref{eq4c-6} may not be saturated for general
pure states $|\psi\rangle_{sd}$ if $N\geq 3$. It is also doubted as
the two terms on its left-hand side (which characterize the wave and
particle nature of a quanton, respectively) can increase or decrease
simultaneously.

\subsection{Distillation and dilution of quantum coherence} \label{sec:4D}

\subsubsection{Standard coherence distillation and dilution}
\begin{figure}
\centering
\resizebox{0.45 \textwidth}{!}{%
\includegraphics{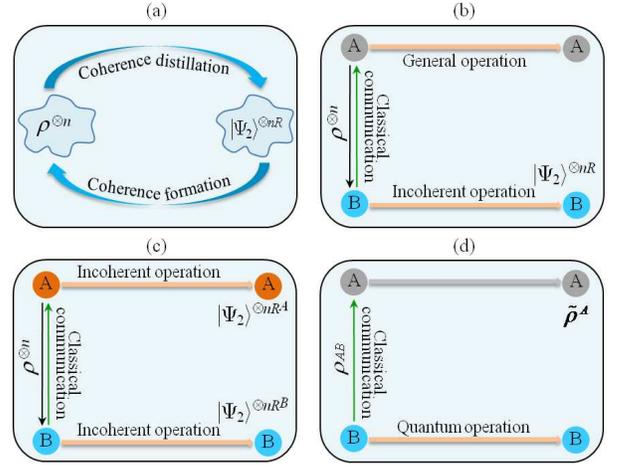}}
\caption{(a) Coherence distillation and coherence formation. (b)
Assisted coherence distillation with general operation on party $A$,
incoherent operation on $B$, and two-way (one-way) classical
communication between them. (c) Coherence distillation with
incoherent operations on both $A$ and $B$ and classical
communication. (d) Remote creation of coherence on party $A$ with
quantum operation on $B$ and classical communication from $B$ to
$A$.}\label{fig:hu4}
\end{figure}

In the same spirit as entanglement distillation and entanglement
formation, one can also consider the tasks of coherence distillation
and coherence formation by incoherent operations $\Lambda$, see Fig.
\ref{fig:hu4}(a). The former corresponds to the transformation of a
general state $\rho$ to the maximally coherent one, i.e., the
distillation of $\rho$ to $\Psi_2= |\Psi_2\rangle \langle\Psi_2|$,
while the latter is the formation of $\rho$ from $\Psi_2$. The
corresponding optimal rate can be recognized as an operational
measure of coherence.

In the asymptotic setting (i.e., infinitely many copies of $\rho$),
\citet{qcd1} defined the distillable coherence as the maximal rate
at which $\Psi_2$ can be obtained from $\rho$, i.e.,
\begin{equation}\label{eq4d-1}
 C_d(\rho)=\sup\{R:\lim_{n\rightarrow \infty\atop\varepsilon\rightarrow 0}
                 (\inf_{\Lambda}\|\Lambda(\rho^{\otimes n})
                 -\Psi_2^{\otimes \lfloor nR\rfloor }\|_1)\leq \varepsilon\},
\end{equation}
and the coherence cost which is the minimal rate at which $\Psi_2$
has to be consumed for formatting $\rho$ is dually to Eq.
\eqref{eq4d-1}, with only the supremum being replaced by the
infimum, and $\Lambda$ acting on $\Psi_2^{\otimes \lceil nR\rceil
}$. The central results are that $C_d(\rho)$ equals to the relative
entropy of coherence $C_r(\rho)$, while $C_c(\rho)$ equals to the
coherence of formation
\begin{equation} \label{cof}
 C_f(\rho)\coloneqq \min_{\{p_i,\psi_i\}}\sum_i p_i S(\Delta [\psi_i]),
\end{equation}
which involves a minimization over all pure state decompositions of
$\rho$ showed in Eq. \eqref{decomp} \cite{cof}, and the additivity
of $C_f$ and $C_r$ implies that both $C_d$ and $C_c$ are additive as
well.

Moreover, different from the possible bound entanglement in a state,
\citet{qcd1} found that there is no bound coherence, that is, there
does not exist quantum state for which its creation consumes
coherence while no coherence could be distilled from it. Thus,
\begin{equation} \label{cof2}
 C_d(\rho)=0\Rightarrow C_c(\rho)=0,
\end{equation}
which reveals that quantum coherence in any state is always
distillable.

\citet{Buprl} considered an one-shot version of coherence
distillation. They defined the relevant coherence cost for
formatting a quantum state $\rho$ under MIO as
\begin{equation}\label{eq4d-15}
  C_{c,\mathrm{MIO}}^{(1),\varepsilon}(\rho)=\inf_{\{\Lambda_{\mathrm{MIO}}\}\atop M\in \mathbb{Z}}
  \{\log_2 M|F[\rho,\Lambda_{\mathrm{MIO}}(\Psi_+^M)]\geq 1-\varepsilon\},
\end{equation}
where $\Psi_+^M=|\Psi_+^M\rangle\langle\Psi_+^M|$ with
$|\Psi_{+}^M\rangle=\sum_{i=1}^M |i\rangle/\sqrt{M}$, and $F(\cdot)$
is the Uhlmann fidelity given in Eq. \eqref{fidelity} [note that it
equals to the square of that adopted by \citet{Buprl}]. Then, they
showed that, for any $\varepsilon>0$, one has
\begin{equation}\label{eq4d-16}
  C_{\max}^{2\sqrt{\varepsilon}}(\rho)\leq C_{c,\mathrm{MIO}}^{(1),\varepsilon}(\rho),
\end{equation}
so the smooth maximum relative entropy of coherence bounds from
below the one-shot coherence cost.

Similar to the one-shot coherence distillation, one can also
consider the one-shot version of coherence dilution, in which the
corresponding coherence cost reads
\begin{equation}\label{eq4d-17}
  C_\mathcal{O}^\varepsilon (\rho)=\inf_{\Lambda\in \mathcal {O}}
  \{\log_2 M|F[\rho,\Lambda(\Psi_+^M)]\geq 1-\varepsilon\},
\end{equation}
where $\mathcal{O}$ is one of the free operations
$\{\mathrm{MIO,DIO,IO, SIO}\}$. To establish an operational
interpretation for the coherence measure, \citet{oneshot} further
introduced an $\varepsilon$-smoothed coherence measure
$C^\varepsilon (\rho)=\min_{\rho'\in B_\varepsilon(\rho)}C(\rho')$,
where $ B_\varepsilon(\rho)=\{\rho'| F(\rho,\rho'\geq
1-\varepsilon)\}$. Based on these preliminaries, they proved that
for any $\varepsilon>0$, we have
\begin{equation}\label{eq4d-18}
 \begin{aligned}
  & C_{\max}^\varepsilon (\rho)\leq C_{\mathrm{MIO}}^\varepsilon (\rho)\leq C_{\max}^\varepsilon (\rho)+1,\\
  & C_{\Delta,\max}^\varepsilon (\rho)\leq C_{\mathrm{DIO}}^\varepsilon (\rho)\leq C_{\Delta,\max}^\varepsilon (\rho)+1, \\
  & C_{\mathrm{MIO}}^\varepsilon (\rho)= C_{\mathrm{SIO}}^\varepsilon (\rho)= C_0^\varepsilon(\rho),
 \end{aligned}
\end{equation}
and in the asymptotic limit, the $\varepsilon$-smoothed coherence
equivalents either to the relative entropy of coherence \cite{coher}
or to the coherence of formation \cite{qcd1}, i.e.,
\begin{equation}\label{eq4d-19}
 \begin{aligned}
  & C_{\mathrm{MIO}}^\infty (\rho)= C_{\mathrm{DIO}}^\infty (\rho)=C_r(\rho),\\
  & C_{\mathrm{IO}}^\infty(\rho)=C_{\mathrm{SIO}}^\infty(\rho)=C_f(\rho).
 \end{aligned}
\end{equation}
where $C_{\mathcal{O}}^\infty (\rho)= \lim_{\varepsilon
\rightarrow0,n\rightarrow \infty} C_{\mathcal {O}}^\varepsilon
(\rho^{\otimes n})/n$. This result, together with that of
\citet{qcd1}, implies that the role of MIO and IO in the asymptotic
scenario of coherence dilution is the same as we also have
$C_{\mathrm{IO}}^\infty (\rho)=C_r(\rho)$.

\subsubsection{Assisted coherence distillation}
In analogy to assisted entanglement distillation, \citet{qcd2}
investigated the task of assisted coherence distillation in the
setting of local quantum-incoherent operations and classical
communication (LQICC), see Fig. \ref{fig:hu4}(b). In this task, two
players, Alice and Bob, share $n$ copies of $\rho^{AB}$, and Alice's
objective is to help Bob to distill as much quantum coherence as
possible. Different from the allowable LOCC in assisted entanglement
distillation, LQICC represents quantum operations $\Lambda_{QI}$
that are general on Alice's side and incoherent on Bob's side. In
this setting, the set $\mathcal {QI}$ of free states called the
quantum-incoherent (QI) states are given by $\chi^{AB}=\sum_i p_i
\rho^A_i \otimes |i^B\rangle\langle i^B|$, with $p_i$ the
probabilities, $\rho^A_i$ arbitrary states for subsystem $A$, and
$\{|i^B\rangle\}$ the incoherent basis for subsystem $B$.

Formally, the generated maximum coherence is called ``coherence of
collaboration" (CoC) for two-way communication, and ``coherence of
assistance" (CoA) in the one-way situation, for which Alice holds a
purifying state and only she is allowed to announce the measurement
results. \citet{qcd2} defined the optimal rate of distillable CoC as
\begin{equation}\label{eq4d-2}
 C_d^{A|B}(\rho)=\sup\{R:\lim_{n\rightarrow \infty}
                 (\inf_{\Lambda_{QI}}\|\Lambda_{QI}(\rho^{\otimes n})
                 -\Psi_2^{\otimes \lfloor nR\rfloor }\|_1)=0\},
\end{equation}
where $C_d^{A|B}$ is upper bounded by the QI relative entropy
$C_r^{A|B}$, i.e.,
\begin{equation} \label{eq4d-v2}
 C_d^{A|B}(\rho^{AB})\leq C_r^{A|B}(\rho^{AB}),
\end{equation}
with equality holding for any pure state, and
\begin{equation}\label{eq4d-3}
 C_r^{A|B}(\rho^{AB})=\min_{\chi^{AB}\in \mathcal {QI}}S(\rho^{AB}\|\chi^{AB})
                     =S(\Delta^B[\rho^{AB}])-S(\rho^{AB}),
\end{equation}
where $\Delta^B[\rho^{AB}]$ denotes dephasing of $\rho^{AB}$ in the
incoherent basis of $B$.

Moreover, when extended to the situation with $N\geq 2$ assistants,
the global operations across all auxiliary systems do not
necessarily outperform the local operations on generating coherence,
e.g., for the initial state $|\Psi\rangle^{A_1\cdots A_N B}$ with
$B$ being a qubit, local operations on $A_1,\cdots, A_N$ together
with classical communication are enough to localize maximum
coherence on $B$.

\citet{qcd2} also proposed quantitative definitions of CoA and the
regularized CoA, which are given respectively, by
\begin{equation}\label{eq4d-4}
 C_a(\rho)=\max_{\{p_i,\psi_i\}}\sum_i p_i C_r(\psi_i),~
 C_a^\infty(\rho)=\lim_{n\rightarrow\infty}\frac{1}{n}C_a(\rho^{\otimes n}),
\end{equation}
and the maximization is taken over the pure state decompositions
$\rho$ showed in Eq. \eqref{decomp}. Moreover, for qubit states
$\rho$, the CoA is showed to be additive, i.e., $C_a(\rho^{\otimes
n})= n C_a (\rho)$.

Notably, there exists some resemblance between CoA of the state
$\rho=\sum_{i,j}\rho_{ij}|i\rangle\langle j|$ and entanglement of
assistance (EoA) of the related maximally correlated state
$\rho_{\rm mc}$ of Eq. \eqref{eq-mcs}, that is,
\begin{equation} \label{EoA}
 C_a(\rho)= E_a(\rho_{\rm mc}),~ C_a^\infty(\rho)= E_a^\infty
(\rho_{\rm mc})=S(\Delta[\rho]).
\end{equation}
Moreover, for pure state $|\psi\rangle^{AB}$, the CoC equals to the
regularized CoA, i.e.,
\begin{equation}\label{eq4d-7}
 C_d^{A|B}(|\psi\rangle^{AB})=C_a^\infty (\rho^B)=S(\Delta[\rho^B]),
\end{equation}
which immediately yields that the maximum extra coherence that Bob
can gain [compared with the standard distillation protocol
\cite{qcd1}] with Alice's assistance equals to the von Neumann
entropy of $\rho^B$.

By replacing LQICC with the local incoherent operations and
classical communication (LIOCC), \citet{qcd3} further studied the
coherence-entanglement tradeoffs in a task similar to \citet{qcd2},
but now both the two parties' operations are restricted to be local
incoherent, see Fig. \ref{fig:hu4}(c). In this new setting, if we
denote by $R^A$ ($R^B$) the rate of coherence formation for Alice
(Bob), and $E^{co}$ that of entanglement formation between Alice and
Bob, then the triple $(R^A, R^B,E^{co})$ is achievable if for every
$\varepsilon>0$ there exists a LIOCC $\Lambda_{II}$ and integer $n$
such that
\begin{equation}\label{eq4d-8}
 \|\Lambda_{II}(\Psi_2^{\otimes\lceil n(R^A+\varepsilon)\rceil}
   \otimes \Psi_2^{\otimes\lceil n(R^B+\varepsilon)\rceil}
   \otimes\Phi_2^{\otimes\lceil n(E^{co}+\varepsilon)\rceil})
   -\rho^{\otimes n}\|_1 \leq \varepsilon,
\end{equation}
where the two $\Psi_2$ belong to Alice and Bob, respectively, while
$\Phi_2=|\Phi_2\rangle \langle\Phi_2|$ with $|\Phi_2\rangle=
(|00\rangle+|11\rangle) /\sqrt{2}$ is shared between them.
Similarly, $(R^A, A^B,E^{co})$ is the achievable
coherence-entanglement distillation triple if
\begin{equation}\label{eq4d-9}
 \|\Lambda_{II}(\rho^{\otimes n})-
   \Psi_2^{\otimes\lfloor n(R^A-\varepsilon)\rfloor}
   \otimes \Psi_2^{\otimes\lfloor n(R^B-\varepsilon)\rfloor}
   \otimes\Phi_2^{\otimes\lfloor n(E^{co}-\varepsilon)\rfloor} \|_1
   \leq \varepsilon.
\end{equation}

For pure states $\Psi=|\Psi\rangle^{AB}\langle\Psi|$, \citet{qcd3}
obtained the possible optimal triples of resource formation
\begin{equation}\label{eq4d-10}
 \begin{aligned}
  & (R^A, A^B,E^{co})=(0,S(B|A)_{\Delta(\Psi)}, S(A)_{\Delta(\Psi)}),\\
  & (R^A, A^B,E^{co})=(S(A)_{\Delta(\Psi)},S(B|A)_{\Delta(\Psi)},E(\Psi)),\\
  & (R^A, A^B,E^{co})=(0, 0, S(AB)_{\Delta(\Psi)}),
 \end{aligned}
\end{equation}
with $S(X)_{\Delta(\Psi)}$ [$S(X|Y)_{\Delta(\Psi)}$] the von Neumann
entropy (conditional entropy) of $\Delta(\Psi)$, and $E(\Psi)=
S(A)_{\Psi}$ the entanglement of $\Psi$. Using monotonicity of a
LIOCC monotone
\begin{equation} \label{eq4d-v10}
 C_L(\rho^{AB})=\min_{\{p_i,\psi_i\}} \sum_i p_i C_L(\psi_i),
\end{equation}
with $\rho^{AB}=\sum_i p_i \psi_i$, $\psi_i = |\psi_i\rangle\langle
\psi_i|$, and
\begin{equation}\label{eq4d-n10}
 C_L(\psi_i)= S(A)_{\Delta(\psi_i)}+ S(B)_{\Delta(\psi_i)}-E(\psi_i),
\end{equation}
\citet{qcd3} also derived the optimal resource distillation tripes
\begin{equation}\label{eq4d-11}
 \begin{aligned}
  & (R^A, A^B,E^{co})=(S(A)_{\Delta(\Psi)}-E(\Psi), S(B)_{\Delta(\Psi)},0),\\
  & (R^A, A^B,E^{co})=(0,S(B|A)_{\Delta(\Psi)},I(A\!:\!B)_{\Delta(\Psi)}),
 \end{aligned}
\end{equation}
where $I(A\!:\!B)_{\Delta(\Psi)}$ is the mutual information of
${\Delta(\Psi)}$.

It is evident that the distillable coherence rate sum $C_D^{\rm
LIOCC}=R^A+R^B$ that can be distilled simultaneously at Alice and
Bob's side is constrained by their shared entanglement. By further
defining two similar quantities $C_D^{\rm Global}$ and $C_D^{\rm
LIO}$, the former with global incoherent operations, and the latter
with local incoherent operations without classical communication,
\citet{qcd3} found that for $\Psi$, the differences
\begin{equation}\label{eq4d-v11}
 \begin{aligned}
 & \delta(\Psi)= C_D^{\rm Global}(\Psi)-C_D^{\rm LIOCC}(\Psi),\\
 & \delta_c(\Psi)=C_D^{\rm LIOCC}(\Psi)-C_D^{\rm LIO}(\Psi),
 \end{aligned}
\end{equation}
are given by
\begin{equation}\label{eq4d-12}
 \delta(\Psi)=E(\Psi)-I(A\!:\!B)_{\Delta(\Psi)},~~
 \delta_c(\Psi)=E(\Psi).
\end{equation}
They describe, respectively, the extra coherence rates that can only
be distilled by nonlocal incoherent operations and by using the data
communicated via a classical channel.

\citet{RCC} considered a similar scenario of collaborative creation
of coherence, see Fig. \ref{fig:hu4}(d). Here, two parties share a
state $\rho^{AB}$, and their aim is to create coherence on $A$ with
the help of quantum operation solely on $B$ and one-way classical
communication from $B$ to $A$. They called this as remote creation
of coherence (RCC), and obtained relations between the created
coherence and entanglement of $\rho^{AB}$. By using the operator-sum
representation $\mathcal {E}(\cdot)= \sum_i E_i(\cdot)E_i^\dag$, and
denoting $\tilde{\rho}^A=\tr_B (\iden_A\otimes\mathcal {E})
\rho^{AB}/p'$ (with $p'=\tr(\iden_A \otimes\mathcal {E})\rho^{AB}$
the probability of getting $\tilde{\rho}^A$), they proved that the
RCC $C(\tilde{\rho}^A)=0$ if and only if $\rho^{AB}=\sum_i p_i
\sum_k q^i_k |k\rangle\langle k| \otimes \rho^B_i$, namely, it is an
incoherent-quantum state.

For the initial pure state $|\psi^{AB}\rangle$ with vanishing
coherence on $A$, the RCC is nonzero if and only if there exists a
basis $\{|i\rangle\}$ which gives $[N, (\langle i|\otimes \iden)
|\psi^{AB}\rangle\langle\psi^{AB}|(|i\rangle\otimes \iden)] \neq 0$,
with $N=\sum_i E_i^\dag E_i\leq \iden$. The amount of RCC measured
by the $l_1$ norm is bounded above by
\begin{equation}\label{eq4d-13}
 C_{l_1}(\tilde{\rho}^A)\leq  \frac{E(|\psi^{AB}\rangle)}{p'}\sqrt{\sum_{j<i}|N_{ji}|^2},
\end{equation}
where $E(|\psi^{AB}\rangle)$ denotes the concurrence of
$|\psi^{AB}\rangle$, and $N_{ji}$s are matrix elements of $N$ under
the Schmidt decomposition basis of $\rho^B=\tr_A(|\psi^{AB}\rangle
\langle\psi^{AB}|)$. Furthermore, if the channel $\mathcal{E}$ is
trace preserving, the average RCC
\begin{equation} \label{eq4d-v13}
 \bar{C}_{l_1}(|\psi^{AB} \rangle)\coloneqq \sum_i p_i C_{l_1}(\tilde{\rho}^A_i),
\end{equation}
with $\tilde{\rho}^A_i=\tr_B [(\iden_A\otimes E_i) |\psi^{AB}\rangle
\langle\psi^{AB}|(\iden_A\otimes E_i)]/p_i$ and $p_i=\tr
[(\iden_A\otimes E_i) |\psi^{AB}\rangle
\langle\psi^{AB}|(\iden_A\otimes E_i)]$, has the following bound
\begin{equation}\label{eq4d-14}
  \bar{C}_{l_1}^{A|B}(|\psi^{AB}\rangle) \leq \frac{d}{2}
  E(|\psi^{AB}\rangle)\bar{C}_{l_1}^{A|B}(|\Phi^{AB}\rangle),
\end{equation}
with $|\Phi^{AB}\rangle$ being the maximally entangled state in the
Schmidt decomposition basis of $|\psi^{AB}\rangle$, and for $d=2$
case, the equality in the above equation holds. This also establish
an operational connection between created coherence of a subsystem
and entanglement of the composite system, although applies only for
the initial pure states.

\subsection{Average coherence of randomly sampled states} \label{sec:4E}
As is known, some measures of quantumness manifest concentration
effect, e.g., the random bipartite pure states sampled from the
uniform Haar measure are typically maximally entangled \cite{avee}.
Along the same line, \citet{avec} studied the coherence properties
of pure states chosen randomly from the uniform Haar measure, and
found that most of them possess almost the same amount of coherence
which are not typically maximally coherent.

For the Haar distributed random pure states $\psi=|\psi\rangle
\langle \psi|$ with dimension $d\geq 3$, they considered the average
coherence of the form
\begin{equation} \label{Haar1}
\bar{C}(\psi)\coloneqq \int d(\psi) C(\psi)=\int
d\mu(U)C(U|1\rangle\langle 1| U^\dag),
\end{equation}
with $U$ being sampled from the uniform Haar distribution, and
$C(\psi)$ can be any faithful measure of coherence.

First, for the relative entropy of coherence, its average over all
$\psi$ was found to be
\begin{equation} \label{Haar2}
 \bar{C}_r(\psi)= H_d-1,
\end{equation}
with $H_d=\sum_{k=1}^d (1/k)$ the $d$th harmonic number, and the
logarithm in Eq. \eqref{eq2a-2} is with respect to natural base
here. The probability for $|C_r(\psi)-(H_d-1)|> \epsilon$ is upper
bounded by $2e^{-{d \epsilon^2}/{36\pi^3\ln 2 \ln ^2 d}}$, hence the
randomly chosen $\psi$ with $C_r(\psi)$ not close to $H_d-1$ is
exponentially small. This is the concentration phenomenon for
relative entropy of coherence. It reveals that most Haar distributed
random $\psi$ have $H_d-1$ amount of coherence, which is solely
determined by the parameter $d$.

Second, for $l_1$ norm of coherence, they found that the mean
classical purity $\bar{P}(\Delta[\psi])$ averaged over the Haar
distributed $\psi$ is given by $2/(d+1)$. The probability for
$|P(\Delta[\psi])-2/(d+1)|>\epsilon$ is $2e^{-{d \epsilon^2}/
{18\pi^3\ln 2}}$, which is also exponentially small for
$\epsilon\rightarrow 0$. Thus by using the upper bound of
$C_{l_1}(\psi)$ given in Eq. \eqref{eq4a-1}, one has
\begin{equation} \label{Haar3}
 \bar{C}_{l_1}(\psi)\leq \sqrt{d(d-1)^2/(d+1)}.
\end{equation}

Finally, to show most of the Haar distributed pure states are not
typically maximally coherent, \citet{avec} calculated the average
trace distance between $\rho^\psi_{\rm diag}$ and the maximally
mixed state $\rho_{\rm mm}$ (which is the optimal $\delta$ for
$|\Psi_d\rangle$). The result shows that
\begin{equation} \label{Haar4}
 \bar{\mathcal{D}} (\rho^\psi_{\rm diag},\rho_{\rm mm})=2(1-1/d)^d,
\end{equation}
which approaches $2/e$ in the limit of $d\rightarrow \infty$. The
probability for a divergence of the amount $\epsilon$ is $2e^{-{d
\epsilon^2}/ {18\pi^3\ln 2}}$, which is arbitrary small for
$\epsilon\rightarrow 0$. This shows that the optimal $\delta$ for
the majority of $C_r(\psi)$ are not $\rho_{\rm mm}$, hence the
random Haar distributed $\psi$ are not maximally coherent.

In fact, for the uniformly distributed pure states, the average
$l_1$ norm of coherence can be obtained. The corresponding
analytical result is derived by \citet{onenorm}, which is given by
\begin{equation} \label{Haar5}
 \bar{C}_{l_1}(\psi)=\frac{(d-1)\pi}{4},
\end{equation}
and there is no concentration phenomenon for it, this is because the
probability for $C_{l_1}(\psi)$ not close to $(d-1)\pi/4$ is given
by $2e^{-{4\epsilon^2}/ {9d\pi^3 \ln 2}}$, which is finite when
$d\rightarrow \infty$. But the scaled $l_1$ norm of coherence
$C_{l_1}(\rho)/(d-1)$ concentrates around $\pi/4$ for very large
values of $d$, and the probability for a divergence of the amount
$\epsilon$ is given by $2e^{-{4(d-1)^2\epsilon^2}/ {9d\pi^3 \ln
2}}$.

Similarly, for $d$-dimensional randomly mixed states sampled from
various induced measures, \citet{avec-mix} considered the average
relative entropy of coherence
\begin{equation} \label{Haar6}
 \begin{split}
 \bar{C}_r(\alpha,\gamma) &\coloneqq \int d\mu_{\alpha,\gamma}(\rho)C_r(\rho)\\
                          &= \int d\mu_{\alpha,\gamma}(U\Lambda U^\dag)C_r(U\Lambda U^\dag),
 \end{split}
\end{equation}
with $\Lambda=\mathrm{diag}\{\lambda_1,\cdots,\lambda_d\}$, and
$U\Lambda U^\dag$ is the isospectral full-ranked density matrices
(i.e., the spectra of $\Lambda$ is nondegenerate), and
$\mu_{\alpha,\gamma}$ is the normalized probability measure on the
set of density matrix $\mathcal{D}(\mathbb{C}^d)$.

For the special case of mixed $\rho$ sampled from induced measures
obtained via partial tracing of the Haar distributed
$dd'$-dimensional ($d'\geq d$) pure states $\psi$, the average
coherence can be further obtained analytically as
$\bar{C}_r(\alpha,\gamma) =(d-1)/2d'$ for $(\alpha,\gamma)=
(d'-d+1,1)$. If $d$ is further restricted to $d\geq 3$, the
probability for $|C_r(\alpha,\gamma)-(d-1)/2d'|> \epsilon$ is
bounded from above by $2e^{-{dd' \epsilon^2}/{144\pi^3\ln 2 \ln ^2
dd'}}$. Hence, nearly all $\rho$ obtained via partial tracing over
the uniformly Haar distributed random pure bipartite states $\psi$
in the Hilbert space $\mathcal{H}_{d'}$ have coherence approximately
equal to the average relative entropy of coherence. These results
were further extended by the same author in a recent work
\cite{avec-mix2}.

\section{Quantum coherence in quantum information} \label{sec:5}

\subsection{Quantum state merging} \label{sec:5A}
For a quantum protocol with two or more parties, e.g., the simplest
case of two players, Alice and Bob, one may wonder how much
coherence is localized (or consumed) at Bob's side, and
simultaneously, how much entanglement is established (or consumed)
for Alice and Bob, after finishing the pre-designed computation
procedure?

\citet{qsm} explored such a problem. They discussed the protocol of
quantum state merging under IO, which they called incoherent quantum
state merging, and is indeed an analog of the standard state merging
with general quantum operations \cite{qsm1}. In this task, Alice,
Bob, and a referee share the state $\rho^{RAB}$. Alice and Bob also
have access to $\Phi_2$ at rate $E$, and Bob has access to $\Psi_2$
at rate $C$. The goal is for them to merge the state of $AB$ on
Bob's side by LQICC, i.e., Alice performs general quantum
operations, while Bob is restricted to IO only.

By denoting $E=E_i-E_t$ and $C=C_i-C_t$, with $E_i$ and $E_t$ ($C_i$
and $C_t$) being the entanglement rate of $AB$ (local coherence of
$B$) before and after the state merging protocol, \citet{qsm} showed
that the entanglement-coherence pair $(E,C)$ is achievable if there
exists $E_i$, $E_t$, $C_i$, $C_t$, and sufficiently large integers
$n$ such that
\begin{equation}\label{eq4e-1}
 \begin{aligned}
   & \|\Lambda_{QI}[\rho_i^{\otimes n}\otimes\Phi_2^{\lfloor(E_i+\delta)n\rfloor}
     \otimes\Psi_2^{\otimes \lfloor (C_i+\delta n)\rfloor}]\\
   & - \rho_t^{\otimes n}\otimes\Phi_2^{\otimes\lceil E_t n\rceil}
    \otimes\Psi_2^{\otimes\lceil C_t n\rceil} \|_1 \leq \varepsilon,
\end{aligned}
\end{equation}
is satisfied for every $\varepsilon>0$ and $\delta>0$. Moreover,
$\rho_i=\rho^{RAB}\otimes |0^{\tilde{B}}\rangle\langle
0^{\tilde{B}}|$, $\rho_t= \rho^{R\tilde{B}B}\otimes
|0^A\rangle\langle 0^A|$, and $|0^{\tilde{B}}\rangle$ is the initial
state of the auxiliary system $\tilde{B}$ (with the same dimension
as $A$) belong to Bob. $E>0$ ($C>0$) corresponds to entanglement
(coherence) consumption in the task of state merging, while $E<0$
($C<0$) corresponds to the reverse situation, i.e., the merging
protocol is achievable for free, with the additional gain of
entanglement (coherence) at rate $|E|$ ($|C|$).

On the above basis, \citet{qsm} found that the sum of $E$ and $C$ is
upper bounded by a nonnegative quantity, i.e.,
\begin{equation}\label{eq4e-2}
 E+C \leq S(\Delta^{AB} [\rho^{RAB}])
             -S(\Delta^B[\rho^{RAB}]),
\end{equation}
where $\Delta^{AB}$ and $\Delta^{B}$ are the same as that in Eq.
\eqref{eq4d-3}. The equality holds for any pure state $\rho^{RAB}$,
for which $(E,C)$ reduces to $(E_0,0)$, with $E_0=S(\bar{\rho}^{AB})
-S(\bar{\rho}^B)$ and $\bar{\rho}^{AB}=\Delta(\rho^{AB})$. It
implies that whenever $E<0$, we must have $C\geq 0$, and vice versa.
Therefore, there is no state merging procedure for which
entanglement and coherence can be gained simultaneously. This can be
recognized as an operational complementarity relation between
entanglement of a bipartite state and quantum coherence of its
reduction.

\subsection{Deutsch-Jozsa algorithm} \label{sec:5B}
The Deutsch-Jozsa algorithm is one of the first quantum algorithms
in quantum information science. It uses quantum coherence as a
resource, and this enables its speedup compared with that of the
classical counterpart \cite{Deutsch}.

By considering a discrete quantum walk version, \citet{DJ-coherence}
studied the Deutsch-Jozsa algorithm. It is performed in three steps:
(\romannumeral+1) let the particle sitting at the edge between 0 and
$A$ (with the state denoting as $|0,A\rangle$) traverses the vertex
$A$, which transforms it to $U_A|0,A\rangle=\sum_{j=0}^N |A,j\rangle
/\sqrt{(N+1)}$; (\romannumeral+2) It goes further from $A$ to $B$
between which there are $N$ paths, and there will be a phase
$e^{i\phi_j}$ ($\phi_j=0$ or $\pi$) being added after traversing the
vertex of the $j$th path, thus the state becomes $(|0,-1\rangle+
\sum_{j=1}^{N} e^{i\phi_j} |j,B\rangle)/ \sqrt{(N+1)}$;
(\romannumeral+3) Finally, the particle traversing the vertex $B$
and it is transformed into
\begin{equation}\label{eq-dj1}
 \begin{split}
  & \frac{1}{\sqrt{N+1}}|-1,-2\rangle+\frac{1}{N+1}\sum_{j,k=1}^{N}
     e^{i\phi_j}e^{2\pi ijk/(N+1)}|B,k\rangle\\
  & +\frac{1}{N+1}\sum_{j=1}^{N} e^{i\phi_j}|B,N+1\rangle.
 \end{split}
\end{equation}

To discuss in a quantitative way how quantum coherence affects
performance of the algorithm, \citet{DJ-coherence} further
introduced a qubit (with the initial state $|0\rangle$) to every
path of the graph. By supposing the qubit state  $|\mu_j\rangle=
\alpha_j|0\rangle_j +\beta_j |1\rangle_j$ after traversing the $j$th
vertex, and defining $|\eta\rangle_j =|\mu_j\rangle\prod_{k\neq j}^N
|0\rangle_k$ and $|\eta\rangle_0 = \prod_{k=0}^N |0\rangle_k$, their
state after passing through the $N$ paths will be
\begin{equation} \label{eq-dj0}
 |\Psi\rangle_{in}=(|0,-1\rangle|\eta_0\rangle+\sum_{j=1}^{N} e^{i\phi_j}
|j,B\rangle|\eta_j\rangle)/\sqrt{(N+1)},
\end{equation}
for which the $l_1$ norm of coherence is given by
$C_{l_1}(|\Psi\rangle_{in})=\sum_{j\neq k}^N
|\langle\eta_k|\eta_j\rangle|/(N+1)$, and the output state after the
vertex $B$ is
\begin{equation}\label{eq-dj2}
 \begin{split}
  |\Psi\rangle_{out}=& \frac{1}{\sqrt{N+1}}|-1,-2\rangle|\eta_0\rangle\\
  & +\frac{1}{N+1}\sum_{j,k=1}^{N} e^{i\phi_j}e^{2\pi ijk/(N+1)}|B,k\rangle|\eta_j\rangle\\
  & +\frac{1}{N+1}\sum_{j=1}^{N}|\eta_j\rangle e^{i\phi_j}|B,N+1\rangle,
 \end{split}
\end{equation}
then the the probability of finding the particle on the edge between
$B$ and $N + 1$ is
\begin{equation}\label{eq-dj3}
 \begin{split}
  p &= |\langle B,N+1|\Psi\rangle_{out}|^2 \\
    &= \frac{1}{(N+1)^2}\sum_{j,k=1}^{N} e^{i(\phi_j-\phi_k)}\langle\eta_k|\eta_j\rangle\\
    &\leq \frac{N}{(N+1)^2}+\frac{C_{l_1}(|\Psi\rangle_{in})}{N+1}.
 \end{split}
\end{equation}

Clearly, the amount of coherence in the system limits our ability to
distinguish between the constant case (i.e., all $\phi_j$ are the
same and thus $p$ takes the maximum value) and the balanced case
(half of $\phi_j$ are zero and half of $\phi_j$ are $\pi$, thus $p$
takes the minimum value).

When one have not detect the particle in the edge between $B$ and
$N+1$, one can guess we have the balanced case. \citet{DJ-coherence}
calculated the error probability for the classical and quantum
Deutsch-Jozsa algorithm after $m$ trials of the discrete quantum
walk experiments, and found they are given respectively, by
\begin{equation}\label{eq-dj4}
 p_{error}^{\mathrm{class}} = \frac{1}{2^m},~
 p_{error}^{\mathrm{quant}} = \frac{1}{2}(1-v)^m,
\end{equation}
with $v=\langle\eta_k|\eta_j\rangle$ is supposed to be positive for
all $j\neq k$. Thus if $v$ is larger than a critical value, the
quantum algorithm always outperforms its classical counterparts.

\subsection{Grover search algorithm} \label{sec:5C}
The Grover search algorithm is another important algorithm in the
developments of quantum information science \cite{Grover}. The
pursue of the reason for the speedup of this algorithm attract
researchers' interest for many years.

For an $N$-qubit database initialized as
\begin{equation}\label{Grover1}
 |\psi_0\rangle=\sqrt{\frac{j}{N}}|X\rangle+ \sqrt{\frac{N-j}{N}}|X^\perp\rangle,
\end{equation}
where $|X\rangle=\sum_{x_s}|x_s\rangle/\sqrt{j}$,
$|X^\perp\rangle=\sum_{x_n}|x_n\rangle/\sqrt{N-j}$, and $j$
represents the number of solutions. To optimize the success
probability, one can perform the Grover operation
\begin{equation} \label{Grover2}
 G=OD,
\end{equation}
with $O=\iden- 2|X\rangle\langle X|$ and $D=2|\psi_0\rangle
\langle\psi_0|-\iden$. After $r$ iterations of the Grover operation
$G$, the initial state $|\psi_0\rangle$ turns to be
\begin{equation}\label{Grover3}
 |\psi_r\rangle\equiv G^r|\psi_0\rangle=\sin\alpha_r|X\rangle+ \cos\alpha_r|X^\perp\rangle,
\end{equation}
with $\alpha_r=(2r+1)\arctan \sqrt{j/(N-j)}$. The success
probability for finding the correct result is $p(r)=\sin^2
\alpha_r$, and the optimal times of search is given by
$r_{\mathrm{opt}}=CI[(\pi-\alpha)/2\alpha]$, with $CI[x]$ denoting
the closest integer to $x$.

\citet{Liusy} calculated the relative entropy and the $l_1$ norm of
coherence for $|\psi_r\rangle$, and found that the success
probability $p(r)$ depends on the amount of quantum coherence
remaining in $|\psi_r\rangle$. To be explicit,
\begin{equation} \label{Grover4}
 \begin{aligned}
 & C_r(|\psi_r\rangle)=H(p)+\log_2 (N-j)+p\log_2\frac{j}{N-j},\\
 & C_{l_1}(|\psi_r\rangle)=\left[\sqrt{jp}+\sqrt{(N-j)(1-p)}\right]^2-1,
 \end{aligned}
\end{equation}
both of which decrease with the increasing value of $p$. Therefore,
the larger the quantum coherence depletion (or equivalently, the
less the remaining quantum coherence in $|\psi_r\rangle$), the
bigger the success probability one can obtain.

\citet{Liusy} also calculated the quantum coherence depletion for
the generalized Grover search algorithm \cite{Grover2}, and found
that the required optimal search time may increase with the
increasing quantum coherence depletion. Moreover, quantum
correlations such as quantum entanglement and QD cannot be directly
related to the success probability or the optimal search time.

\subsection{Deterministic quantum computation with one qubit} \label{sec:5D}
The DQC1 algorithm is the first algorithm that shows quantum
computation can outperform those of the classical computation even
without entanglement \cite{DQC1-2,DQC1-3}. The standard DQC1
algorithm starts with an initial product state $|0\rangle\langle
0|\otimes (\iden/2^n)$, and then it was transformed into
\begin{equation}\label{eq-dqc1}
\tilde{\rho}^{AR}= \frac{1}{2}\left(\iden_2\otimes
\frac{\iden_{2^n}}{2^n}+|0\rangle\langle 1|\otimes
\frac{U^\dag}{2^n}+|1\rangle\langle 0|\otimes \frac{U}{2^n}\right),
\end{equation}
after performing a Hadamard operation on the first qubit, who then
served as the control qubit when a controlled unitary operation $U$
is performed on the target qubits in the maximally mixed state
$\iden_{2^n}/2^n$ \cite{DQC1}. The goal of this algorithm is to
estimate the normalized trace of $U$.

As the reduced states of the control qubit after the series
operations is given by
\begin{equation}\label{eq-dqc2}
 \tilde{\rho}^A=\frac{1}{2}
              \left(\begin{array}{cc}
              1  &   \frac{\tr U^\dag}{2^n} \\
              \frac{\tr U}{2^n}    &   1
              \end{array}\right),
\end{equation}
the estimation can be finalized by measurements of the ancilla in an
appropriate basis, i.e., $\langle \sigma_x+ i\sigma_y
\rangle_{\tilde{\rho}_A}=\tr U/ 2^n$.

For the above fashion of DQC1, several works have been undertaken to
understand the origin of its superiority over the classical
algorithm \cite{DQC1-mid,DQC1-qd,GQD}. Recently, \citet{Mile}
provided a further viewpoint of its superiority from the perspective
of quantum coherence. By choosing the computational basis for the
ancilla qubit and the the eigenbasis of $U$ for the register as the
reference basis, they studied how the interplay between coherence
consumption and creation of QD works in DQC1, and showed that
\begin{equation} \label{eq-dqc3}
 \bar{D}(\tilde{\rho}^{AR})\leq \delta C(\rho^A), ~
 \bar{D}_{R|A}(\tilde{\rho}^{AR})\leq \delta C(\rho^A),
\end{equation}
which are direct consequences of Eqs. \eqref{eq3b-4} and
\eqref{eq3b-as4}. When being measured by the relative entropy, the
coherence consumption can be obtained (note that $\rho^A$ is
maximally coherent) from Eq. \eqref{eq-dqc1} as
\begin{equation} \label{eq-dqc4}
 \delta C(\rho^A)=C_r(\rho^A)-C_r(\tilde{\rho}^A)=H_2\left(\frac{1-|\tr U|/2^n}{2}\right),
\end{equation}
where $H_2(\cdot)$ is the binary Shannon entropy function. It shows
that the speedup of this algorithm always corresponds to the
consumption of quantum coherence in the ancilla. When there is no
coherence to be consumed, we must have $|\tr U|=2^n$, and thus
$U=e^{i\phi}\iden$ for some $\phi$.

By considering a duplication of the DQC1 protocol termed as nonlocal
deterministic quantum computation with two qubits (NDQC2), i.e., the
collaborative task of estimating the product of normalized traces of
two unitaries without obtaining the individual trace value of each
unitary, \citet{NDQC2} found that its computational advantage can be
achieved with quantum states that have no quantum entanglement and
QD. To interpret this phenomenon, they introduced an operational
definition of nonclassical correlations, that is, a state
$\rho^{AB}$ is said to be nonclassical if it enables a collaborative
task only using correlated inputs and measurement results of
correlations more efficiently than any classical algorithm. Based on
this framework, they defined
\begin{equation} \label{eq-dqc5}
 C^{\mathrm{net}}(\rho^{AB})=C(\rho^{AB})-C(\rho^A)-C(\rho^B),
\end{equation}
and suggested that this quantity can be used for interpreting the
efficiency of the NDQC2 protocol, as its quantum advantage is
achieved only when $C^{\mathrm{net}}(\rho^{AB})>0$.

For the relative entropy of coherence, from Eq. \eqref{eq3d-1} it is
clearly that $C_r^{\mathrm{net}}(\rho^{AB})\geq 0$, and it takes the
maximum for the maximally coherent states of the form of Eq.
\eqref{Peng}. Moreover, $\rho^{AB}$ is a classical-classical state
if and only if $C^{\mathrm{net}} (\rho^{AB})=0$ for certain
reference bases. Operationally, when there are no local coherence,
i.e., when $C(\rho^A)=C(\rho^B)=0$, two spatially separated parties
(Alice and Bob) cannot distil quantum coherence on neither sides
using LICC if and only if $C_r^{\mathrm{net}}(\rho^{AB})=0$. This
shows another physical implication of the net global coherence as a
primitive property of quantum systems which is distinct from those
captured by entanglement or QD.

Based on the aforementioned facts, \citet{NDQC2} also gave a basis
dependent characterization of nonclassical states, that is, a state
$\rho^{AB}$ is said to be nonclassical if and only if
\begin{equation}\label{eq-dqc6}
 C_{r,\max}^{\mathrm{net}}(\rho^{AB})= \max_{\{|i\rangle_A|j\rangle_B\}}
                                       C_r^{\mathrm{net}}(\rho^{AB})
                                     >0,
\end{equation}
and it vanishes only for the product states $\rho^{AB}=\rho^A\otimes
\rho^B$. On the contrary, as $C_r^{\mathrm{net}}(\rho^{AB})=
I(\rho^{AB})- I(\Delta[\rho^{AB}])$, then when it is minimized over
the reference bases, we obtain the symmetric discord \cite{RQC}, see
Eq. \eqref{eq3-9}.

\subsection{Quantum metrology} \label{sec:5E}
Considering an explicit metrology task, i.e., the phase
discrimination (PD) game. In this game, a particle in the state
$\rho\in \mathcal{D} (\mathbb{C}^d)$ passes through a black box,
after which an unknown phase was encoded to it as $U_\phi\rho
U_\phi^\dag$, with $U_\phi=\sum_{j=0}^{d-1} e^{ij\phi}|j\rangle
\langle j|$, $\phi\in \mathbb{R}$, and $\{|j\rangle\}$ being the
reference basis. For a collection of pairs $\Theta=
\{p_k,\phi_k\}_{k=0}^{m-1}$, the goal of the PD game is to predict
the phases $\{\phi_k\}$ with success probability as high as
possible. In general, the optimal probability can be obtained by
optimizing over all measurements $\{M_k\}$, and it is given by
\begin{equation} \label{eq-metrology}
 p_\Theta^{\mathrm{succ}}(\rho)=\max_{\{M_k\}}\sum_k p_k \tr[U_{\phi_k}\rho
U_{\phi_k}^\dag \rho],
\end{equation}

\citet{meas6} showed that for the above metrology task, the optimal
probability can be linked to RoC of the state $\rho$, i.e.,
\begin{equation} \label{metro1}
 \max_{\Theta}\frac{p_\Theta^{\mathrm{succ}}(\rho)}{p_\Theta^{\mathrm{succ}}(\mathcal{I})}
 =1+C_R(\rho),
\end{equation}
where the maximum is achieved for $\Theta^\star= \{1/d,2\pi
k/d\}_{k=0}^{d-1}$, and $p_\Theta^{\mathrm{succ}}(\mathcal{I})$ is
the corresponding classical probability obtained only by guessing.
Therefore, $C_R(\rho)$ quantifies the quantum advantage of the PD
task, thus suggests a prominent role of RoC in quantum information
processing.

\citet{Yucsskw} also investigated a similar metrology task. The
difference is that they linked the skew-information-based coherence
measure $C_{sk}(\rho)$ in Eq. \eqref{eq2f-6} to uncertainty of the
estimated phase. By using the quantum Cram\'{e}r-Rao bound
\begin{equation} \label{metro2}
 (\delta\phi_k)^2 \geq \frac{1}{N \mathcal{F}(\rho_\phi)},
\end{equation}
with $\mathcal{F}(\rho_\phi)$ being the quantum Fisher information
given in Eq. \eqref{QFI}, they showed that
\begin{equation} \label{metro3}
 \frac{1}{4N C_{sk}(\rho)}\leq \sum_k (\delta\phi_k^\star)^2 \leq \frac{1}{8N C_{sk}(\rho)},
\end{equation}
where $\delta\phi_k^\star$ denotes the optimal variance. It shows
that the measurement precision can be increased by increasing RoC of
the state $\rho$.

By considering the subchannel discrimination task which is a
generalization of the PD task, \citet{Buprl} provided an operation
interpretation for the maximum relative entropy of coherence defined
in Eq. \eqref{eq-mre1}. Here, the subchannel refers to a map that is
linear completely positive and trace nonincreasing, and an
instrument $\mathfrak{I}$ for a channel $\mathcal{E}= \sum_a
\mathcal{E}_a$ is a collection of subchannels $\{\mathcal{E}_a\}$.
The optimal probability of successfully discriminating the
subchannels in $\mathfrak{I}$ reads
\begin{equation} \label{metro4}
 P_\mathfrak{I}^{\mathrm{succ}}(\rho)= \max_{\{M_k\}}\sum_a \tr[\mathcal{E}_a(\rho)M_k],
\end{equation}
where the optimization is taken over the POVM $\{M_k\}$. If we are
restricted to the set of incoherent states, the resulting optimal
probability turns out to be $P_\mathfrak{I}^{\mathrm{succ}}(\mathcal
{I})= \max_{\sigma\in \mathcal {I}}P_\mathfrak{I}^{\mathrm{succ}}
(\sigma)$. \citet{Buprl} showed that the optimal maximum advantage
achievable in subchannel discrimination can be characterized by the
maximum relative entropy of coherence. To be precise, we have
\begin{equation} \label{metro5}
 2^{C_{\max}(\rho)}=\max_{\mathfrak{I}} \frac{P_{\mathfrak{I}}^{\mathrm{succ}}(\rho)}
                                      {P_{\mathfrak{I}}^{\mathrm{succ}}(\mathcal{I})},
\end{equation}
which is very similar to Eq. \eqref{eq-metrology} as $C_\mathrm{max}
(\rho)$ is connected to $C_R(\rho)$ via $C_\mathrm{max}(\rho)=\log_2
[1+C_R(\rho)]$, see Eq. \eqref{eq-mre2}.

\section{Quantum correlations and coherence under quantum channels} \label{sec:6}
As a precious physical resource for implementing quantum computation
and communication tasks that are otherwise impossible classically,
and due to the obvious fact that nearly all quantum systems are
inevitably interact with their surroundings which may cause
decoherence and other negative effects, the study of GQD and quantum
coherence, in particular, the control and maintenance of them in
noisy environments, is of equal importance to the study of other
similar problems such as quantum correlation measures
\cite{RMP,Licf}.

\subsection{Frozen phenomenon of QD and quantum coherence} \label{sec:6A}

\subsubsection{Freezing of quantum discord}
\citet{frozen-epl} investigated the family of local quantum channels
under the action of which the QD is preserved for all bipartite
states. By using a result of \citet{Petz} which says that
\begin{equation} \label{free1}
 S(\rho\|\sigma)= S(T[\rho]\|T[\sigma]),
\end{equation}
if and only if the map $\rho\mapsto T[\rho]$ and $\sigma\mapsto
T[\sigma]$ are invertible, they showed that the
mutual-information-based QD is frozen for all states if and only if
the channels are invertible. Explicitly, by denoting $\Lambda_A$
($\Lambda_B$) the quantum channel acting on party $A$ ($B$), then
\begin{equation} \label{free2}
 D_A(\rho^{AB})=D_A(\Lambda_A\otimes\Lambda_B [\rho^{AB}]),
\end{equation}
if and only if there exists $\Lambda_A^*$ and $\Lambda_B^*$ such
that
\begin{equation}\label{fro-01}
 \begin{aligned}
  &(\Lambda_A^*\otimes\Lambda_B^*)(\Lambda_A\otimes\Lambda_B)[\rho^{AB}]=\rho^{AB},\\
  &(\Lambda_A^*\otimes\Lambda_B^*)(\Lambda_A\otimes\Lambda_B) [\rho^A\otimes\rho^B]=
  \rho^A\otimes\rho^B.
\end{aligned}
\end{equation}

Moreover, for a distance measure of two states that is monotonic
under the action of quantum channel, the corresponding GQD defined
based on it is frozen if and only if the local quantum channels
$\Lambda_A$ and $\Lambda_B$ are invertible. The related distance
measures include those based on the trace norm and Uhlmann fidelity.

For certain quantum channels, the QD may be frozen for a restricted
family of states. \citet{frozen-pra} studied such a problem. They
considered the phase damping channel whose action on a state can be
described by $\Lambda_{pd}(\rho)=\sum_i E_i\rho E_i^{\dag}$, with
\begin{equation}\label{eq-pd}
  E_0={\rm diag}\{1,p(t)\},~
  E_1={\rm diag}\{0,\sqrt{1-p^2(t)}\},
\end{equation}
being the Kraus operators, and $p(t)$ a time-dependent parameter
containing the information of the channel. For the initial
Bell-diagonal states $\rho^{\rm Bell}$ of Eq. \eqref{eq-bell} with
one subsystem subjecting to the channel $\Lambda_{pd}$, they
obtained necessary and sufficient conditions for freezing QD, which
are given in terms of the triple $(c_1,c_2,c_3)$. Explicitly, the QD
in $\rho^{\rm Bell}$ is frozen if and only if
\begin{equation}\label{fro-02}
 \begin{aligned}
  &c_2=-c_1c_3,~ |c_1| > |c_3| \\
  \rm {or}~ & c_1=-c_2c_3, ~ |c_2| >|c_3|.
\end{aligned}
\end{equation}
The above condition is also of special importance for studying the
universal freezing of geometric quantum correlations
\cite{frozen-universal}. Besides these, they also generalized their
results to an extended family of two-qubit states
\begin{equation}\label{free3}
  \rho=\rho^{\rm Bell}+\frac{1}{4}(c_{12}\sigma_1\otimes\sigma_2+ c_{21} \sigma_2\otimes
\sigma_1),
\end{equation}
and obtained a similar necessary and sufficient conditions.

\citet{frozen-ohmic} also studied the frozen phenomenon of QD in
dephasing reservoir. The difference is that they considered the
explicit Ohmic-type spectrum given by
\begin{equation} \label{free4}
  J(\omega)=\omega^s \omega_c^{1-s}e^{-\omega/\omega_c},
\end{equation}
with $\omega_c$ being cutoff frequency of the reservoir, and it is
said to be sub-Ohmic if $0<s < 1$, Ohmic if $s = 1$, and super-Ohmic
if $s > 1$. For a subset of the initial Bell-diagonal state
\begin{equation} \label{free5}
 \rho^{\rm Bell}_{\rm sub}=\frac{1+c}{2}|\Psi^{\pm}\rangle
\langle\Psi^{\pm}|+\frac{1-c}{2}|\Phi^{\pm}\rangle
\langle\Phi^{\pm}|,
\end{equation}
with $|\Psi^{\pm}\rangle=(|00\rangle\pm |11\rangle)/\sqrt{2}$ and
$|\Phi^{\pm}\rangle=(|01\rangle\pm |10\rangle)/\sqrt{2}$, they
obtained the expression of the evolved QD, and found that if
$e^{-\Lambda(\bar{t})}=c$ [$\Lambda(t)$ is the dephasing factor],
then there will be a transition from classical decoherence to quantum
decoherence. But if there is no solution for $e^{-\Lambda(\bar{t})}
=c$, the QD will be frozen forever, with the time-invariant value
\begin{equation} \label{free6}
 D_A(\rho^{\rm Bell}_{\rm sub})=\frac{1+c}{2}\log_2(1+c) +\frac{1-c}{2} \log_2 (1-c).
\end{equation}
The $(s,c)$ region for which the frozen condition is satisfied is
determined by temperature of the reservoir. For the zero-temperature
case, they obtained numerically the corresponding $(s,c)$ region,
which shrinks with the increase of $c$ and vanishes when $c\gtrsim
0.16$.

For two qubits prepared initially in the Bell-diagonal states
described by the triple $(c_1,c_2,c_3)$, there may be universal
freezing of GQD defined based on the distance measures $\mathcal{D}$
of quantum states that satisfy the following conditions:
(\romannumeral+1) contractivity under CPTP maps, (\romannumeral+2)
invariant under transposition, and (\romannumeral+3) convexity under
mixing of states. Distance of this type include the relative
entropy, the squared Bures distance, the squared Hellinger distance,
and the trace distance.

From the above conditions, \citet{frozen-universal} considered the
initial Bell-diagonal state $\rho^{\rm Bell}$ of Eq. \eqref{eq-bell}
with the triple $(c_1,-c_1c_3,c_3)$, and proved that its distance to
$(c_1,0,0)$ [$(0,0,c_3)$] is independent of $c_1$ ($c_3$). Moreover,
one of the closest classical state to $\rho^{\rm Bell}$ is still a
Bell-diagonal state $(s_1,s_2,s_3)$ with however only one of $s_k$
is nonzero, and for the special case $c_2=-c_1c_3$, the closest
classical state further reduces to $(c_1,0,0)$ if $|c_1|\geqslant
|c_3|$, and $(0,0,c_3)$ otherwise. This extends the results of Eq.
\eqref{fro-02}. As it shows Eq. \eqref{fro-02} holds for general
distance measure of states satisfying the above three conditions.

Based on these formulas, \citet{frozen-universal} found that when
the two qubits are subject to independent phase flip (similar for
bit flip and bit-phase flip) channels, the GQDs satisfying the above
three conditions will be frozen in the time interval
$t<t^*=-(1/2\gamma) \ln(|c_3(0)|/ |c_1(0)|)$. As this conclusion
depends only on the proposed properties of distance measures of
states, it shows the universal freezing of geometric quantum
correlations.

\citet{traceqd1} studied the trace norm of discord for a two-qubit
system (initially prepared in the Bell-diagonal state) passes
through the local bit flip, phase flip, bit-phase flip, and
generalized amplitude damping channels. Through detailed analysis
with different initial state parameters $(c_1,c_2,c_3)$, they found
that the trace norm of discord exhibits the phenomenon of freezing
behavior during its evolution process. \citet{traceqd2} discussed
the trace norm of discord with the locally applied phase-flip
channels and random external fields and observed the freezing
phenomenon. They also compared dynamics of the total and classical
classical correlations defined via the trace norm, see Eq.
\eqref{eq1a-5}. Moreover, the trace norm, Bures distance, and
Hellinger distance measure of GQD for two non-interacting qubits
subject to two-sided and one-sided thermal reservoirs have also been
investigated \cite{husun}. In fact, the frozen phenomenon of various
GQDs were proved to be universal by \citet{bures2} and
\citet{frozen-universal}.

\subsubsection{Freezing of quantum coherence}
For a $N$-qubit quantum system subject to local independent and
identical decohering environments, \citet{froz1} studied decay
dynamics of coherence and provided important insights between them
and the discordlike correlation measures. The extension of
two-qubit Bell-diagonal states, i.e.,
\begin{equation}  \label{cfree1}
 \rho^{\rm Bell}_N= \frac{1}{2^N}\left(\iden_2^{\otimes N}+\sum_{i=1}^3 c_i
                    \sigma_i^{\otimes N}\right),
\end{equation}
are also described by the triple $(c_1,c_2,c_3)$. For a system with
even number of qubits, \citet{froz1} found that if
\begin{equation} \label{cfree2}
 c_2= (-1)^{N/2}c_1 c_3,
\end{equation}
then all bona fide distance-based coherence measures will be
permanently frozen for local bit flip channel (similar result can be
obtained for local bit-phase flip channel by exchanging $c_1$ and
$c_2$). These include the relative entropy of coherence for general
even $N$, and the trace norm of coherence for $N=2$.

Moreover, for general one-qubit state (i.e., $\rho^{\rm Bell}_N$
with $N=1$) subject to bit flip channel, the $l_1$ norm of coherence
is frozen forever if $c_2=0$, while for the two-qubit state of Eq.
\eqref{gqd-ad1} with the elements of $T$ vanishing for all
non-diagonal elements, it is frozen when the parameters
\begin{equation} \label{cfree3}
 x_2=y_2=0,~T_{22}=uT_{11},~~(-1\leq u\leq 1).
\end{equation}

Experimentally, the freezing phenomenon for relative entropy of
coherence \cite{coher}, fidelity-based measure of coherence
\cite{meas2}, and trace norm of coherence \cite{froz1} for two and
four qubits exposing to the phase damping channel were observed in
an nuclear magnetic resonance system \cite{expfreeze}.

If an incoherent operation satisfy not only $K_i\mathcal{I}
K_i^\dag\subset \mathcal{I}$, but also the additional constraint
$K_i^\dag \mathcal{I} K_i\subset \mathcal{I}$ for all $K_i$, then it
is said to be strictly incoherent \cite{qcd1}. Their Kraus operators
contain at most one nonzero entry in each row and each column, and
incoherent channels of such type cover the paradigmatic source of
noises in quantum information science, e.g., the bit flip, phase
flip, bit-phase flip, depolarizing, amplitude damping, and phase
damping channels.

By restricting to strictly incoherent channels, \citet{froz2}
established a measure-independent freezing condition of coherence,
which states that for any initial state of a system, all measures of
its coherence are frozen if and only if its relative entropy of
coherence is frozen. The proof for this claim comprises two
essential steps. First, if $\delta^\star$ is the closest incoherent
state to $\rho$ in the definition of $C_r(\rho)$, then
$\Lambda[\delta^\star]$ is the closest state to $\Lambda[\rho]$ for
$C_r(\Lambda[\rho])$. Second, if the channel maps $\rho(0)$ to
$\rho(t)$, i.e., $\Lambda[\rho(0)]=\rho(t)$, then one can always
construct an incoherent operation which gives the map $\mathcal
{R}[\rho(t)]=\rho(0)$.

For a system of $N$ qubits interacting independently with $N$ bit
flip (not necessary to be identical) channels, \citet{froz2} further
identified two families of states for which all measures of quantum
coherence are frozen, they are given respectively by:

(1) $|\varphi^{\pm}_l\rangle=(|l\rangle\pm |\bar{l}\rangle)
/\sqrt{2}$, with the sequences $l=l_1 l_2 \cdots l_N$,
$\bar{l}=\bar{l}_1 \bar{l}_2 \cdots \bar{l}_N$, $l_1=0$, $l_{i\neq
1}=\{0,1\}$, $\bar{l}_i=1-l_i$.

(2) $\rho=\sum_l p_l [p|\varphi^{+}_l\rangle
\langle\varphi_l^{+}|+(1-p)|\varphi^{-}_l \rangle\langle
\varphi_l^{-}|]$, with $0\leq p\leq 1$, and $p_l$ being any
probability distribution.

By decomposing state $\rho$ as Eq. \eqref{eq2d-4}, and using a
transformation matrix $T$ to describe the action of $\mathcal {E}$,
i.e.,
\begin{equation}\label{eq-master}
 \mathcal{E}^\dag (X_i)=\sum_j T_{ij}X_j,
\end{equation}
\citet{fac3} also derived a condition for freezing the $l_1$ norm of
coherence. They found that when $T_{k0}=0$ for $k\in
\{1,2,\ldots,d^2-d\}$, and $T^S$ (the submatrix of $T$ consisting
$T_{ij}$ with $i$ ranging from 1 to $d^2-d$ and $j$ from 1 to
$d^2-1$) is a rectangular block diagonal matrix, with the main
diagonal blocks
\begin{equation}\label{eq5a-1}
 T^S_r =\left(\begin{array}{cc}
              T_{2r-1,2r-1}  &   T_{2r-1,2r} \\
              T_{2r,2r-1}    &   T_{2r,2r}
              \end{array}\right)
              ~(r\in \{1,\dots,d_0\}),
\end{equation}
being orthogonal matrices, i.e., $(T^S_r)^T T^S_r = \iden_2$, the
$l_1$ norm of coherence for $\rho^{\hat{n}}$ will be frozen during
the entire evolution. Here, $\rho^{\hat{n}}$ represents states with
the characteristic vectors $\vec{x}$ [see Eq. \eqref{eq2d-4}] along
the same or completely opposite directions but possessing different
lengths.

\subsection{Enhancing the quantum resources via quantum operations} \label{sec:6B}
Since QD and quantum coherence are both quantum properties of
quantum states, the ability of a quantum channel to create and/or
enhance strength of QD or quantum coherence is related to the
quantumness of the channel. It is then of interest to study whether
a channel has the ability to create quantum resources, and how many
quantum resources the channel can create or enhance.

\subsubsection{Creation of quantum discord from classical states}
When a bipartite system is coupled to a common bath, it was proved
that a Markovian dissipative quantum channel can generate QD from
some bipartite product states if and only if it cannot be reduced to
individual decoherence channels independently acting on each qudit
\cite{create-collective}. Further, if the subsystems initially share
classical correlations, even local operations can create QD.

The local creation of quantum correlations was first studied by
\citet{create-prl}. They investigated the completely decohering (or
semiclassical) channel $\Lambda_{sc}$ described by
\begin{equation} \label{semic}
 \Lambda_{sc}(\rho)=\sum_k p_k(\rho)|k\rangle\langle k|,
\end{equation}
and the unital channel $\Lambda_{u}$ which keeps the maximally mixed
state invariant. For a single qubit of a multiqubit system subject
to a channel, they proved strictly that $\Lambda_{sc}$ and
$\Lambda_u$ are the only two types of channels that cannot create
quantum correlations. Equivalently, for qubit systems, a necessary
and sufficient condition for a local channel to create quantum
correlation is that the channel is neither completely decohering nor
unital. The geometric quantum correlations defined based on
contractive distance measures of states are further showed to be
nonincreasing under local semiclassical channels and local unital
channels. However, this result does not hold for states with higher
dimensions. For multi-qudit states with dimension $d\geq 3$, even
unital channels can create quantum correlation.

\citet{create-huxy} defined commutativity-preserving channels
$\Lambda_{cp}$ as those preserve the commutativity of any two input
states, that is,
\begin{equation} \label{eq-cp}
 [\rho, \sigma]=0\Rightarrow [\Lambda_{cp}(\rho),\Lambda_{cp}(\sigma)]=0.
\end{equation}
For any finite-dimensional multipartite systems, they proved that a
channel $\Lambda$ acting locally on $B$ of a quantum-classical state
can create quantum correlation if and only if $\Lambda\notin
\Lambda_{cp}$. For qubit case, a commutativity-preserving channel is
either a completely decohering channel or a unital channel. For
qutrit case, a commutativity-preserving channel is either a
completely decohering channel or an isotropic channel, which is
defined as
\begin{equation}\label{cre-01}
 \Lambda_{iso}(\rho)=p\Gamma(\rho)+(1-p) \frac{\iden}{d},
\end{equation}
with $p$ being the parameter for ensuring CPTP of $\Lambda_{iso}$,
and $\Gamma$ is either a unitary operation [$-1/(d-1)\leqslant p
\leqslant 1$] or is unitarily equivalent to transpose [$-1/(d-1)
\leqslant p \leqslant 1/(d+1)$].

\citet{create-huxy} also conjectured that for systems with dimension
higher than 3, a commutativity-preserving channel is also either
isotropic or completely decohering. \citet{create-jpa} further gave
an affirmative answer to this conjecture. They proved that $\Lambda$
acting on party $B$ of a system cannot create QD (i.e.,
$D_B(\rho_{AB})=0 \Rightarrow D_B(\iden\otimes
\Lambda[\rho_{AB}])=0$) if and only if it is either a completely
decohering channel or a nontrivial isotropic channel
($\Lambda_{iso}$ with $p\neq 0$). Channels of these types are also
showed to preserve commutativity and normality of quantum states.
Furthermore, $\Lambda_B$ which yields $D_B(
\rho_{AB})=0\Leftrightarrow D_B(\Lambda_B[\rho_{AB}])=0$ are
restricted only to the nontrivial isotropic channels.

\citet{create-pra} considered the local creation of QD by a
Markovian amplitude-damping channel described by
$\Lambda_{ad}(\rho)=\sum_i E_i\rho E_i^{\dag}$, with
\begin{equation}\label{cre-02}
 E_0=|0\rangle\langle 0|+ \sqrt{1-p(t)} |1\rangle\langle 1|,~
 E_1=\sqrt{p(t)}|0\rangle \langle 1|.
\end{equation}
For the initial state
\begin{equation}\label{cre-03}
\rho =\frac{1}{2}(|0\rangle \langle 0| \otimes \tau_{0}+|1\rangle
\langle 1|\otimes\tau_{1}),
\end{equation}
with the length of local Bloch vectors for $\tau_{0}$ and $\tau_{1}$
equal to each other (i.e., $|\vec{s}_0|=|\vec{s}_1|=s$), they
obtained
\begin{equation}\label{cre-04}
 \begin{aligned}
  D_B(\rho)= & h\left(\frac{1+s|\cos(\varphi/2)|}{2}\right)+
               h\left(\frac{1+s|\sin(\varphi/2)|}{2}\right)\\
             & -h\left(\frac{1+s}{2}\right)-1,
 \end{aligned}
\end{equation}
where $\varphi$ is the angle between $\vec{s}_0$ and $\vec{s}_1$.
Using this formula, they showed rigorously that $\Lambda_{ad}$
acting on $B$ can create QD from $\rho$. In fact, it is easy to
check that $\Lambda_{ad}$ is neither a completely decohering nor a
unital channel, hence the local creation of QD in the present case
is easy to understand from the result of \citet{create-huxy}.

Now we have reviewed the conditions on the quantum channels which
has the ability to increase quantum correlations. An equally
important problem is to characterize the quantum states whose
quantum correlations can be increased locally. \citet{increase-huxy}
studied this problem by employing the tool of quantum steering
ellipsoids \cite{qse}. They considered the amplitude damping channel
acting on qubit $B$ of a Bell-diagonal state of Eq. \eqref{eq-bell}.
For such a state, both $\mathcal E_A$ and $\mathcal E_B$ are unit
spheres shrunk by $c_1,\ c_2$ and $c_3$ in the $x,\ y$ and $z$
direction, respectively. It is observed that, the local increase of
discord occurs when $|c_1|\gg |c_2|,|c_3|$. An interesting
consequence is that, the local quantum operation can increase the QD
of an entangled state.

\subsubsection{Enhancing the coherence via quantum operations}
Quantum coherence measures should be monotonically decreasing under
IO, which are a strict subset of non-coherence-generating (NC)
channels. The behavior of different measures of coherence under the
action of NC channels was studied by \citet{ncgc}. While the
relative entropy of coherence was proved monotone under all NC
channels, the coherence of formation $C_f$ can be increased by some
NC channels. An example was presented that $C_f$ of a two-qubit
state is increased when a NC qubit channel is acting on one of the
two qubits. Here, the NC channel is chosen as $\Lambda(\cdot)=
E_1(\cdot)E_1^\dagger+E_2(\cdot)E_2^\dagger$, with
\begin{equation} \label{ncchannel}
E_1=\frac{1}{2}\left(\begin{array}{cc}
1 & 0\\
-1 & \sqrt2
\end{array}\right),~
E_2=\frac{1}{2}\left(\begin{array}{cc}
1 & \sqrt2\\
1 & 0
\end{array}\right),
\end{equation}
and the two-qubit input state is $\Psi^+\equiv |\Psi^+\rangle
\langle\Psi^+|$ with $|\Psi^+\rangle=(|00\rangle+|11\rangle)/
\sqrt{2}$. It was checked that the $C_f$ of the output state is
strictly larger than the input state, i.e., $C_f(\iden\otimes
\Lambda(\Psi^+))>1=C_f(\Psi^+)$. Interestingly, the channel
$\Lambda(\cdot)$ can never increase $C_f$ of a single-qubit state,
so the ability of this channel to increase coherence is enhanced
when extending to composed Hilbert space. The reason for this
enhancement is that, the local NC operation turn the quantum
correlation into the local coherence, and meanwhile increase the
quantum coherence of the total state.

Although the amount of coherence for a state cannot be enhanced
under IO by definition, this does not prevent us from obtaining
probabilistically a postmeasurement state with enhanced coherence
when selective measurements are allowed, e.g., by retaining those
$\rho_n=K_n\rho K_n^\dagger/p_n$ [$p_n=\tr (K_n\rho K_n^\dagger)$]
that satisfy $C(\rho_n)> C(\rho)$ and discarding the other $\rho_n$,
one can obtain a mixed state $\sum_{n|C(\rho_n)> C(\rho)}p_n\rho_n$
with enhanced coherence. \citet{SSIO} considered one such problem.
By taking the $l_1$ norm of coherence as a measure and considering
the stochastic strictly incoherent operation $\Lambda_s$ whose Kraus
operators $\{K_n\}_{n=1}^L$ (a subset of SIO) fulfilling
$\sum_{n=1}^L K_n^\dagger K_n\leq \iden$, they obtained the maximum
attainable coherence for the postmeasurement state
$\Lambda_s[\rho]$, which reads
\begin{equation} \label{eq-ssio01}
 \max_{\Lambda_s} C_{l_1}(\Lambda_s[\rho])=\lambda_{\max}
 \left(\rho_{\mathrm{diag}}^{-1/2}|\rho|\rho_{\mathrm{diag}}^{-1/2} \right)-1,
\end{equation}
where $\lambda_{\max}(\cdot)$ is the largest eigenvalue of
$\rho_{\mathrm{diag}}^{-1/2}|\rho|\rho_{\mathrm{diag}}^{-1/2}$, and
$|\rho|$ is a matrix obtained from $\rho$ by taking absolute values
to all its elements. \citet{SSIO} also constructed the Kraus
operator and the corresponding optimal probability for obtaining Eq.
\eqref{eq-ssio01}. If $\rho$ is irreducible, by denoting
$|\varphi_{\max} \rangle=(\varphi_1, \varphi_2, \ldots,
\varphi_d)^T$ ($d= \dim\rho$) the eigenvector corresponding to the
largest eigenvalue of $\rho_{\mathrm{diag}}^{-1/2}|\rho|
\rho_{\mathrm{diag}}^{-1/2}$ and $U_{\mathrm{in}}$ an arbitrary
incoherent unitary matrix, one has
\begin{equation} \label{eq-ssio02}
 \begin{aligned}
  & K'= \min_i \frac{\sqrt{\rho_{ii}}}{\varphi_i} U_{\mathrm{in}}\mathrm{diag}
        \left( \frac{\varphi_1}{\sqrt{\rho_{11}}}, \frac{\varphi_2}{\sqrt{\rho_{22}}},
        \ldots,\frac{\varphi_d}{\sqrt{\rho_{dd}}}\right), \\
  & p_{\max}(\rho)= \min_i \frac{\rho_{ii}}{\varphi_i^2},
 \end{aligned}
\end{equation}
and if $\rho$ is reducible, i.e., it can be transformed by a
permutation matrix into $p_1\rho_1 \oplus p_2\rho_2\oplus \ldots
\oplus p_n\rho_n \oplus \bold{0}$, one has
\begin{equation} \label{eq-ssio03}
 \begin{aligned}
  & K'= U_{\mathrm{in}}(K'_1\oplus K'_2 \oplus\ldots\oplus K'_n \oplus\bold{0}), \\
  & p_{\max}(\rho)= \sum_{\lambda_{\max}^\alpha
                  = \lambda_{\max}}p_\alpha p_{\max}(\rho_\alpha),
 \end{aligned}
\end{equation}
with
\begin{equation}  \label{eq-ssio04}
  K'_\alpha=\min_i \frac{\sqrt{\rho_{ii}^\alpha}}{\varphi_i^\alpha}
            \mathrm{diag} \left( \frac{\varphi_1^\alpha}{\sqrt{\rho_{11}^\alpha}},
            \frac{\varphi_2^\alpha}{\sqrt{\rho_{22}^\alpha}},\ldots,\frac{\varphi_d^\alpha}
            {\sqrt{\rho_{dd}^\alpha}}\right).
\end{equation}

\subsubsection{Energy cost of creating quantum coherence}
As it has been shown in the above sections, quantum operations
acting on an incoherent state can map it to be incoherent. The
amount of quantum coherence in a non-maximally coherent state can
also be enhanced via some quantum operations. When the quantum
operations are restricted to be unitary, it has been shown by
\citet{RQC}, \citet{maxco}, and \citet{Yucs} that the maximal
achieved relative entropy of coherence is $\log_2 d-S(\rho)$ for any
initial state $\rho$, see Eq. \eqref{eq-maxrc}. When $\rho$ is
incoherent, this is also the maximal coherence created by unitary
operation.

In a similar manner, \citet{energy} also considered the maximal
creation of quantum coherence. They considered the initial state to
be the thermal state of a system, and adopted eigenbasis
$\{|j\rangle\}$ of the system Hamiltonian $\hat{H}$ as the reference
basis. The initial thermal state $\rho^T=e^{-\hat{H}/T}$ ($T$ is the
temperature) before acting the unitary operation is incoherent. The
maximum amount of relative entropy of coherence created by using
unitary operations thus has the same form as Eq. \eqref{eq-maxrc},
i.e., $C_r^{\rm max}(\rho^f)=\log_2 d-S(\rho^T)$, where
$\rho^f=U\rho^T U^\dag$ denotes the output state after performing
the unitary operation.

To construct the corresponding optimal unitary operations,
\citet{energy} used the maximally coherent basis
$\{|\phi_j\rangle\}$ which are very similar to that of the maximally
entanglement basis (e.g., for the two-qubit case they are the four
Bell states). Here, all $\{|\phi_j\rangle\}$ have maximal value of
coherence, and they are orthogonal to each other. To be explicit,
they can be written as
\begin{equation} \label{eq-mcb1}
 |\phi_j\rangle=\frac{1}{\sqrt{d}}\sum_{m=0}^{d-1} e^{i\frac{2\pi jm}{d}}|m\rangle,
\end{equation}
which is in fact a map of $\mathbb{Z}= \sum_m e^{2\pi i m/d}
|m\rangle\langle m|$ on the maximally coherent state
$|\Psi_d\rangle$, and $i$ in the superscript is the imaginary unit.
Then they gave an unitary operation
\begin{equation}  \label{eq-mcb2}
 U=\sum_{j=0}^{d-1} |\phi_j\rangle\langle j|,
\end{equation}
for which the output state after the action of it is given by
\begin{equation}  \label{eq-mcb3}
 \rho^f=\frac{1}{Z}\sum_{j=0}^{d-1} e^{-E_j/T}|\phi_j\rangle\langle\phi_j|,
\end{equation}
thus $S(\rho^f_\mathrm{diag})=\log_2 d$.

On the other hand, during these processes of coherence creation and
coherence enhancement, a supply of external energy is needed. Hence,
it is natural to inquire if there are quantitative connections
between the created quantum coherence and the amount of energy cost?
In general, the energy cost is given by $\Delta E=\tr(U\rho U^\dag
\hat{H})- \tr(\rho\hat{H})$.

For the case of initial thermal state $\rho^T$, by using the fact
that $\tr(\rho^f \hat{H})=\tr(\rho^f_\mathrm{diag} \hat{H})$ as
$\hat{H}$ is diagonal, and the maximum entropy principle which says
that the thermal state has maximum entropy among all states with a
fixed average energy \cite{mep1,mep2}, one can show that with
limited energy cost $\Delta E$, the maximum created relative entropy
of coherence is bounded from above by
\begin{equation} \label{eq-bound}
 C_r^{\max}(\Delta E)\leq S(\rho^{T'})-S(\rho^T),
\end{equation}
where $\rho^{T'}$ is the thermal state at the higher temperature
$T'$ such that
\begin{equation} \label{eq-bound2}
 \Delta E=\tr(\rho^{T'}\hat{H})-\tr(\rho^T\hat{T}).
\end{equation}

To obtain the maximal coherence with limited energy $\Delta E$,
i.e., to saturate the upper bound of Eq. \eqref{eq-bound}, one
should find an optimal $U$ such that the diagonal part of $\rho^f$
equals to $\rho^{T'}$. \citet{energy} proved strictly that there
always exists such an (real) unitary.  The derivation of such an
unitary for single-qubit state is easy, but for higher dimensional
case it turns out to be very complicated.

For multipartite system, \citet{energy} also compared the amounts of
quantum coherence and quantum total correlations (measured by the
quantum mutual information) by using the same unitary operations.
For the noninteracting system described by the Hamiltonian
$\hat{H}=\sum_{k=1}^N \hat{H}_k$ ($\hat{H}_k$ is the Hamiltonian for
subsystem $k$), and starting from the initial product thermal states
$\prod_k \otimes \rho_k^{T}$ of each subsystems, the maximum created
quantum mutual information with limited energy $\Delta E$ is given
by
\begin{equation} \label{eq-energy}
 I^{\max}(\Delta E)=\sum_k [S(\rho_k^{T'})-S(\rho_k^T)],
\end{equation}
where the optimal unitary transforms the initial state to a final
state $\rho^f$ whose all marginals are thermal states at a higher
temperature $T'$, i.e., $\rho^{T'}_k=e^{-\hat{H}_k/T'}/\tr
e^{-\hat{H}_k/T'}$.

As $\rho^f_\mathrm{diag}$ and the products of the marginals $\prod_k
\otimes\rho^{T'}_k$ have the same average energy, the maximum
entropy principle implies that $S(\rho^f_\mathrm{diag})\leq
S(\prod_k \otimes \rho_k^{T'})$. Hence, when the maximum correlation
is created among the multipartite system, the corresponding
coherence is upper bounded by it. Contrary, if maximum coherence is
created in the multipartite system, then the diagonal part of the
output $\rho^f$ will be a thermal state at temperature $T'$, and it
has the same average energy with the products of the marginals
$\prod_k\otimes\rho_k^f$ (with $\rho^f_k=\tr_{l\neq k}\rho^f$) due
to the product structure of $\hat{H}$, the maximum entropy principle
implies $S(\rho^f_\mathrm{diag})\geq \sum_k S(\rho^f_k)$. Hence, in
this case the created correlation turns to be bounded from above by
the maximum created coherence. As for the problem of whether the
maximum quantum coherence and correlation can be created
simultaneously, the study of the two-qubit case shows that the
answer to this may be negative \cite{energy}.

\subsection{Resource creating and breaking power} \label{sec:6C}
When a quantum channel has the ability to create or enhance quantum
resources, it is of interest to quantify this ability. This
quantification can be regarded an intrinsic property of the channel.
As a dual problem, the power of a channel to decrease or destroy the
quantum resources, also attracts some research interest.

\subsubsection{Quantum correlating power}
The quantum correlating power (QCP) is defined as the maximum amount
of quantum correlations that can be created when the channel acts
locally on one party of a multipartite system
\cite{PhysRevA.87.032340}, i.e.,
\begin{equation} \label{qcp1}
 \mathcal{Q}(\Lambda)\coloneqq \max_{\rho\in \mathcal{CQ}} Q(\Lambda\otimes \iden(\rho)),
\end{equation}
where $Q$ is a bona fide measure of quantum correlation, and
$\mathcal {CQ}$ denotes the set of classical-quantum states.

The QCP is an intrinsic attribute of a channel, which quantifies the
channels's ability to create quantum correlations. In its
definition, the maximization is taken over the set of
quantum-classical states. The input states that correspond to the
maximization are called the optimal input states, which are proved
to be in the set of classical-classical (CC) states
\begin{equation}  \label{qcp2}
 \CC=\left\{\rho|\rho=\sum_i p_i|\psi_i\rangle_A
     \langle\psi_i|\otimes|\phi_i\rangle_B\langle\phi_i|\right\}.
\end{equation}
where $\{|\psi_i\rangle_A\}$ and $\{|\phi_i\rangle_B\}$ are
orthogonal basis of $\mathcal H_A$ and $\mathcal H_B$, respectively.
The proof can be easily sketched. For any output state $\rho'$ that
corresponds to a general QC input state, one can find a CC state
whose corresponding output state $\rho$ can be transformed to
$\rho'$ by a local channel on $B$, i.e., $\rho'=\iden\otimes
\lambda_B (\rho)$. From the contractivity of the measure $Q$ under
CPTP map, we have $Q(\rho)\geq Q(\rho')$. Hence the definition of
QCP can be optimized to
\begin{equation}  \label{qcp3}
\mathcal Q(\Lambda)\coloneqq \max_{\rho\in\CC}
Q(\Lambda\otimes\iden(\rho)).
\end{equation}

A channel with larger amount of QCP is more quantum, in the sense of
the ability to create quantum correlations. Hence it is of interest
to find out the channels with the most QCP. It can be proved that,
the local single-qubit channel which maximum QCP can be found in the
set of the following channels
\begin{equation}  \label{qcp4}
\mathcal{MP}=\left\{\Lambda|\Lambda(\cdot)=\sum_{i=0}^1
|\phi_i\rangle\langle\alpha_i|(\cdot)|\alpha_i\rangle\langle\phi_i|\right\},
\end{equation}
where $|\phi_0\rangle$ and $|\phi_1\rangle$ are two nonorthogonal
pure states.

\citet{PhysRevA.88.012315} also studied the superadditivity of QCP.
Its says that two zero-QCP channels can constitute a positive-QCP
channel. The phase damping channel $\Lambda_{pd}$ was used as an
example to show how this property works. The corresponding Kraus
operators are given in Eq. \eqref{eq-pd}. $\Lambda_{pd}$ is unital
and thus $\mathcal {Q}(\Lambda_{pd})=0$. For a four-qubit initial
state shared between Alice ($AA'$) and Bob ($BB'$)
\begin{equation}  \label{qcp5}
\rho_{AA'BB'}=\frac14\sum_{i,j}|ij\rangle_{AA'}\langle
ij|\otimes|\psi_{ij}\rangle_{BB'}\langle \psi_{ij}|,
\end{equation}
where
\begin{equation}  \label{qcp6}
 \begin{aligned}
   & |\psi_{00}\rangle=\frac{1}{\sqrt{2}}(|00\rangle+|11\rangle),~
     |\psi_{11}\rangle= \frac{1}{\sqrt{2}}(|0+\rangle+|1-\rangle),\\
   & |\psi_{01}\rangle=\frac{1}{\sqrt{2}}(|01\rangle-|10\rangle),~
     |\psi_{10}\rangle=\frac{1}{\sqrt{2}}(|0-\rangle-|1+\rangle).
 \end{aligned}
\end{equation}
Since $|\psi_{ij}\rangle$ are orthogonal to each other, the quantum
correlation on Bob is zero. After the action of $\Lambda_{pd}^B
\otimes \Lambda_{pd}^{B'}$ on $B$ and $B'$, the output state becomes
$\rho'_{AA'BB'} =\iden_{AA'}\otimes\Lambda_{pd}^B \otimes
\Lambda_{pd}^{B'}(\rho_{AA'BB'})$. Because $[\Lambda_{pd}\otimes
\Lambda_{pd} (\psi_{00}), \Lambda_{pd}\otimes\Lambda_{pd}
(\psi_{11})] =\frac{1}{8}\tilde{i} p\sqrt{1-p} (\sigma^y\otimes
\sigma^z +\sigma_z\otimes\sigma^y)\neq0$, the output state
$\rho'_{AA'BB'}$ is not a QC state. Therefore, the quantum
correlation on Bob's qubits $BB'$ is created by the channel
$\Lambda_{pd}^B \otimes\Lambda_{pd}^{B'}$.

\subsubsection{Cohering and decohering power}
For a system traversing a quantum channel $\mathcal{E}$, the amount
of coherence contained in it may be increased or decreased. Building
upon this fact and in the same spirit as defining entangling power
and discording power [see \citet{dpower} and references therein],
one can also consider the ability of $\mathcal{E}$ on producing or
destroying coherence, and introduce the concepts of cohering and
decohering power of $\mathcal{E}$.

\citet{cpower} defined the cohering power of $\mathcal{E}$ as the
maximum coherence (measured in some way) that it can produce from
the full set $\mathcal{I}$ of incoherent states, and the decohering
power as the maximum amount of coherence lost after all the
maximal-coherence-value states $\rho^{\rm mcs} \in\mathcal{M}$
\cite{mcvs} passing through this channel. To be precise,
\begin{equation}\label{eq5b-3}
 \begin{aligned}
  & {\rm CP}(\mathcal{E})=\max_{\delta\in\mathcal{I}}C(\mathcal{E}[\delta]),\\
  & {\rm DP}(\mathcal{E})=C(\rho^{\rm mcs})-\min_{\rho^{\rm mcv}
               \in \mathcal{M}}C(\mathcal{E}[\rho^{\rm mcs}]),
 \end{aligned}
\end{equation}
and the optimization can in fact be restricted to pure states, in
particular, for ${\rm CP}(\mathcal{E})$ the maximization can be
taken with only the basis states $\{|k\rangle\}$.

Focusing on single-qubit states and coherence measured by the $l_1$
norm and skew information, they calculated ${\rm CP}(\mathcal{E})$
and ${\rm DP}(\mathcal{E})$ for the depolarizing and bit flip
(similarly for bit-phase and phase flip) channels, and showed that
the unitary channel has equal cohering and decohering power in any
basis, i.e., ${\rm CP}(U)={\rm DP}(U)$. Moreover, for $N$ qubits
subjecting to $N$ independent unitary channels, the cohering power
\begin{equation} \label{eq5b-n3}
 {\rm CP}(\otimes_{i=1}^N U_i)= \prod_{i=1}^N [{\rm CP}(U_i)+1]-1,
\end{equation}
while the decohering power is bounded from below by
\begin{equation} \label{eq5b-n4}
 {\rm DP}(\otimes_{i=1}^N U_i)\geq 2^N-\prod_{i=1}^N [2-{\rm DP}(U_i)],
\end{equation}
and apart from the very special case of ${\rm DP}(U_i)=0$, $\forall
U_i$, ${\rm DP}(\otimes_{i=1}^N U_i)$ approaches $2^N-1$ when
$N\rightarrow\infty$, hence the coherence in this state will be
completely deteriorated for infinitely large $N$.

\citet{Bukf} obtained the analytical solutions of the cohering
power. First, when the coherence is measured by the $l_1$ norm, they
showed that
\begin{equation} \label{eq5b-n5}
 {\rm CP}_{l_1}(U)= \|U\|_{1\rightarrow1}^2-1,
\end{equation}
where $\|U\|_{1\rightarrow1}=\max_{1\leq j\leq d}\{\sum_{i=1}^d
|U_{ij}|\}$ is the matrix norm, and for $N$-qubit system, the
Hadamard gate $H^{\otimes N}$ [with $H=(\sigma_x+\sigma_z)/
\sqrt{2}$] was showed to have maximum cohering power. When we adopt
the relative entropy of coherence, it is given by
\begin{equation} \label{eq5b-n6}
 {\rm CP}_r (U)= \max_{1\leq j\leq d}H(|U_{1j}|^2,\ldots,|U_{dj}|^2),
\end{equation}
where $H(p_1,\ldots,p_d)$ denotes the Shannon entropy.

For general quantum channel $\mathcal{E}$, although there is no
analytical solutions for the cohering power, they were showed to
satisfy the additivity relation
\begin{equation} \label{eq5b-n7}
 \begin{aligned}
  & {\rm CP}_{l_1}(\mathcal{E}_1 \otimes \mathcal{E}_2)+1=({\rm
    CP}_{l_1} (\mathcal{E}_1)+1)({\rm CP}_{l_1}(\mathcal{E}_2)+1),\\
  & {\rm CP}_{r} (\otimes_{i=1}^N \mathcal {E}_i)=
    \sum_{i=1}^N {\rm CP}_{r}(\mathcal {E}_i).
 \end{aligned}
\end{equation}

Moreover, one may consider a slightly different definition of
cohering power
\begin{equation} \label{eq5b-n8}
 {\rm CP}^{(\rho)}(\mathcal{E})= \max_{\rho\in \mathcal {D}_\mathcal
{H}} \{C(\mathcal{E}[\rho])-C(\rho)\},
\end{equation}
which characterizes the maximum enhancement of quantum coherence
after the action of the channel $\mathcal{E}$, it was found that for
the 2-dimensional system, the two different cohering powers measured
by the $l_1$ norm are always the same for any unitary channel, but
when $d \geq 3$ or when the coherence is measured by the relative
entropy, they can be different.

On the other hand, we know that the action of $\mathcal{E}$ on
$\rho$ can be implemented by an IO $\Lambda$ on the product state of
$\rho$ and an ancillary state $\sigma$, i.e., $\Lambda(\rho\otimes
\sigma)=\mathcal{E}(\rho)\otimes\sigma'$ [see, e.g., the work of
\citet{coher}]. Start from this point of view, \citet{Bukf} further
gave an interpretation of the cohering power. To be explicit, they
showed that the minimal amount of coherence of $\sigma$ is just the
cohering power of $\mathcal{E}$, i.e., $C(\sigma)\geq {\rm CP}
(\mathcal {E})$.

\citet{cpower2} also explored the cohering and decohering powers of
various typical channels, with however the coherence being measured
by the relative entropy. These include the amplitude damping, phase
damping, depolarizing, as well as the bit flip, bit-phase flip and
phase flip channels. They also found that the cohering power can be
enhanced by applying weak measurement and reversal operation to the
qubit.

For the HS norm measure of quantum coherence, \citet{cgp} discussed
the cohering power of the unitary and unital channels. As the HS
norm is not an monotonic quantity under general CP maps, they
restricted in their work only to those of the unital incoherent CP
maps, under the action of which the HS norm of coherence is
monotonically decreasing. By denoting $\Delta(\rho)$ the full
dephasing of $\rho$ in a given basis [see Eq. \eqref{eq2d-v2}] and
$\tilde{\Delta}=\iden-\Delta$ the the complementary projection of
$\Delta$, they defined the cohering power of the quantum channel as
the average coherence generated from an uniformly distributed
incoherent states, i.e.,
\begin{equation} \label{eq5b-n9}
 C_{av}(\mathcal{E})=\langle C_{l_2}(\mathcal{E}_{\rm off}[\psi])
 \rangle_\psi,
\end{equation}
where $\mathcal{E}_{\rm off}=\tilde{\Delta}\mathcal{E}\Delta$, and
the average is taken over the ensemble of pure states
$\psi=|\psi\rangle\langle\psi|$ sampled randomly from the uniform
Haar measure. That is to say, the uniform ensemble of incoherent
states are generated by dephasing $\{\psi\}$.

When the channel is unitary, they found that the cohering power can
be obtained analytically. For a $d$-dimensional system, it is given
by
\begin{equation} \label{eq5b-n10}
 C_{av}(U)=\frac{1}{d+1}\left(1-\frac{1}{d}\sum_{i,j}|\langle
 i|U|j\rangle|^4\right),
\end{equation}
where the upper bound $C_{av}^{\max}(U)=(d-1)/d(d+1)$ is achieved
when the basis $\{|i\rangle\}$ and $\{U|i\rangle\}$ are mutually
unbiased, and the lower bound $C_{av}(U)=0$ is achieved when $U$ is
an incoherent operation, i.e., $[U,\Delta]=0$.

Similarly, when the channel $\mathcal{E}$ is unital, i.e.,
$\mathcal{E}(\iden/d) =\iden/d$, with $\{A_k\}$ being the
corresponding Kraus operators, the cohering power is given by
\begin{equation} \label{eq5b-n11}
 C_{av}(\mathcal{E})=\frac{1}{d(d+1)}\sum_{i,l\neq m}\Big|\sum_k
 (A_k)_{li}(A_k)^*_{mi}\Big|^2,
\end{equation}
but now $C_{av}(\mathcal{E})=0$ does not always implies $[\mathcal
{E},\Delta]=0$, that is to say, the cohering power
$C_{av}(\mathcal{E})$ for unital channel is not faithful.

In fact, as the unitary operation $U$ is a subset of the unital
operation, $C_{av}(\mathcal{E})$ also covers the result of
$C_{av}(U)$. Moreover, the above equation is equivalent to
\begin{equation} \label{eq5b-n12}
 C_{av}(\mathcal {E})= \frac{1}{d+1}\{\tr[S\tilde{\omega}(\mathcal
{E})]-\tr[S\omega(\mathcal {E})]\},
\end{equation}
where $\tilde{\omega}(\mathcal{E})=\mathcal{E}^{\otimes
2}(\rho_{\mathrm{mm}})$, $\omega(\mathcal{E})=(\Delta
\mathcal{E})^{\otimes 2}[\rho_{\mathrm{mm}}]$, $\rho_{\mathrm{mm}}=
\iden/d$, and $S=\sum_{ij}|ij\rangle\langle ji|$ is the
\textsc{swap} operator.

\citet{cgp1} also examined power of the dephasing channel
$\Delta_{B'}$ described by the projector $B'=\{|i'\rangle\langle
i'|\}$ (i.e., $\Delta_{B'}[\rho]=\sum_{i'} \langle i'|\rho
|i'\rangle |i'\rangle\langle i'|$) on generating quantum coherence
defined with respect to the basis $B=\{|i\rangle\langle i|\}$. To be
explicitly, they defined the cohering power as
\begin{equation} \label{eq5b-n13}
 C^B(\Delta_{B'})=\int d\mu_{\mathrm{unif}} (\delta)
                  C^B\left(\Delta_{B'}[\delta]\right),
\end{equation}
where $d\mu_{\mathrm{unif}} (\delta)$ denotes the uniform measure in
the $(d-1)$-dimensional simplex. $C^B(\Delta_{B'})$ is in fact the
average coherence generated from uniformly distributed incoherent
states $\delta\in \mathcal {I}$. When using the relative entropy of
coherence as a quantifier, they proved that
\begin{equation} \label{eq5b-n14}
 C_r^B(\Delta_{B'})= \tilde{Q}(X_U X_U^T)-\tilde{Q}(X_U),
\end{equation}
and $C_r^B(\Delta_{B'})=0$ if and only if the dephasing operators
$\Delta_{B'}$ and $\Delta_{B}$ commute. Here, $X_U$ is bistochastic
with elements $(X_U)_{ij}=|\langle i|U|j\rangle|^2$, $U$ is the
unitary operator ensures $|i'\rangle= U|i\rangle$, $\forall i$, and
$\tilde{Q}(X)=\sum_j Q(\bm{p}_j)/d$, with $Q(\bm{p}_j)$ being the
subentropy of the column vector $\bm{p}_j$ with elements
$(\bm{p}_j)_i=(X_U)_{ij}$.

When one uses the HS norm of coherence, the power turns out to be
\cite{cgp1}
\begin{equation} \label{eq5b-n15}
 C_{l_2}^B(\Delta_{B'})= \frac{1}{d(d+1)}\tr [X_U X_U^T(1-X_U X_U^T)],
\end{equation}
and it is bounded from above by $(d-1)/4d(d+1)$, which is just
one-quarter of the maximum $C_{av}^{\max}(U)$.

\subsubsection{Coherence-breaking channels}
\citet{breaking} investigated coherence breaking channels (CBC)
which were defined as those of the incoherent channels who destroy
completely the coherence of any input state. They also discussed the
selective CBC for which the Kraus operators $\{K_n\}$ give $K_n\rho
K_n^\dag \in \mathcal{I}$, and found that they are equivalent to
CBC, i.e., the two sets $\mathcal {S}_{\rm cbc}=\mathcal{S}_{\rm
scbc}$. The CBC are subsets of the entanglement-breaking channels
\cite{breaking-e} and quantum-classical channels.

When a channel $\Phi\in \mathcal {S}_{\rm cbc}$, then
$\Phi(|i\rangle\langle j|)$ is diagonal for any two incoherent basis
states $|i\rangle$ and $|j\rangle$, and the action of $\Phi$ on
$\rho$ can be written as
\begin{equation} \label{CBC01}
 \Phi(\rho)=\sum_i |i\rangle\langle i| \tr(\rho F_i),
\end{equation}
with $\{F_i\}$ being the set of positive semidefinite operators
satisfying $\sum_i F_i=\iden$. For the special case of single-qubit
state $\rho=(\iden+\vec{r}\cdot\vec{\sigma})/2$, the channel
\begin{equation} \label{CBC02}
 \Phi(\rho)=\frac{1}{2}[\iden+(M\vec{r}+\vec{n}) \cdot\vec{\sigma}],
\end{equation}
belongs to CBC if the nonzero elements of $M$ and $\vec{n}$ lie only
in the third row of them.

\citet{breaking} further introduced a notion which they termed as
coherence-breaking index. It concerns the iterative actions of an
incoherent quantum channel $\Phi$ on a given state, and can be
defined explicitly as
\begin{eqnarray}\label{eq5b-add}
 n(\Phi)=\min\{n\geqslant 1\!: \Phi^n\in\mathcal {S}_{\rm cbc}\},
\end{eqnarray}
that is, $n(\Phi)$ characterizes the minimum number of iterations of
$\Phi$ such that $\Phi^n(\rho)\in \mathcal{I}$ for any $\rho$.
Clearly, if $\Phi$ is already a CBC, then $n(\Phi)=1$, while for the
case of $n(\Phi)= \infty$, $\Phi^n$ is not a CBC.

For the single-qubit case, they also investigated the sudden death
phenomenon for the $l_1$ norm of coherence by using the result of
\citet{fac3}. Explicitly, the occur of coherence sudden death is
only determined by forms of the incoherent channel and is
independent of the  initial state. But this does not apply to high
dimensional states.

\subsection{Evolution equation of quantum correlation and coherence} \label{sec:6D}
Refereing to various measures of quantumness in open system, the
search of certain dynamical law governing their evolution is of
practical significance, as this can simplify the assessment of their
robustness against decoherence. For entanglement measured by
concurrence, its evolution was found to obey a factorization law for
the initial two-qubit states and arbitrary quantum channels
\cite{fac1,fac2}, we review here the similar problems for geometric
quantum correlations and quantum coherence.

\subsubsection{Evolution equation of geometric quantum correlation}
Similar to the evolution equation of entanglement measured by
concurrence \cite{fac1}, \citet{fac4} found that when a bipartite
system traverses the local quantum channel, the evolution of GQD may
demonstrate a factorization decay behavior.

For a bipartite state $\rho$ decomposed as
\begin{eqnarray}\label{eq1}
 \rho = -\frac{1}{d_A d_B} \iden_{AB}
        +\rho_A\otimes \frac{1}{d_B}\iden_B+\frac{1}{d_A}
        \iden_A\otimes\rho_B+\rho_{co},
\end{eqnarray}
where the reduced states $\rho_A = \tr_B \rho$, $\rho_B=\tr_A \rho$,
and the traceless `correlation operator' $\rho_{co}$ are give by
\begin{equation}\label{eq2}
 \begin{split}
  & \rho_A = \frac{1}{d_A}\iden_A+\vec{x}\cdot\vec{X},~
    \rho_B = \frac{1}{d_B}\iden_B+\vec{y}\cdot\vec{Y},\\
  & \rho_{co} = \sum_{i=1}^{d_A^2-1}\sum_{j=1}^{d_B^2-1}t_{ij} X_i \otimes Y_j,
 \end{split}
\end{equation}
they proved that when the channel $\mathcal {E}$ gives $\mathcal
{E}(\varrho)= q(t)\varrho$, with $\varrho = \vec{x} \cdot \vec{X}
\otimes \iden_B/d_B+\rho_{co}$, the evolution of $D_p(\mathcal
{E}[\rho])$ fulfills the factorization decay behavior
\begin{eqnarray}\label{eq6}
 D_p(\mathcal{E}[\rho]) = |q(t)|^p D_p(\rho),
\end{eqnarray}
which is solely determined by the product of the initial $D_p(\rho)$
and a channel-dependent factor $|q(t)|$, and
\begin{eqnarray}\label{eq3}
 D_p(\rho) = \mathop{\rm opt}_{\Pi^A\in \mathcal {M}}
             \parallel\rho-\Pi^A(\rho)\parallel_p^p,
\end{eqnarray}
with opt representing the optimization over some class $\mathcal
{M}$ of the local measurements $\Pi^A=\{\Pi_k^A\}$ acting on party
$A$.

\begin{figure}
\centering
\resizebox{0.41 \textwidth}{!}{%
\includegraphics{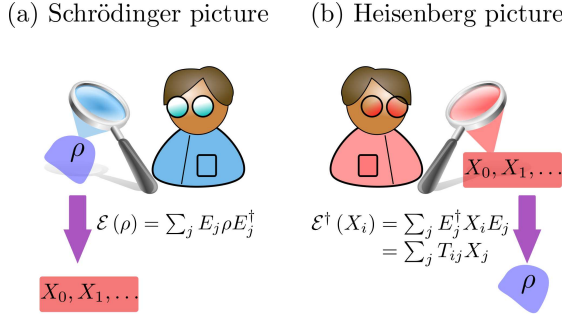}}
\caption{The action of $\mathcal {E}$ on a state. In the
Schr\"{o}dinger picture (a), the state $\rho$ evolves in time while
the observables $X_i$ are time-independent. But when we turn to the
Heisenberg picture (b), the case is opposite, with which we obtain
the condition for $\rho$ obeying the evolution equation for quantum
coherence.}\label{fig:hu3}
\end{figure}

By turning to the Heisenberg picture to describe the action of
$\mathcal {E}$ (with the Kraus operators $\{E_j\}$), i.e., $\mathcal
{E}^\dag(X_i)=\sum_j E_j^\dag X_i E_j$ (see Fig. \ref{fig:hu3}),
they identified the family of states for which the factorization
relation holds. Explicitly, if $\mathcal {E}_1^\dag (X_i)=q_A X_i$
for all $\{X_i\}$, and $\mathcal {E}_2^\dag (Y_j) = q_B Y_j$ for all
$\{Y_j\}$, then Eq. \eqref{eq6} holds for the families of $\rho$
with
\begin{equation}\label{eq-c1}
 \begin{aligned}
  (1)&~ \mbox{arbitrary}~ \rho_A,~ \rho_B,~ \rho_{co} ~~
        (\mbox{for \,} \mathcal {E}_1 \otimes \iden_B), \\
  (2)&~ \rho_A = \frac{1}{d_A}\iden_A, \mbox{or~} \rho_{co}=0 ~~
        (\mbox{for \,} \mathcal {E}_1 \otimes \mathcal {E}_2~ \mbox{with~} \mathcal {E}_2\neq
        \iden_B),
 \end{aligned}
\end{equation}
while $\mathcal {E}_1^\dag (X_k) = q_A X_k$ only for $\{X_k\}$ with
$k = \{k_1,\ldots,k_\alpha\}$ ($\alpha < d_A^2-1$), and $\mathcal
{E}_2^\dag (Y_l) = q_B Y_l$ only for $\{Y_l\}$ with
$l=\{l_1,\ldots,l_\beta\}$ ($\beta < d_B^2-1$), Eq. \eqref{eq6}
holds for the families of $\rho$ with
\begin{equation}\label{eq-c2}
 \begin{aligned}
  (1)&~ \rho_A = \rho_A^{(1)},~ \rho_{co} = \rho_{co}^{(1)} ~~
        (\mbox{for \,} \mathcal{E}_1 \otimes \iden_B),\\
  (2)&~ \rho_A = \frac{1}{d_A}\iden_A,~ \rho_{co} = \rho_{co}^{(2)},~
        \mbox{or~} \rho_{co} = 0~
        (\mbox{for \,} \iden_A \otimes \mathcal {E}_2),\\
  (3)&~ \rho_A = \frac{1}{d_A}\iden_A,~ \rho_{co} = \rho_{co}^{(3)},~
        \mbox{or~} \rho_A = \rho_A^{(1)},~\rho_{co} = 0 \\
     &~ (\mbox{for \,} \mathcal {E}_1 \otimes \mathcal {E}_2),
 \end{aligned}
\end{equation}
where
\begin{equation}\label{eq-cc12}
 \begin{split}
  & \rho_A^{(1)} = \frac{1}{d_A}\iden_A +
                   \sum_{k=k_1}^{k_\alpha} x_k X_k,~
    \rho_{co}^{(1)} = \sum_{k=k_1}^{k_\alpha}\sum_{j=1}^{d_B^2-1}
                   t_{k j} X_k \otimes Y_j,\\
  & \rho_{co}^{(2)} = \sum_{i=1}^{d_A^2-1}\sum_{l=l_1}^{l_\beta}
                   t_{i l} X_i \otimes Y_l,~
    \rho_{co}^{(3)} = \sum_{k=k_1}^{k_\alpha}\sum_{l=l_1}^{l_\beta}
                   t_{k l} X_k \otimes Y_l.
 \end{split}
\end{equation}

Besides the above statements, they also discussed the case of
symmetric GQD, the families of states for which Eq. \eqref{eq6}
holds are similar to those of Eqs. \eqref{eq-c2} and
\eqref{eq-cc12}, and we do not list them here again.

\subsubsection{Evolution equation of quantum coherence}
For quantum coherence measured by $l_1$ norm, \citet{fac3} explored
its evolution for a $d$-dimensional system traversing the quantum
channel $\mathcal {E}$. As for any master equation which is local in
time, whether Markovian, non-Markovian, of Lindblad form or not, one
can always construct a linear map which gives $\rho(t)= \mathcal {E}
(\rho(0))$ (the opposite case may not always be true), and the
linear map can be expressed in the Kraus-type representations
\cite{me}. If the map $\mathcal{E}$ is CPTP, then one can explicitly
construct the Kraus operators $\{E_\mu\}$ such that $\mathcal
{E}(\rho)= \sum_\mu E_\mu \rho E_\mu^\dag$.

For $\rho$ of Eq. \eqref{eq2d-4}, one can turn to the Heisenberg
picture to describe $\mathcal{E}$ via the map
\begin{eqnarray} \label{eq-ee1}
 \mathcal{E}^\dag (X_i)= \sum_\mu E_\mu^\dag X_i E_\mu,
\end{eqnarray}
which gives $ x'_i= \tr(\rho\mathcal{E}^\dag [X_i])$. As any
Hermitian operator $\mathcal{O}$ on $\mathbb{C}^{d \times d}$ can
always be decomposed as $\mathcal{O}=\sum_{i=0}^{d^2-1} r_i X_i$
($r_i \in \mathbb{R}$), $\mathcal{E}^\dag (X_i)$ can be further
characterized by the transformation matrix $T$ defined in Eq.
\eqref{eq-master}, namely, $\mathcal {E}^\dag (X_i)=
\sum_{j=0}^{d^2-1} T_{ij}X_j$, where $T_{ij}=\tr(\mathcal{E}^\dag
[X_i]X_j)/2$, and $X_0=\sqrt{2/d}\iden_d$. Clearly, $T_{00}=1$, and
$T_{0j}=0$ for $j \geq 1$. This further gives
\begin{eqnarray} \label{eq-ee2}
 x'_i=\sum_{j=0}^{d^2-1} T_{ij} x_j.
\end{eqnarray}

By classifying state $\rho$ of Eq. \eqref{eq2d-4} into different
families: $\rho= \{\rho^ {\hat{n}}\}$, with $\rho^{\hat{n}}=
\iden_d/d+ \chi \hat{n}\cdot\vec{X}/2$ ($\hat{n}$ is a unit vector
in $\mathbb{R}^{d^2-1}$, and $\chi$ is smaller than
$\sqrt{2(d-1)/d}$ as $\tr(\rho^{\hat{n}})^2=|\chi|^2/2+ 1/d$), one
can found that if $T_{k0}=0$ for $k \in \{1,2,\ldots,d^2-d\}$, then
the evolution of $C_{l_1}(\mathcal{E}[\rho^{\hat{n}}])$ obeys the
factorization relation
\begin{eqnarray}\label{eq5c-1}
 C_{l_1}(\mathcal {E}[\rho^{\hat{n}}])= C_{l_1}(\rho^{\hat{n}})
                        C_{l_1}(\mathcal {E}[\rho_p^{\hat{n}}]),
\end{eqnarray}
with
\begin{eqnarray} \label{eq-ee3}
 \rho_p^{\hat{n}}= \frac{1}{d}\iden +\frac{1}{2}\chi_p\hat{n}
\cdot\vec{X},
\end{eqnarray}
being the probe state, and $\chi_p = 1/\sum_{r=1}^{d_0} (n_{2r-1}^2+
n_{2r}^2)^{1/2}$.

As a corollary of the above equation, one can also show that if the
operator $A=\sum_\mu E_\mu E_\mu^\dag$ is diagonal, then the
evolution of $C_{l_1}(\mathcal{E}[\rho^{\hat{n}}])$ is governed by
Eq. \eqref{eq5c-1}. Moreover, if more restrictions are imposed on
the quantum channel, e.g., if a channel $\mathcal{E}$ yields
$\mathcal{E} ^\dag (X_k)= q(t) X_k$ for $\{X_k\}_{k= k_1, \ldots,
k_\beta}$ ($\beta \leq d^2-d$), with $q(t)$ containing information
on $\mathcal{E}$'s structure, then
\begin{equation}\label{eq5c-2}
 C_{l_1}(\mathcal{E}[\rho])=|q(t)| C_{l_1}(\rho),
\end{equation}
holds for all the
\begin{equation}\label{eq5c-3}
 \rho = \frac{1}{d}\iden_d +\frac{1}{2}\sum_{k=k_1}^{k_\beta}
         x_k X_k+\frac{1}{2}\sum_{l=d^2-d+1}^{d^2-1} x_l X_l.
\end{equation}
The channel $\mathcal{E}$ satisfying the requirements includes the
Pauli channel (covers bit flip, phase flip, bit-phase flip, phase
damping, and depolarizing channels) and Gell-Mann channel given by
\citet{fac3}, and the generalized amplitude damping channel. They
also constructed a quantum channel $\mathcal{E}_{\rm G}$ for which
$C_{l_1}(\mathcal{E}_{\rm G}[\rho])$ obeys Eq. \eqref{eq5c-2} for
arbitrary initial state.

\subsection{Preservation of GQD and quantum coherence} \label{sec:6E}
Along with the similar line for exploring quantum entanglement and
entropic discord dynamics in open quantum systems, some works have
also been devoted to the study of dynamical behaviors of various
GQDs and coherence monotones. Apart from those focused on
identifying freezing conditions discussed above, the others are
aimed at seeking flexible methods to control their evolution. We
summarize the key results in this section, mainly for the qubit
states and typical noisy sources in quantum information processing.

\citet{orderd1} discussed robustness of the HS norm of discord for
two qubits coupled to a multimode vacuum electromagnetic field, and
found that the robustness can be enhanced if appropriate local
unitary operations were performed on the initial state of the
system. \citet{hutian} discussed trace norm of discord and Bures
norm of discord for a two-qubit system subject to independent and
common zero-temperature bosonic structured reservoirs. The results
showed that the two GQDs can be preserved well or even be improved
and generated by the noisy process of the common reservoir. If one
can detuning the transition frequency of the qubits to large enough
values, the long-time preservation of these two GQDs in independent
reservoirs can also be achieved. Moreover, it was found that the
decay rates of GQD can be retarded apparently by properly choosing
the Heisenberg type interaction of two qubits when they are embedded
in two independent Bosonic structured reservoirs \cite{Lizhao}.

If the noisy channel (or reservoir) coupled to the central system is
non-Markovian, the backflow of information from the reservoir to the
system can induce damped oscillation behaviors of the GQD. For two
qubits subject to bosonic structured reservoirs with Lorentzian and
Ohmic-like spectra, the relation between behaviors of GQDs and the
extent of non-Markovianity of the reservoir have been studied
\cite{hulian}. By analyzing their dependence on a factor whose
derivative signifies the (non-)Markovianity of the dynamics, it was
demonstrated that the non-Markovianity induced by the backflow of
information from the reservoirs to the system enhances the GQDs in
most of the parameter regions.

For a single qubit subjecting to pure dephasing channel with the
Ohmic-like spectral densities, \citet{trap1} compared the coherence
evolution behaviors (measured by the $l_1$ norm and relative
entropy) with different system and bath parameters. They found that
the initial system-bath correlations are preferable for realizing
long-lived coherence in the super-Ohmic baths, and the region of
coherence trapping is enlarged with increasing the correlation
parameter. For the given initial state with equal amplitudes, they
obtained numerically the optimal Ohmicity parameter $\mu\doteq 1.46$
for the most efficient coherence trapping, which is independent of
the coupling constant and the correlation parameter.

The atomic system is also an important candidate for various quantum
information processing tasks. For the static polarizable two-level
atoms interacting with a fluctuating vacuum electromagnetic field,
\citet{LiuTian2016a} explored the coherence dynamics measured by the $l_1$
norm and relative entropy, both for an initial product state of the
atoms and the field. The results show that for the initial one-qubit
pure states and two-qubit Bell-diagonal states, the coherence cannot
be protected for non-boundary electromagnetic field. Contrarily,
when there is a reflecting boundary, the coherence will be trapped
if the atom is close to the boundary and transversely polarized. The
coherence can also be protected to some extent for other specific
polarization directions. All these show that the coherence behavior
is position and polarization dependent.

\section{Quantum coherence and GQD in many-body systems} \label{sec:7}
Quantum coherence can be regarded as a fundamental property in
quantum realm. The many-body systems of condensed matter physics
possess various quantum characteristics which may have no classical
analogue. In this sense, the exploring of quantum coherence in
many-body systems may lead some intriguing connections and may also
result in developments in both research areas. We next start from a
interesting concept in condensed matter physics.

\subsection{Off-diagonal long-range order and $l_1$ norm of coherence} \label{sec:7A}
In theory of superconductivity, one of the well-known properties is
the off-diagonal long-range order (ODLRO). Apparently, this property
is related with $l_1$ norm quantum coherence measure, which uses the
summation of the off-diagonal elements norm of a (reduced) density
matrix quantifying coherence.

Let us consider a $\eta$-pairing state as an example. We start from
the Hamiltonian of the Hubbard model,
\begin{equation}
 \begin{aligned}
  H =& -\sum _{\sigma ,\langle i,j\rangle }\left( c_{j,\sigma}^{\dagger }
      c_{k,\sigma }+c_{k,\sigma }^{\dagger }c_{j,\sigma}\right)\\
     & +U\sum _{j=1}^L\left( n_{j\uparrow }-\frac {1}{2}\right)\left(
       n_{j\downarrow }-\frac {1}{2}\right) ,
\end{aligned}
\end{equation}
where $\sigma =\uparrow ,\downarrow $, and $\langle j,k\rangle $ is
considered as a pair of the nearest-neighboring sites, $c_{j\sigma
}^{\dagger }$ and $c_{j\sigma }$ are the creation and annihilation
operators of fermions. The $\eta $-pairing operators at lattice site
$j$ are defined as
\begin{equation}
 \eta _j=c_{j\uparrow}c_{j\downarrow}, ~
 \eta^{\dagger}_j=c_{j\downarrow}^{\dagger}c_{j\uparrow}^{\dagger}, ~
 \eta_j^z=-\frac {1}{2}n_j+\frac {1}{2},
\end{equation}
and they constitutes a SU(2) algebra. The $\eta $ operators are
defined as $\eta =\sum \eta_j$ and $\eta^{\dagger}= \sum
\eta_j^{\dagger}$. The $\eta $-pairing state is defined as
\cite{CNYang1989,EsslerKorepin,FanLloyd}
\begin{eqnarray}
 |\Psi \rangle =(\eta^{\dagger})^N|\mathrm{vac}\rangle
\end{eqnarray}
where $|{\rm vac}\rangle $ is the vacuum state, and  $|\Psi \rangle$
is an eigenstate of the Hubbard model. We can find that the $\eta
$-pairing state is actually the completely symmetric state with $N$
sites filled while the other $L-N$ sites unfilled up to a
normalization factor. The ODLRO of this $\eta$-pairing state is
shown as
\begin{eqnarray}
 C_{odlro}=\frac {\langle \Psi |\eta _k^{\dagger }\eta_l|\Psi \rangle}
           {\langle \Psi |\Psi \rangle }=\frac {N(L-N)}{L(L-1)},
           ~~k\not=l.
\end{eqnarray}
The off-diagonal element $C_{odlro}$ is a constant which does not
depend on the distance $|k-l|$, in particular when $|k-l|\rightarrow
\infty $.

We may find that the density matrix $\rho =|\Psi\rangle\langle\Psi
|$ of the $\eta$-pairing state is a $L$-qubit state, the quantity
$C_{odlro}$ corresponds to one off-diagonal element of $\rho $. The
$l_1$ norm of $\eta$-pairing state can be calculated as
\begin{eqnarray}
 C_{l_1}(\rho)= L(L-1)C_{odlro}.
\end{eqnarray}
In this sense, the ODLRO is directly related with $l_1$ norm of
coherence. As we mentioned that the coherence measure depends on a
specified basis, here the definition of ODLRO naturally provides the
basis by which the quantum coherence can be quantified.

By using this example, we try to show that further perspective study
can be expected concerning about the quantum coherence in many-body
systems.

\subsection{Quantum coherence of valence-bond-solid state} \label{sec:7B}
Haldane conjectured that antiferromagnetic spin chains will be
gapless for half-odd-integer spins and gapped for integer spins
\cite{Haldaneconj1,Haldaneconj2}. The Affleck-Kennedy-Lieb-Tasaki
(AKLT) model \cite{AKLT1,AKLT2} is a spin-1 chain in bulk and
spin-$1/2$ at the two ends, which agrees with the Haldane
conjecture. The ground state of AKLT model is known as the
valence-bond-solid (VBS) state. The Hamiltonian of the AKLT model is
written as
\begin{equation}
 H=\sum _{j=1}^{N-1}\left(\vec {S}_j\cdot \vec {S}_{j+1}
   +\frac{1}{3}(\vec {S}_j\cdot \vec {S}_{j+1})^2\right)
   +\pi _{0,1}+\pi_{N,N+1},
\end{equation}
where $\vec {S}$ is the spin-1 operator in bulk, $\pi$ describes the
interaction of spin-1 in bulk and spin-{1/2} at one end.

\begin{figure}
\centering
\resizebox{0.46 \textwidth}{!}{%
\includegraphics{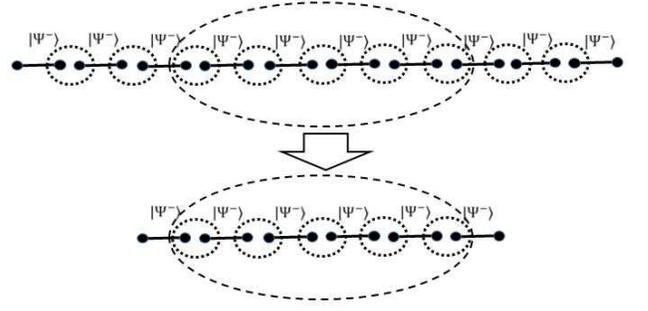}}
\caption{Schematic picture of the VBS state for the AKLT model. The
VBS state is constructed by a series of projections, represented as
dashed circles, each acts on a pair of ends, represented as filled
dots, of two singlet states. It can be shown that entropy of a
subsystem of the VBS state, represented as large dashed circle, only
depends on the size of the subsystems itself, so the VBS state can
be chosen to contain the studied subsystem directly connected with
two ends, as shown in lower part of the picture.}\label{fig:fan}
\end{figure}

The ground state, VBS state, is written as
\begin{eqnarray}
|G\rangle =(\otimes_{j=1}^NP_{j \bar{j}})|\Psi^-\rangle _{\bar
{0}1}|\Psi^-\rangle_{\bar {1}2} \cdots |\Psi^-\rangle_{\bar{N}N+1},
\end{eqnarray}
where $|\Psi ^-\rangle =(|\uparrow \downarrow \rangle -|\downarrow
\uparrow\rangle)/\sqrt{2}$ is a singlet state which corresponds to
the operator form, $(a_{\bar {i}}^{\dagger }b_i^{\dagger }-b_{\bar
{i}}^{\dagger }a_i^{\dagger })|{\rm vac}\rangle$, where
$a^{\dagger}$ and $b^{\dagger }$ are bosonic creation operators. The
projector $P$ maps a two-qubit state, which is a four-dimensional
Hilbert space, on a symmetric subspace which is three-dimension for
spin-1 operator $\vec{S}$. So the VBS state is constructed by a
chain of singlet states under the projection of $P$ at the bulk
sites and leaves the spin-$1/2$ at two ends (see Fig.
\ref{fig:fan}).

By using the teleportation technique sequentially \cite{FKR}, the
VBS state takes a form like the following
\begin{eqnarray}
 |G\rangle &=& \frac {1}{3^{N/2}}\sum _{\alpha _j=1}^3|\alpha_1\rangle
               \cdots |\alpha _N\rangle \nonumber \\
            && \times\left[\iden\otimes(\sigma_{\alpha_N}\cdots\sigma_{\alpha_1})\right]
               |\Psi^-\rangle_{\bar{0},N+1},
\end{eqnarray}
where $\iden$ is the identity operator, $\sigma_1$, $\sigma_2$,
$\sigma_3$ are Pauli matrices. It is proved that the reduced density
matrix of continuously $L$ bulk spins is invariant which does not
depend on its position in the spin chain, it can be written as
\begin{eqnarray}
\rho_L=\frac {1}{3}\sum _{\alpha,\alpha '}f_{\alpha \alpha '}
|\alpha_1\rangle \langle \alpha_1'|\cdots|\alpha_L\rangle \langle
\alpha'_L|,
\end{eqnarray}
where parameters $f_{\alpha \alpha'}$ can be determined as
\begin{equation}
 f_{\alpha\alpha'}=\tr(\iden\otimes V_{\alpha })|\Phi^-\rangle
                   \langle\Psi^-|(I\otimes V_{\alpha'}^{\dagger}),
\end{equation}
with $V_{\alpha }= \sigma_{\alpha_L}\cdots \sigma_{\alpha_1}$, and
$V_{\alpha'}= \sigma_{\alpha'_L}\cdots \sigma_{\alpha'_1}$.

Here we consider the measure of relative entropy of coherence. Due
to the form of $f_{\alpha \alpha'}$, one may find that the diagonal
matrix $\rho ^{diag}_L$ of $\rho _L$ is a completely mixed state
with tensor product of $L$ identities and a normalization factor,
so we have
\begin{eqnarray} \label{volumepart}
S(\rho_L^{\rm diag})= L\log_2 3.
\end{eqnarray}
This quantity corresponds to the volume quantity of the bulk $L$
spins. On the other hand, we know \cite{FKR} that the von
Neumann entropy of $\rho _L$ equals to the von Neumann entropy of
the state of two ends which is a Werner state
\begin{equation}\label{areapart}
 \begin{aligned}
  & S(\rho_L)=S(\tilde\rho_L), \\
  & \tilde{\rho}_L= \frac{1}{4}(1-p_L)\iden+ p_L|\Psi^-\rangle \langle \Psi^-|,
 \end{aligned}
 \end{equation}
where $p_L=(-1/3)^L$. We know that the entropy of $L$ bulk spins,
$S(\rho_L)$, reaches to a constant $2$ exponentially fast in terms
of the number of bulk spins $L$.

With combination of those analysis, we can find that the relative
entropy of coherence of bulk $L$ spins,
\begin{eqnarray}
C_r(\rho_L)= S(\rho_L^{\rm diag})-S(\rho_L).
\end{eqnarray}
By substituting the results of Eqs.
(\ref{volumepart}-\ref{areapart}) into the definition, the coherence
of the VBS state can be obtained straightforwardly. We would like to
remark that this result is independent of the basis chosen because
of the complete mixed form of $\rho _L^{\rm diag}$ is invariant for
different bases.

For gapped one-dimensional system, the above results can be
interpreted as,
\begin{eqnarray}
 C_r(\rho_L)={\rm volume}-{\rm constant}.
\end{eqnarray}
Let us point out that the ``constant'' term corresponds to the area
law \cite{HammaArea,PlenioRev}.

Perspectively, it is worth exploring whether the relative entropy of
coherence can be generally written as the difference between volume
of the studied subsystem and the boundary term of area law,
\begin{eqnarray}
 C_r= {\rm volume}-{\rm area~law~(boundary)}.
\end{eqnarray}
Further evidences of this expectation are necessary. Recently, it is
shown that the volume effect can be found for the \emph{XY} model,
where the factor for the volume term is also important which can be
used to distinguish different quantum phases \cite{WangZhengAn}.

In the seminal work of \citet{VLRK}, two different behaviors of
entanglement entropy for one-dimensional gapped and gapless models
were proposed. For gapless models like critical spin chains, the
entanglement of the ground state for a bulk of spins grows
logarithmically in number of particles in the bulk. The prefactor of
the logarithm term is related with the central charge of the
conformal field theory. The same scaling behavior holds also for
mean entanglement at criticality for a class of strongly random
quantum spin chains \cite{RefaelMoore}. For a gapped model, the
entanglement approaches a constant bound. Additionally, the
topological entanglement entropy and entanglement spectrum are
studied based on the theory of entanglement \cite{HammaArea,KP,
LW,LH}. The entanglement entropy for a pure state is defined as the
von Neumann entropy of the reduced density matrix of its subsystem.
R\'enyi entropies parametrized by a parameter $\alpha$ are the
generalizations of the von Neumann entropy. So topological
entanglement entropy can also be generalized as topological
entanglement R\'enyi entropies \cite{Flammia}. However, it is found
that all topological R\'enyi entropies are the same, which is due to
the fact that R\'enyi entropies are additive and the studied density
matrix takes a product form. In correspondence, those results can be
further explored from point of view of quantum coherence. There are
some works about the characteristics of quantum phase transitions by
quantum coherence, as presented next.

\subsection{Quantum coherence and correlations of localized and thermalized states} \label{sec:7C}
Quantum dynamics of isolated quantum systems far from equilibrium
has recently been extensively studied \cite{Eisert2015}. By
principles of statistical mechanics, it is known that the
non-equilibrium state will evolve to a thermalized state which is
ergodic \cite{Thermalization-Deutsch,Thermalization-Srednicki,
Thermalization-Rigol}, and no quantum correlation is expected to
exist. For an isolated quantum system initially in a pure state, the
time evolution is unitary transformation which keeps the system in a
pure state. The thermalization means that the reduced density matrix
of a subsystem, which is relatively small compared with the whole
system, takes the form of a thermal state, $\rho^S= e^{-\beta
H^{S}}/Z$, where $\beta$ is the inverse of temperature, $H^{S}$ is
the Hamiltonian of the studied subsystem \emph{S}, and $Z= \tr
e^{-\beta H^{S}}$ is the partition function.

On the other hand, it is pointed out that the disorder may prevent
the system from thermalizing, resulting in localized state. In
general, there are two different types of localization, the
single-particle localization in name of Anderson localization
\cite{Anderson58} and the many-body localization
\cite{MBL-Altshuler,MBL-Polyakov}. The many-body localization is
induced by competition between interactions and disorder, in
contrast, Anderson localization is only due to disorder but without
interaction. Besides, thermalization cannot happen for integrable
models because of the constraints imposed by infinite number of
conserved quantities. There are various signatures in characterizing
the thermalized states and localized states. Here we will review the
properties of quantum coherence and quantum correlations for those
states.

\subsubsection{Entanglement entropy}
The mechanism of thermalization is based on the eigenstate
thermalization hypothesis, and the thermal state is ergodic
\cite{Thermalization-Rigol,AltmanVosk,NandkishoreHuse}. Those facts
lead to that the thermalized state $\rho^{S}$ takes a diagonal form,
so state $\rho^{S}$ possesses neither quantum coherence, nor quantum
correlations. On the other hand, the von Neumann entropy of the
thermal state, $S(\rho ^{S})=-\tr(\rho^{S}\log_2\rho^{S})$ satisfies
the volume law, implying that it is proportional to the number of
the particles $L$ of the subsystem \emph{S}. Considering that the
isolated system is always in pure state, the von Neumann entropy of
the reduced density matrix is the entanglement entropy itself. Thus
in thermal phase, the entanglement entropy satisfies the volume law.
If the initial state of the system is a highly excited state such as
N\'eel state, the thermalized state approaches to the completely
mixed state corresponding to infinite temperature, so entanglement
entropy approaches $L$, corresponding to particle number of the
subsystem for spin-1/2 particles. Note that $L$ is the upper bound
of the entanglement entropy.

The quantum dynamics of localizations, both Anderson localization
and many-body localization, and thermalization can be well
characterized by behaviors of entanglement entropy. Suppose the
initial state is a product state like N\'eel state, which is also a
highly excited state, the initial entropy will be zero for the
subsystem \emph{S}. For thermalization, the entropy will increase
quickly and approaches its upper bound. For Anderson localization,
similarly, the entropy will quickly saturate its bound but the bound
is much smaller than that of the thermalized state. In contrast,
many-body localization is a consequence of the competition between
particle interactions and disorder. The localized state breaks the
ergodicity and eigenstate thermalization hypothesis. The
entanglement entropy does not obey the volume law. Instead, the
entropy for the stationary state demonstrates a long time slow
increase characterized as logarithmic increasing in time
\cite{MBL-Log1,MBL-Log2,MBL-Log3,MBL-Log4, MBL-entropyincrease1,
MBL-entropyincrease2}, or algebraic with power-law interactions
\cite{MBL-entropyincrease3}. Reminding that the coherence can be
quantified as the diagonal entropy subtracting the von Neumann
entropy, the coherence of the subsystem will demonstrate decrease
for many-body localization.

Both Anderson localization and many-body localization have been
realized experimentally. In system of trapped ions with long-range
interactions, the growth of entanglement is shown by measuring the
quantum Fisher information \cite{MBL-experiment}. Recently, the
entanglement entropy logarithmic increase in time of many-body
localization is successfully demonstrated in a 10-qubit
superconducting quantum simulation based on single-shot state
tomography measurement \cite{XuKai18}. The many-body localization
and thermalization can also be distinguished by energy spectrum of
the system. In thermal phase, the energy levels of the system tend
to repel one another and their statistics are Wigner-Dyson distribution,
while for many-body localized state, the energy levels show a
Poisson statistics \cite{energy-statistics1,energy-statistics2,
energy-statistics3}. These phenomena are also demonstrated
experimentally \cite{energy-statistics4}.

\subsubsection{Entanglement spectrum}
The entanglement entropy of a ground state is just one quantity
based on entanglement spectrum \cite{LH} due to Schmidt
decomposition for a pure state. The entanglement spectrum possesses
more information which may be invisible for entanglement entropy.
For a ground state $|G\rangle$, the reduced density matrix for a
bulk of $L$ sites is written as $\rho_L$, which can be rewritten as
\begin{eqnarray}\label{entanglementHamiltonian}
 \rho_L= e^{-H_\mathrm{E}}.
\end{eqnarray}
The entanglement spectrum is the energy spectrum of the so-defined
entanglement Hamiltonian $H_\mathrm{E}$. It is shown that there is
one-to-one correspondence between low-energy edge states of the
system with open boundary condition and the low-lying eigenstates of
the entanglement Hamiltonian \cite{Fidkowski,QiXiaoLiang}.

The entanglement spectrum of the ground state for a topological
Chern insulator with disorder exhibits level repulsion, which is
consistent with Wigner-Dyson distribution. This result in addition
with energy spectrum and Chern number can be used to describe
transition of Chern insulator to Anderson-insulator \cite{Prodan}.
The many-body localization and thermalization and the periodically
driven systems, which are known as Floquet systems, can be
characterized by entanglement spectrum \cite{GNRegnault}. The level
statistics of the entanglement spectrum in the thermalizing phase is
governed by an appropriate random matrix ensemble. The Floquet
entanglement spectrum has similar results showing a result beyond
eigenstate thermalization hypothesis. In many-body localized phase,
the entanglement spectrum shows level repulsion and obeys a
semi-Poisson distribution. Also, the dynamical many-body
localization is observed in an integrable system with periodically
driven \cite{Keser16}. Perspectively, the study of quantum
benchmarks such as quantum coherence, entanglement may be performed
for those systems and phases like Floquet topological insulators
induced by disorder \cite{Titum15}.

\subsection{Quantum coherence and quantum phase transitions} \label{sec:7D}
Quantum phase transition (QPT) describes an abrupt change for
properties of the ground state of a many-body system driven by its
quantum fluctuations. It is a purely quantum process and is caused
by variation of the system parameters of the Hamiltonian, such as
the spin coupling and external magnetic field \cite{QPTbook}. With
the development of quantum entanglement theory, it is natural to
study QPT of a many-body system from the point of view of
entanglement. Indeed, it has been found that the singularity and
extreme point of entanglement or its derivative can be used for
detecting QPTs. An overview for the related progress can be found in
the work of \citet{RMP2} and \citet{ZengBei}.

As quantum coherence measures defined within the framework of
\citet{coher} are also quantitative characterizations of the quantum
feature of a system, they are hoped to play a role in studying
quantum phase transitions (QPTs) of the many-body systems. We review
briefly in this section some main progress for such studies.

\citet{qptc1} demonstrated role of the coherence susceptibility on
studying QPTs at both absolute zero and finite temperatures. Here,
the coherence susceptibility is defined as the first derivative of
the relative entropy of coherence $C_r(\rho)$, that is,
\begin{eqnarray}
 \chi^{\rm co}= \frac{\partial C_r(\rho)}{\partial \lambda},
\end{eqnarray}
with $\lambda$ being a characteristic parameter of the system
Hamiltonian. For the transverse Ising model with the Hamiltonian
$H_I$, the spin-1/2 Heisenberg \emph{XX} model with $\hat{H}_{XX}$,
and the Kitaev honeycomb model with $\hat{H}_K$, described by
\begin{eqnarray} \label{Ising}
 \begin{aligned}
  & \hat{H}_I= -\sum_{i=1}^N \sigma_i^z \sigma_{i+1}^z-\lambda\sum_{i=1}^N \sigma_x, \\
  & \hat{H}_{XX}=-\frac{1}{2}\sum_{i=1}^N(\sigma_i^x \sigma_{i+1}^x +
                 \sigma_i^y \sigma_{i+1}^y)- \lambda\sum_{i=1}^N \sigma_i^x,\\
  & \hat{H}_K=- \sum_{\alpha=\{x,y,z\}}J_\alpha \sum_{i,j\in \alpha-{\rm links}}
             \sigma_i^\alpha \sigma_j^\alpha,
  \end{aligned}
\end{eqnarray}
where $\lambda$ is the strength of the external magnetic field in
units of the interaction energy, they showed that apart from the
figure of merit that this method requires no prior knowledge of
order parameter (the same as those based on entanglement and
discord), the coherence susceptibility pinpoints not only the exact
QPT points via its singularity with respect to $\lambda$, but also
the temperature frame of quantum criticality. In particular, the
latter has been considered to be a superiority of the coherence
susceptibility method.

\citet{qptc2} showed validity of the Skew-information-based
coherence measure $\mathsf{I} (\rho,K)$ (will be called the $K$
coherence for brevity) and its lower bound $\mathsf{I}^L (\rho,K)$
on studying QPTs \cite{meas8}. They considered the spin-1/2
Heisenberg \emph{XY} model described by the Hamiltonian
\begin{eqnarray}
 \begin{aligned}
 \hat{H}=& -\frac{\lambda}{2} \sum_i [(1+\gamma) \sigma_i^x \sigma_{i+1}^x
         +(1-\gamma)\sigma_i^y\sigma_{i+1}^y]\\
         & - \sum_i\sigma_i^z,
 \end{aligned}
\end{eqnarray}
with $0\leq \gamma\leq 1$ being the anisotropy parameter, and
$\lambda$ strength of the inverse magnetic field. They calculated
the single-spin coherence $\mathsf{I}(\rho, \sigma^\beta)$, two-spin
local coherence $\mathsf{I}(\rho, \sigma^\beta\otimes\iden_2)$
($\beta=x,y,z$), and their lower bounds. The numerical results show
that the divergence of the first derivatives of $\mathsf{I}
(\rho,\sigma^x)$ and $\mathsf{I} (\rho,\sigma^{x,y,z} \otimes
\iden_2)$ (including their lower bounds) with respect to $\gamma$
pinpoint exactly the transition point $\gamma_c=1$ [$\mathsf{I}
(\rho,\sigma^y\otimes \iden_2)$ fails for the special case
$\gamma=0.5$], while the derivatives of $\mathsf{I} (\rho,
\sigma^{x,y,z}\otimes \iden_2)$ also detect the factorization point
$\lambda_f\sim 1.1547$. Moreover, the performance of
$\mathsf{I}^L(\rho,\sigma^x)$ in detecting QPTs at relatively high
temperatures outperforms that of $\mathsf{I}(\rho,\sigma^x)$ for the
considered model. A review of these results in addition with some
quantum correlations are presented in \cite{Entropy}.

\citet{qptc3} studied quantum coherence measured by the WY skew
information on diagnosing critical points of the spin-1/2 transverse
field \emph{XY} model with the \emph{XZY$-$YZX} type of three-spin
interactions. The Hamiltonian is given by
\begin{equation}\label{eq6-1}
 \begin{aligned}
 \hat{H}=&-\sum_{i=1}^N \bigg[\frac{1+\gamma}{2}\sigma_i^x\sigma_{i+1}^x
          +\frac{1-\gamma}{2}\sigma_i^y\sigma_{i+1}^y+h\sigma_l^z \\
         &+\frac{\alpha}{4} (\sigma_{i-1}^x\sigma_i^z\sigma_{i+1}^y
         -\sigma_{i-1}^y\sigma_i^z\sigma_{i+1}^x)\bigg].
  \end{aligned}
\end{equation}

By examining the single-spin $\sigma^{x,y,z}$ coherence [i.e.,
$K=\sigma^x$, $\sigma^y$, or $\sigma^z$ in Eq. \eqref{eq2f-1}], the
two-spin local $\sigma^{x,y,z}$ coherence [i.e., $K=\sigma^x\otimes
\iden_2$, $\sigma^y\otimes \iden_2$, or $\sigma^z\otimes \iden_2$],
and their lower bounds $\mathsf{I}^L(\rho,K)$ \cite{meas8},
\citet{qptc3} found that if the three-spin interaction $\alpha=0$
and the external magnetic field $h<1$, the single-spin
$\sigma^{x,y,z}$, two-spin local $\sigma^z$ coherence, and their
lower bounds are extremal at the critical point $\gamma_c=0$ of
anisotropy transition. But the two-spin local $\sigma^x$
($\sigma^y$) coherence and its lower bound decrease (increase) with
the increasing $\gamma$. Their first derivative with respect to
$\gamma$ are minimal (maximum) at the critical point $\gamma_c=0$,
and show scaling behaviors with respect to $\log N$, i.e., $d
\mathsf{Q} /d\gamma=a_1+a_2\log_2 N$, where $\mathsf{Q}=\mathsf{I}
(\rho,K)$ or $\mathsf{I}^L(\rho,K)$, and $a_1$ and $a_2$ are the
system-dependent parameters.

When the three-spin interaction is introduced, there will be a
gapless phase in the range $h\in[h_{c1},h_{c2}]$ for
$\gamma<\alpha$. The system undergoes two QPTs (second order
transitions) with increasing $h$, the first from gapped phase to
gapless phase when $h$ increases from $h<h_{c1}$ to $h>h_{c1}$, and
the second from the gapless phase to gapped phase when $h$ increases
from $h<h_{c2}$ to $h>h_{c2}$. For this case, it was found that both
the single-spin and two-spin local $\sigma^{x,y,z}$ coherence and
their lower bounds are affected by the existence of $\alpha$ only in
the gapless phase, and the two critical points $h_{c1}$ and $h_{c2}$
of the gapless phase can be pinpointed by the extremal points of
their first derivatives. But different from that of $\alpha=0$,
there are no size effect of the corresponding derivatives of
coherence around the critical points, that is, the derivatives for
different $N$ are almost the same.

\citet{qptc4} also showed effectiveness of the two-spin local
$\sigma^\beta$ ($\beta=x,y,z$) coherence $\mathsf{I}^L
(\rho,\sigma^\beta)$ in detecting QPTs of different physical
systems. For the \emph{XY} model with transverse magnetic fields and
\emph{XZX+YZY} type of three-spin interactions [the Hamiltonian is
similar as that of Eq. \eqref{eq6-1}, with only the terms in the
second line being replaced by $\alpha(\sigma_{i-1}^x \sigma_i^z
\sigma_{i+1}^x+ \sigma_{i-1}^y\sigma_i^z\sigma_{i+1}^y)$]. Contrary
to the case of $h=0.5$ studied in \citet{qptc3}, when $h=\alpha=0$,
it was found that while the extremal of $\mathsf{I}^L (\rho,
\sigma^z\otimes \iden_2)$ can pinpoint the critical points of QPT at
$\gamma_c=0$, $\mathsf{I}^L (\rho,\sigma^x\otimes \iden_2)$
[$\mathsf{I}^L (\rho,\sigma^y\otimes \iden_2)$] increases
(decreases) with $\gamma$, and its first derivative with respect to
$\gamma$ is maximal (minimal) at the first-order QPT point
$\gamma_c=0$. For the second-order QPT at $h_c=1$, the derivative of
$\mathsf{I}^L (\rho,\sigma^x\otimes \iden_2)$ with respect to $h$ at
$h_c=1$ show a size-dependent scaling behavior, which implies that
it will be divergent in the thermodynamic limit. Even at finite
temperature, $\mathsf{I}^L (\rho,\sigma^x\otimes \iden_2)$ and its
first derivative can also detect the second order QPT at
$\alpha=0.5$ for $\gamma= 0.5$ and $h=0$. Moreover, they also showed
that the two-spin local coherence can detect QPTs for the
one-dimensional half-filled Hubbard model with both on-site and
nearest-neighboring interactions and topological phase transition
for the Su-Schrieffer-Heeger model.

The amount of WY skew information $\mathsf{I}(\rho,K)$ is determined
by the observable one chooses. \citet{luoprl} introduced a quantity
\begin{equation}
Q(\rho)=\sum_i \mathsf{I}(\rho,X_i)
\end{equation}
where $\{X_i\}$ is the set of observables which constitute an
orthonormal basis, and proved it to be independent of the choice of
$\{X_i\}$ \cite{luopra}. Based on this, \citet{qptc5} explored its
role in detecting critical points of QPTs and the factorization
transition of the spin model. For the density matrix $\rho^{AB}$ of
two neighboring spins, they calculated
\begin{equation}
 F(\rho^{AB})=Q_A(\rho^{AB})- Q_A(\rho^A\otimes\rho^B),
\end{equation}
with $Q_A(\rho^{AB})=\sum_i \mathsf{I} (\rho^{AB},X_i\otimes
\iden_B)$, and likewise for $Q_A(\rho^A\otimes\rho^B)$. For the
\emph{XY} model with transverse magnetic fields, their results show
that the second-order QPT from the ferromagnetic to the paramagnetic
phase (Ising transition) can be detected by the minimum of the first
derivative of $F(\rho^{AB})$ with respect to $h$. At the vicinity of
the transition point $h_c=1$, $\partial F(\rho^{AB})/\partial h$
shows a size-dependent scaling behavior, and is logarithmic
divergent in the thermodynamic limit. On the other hand, the
first-order transition from a ferromagnet with magnetization in the
$x$ direction to one with magnetization in the $y$ direction
(anisotropy transition) at $\gamma_c=0$ can be detected directly by
the minimum of $F(\rho^{AB})$, but its first derivative with respect
to $\gamma$ is continuous and size-independent. Moreover, it was
found that both $\partial F(\rho^{AB})/\partial h$ and $\partial
F(\rho^{AB}) /\partial \gamma$ are discontinuous along the curve
$h^2+\gamma^2 =1$. The emergence of the discontinuity pinpoints the
factorization transition for ground states of the considered system,
and has its roots in the elements of $\rho^{1/2}$.

Compared with the spin-1/2 models, various high-spin systems show
richer phase diagrams. \citet{qptc6} considered the spin-1
\emph{XXZ} model and bilinear biquadratic model, with the
Hamiltonian
\begin{equation}\label{eq6-2}
 \begin{aligned}
  & \hat{H}_{XXZ}=\sum_i(S_i^x S_{i+1}^x+S_i^y S_{i+1}^y+\Delta S_i^z S_{i+1}^z), \\
  & \hat{H}_{BB}=\sum_i [\cos\theta (\bold{S}_i \cdot\bold{S}_{i+1})
                 +\sin\theta(\bold{S}_i\cdot \bold{S}_{i+1})^2]
  \end{aligned}
\end{equation}
where $\bold{S}_i= (S_i^x,S_i^y,S_i^z)$ are spin-1 operators. For
the spin-1 \emph{XXZ} model, the relative entropy and $l_1$ norm of
coherence for two neighboring spins, and the local two-spin $S^x$
and $S^y$ coherence cannot detect the Kosterlitz-Thouless QPT at
$\Delta_{c2} \approx 0$, while their inflection points detect the
Ising type second-order QPT at $\Delta_{c2}\approx 1.185$. Moreover,
the extremum of single-spin $S^x$ coherence pinpoints the SU(2)
symmetry point $\Delta= 1$. For the spin-1 bilinear biquadratic
model, the single spin density matrix is diagonal in $S_z$ basis for
all values of the anisotropy parameter, so all coherence measures
are zero. On the other hand, the transition point can be identified
by mutual information and discord, which coincidences to both the
infinite order Kosterlitz-Thouless transition and the SU(3) symmetry
point $\theta= 0.25\pi$.

\subsection{GQD and quantum phase transition} \label{sec:7E}
The singularity or extreme point of QD can be used for detecting
QPTs. \citet{qptd} studied one such problem. They considered a
general Heisenberg \emph{XXZ} model with the Hamiltonian
\begin{equation}\label{qpt-01}
 \hat{H}= J \sum_{i=1}^N (\sigma_i^x \sigma_{i+1}^x+\sigma_i^y \sigma_{i+1}^y
          +\Delta\sigma_i^z \sigma_{i+1}^z),
\end{equation}
and by setting $J=1$, they calculated QD of Eq. \eqref{new1-5} as
well as its first and second order derivatives for thermal states of
the neighboring spins, and showed that it can efficiently detect the
QPT points $\Delta=\pm 1$ for this model even at finite temperature,
while the entanglement measured by entanglement of formation does
not. This shows potential role of QD in investigating QPT. In
particular, it is very important for experimental characterization
of QPTs as in principle one is unable to reach a zero temperature
experiment.

\citet{qpt-qd} studied QD for ground states of the transverse Ising
and Heisenberg \emph{XXZ} model, and found that the amount of QD
increases close to the QPT points. Indeed, there are many other
related works discussing role of the QD defined in Eq.
\eqref{new1-5} in detecting QPTs, and we refer to the work of
\citet{RMP} for a detailed overview in this respect. In what
follows, we focus on role of GQDs on studying QPTs in various
many-body systems.

\citet{togd} examined ground state properties of the Heisenberg
\emph{XXZ} model Eq. \eqref{qpt-01} by setting $J=-1$. By employing
the trace norm of discord as a quantifier of correlation, they found
that $D_T(\rho)$ defined in Eq. \eqref{eq-tdd} as well as
$C_T(\rho)$ and $T_T(\rho)$ defined in Eq. \eqref{eq1a-4} detects
successfully the first-order phase transition at $\Delta=1$. On the
other hand, the infinite-order QPT at $\Delta=-1$ can only be
detected by the classical correlation $C_T(\rho)$, while $D_T(\rho)$
and $T_T(\rho)$ failed. This seems to casting a doubt on the
usefulness of GQD, but for other many-body systems it may work
effectively for detecting phase transition points.

By using the quantum renormalization group method, \citet{qpt08}
studied HS norm of discord for ground states of the Heisenberg
\emph{XXZ} model with Dzyaloshinskii-Moriya (DM) interaction. The
Hamiltonian reads
\begin{equation}\label{qpt-02}
 \begin{aligned}
  \hat{H}= & \frac{J}{4}\sum_{i=1}^N [\sigma_i^x \sigma_{i+1}^x
             +\sigma_i^y \sigma_{i+1}^y+\Delta\sigma_i^z \sigma_{i+1}^z \\
           &+ D(\sigma_i^x\sigma_{i+1}^y-\sigma_i^y\sigma_{i+1}^x)].
 \end{aligned}
\end{equation}
Their calculation shows that the HS norm of discord can effectively
characterize the QPT point $\Delta_c= \sqrt{1+D^2}$ separating the
spin-fluid phase and the N\'{e}el phase.

\citet{qpt01} also studied quantum correlations for the ground state
properties of several three different spin models, but they used the
trace norm of discord. First, for the \emph{XXZ} model given in Eq.
\eqref{qpt-02}, they found that the trace norm of discord detects
successfully the QPT point $\Delta_c$. Second, for the Ising model
with DM interaction,
\begin{equation}\label{qpt-03}
 \hat{H}= \frac{J}{4}\sum_{i=1}^N [\sigma_i^z \sigma_{i+1}^z
          + D(\sigma_i^x\sigma_{i+1}^y-\sigma_i^y\sigma_{i+1}^x)],
\end{equation}
it was showed that the trace norm of discord can also be used to
detect the critical point $D=1$ which separates the antiferromagnet
phase and chirality phase. \citet{qpt01} also considered the
Heisenberg \emph{XXZ} model with staggered DM interaction, with the
Hamiltonian being given by
\begin{equation}\label{qpt-04}
 \begin{aligned}
    \hat{H}= & \frac{J}{4}\sum_{i=1}^N [(1+\Delta)\sigma_i^x \sigma_{i+1}^x
               -(1-\Delta)\sigma_i^y \sigma_{i+1}^y \\
             &+ D(\sigma_i^x\sigma_{i+1}^y+\sigma_i^y\sigma_{i+1}^x)],
 \end{aligned}
\end{equation}
and their result showed that the trace norm of discord also detects
successfully the region $|\Delta|\leq \sqrt{1+D^2}$ in which the
system is in the N\'{e}el phase. Concerning entanglement properties
of this model, we refer to the work of \citet{Mafw}

For the \emph{XX} model with transverse magnetic field as showed in
Eq. \eqref{Ising}, \citet{qpt03} found that the trace norm of
discord can also effectively characterize the second-order QPT
occurs at $\lambda_c=1$ which separates the ferromagnetic and
paramagnetic phases. \citet{qpt04} studied QPT in an
Ising-\emph{XXZ} diamond model. By analyzing scaling behavior of the
trace norm of discord for the thermal state, they found that around
the critical lines, its first-order derivative exhibits a maximal at
finite temperature and diverges when $T\rightarrow 0$.

Moreover, \citet{qpt05} studied the problem of many-body
localization (MBL) in a spin-1/2 Heisenberg model with random
on-site disorder of strength $h$. The Hamiltonian is
\begin{equation}\label{qpt-05}
  \hat{H}= \frac{1}{2}\sum_{i=1}^N [J(\sigma_i^x \sigma_{i+1}^x
           +\sigma_i^y \sigma_{i+1}^y+\sigma_i^z \sigma_{i+1}^z)
           +h_i \sigma_i^z],
\end{equation}
where $h_i$ are uniformly distributed random numbers in the interval
$[-h,h]$. They founded that the derivatives of the trace norm of
discord of Eq. \eqref{eq-tdd} and the geometric classical and total
correlations of Eq. \eqref{eq1a-4} give the range $h_c/J \in [3,4]$
for the MBL critical point. This estimate is in accordance with the
result $h_c/J\sim 3.8$ given in the literature \cite{qpt06,qpt07}.

\section{Quantum correlations and coherence in relativistic settings} \label{sec:8}
Since the early 20th century, much efforts have been put forward to
bridge the gap between quantum mechanics and relativity theory,
which are two fundamentals of modern physics. The reconciliation
between them gives birth for quantum field theory (QFT), and several
predictions have been made based on this theory. A fundamental
prediction in QFT is that the particle content of a quantum field is
observer dependent, such a phenomenon is named Unruh effect
\cite{unruh1976,unruhreview}. Again, the phenomenon of a quantum
field is in vacuum state as observed by a freely falling observer of
an eternal black hole, while it is a thermal state for a observer
who hovers outside the event horizon the black hole. Such a
phenomenon is named Hawking effect. The study of quantum correlation
in a relativistic framework is not only helpful to understand some
of the key questions in quantum information theory, but also plays
an important role in the black hole entropy and black hole
information paradox \cite{Hawking76,Terashima}. Following the
pioneering work of \citet{SRQIT1}, many authors have studied quantum
correlations in relativistic setting from different aspects.

\subsection{Quantum correlations for free field modes} \label{sec:8A}
For a free mode scalar field, the dynamics of the field obeys the
Klein-Gordon (KG) equation \cite{Birrelldavies}
\begin{eqnarray}\label{K-G Equation}
 \frac{1}{\sqrt{-g}}\frac{{\partial}}{\partial x^{\mu}}
 \left(\sqrt{-g}g^{\mu\nu}\frac{\partial\phi}{\partial x^{\nu}}\right)=0,
\end{eqnarray}
where $g$ is the determinant of the metric $g_{\mu\nu}$
\cite{wald94}. Similarly, the motion equation of a Dirac field
$\Psi$ in a background reads
\begin{eqnarray}\label{DiracEquation}
 i\gamma^{\mu}(x)\left(\frac{\partial}{\partial x^{\mu}}- \Gamma_{\mu}\right)\Psi=m\Psi,
\end{eqnarray}
where the background-dependent Dirac matrices $\gamma^{\mu}(x)$
relate to the matrices in flat space through $\gamma^{\mu}(x)=
e^{\mu}_{a}(x)\bar{\gamma}^{a}$, and $\bar{\gamma}^{a}$ are the
flat-space Dirac matrices. Here,
\begin{equation}
 \Gamma_\mu = \frac{1}{8} [\gamma^\alpha,\gamma^\beta]e_\alpha^\nu
e_{\beta\nu;\mu},
\end{equation}
are the spin connection coefficients. Throughout this section we set
$G=c=\hbar=\kappa_{B}=1$.

The field (either scalar field or Dirac field) can be quantized in
terms of a complete set of modes $u_{k} (x, \eta)$, which is an
orthonormal basis of solutions of the scalar field (or Dirac field).
That is,
\begin{equation}
 \Phi(\Psi)= \int d^3 k (a_{k} u_{k} + a^{\dagger}_{k} u^*_{k}),
\end{equation}
where $\Phi$ denotes the scalar field and $\Psi$ denotes the Dirac
field, $k$ is the wave vector labeling the modes and for massless
fields $\omega=|k|$.

For a scalar field, the positive and negative frequency modes
satisfy the canonical commutation relations $[a_{k},
a^{\dagger}_{k'}]= \delta^3({k}-{k}')$, while for the Dirac fields
the anti-commutation relations $\{ a_{k}, a^{\dag}_{k'}\}=
\delta^3(k-k')$ should be satisfied. The annihilation operators
$a_{k}$ define the vacuum state $\vert0\rangle$ through $a_{k}
\vert0\rangle=0,\,\,\forall\, k$. A different inequivalent choice of
modes $\{ \tilde{u}_k \}$ might exist which satisfies the same
equation of motion in different spacetime background. For example,
the appropriate coordinates to describe the accelerated observer's
motion is the Rindler coordinates $(\eta,\xi)$, which is given by
the transformation
\begin{equation}\label{Rindlerc}
 t= a^{-1}e^{a\xi}\sinh(a\eta),~
 x=a^{-1}e^{a\xi}\cosh(a\eta).
\end{equation}

Solving the KG equation or Dirac equation in the Rindler
coordinates, we obtain some sets of positive frequency modes
propagating in the regions I and II of the Rindler spacetime,
respectively. For free scalar fields, the positive frequency modes
can be used to expand the field as \cite{Schuller-Mann}
\begin{equation}\label{Firstexpand}
\Phi= \int d\omega[\hat{a}^{I}_{\omega}\Phi^{+}_{{\omega},\text{I}}
     +\hat{b}^{I\dag}_\omega \Phi^{-}_{{\omega},\text{I}}
     +\hat{a}^{II}_{\omega}\Phi^{+}_{{\omega},\text{II}}
     +\hat{b}^{II\dag}_{\omega}\Phi^{-}_{{\omega},\text{II}}],
\end{equation}
where $\hat{a}^{I}_{\omega}$ and $\hat{b}^{I\dag}_{\omega}$ are the
bosonic annihilation and anti-boson creation operators in the
Rindler region $I$, and $\hat{a}^{II}_{\omega}$ and
$\hat{b}^{II\dag}_{\omega}$ are the bosonic annihilation and
creation operators in the region $II$.

The quantum field theory for Dirac fields is constructed by
expanding the field in terms of the positive and negative frequency
modes \cite{Alsing2006}
\begin{equation}\label{Firstexpandd}
 \Psi= \int d\mathbf{k}[\hat{c}^{I}_{\mathbf{k}}\Psi^{I+}_{\mathbf{k}}
       +\hat{d}^{I\dag}_{\mathbf{k}}\Psi^{I-}_{\mathbf{k}}
       +\hat{c}^{II}_{\mathbf{k}}\Psi^{II+}_{\mathbf{k}}
       +\hat{d}^{II\dag}_{\mathbf{k}}\Psi^{II-}_{\mathbf{k}}],
\end{equation}
where $\hat{c}^{I}_{\mathbf{k}}$ and $\hat{d}^{I\dag}_{\mathbf{k}}$
are the fermionic annihilation and creation operators acting on the
state in region $I$, and $\hat{c}^{II}_{\mathbf{k}}$ and
$\hat{d}^{II\dag}_{\mathbf{k}}$ are the fermionic operators in the
region $II$. The above positive and negative frequency modes are
defined in terms of the future-directed timelike Killing vector in
different regions, in Rindler region $I$ the Killing vector is
$\partial_\eta$ and in the region $II$ the Killing vector is
$\partial_{-\eta}$.

After some calculations, the Minkowski vacuum is found to be an
entangled two-mode squeezed state for a free scalar field
\begin{equation}\label{vacuums}
 |0_\omega\rangle_\text{M}=\frac{1}{\cosh r_\omega^{2}}
     \sum_{n=0}^{\infty}\tanh r_\omega^{n}|nn\rangle_{\omega},
\end{equation}
where $\cosh r=(1-e^{-2\pi\omega/a})^{-1/2}$ and $a$ is Rob's
acceleration. For a free Dirac field, the Minkowski vacuum has the
following form
\begin{equation}\label{Dirac-vacuum1}
 |0_{\mathbf{k}}\rangle_{M}= \cos r|0_{\mathbf{k}}\rangle_{I}|0_{-\mathbf{k}}\rangle_{II}+\sin r
                             |1_{\mathbf{k}}\rangle_{I}|1_{-\mathbf{k}}\rangle _{II},
\end{equation}
where $\cos r=(e^{-2\pi\omega/a}+1)^{-1/2}$.

\subsubsection{Quantum entanglement}
\citet{Schuller-Mann} studied quantum entanglement between two free
bosonic modes as observed by two relatively accelerated observers.
They found that the quantum entanglement is an observer-dependent
quantity in noninertial frames. A maximally entangled initial state
in an inertial frame becomes less entangled under the influence of
the Unruh effect. In the infinite acceleration limit, the
distillable entanglement for the final state of the scalar field
vanishes. \citet{Alsing2006} studied the entanglement between two
free modes of a Dirac field in noninertial frames. They found that
entanglement between the Dirac modes is destroyed by the Unruh
effect. Differently, the entanglement of the fermionic modes
asymptotically reaches a nonzero minimum value in the infinite
acceleration limit.

\citet{LingYi} studied entanglement of the electromagnetic field in
a noninertial reference frame. They employed the photon helicity
entangled state and found that the logarithmic negativity of the
final state remains the same as those in the inertial reference
frame, which is completely different from that of  the particle
number entangled state. \citet{Panqiyuan} investigated the
entanglement between two modes of free scalar and Dirac fields. They
proved that the different behavior of the field modes is owing to
the in equivalence of the quantization of the free field modes in
the Minkowski and the Rindler coordinates. In the
infinite-acceleration limit, the mutual information equals to the
half mutual information of the initial state, which is independent
of the initial state and the type of field.

\citet{Adesso2007} studied the distribution of entanglement between
modes of a free scalar field from the perspective of observers in
uniform acceleration. We consider a two-mode squeezed state of the
field from an inertial perspective, and analytically study the
degradation of entanglement due to the Unruh effect, in the cases of
either one or both observers undergoing uniform acceleration. The
effect of Unruh effect on a quantum radiation can be described by a
two-mode squeezing operator acting on the input state of the quantum
system. In the phase space the symplectic phase-space
representation, $S_{B, \bar B}(r)$ for the two-mode squeezing
transformation is \cite{Adesso2007}
\begin{eqnarray}\label{cmtwomode}
 S_{B,\bar B}(r)=\left(
                   \begin{array}{cc}
                     \cosh r \iden_2 & \sinh r Z_2 \\
                     \sinh r Z_2 & \cosh r \iden_2 \\
                   \end{array}
                 \right),
\end{eqnarray}
where $\iden_2$ is a $2\times 2$ identity matrix and $Z_2=
\mathrm{diag} \{1,-1\}$. After the transformation, the final state
of the entire three-mode system is given by the covariance matrix
\cite{Adesso2007}
\begin{eqnarray}\label{in34}
\nonumber\sigma^{\rm }_{AB \bar B}(s,r) &=&
                 \big[\iden_A \oplus  S_{B,\bar B}(r)\big]
                 \big[\sigma^{\rm (M)}_{AB}(s) \oplus \iden_{\bar B}\big]
                 \big[\iden_A \oplus  S_{B,\bar B}(r)\big]\\
 &=& \left(
       \begin{array}{ccc}
         \mathcal{\sigma}_{A} & \mathcal{E}_{AB} & \mathcal{E}_{A\bar B} \\
         \mathcal{E}^{\sf T}_{AB} &  \mathcal{\sigma}_{B} & \mathcal{E}_{B\bar B} \\
         \mathcal{E}^{\sf T}_{A\bar B} & \mathcal{E}^{\sf T}_{B\bar B} &  \mathcal{\sigma}_{\bar B} \\
       \end{array}
     \right),
\end{eqnarray}
where the diagonal elements are
\begin{equation}
 \begin{aligned}
  & \mathcal{\sigma}_{A}= \cosh(2s)\iden_2, \\
  & \mathcal{\sigma}_{B}=[\cosh(2s) \cosh^2(r) + \sinh^2(r)]\iden_2, \\
  & \mathcal{\sigma}_{\bar B}=[\cosh^2(r) + \cosh(2s) \sinh^2(r)]\iden_2,
 \end{aligned}
\end{equation}
and the non-diagonal elements have the following forms:
\begin{equation}
 \begin{aligned}
  & \mathcal{E}_{AB}=[\cosh(r) \sinh(2s)]Z_2, \\
  & \mathcal{E}_{B\bar B}=[\cosh^2(s) \sinh(2r)]Z_2, \\
  & \mathcal{E}_{A\bar B}=[\sinh(2s) \sinh(r)]Z_2.
 \end{aligned}
\end{equation}
It was found that for two observers undergoing the finite
acceleration, the entanglement vanishes between the lowest-frequency
modes. The loss of entanglement is precisely explained as a
redistribution of the inertial entanglement into the multipartite
quantum correlations among accessible and inaccessible modes from a
noninertial perspective. The classical correlations are also lost
from the perspective of two accelerated observers but conserved if
one of the observers remains inertial.

\citet{Leon2009} investigated the effect of Unruh effect on spin and
occupation number entanglement of Dirac fields in the noninertial
frame. They analyzed spin Bell states and occupation number
entangled state in a relativistic setting, obtained their
entanglement dependence on the acceleration. They showed that the
acceleration produces a qubit$\times$four-level quantum system state
for the spin case, while there is always qubit$\times$qubit for the
spinless case despite their apparent similitude. The entanglement
degradation in the spin case is greater than in the spinless case.
They as well introduced a procedure to consistently erase the spin
information from the system and preserving occupation numbers at the
same time. \citet{MannVillalba09} studied the speeding-up
degradation of entanglement as a function of acceleration for the
free scalar field in an accelerated frame.

\citet{Moradi} studied the distillability of entanglement of
bipartite mixed states of two modes of a free Dirac field  in
accelerated frames. It was showed that there are some certain value
of accelerations which will change the state from a distillable one
into separable one. \citet{Doukas09} studied the loss of spin
entanglement for accelerated electrons in electric and magnetic
fields by using an open quantum system. They found that the proper
time for the extinguishment of entanglement is proportional to the
inverse of the acceleration cubed at high Rindler temperature.
\citet{Ostapchuk09} studied the generation of entangled fermions by
accelerated measurements on the vacuum. \cite{Aspachs:2010} find
that the Unruh-Hawking effect acts on a quantum system as a bosonic
amplification channel.

\citet{wangjing2010} studied the dynamics of quantum entanglement
for Dirac field when the field interacts with noise environment in
noninertial frames. They found that the decoherence induced by the
noise environment  and loss of the entanglement generated by the
Unruh effect will influence each other remarkably. In the case of
the total system  interact with noise environment, the sudden death
of entanglement may appear for any acceleration. However,  sudden
death may only occur when the acceleration parameter is greater than
a critical point when only Rob's qubit under decoherence.

\citet{Hwang2011} examined the entanglement of a tripartite of
scalar field when one of the three parties is moving with uniform
acceleration. The tripartite entanglement exhibits a decreasing
behavior but does not completely vanish in the infinite acceleration
limit, which is different from the behavior of bipartite
entanglement. This fact indicates that the quantum information
processing tasks using tripartite entanglement may be possible even
if one of the parties approaches to the horizon of the Rindler
spacetime.

\citet{wang2011} investigated tripartite entanglement of a fermionic
system when one or two subsystems accelerated. They found that all
the one-tangles decrease with increasing acceleration but never
reduce to zero for any acceleration, which is different from the
scalar case of scalar field. It was shown that the system has only
tripartite entanglement when one or two subsystems with accelerated
motion, which means that the acceleration does not effect the
entanglement structure of the quantum states. The tripartite
entanglement of the case of two observers accelerated decreases much
quicker than the one-observer-accelerated case.

\citet{Olson11} studied quantum entanglement between the future
region and the past region in the quantum vacuum of the Rindler
spacetime. The massless free scalar  fields within the future and
past light cone was quantized as independent systems. The initial
vacuum between the future and past regions became an entangled state
of these systems, which exactly mirrors the prepared entanglement
between the space-like separated Rindler wedges. This lead to the
notion of time-like entanglement. They described an detector which
would exhibit thermal response to the vacuum and discussed the
feasibility of detecting the Unruh effect.

\citet{wangjing2012} discussed the system-environment dynamics of
Dirac fields for amplitude damping and  phase damping  channels  in
noninertial systems. They found that the thermal fields generated by
the Unruh thermal bath promotes the sudden death of entanglement
between the subsystems while postpone the sudden birth of
entanglement between the environments. However, no entanglement was
generated between the system and environment  when the system
coupled with the phase damping environment.

\citet{Montero13} argued that in the infinite acceleration limit,
the entanglement in a bipartite system of the fermionic field must
be independent of the choice of Unruh modes. Therefore, to compute
field entanglement in relativistic quantum information, the tensor
product structures should be modified to give rise to physical
results.

\citet{Khan14} studied the dynamics of tripartite entanglement for
Dirac fields through linear contraction criterion in the noninertial
frames. It is found that the entanglement measurement is not
invariant with respect to the partial realignment of different
subsystems if one observer is accelerated case, while it is
invariant in the two observers accelerated case. It is shown that
entanglement are not generated by the acceleration of the frame for
any bipartite subsystems. Unlike the bipartite states, the genuine
tripartite entanglement does not completely vanish in both one
observer accelerated and two observers accelerated cases even in the
limit of infinite acceleration.

\citet{Dai15JHEP} discussed the entanglement of two accelerated
Unruh-Wald detectors which couple with real scalar fields. The found
that the bipartite entanglement of the two qubits suddenly dies when
the acceleration of one or more qubits are large enough, which is a
result of Unruh thermal bath. \citet{Dai16PRD} studied the
entanglement of three accelerated qubits, each of them is locally
coupled with the real scalar field, without causal influence among
the qubits or among the fields. The obtained how the entanglement
depends on the accelerations of the three qubits and found that all
kinds of entanglement would  sudden death if at least two of three
qubits have large enough accelerations.

\citet{Metwally} studied the possibility of recovering the
entanglement of accelerated qubit and qutrit systems by using
weak-reverse measurements. It is found that the accelerated coded
local information in the qutrit system is more robust than that
encoded in the qubit system. In addition, the non-accelerated
information in the qubit system is not affected by the local
operation compared with that depicted on qutrit system.

\subsubsection{Discordlike correlations}
\citet{Datta09} discussed the QD between two free modes of a scalar
field which are observed by two relatively accelerated observers. It
was showed that  finite amount of QD exists in the regime where
there is no distillable entanglement due to the Unruh effect. In
addition, they provided evidence for a nonzero amount of QD in the
limit of infinite acceleration. \citet{Martin2010c} studied the
behavior of classical and quantum correlations in a spacetime with
an event horizon, comparing fermionic with bosonic fields. They
showed the emergence of conservation laws for classical correlations
and quantum entanglement, pointing out the crucial role that
statistics plays in the entanglement tradeoff across the horizon.

\citet{wang2010c} investigated the distribution of classical
correlations and QD of Dirac modes among different regions in a
noninertial frames. They found that for the Dirac field, the
classical correlation decreases with increasing acceleration, which
is different from the scalar field case where the classical
correlation is independent of the acceleration.

\citet{Ramzan} studied the dynamics of GQD and MIN for noninertial
observers at finite temperature. It was found that the GQD can be
used to distinguish the Bell, Werner, and general type initial
quantum states. In addition, sudden transition in the behavior of
GQD and MIN depends on the mean photon number of the local
environment. In the case of environmental noise is introduced in the
system, this effect becomes more prominent. In the case of
depolarizing channel, the environmental noise is found to have
stronger affect on the dynamics of GQD and MIN as compared to the
Unruh effect. \citet{qiang15} investigated the distribution of GQD
among all possible bipartite divisions of a tripartite system for
the free Dirac field modes in noninertial frames. As a comparison,
they also discussed the geometric measure of entanglement for the
same quantum state.

\subsubsection{Quantum coherence}
\citet{chencoherence} investigated the behavior of quantum coherence
for free scalar and Dirac  modes as detected by accelerated
observers. They showed that the relative entropy of coherence is
destroyed as increasing acceleration of the detectors. In addition,
the shared coherence between the accelerated observers vanishing  in
the infinite acceleration limit for the scalar field, but tends to a
non-vanishing value for the Dirac field.

\citet{huangcoh} studied the freezing condition of coherence for
accelerated free modes in a relativistic setting beyond the
single-mode approximation. They also discussed the behavior of
cohering power and decohering power under the Unruh channel. It was
found that the quantum coherence can be distributed between
different modes, but the coherence lost in the particle mode sector
is not transferred entirely to the antiparticle mode sector. They
also demonstrated that the robustness of quantum coherence are
better than entanglement under the influence of Unruh effect.

\citet{LiuTian2016a,LiuTian2016b} investigated the dynamics of
quantum coherence of two-level atoms interacting with the
electromagnetic field in the absence and presence of boundaries.
They found that for the two-level systems, the quantum coherence
cannot be protected from noise without boundaries. However, in the
presence of a boundary, the insusceptible of the quantum coherence
can be fulfilled when the atoms is close to the boundary and is
transversely polarizable. In addition, in the presence of two
parallel reflecting boundaries, for some special distances  the
quantum coherence of atoms can be shielded from the influence of
external environment when the atoms have a parallel dipole
polarization at arbitrary location between these two boundaries.

\subsection{Free field modes beyond the single-mode approximation} \label{sec:8B}
\citet{SMA2009} introduced an arbitrary number of accessible modes
when analyzing the Unruh effect on bipartite entanglement
degradation. \citet{Bruschi2010} performed that an inertial observer
has the freedom to create excitations in any accessible modes
$\Omega_j, \forall j$ rather than a typical mode. Therefore, one
cannot maps a single-frequency Minkowski mode into a set of single
frequency Rindler modes in an accelerated setting
\cite{Bruschi2010}. That is, the single-mode approximation should be
relaxed in a general setting. To overcome the shortage of the
single-mode approximation, one should employ the Unruh basis which
provides an intermediate step between the Minkowski modes and
Rindler modes. The relations between the Unruh and the Rindler
operators are
\begin{equation}\label{Unruhop}
\begin{aligned}
   & C_{\omega,\text{\text{R}}}=\left(\cosh r_{\omega}\, \hat{a}_{{\omega},\text{I}}-\sinh r_{\omega}\, \hat{b}^\dagger_{{\omega},\text{II}}\right),\\
   & C_{\omega,\text{\text{L}}}=\left(\cosh r_{\omega}\, \hat{a}_{{\omega},\text{II}}-\sinh r_{\omega}\, \hat{b}^\dagger_{{\omega},\text{I}}\right),\\
   & D^\dagger_{\omega,\text{\text{R}}}=\left(-\sinh r_{\omega}\, \hat{a}_{{\omega},\text{I}}+\cosh r_{\omega}\, \hat{b}^\dagger_{{\omega},\text{II}}\right),\\
   & D^\dagger_{\omega,\text{\text{L}}}=\left(-\sinh r_{\omega}\, \hat{a}_{{\omega},\text{II}}+\cosh r_{\omega}\, \hat{b}^\dagger_{{\omega},\text{I}}\right),
 \end{aligned}
\end{equation}
where $\sinh r_{\omega}=(e^{2\pi\omega/a}-1)^{-1/2}$. The Unruh
modes are positive-frequency combinations of plane waves in the
Minkowski spacetime, but enjoy an important property: they are
mapped into single frequency Rindler modes.

For the free scalar fields, the generic Rindler Fock state
$|nm,pq\rangle_{\omega}$ describing both boson and antiboson is

\begin{align}\label{shortnot}
 |nm,pq\rangle_{\omega}\coloneqq \frac{\hat{a}_{{\omega},\text{out}}^{\dagger n}}{\sqrt{n!}}
 \frac{\hat{b}^{\dagger m}_{\omega,\text{in}}}{\sqrt{m!}}\frac{\hat{b}^{\dagger p}_{\omega,\text{out}}}{\sqrt{p!}}
 \frac{\hat{a}^{\dagger q}_{\omega,\text{in}}}{\sqrt{q!}}|0\rangle_S,
\end{align}
where the $\pm$ sign is the notation for boson and antiboson,
respectively. This allows us to rewrite the Unruh vacuum as
\cite{Fabbri,Bruschi2}
\begin{equation}\label{vacuumba}
 |0_\omega\rangle_\text{U}=\frac{1}{\cosh r_\omega^{2}}\sum_{n,m=0}^{\infty}\tanh r_\omega^{n+m}|nn,mm\rangle_{\omega},
\end{equation}
where $|0_\omega\rangle_\text{U}$ is a shortcut notation used to
underline that each Unruh mode $\omega$ is mapped into a single
frequency Rindler mode $\omega$.

One particle Unruh states are defined as $|1_{j}\rangle^+_{\text{U}}
=c_{\omega,\text{U}}^\dagger|0\rangle_\text{H}$, $|1_{j}
\rangle^-_{\text{U}}=d_{\omega,\text{U}}^\dagger|0\rangle_\text{H}$,
where $|0\rangle_\text{H}$ denotes the Hartle-Hawking vacuum. The
particle and antiparticle creation operators for Unruh modes are
defined as
\begin{equation}\label{creat}
 c_{\omega,\text{U}}^\dagger= q_{\text{R}}C^\dagger_{\omega,\text{R}}+q_{\text{L}}C^\dagger_{\omega,\text{L}},~
 d_{\omega,\text{U}}^\dagger= q_{\text{R}}D^\dagger_{\omega,\text{R}}+q_{\text{L}}D^\dagger_{\omega,\text{L}},
\end{equation}
where $q_\text{\text{R}}$ and $q_\text{\text{L}}$ satisfy
$|q_\text{\text{R}}|^2+|q_\text{\text{L}}|^2=1$. The operator
$c_{\omega,\text{U}}^\dagger$ in Eq. \eqref{creat} means the
creation of a pair of particles \cite{Bruschi2,jieci1}, i.e., a
boson with mode $\omega$ in the Rindler region {\it I} and an
antiboson in the Rindler region {\it II}. Similarly, the creation
operator $d_{\omega,\text{U}}^\dagger$ denotes that an antiboson and
a boson are created in Rindler region {\it I} and {\it II},
respectively.

\citet{SMA2009} introduced an arbitrary number of accessible modes
for the Dirac field. They proved that under the single-mode
approximation a fermion only has a few accessible levels due to
Pauli exclusion principle, which is different from the bosonic
fields which has infinite number of excitable levels. This was
argued to justify entanglement survival in the fermionic case at the
infinite acceleration limit under the single-mode approximation. By
relaxing the single-mode approximation, entanglement loss for the
Dirac field mode is limited, which comes from fermionic statistics
through the characteristic structure. In addition, the surviving
entanglement in the infinite acceleration limit is found to be
independent of the the type of fermionic field and the number of
considered accessible modes.

\citet{Bruschi2010} addressed the validity of the single-mode
approximation and discussed the behavior of Unruh effect beyond the
single-mode approximation. They argued that the single-mode
approximation is not valid for arbitrary states in a relativistic
setting. In addition, some corrections to previous studies on
relativistic quantum information beyond the single-mode
approximation are performed both for the bosonic and fermionic
cases. They also exhibited a sequence of wave packets where such
approximation is justified subject to the peaking constraints which
set by some appropriate Fourier transforms.

\citet{Bruschi2} analyzed the tradeoff of quantum entanglement
between particle and anti-particle modes of a charged bosonic field
in a noninertial frame beyond the single-mode approximation. They
found that the redistribution of entanglement between bosonic and
antibosonic modes does not prevent the entanglement from vanishing
in the limit of infinite acceleration. That is, they included
antiparticles in the study of bosonic entanglement by analyzing the
charged bosonic case and find that mode entanglement always vanishes
in this limit. This supports the conjecture that the main
differences in the behavior of entanglement in the bosonic field
mode and fermionic field mode case are due to the difference between
the Bose-Einstein statistics and the Fermi-Dirac statistics.

\citet{Brown2012} demonstrated that quantum correlations measured by
the GQD decays to zero in the limit of infinite acceleration, which
is in contrast with previous research showing that the degradation
of QD vanish in this limit. They argued that the usable quantum
correlations measured by GQD in the large acceleration regime appear
severely limited for any protocols. In addition, vanishing of the
GQD implies a significant limitation on the usable quantum
correlations for large accelerations.

\citet{Tian13} studied the MIN for both Dirac and Bosonic fields in
non-inertial frames  beyond the single-mode approximation. They
found that two  different behaviors exist between the Dirac and
scalar  fields are: (i) the MIN for Dirac fields persists for any
acceleration, while for Bosonic fields this quantity does decay to
zero in the limit of infinite acceleration; (ii) the dynamic
behaviors of the MIN for scalar fields is quite different from the
Dirac fields case in the accelerated frame. In addition, the MIN is
found to be more general than the quantum nonlocality related to
violation of Bell inequalities.

\citet{Richter} studied the entanglement of Unruh modes shared by
two accelerated observers and find some differences in the
robustness of entanglement for these states under the effect of
Unruh thermal bath. For the initial state prepared in Bell states of
free  bosonic and a fermionic modes, they found that the states
$\Psi_\pm$ are entangled for any finite accelerations. However, the
states $\Phi_\pm$ exists entanglement sudden death for some finite
accelerations due to the effect of Unruh radiation. They also
considered the differences in the behavior of entanglement for
fermionic modes and discussed the role that is played by particle
statistics. These results suggest that the degradation of
entanglement in noninertial frames strongly depends on the
occupation patterns of the constituent states.

\subsection{Curved spacetime and expanding universe} \label{sec:8C}

\subsubsection{In the background of a black hole}
As discussed in Refs. \cite{Schuller-Mann,Panqiyuan2008}, the role
of a Rindler observer in the accelerated frame corresponds to a
Schwarzschild observer in the background of a black hole
\cite{Fabbri}. In addition, it was found that the effect of Hawking
radiation of the black hole on a quantum system can be described by
a bosonic amplification channel \cite{Aspachs:2010}. In this case,
we assume Alice stays stationary at an asymptotically flat region of
an external black hole, and Bob is a Schwarzschild observer who
hovers near the event horizon of a black hole. The spacetime
background near a static and asymptotically flat Schwarzschild black
hole, is described by
\begin{equation}\label{matric}
 \begin{split}
  ds^2 =&-(1-\frac{2M}{r}) dt^2+(1-\frac{2M}{r})^{-1} dr^2 \\
        &+r^2(d\theta^2 +sin^2\theta d\varphi^2),
 \end{split}
\end{equation}
where $M$ represents the mass of the black hole.

Solving the KG equation or Dirac equation near the event horizon of
the black hole, one can obtain a set of positive frequency modes
propagating in the exterior and interior regions of the event
horizon. Here, we introduce the quantization of Dirac fields, in
this case the positive (fermions) frequency solutions are found to
be
\begin{equation}\label{inside and outside mode}
\Psi^{+}_{I,\mathbf{k}}=\mathcal {G}e^{-i\omega \mathcal {U}},~
\Psi^{+}_{II,\mathbf{k}}=\mathcal {G}e^{i\omega \mathcal {U}},
\end{equation}
where $\mathcal {U}=t-r_{*}$ and $\mathcal{G}$ is a 4-component
Dirac spinor, $\mathbf{k}$ is the wave vector used to label the
modes and for massless Dirac field we have $\omega=|\mathbf{k}|$.

In terms of these basis, the Dirac field $\Psi$ can be expanded as
\begin{equation}\label{Firstexpand02}
  \Psi =\int d\mathbf{k}[\hat{a}^{out}_{\mathbf{k}}\Psi^{+}_{out,\mathbf{k}}
         +\hat{b}^{out\dag}_{\mathbf{-k}}\Psi^{-}_{out,\mathbf{k}}
        + \hat{a}^{in}_{\mathbf{k}}\Psi^{+}_{in,\mathbf{k}}
         +\hat{b}^{in\dag}_{\mathbf{-k}}\Psi^{-}_{in,\mathbf{k}}],
\end{equation}
where $\hat{a}^{out}_{\mathbf{k}}$ and
$\hat{b}^{out\dag}_{\mathbf{k}}$ are the fermionic annihilation and
antifermion creation operators acting on the state of the exterior
region of the black hole, and $\hat{a}^{in}_{\mathbf{k}}$ and
$\hat{b}^{in\dag}_{\mathbf{k}}$ are the operators acting on the
state in the interior region of the black hole. These operators
$\hat{a}^{out}_{\mathbf{k}}$ satisfy the canonical anticommutation
relations
\begin{equation}
 \begin{split}
  & \{\hat{a}^{out}_{\mathbf{k}}, \hat{a}^{out}_
    {\mathbf{k'}}\}=\delta_{\mathbf{k}\mathbf{k'}},\\
  & \{\hat{a}^{out}_{\mathbf{k}},\hat{a}^{out\dagger}_{\mathbf{k'}}\}
    =\{\hat{a}^{out\dagger}_{\mathbf{k}},\hat{a}^{out\dagger}_{\mathbf{k'}}\}=0
 \end{split}
\end{equation}
where $\{.,.\}$ denotes the anticommutator.

Making analytic continuation for Eq. \eqref{inside and outside mode}
according to the suggestion of Damour-Ruffini \cite{D-R}, a set of
Kruskal modes is obtained. The Kruskal modes can be used to define
the Hartle-Hawking vacuum, corresponding to Minkowski vacuum in an
inertial frame \cite{Fabbri}. These two sets of operators are
related to each other by the Bogoliubov transformation
\begin{gather} \label{Bogoliubov transformation}
 \tilde{a}=\int_{k'} dk'\left[\alpha_{k k'} a_{k'}+\beta^*_{k k'} a_{k'}^{\dagger}\right],
\end{gather}
where $\alpha_{k k'}$ and $\beta_{k k'}$ are the Bogoliubov
coefficients, which encode information about the spacetime. To
quantize the Dirac filed beyond the single-mode approximation
\cite{SMA2009,Bruschi2010}, we construct a different set of
operators in the inside and outside regions of the black hole, which
are
\begin{equation}\label{Dirac-ex}
 \begin{aligned}
  & \tilde{c}^\dagger_{\mathbf{k},R}= \cos r \hat{a}^{out\dagger}_{\mathbf{k}}- \sin r \hat{b}^{in\dagger}_{\mathbf{-k}} \\
  & \tilde{c}^\dagger_{\mathbf{k},L}= \cos r \hat{a}^{in\dagger}_{\mathbf{k}}- \sin r \hat{b}^{out\dagger}_{\mathbf{-k}},
 \end{aligned}
\end{equation}
where
\begin{equation}
\cos r = (e^{-8\pi\omega M}+1)^{-1/2},~ \sin r  = (e^{8\pi\omega
M}+1)^{-1/2}.
\end{equation}
A relevant set of annihilation operators can be constructed in a
analogous way. These modes with subscripts L and R are left and
right Unruh modes. After some calculations, the Hartle-Hawking
vacuum is found to be $|0\rangle_H = \bigotimes_{\mathbf{k}}
|0_{\mathbf{k}}\rangle_{K}$, where
\begin{equation}\label{Dirac-vacuum}
 \begin{aligned}
  |0_{\mathbf{k}}\rangle_{K}=&\cos^2 r|0000\rangle -\sin r |0011\rangle \\
      &+\sin r\cos r|1100\rangle-\sin^2 r|1111\rangle.
 \end{aligned}
\end{equation}
In the last-written equation
\begin{equation}
 |mnm'n'\rangle=|m_{\mathbf{k}} \rangle^{+}_{out}
|n_{-\mathbf{k}}\rangle^{-}_{in} |m'_{-\mathbf{k}}\rangle^{-}_{out}
|n'_{\mathbf{k}} \rangle^{+}_{in},
\end{equation}
with $\{|n_{-\mathbf{k}}\rangle^{-}_{in}\}$ and $\{|n_{\mathbf{k}}
\rangle^{+}_{out}\}$ being the orthonormal bases of the inside and
outside the event horizon of the black hole, and the $\{+,-\}$ is
used to indicate the fermion and antifermion vacuum states.

\citet{Panqiyuan2008} discussed the effect of the Hawking
temperature of a static and asymptotically flat black hole on the
entanglement and teleportation for the free scalar modes. It was
demonstrated that the fidelity of teleportation decreases as the
Hawking temperature of the black increases, which indicates the
thermal bath induced by the Hawking radiation destroys the quantum
channel. The final state are absent of any distillable entanglement
in the infinite Hawking temperature limit, which corresponds to the
case of the black hole evaporating completely.

\citet{gesang} studied the dynamics of entanglement and the fidelity
of teleportation in the background of a rotating black hole with
extra dimensions. The metric of a $d$-dimensional black hole is
given by
\begin{equation}\label{schw}
 \begin{aligned}
   ds^2=&-\left[1-\Bigl(\frac{r_{h}}{r} \Bigr)^{d-3}\right]dt^2
        +\left[1- \Bigl(\frac{r_{h}}{r}
        \Bigr)^{d-3}\right]^{-1}dr^2\\
        &+r^2d\Omega^2_{d-2},
 \end{aligned}
\end{equation}
where $r_{h}$ is the event horizon of the black hole with area
$A_{d}=r^{d-2}_{h}\Omega_{d-2}$, and $\Omega_{d-2}$ is the volume of
a unit $(d-2)$-sphere. From Eq. \eqref{schw}, one can obtain the
mass of the $d$-dimensional black hole, which is
\begin{equation}
\label{mbh} M=\frac{(d-2)r^{d-3}_{h}\Omega_{d-2}}{16\pi G_{d}},
\end{equation}
for the $d$-dimensional Newton's constant $G_{d}$. They discussed
how the extra dimensions, the black hole's mass and angular momentum
parameter, and mode frequency would influence the behavior of
quantum entanglement and fidelity in the curved spacetime. They
showed that a maximally entangled initial  state which is prepared
in an inertial frame becomes less entangled in the curved space due
to the Hawking radiation. In addition, the degree of entanglement
and fidelity of quantum teleportation were found to be degraded with
increasing extra dimension parameter and  surface gravity of the
black hole.

\citet{wang2009plb} studied quantum entanglement of the coupled
massive scalar field in the spacetime of a
Garfinkle-Horowitz-Strominger dilation black hole. The metric for a
Garfinkle-Horowitz-Strominger dilation black hole spacetime is
\cite{Horowitz}
\begin{equation}\label{gem1}
 \begin{aligned}
  ds^2=&- \left(\frac{r-2M}{r-2\alpha}\right)dt^2+\left(\frac{r-2M}{r-2\alpha}\right)^{-1} dr^2 \\
       &+ r(r-2\alpha)d \Omega^2,
 \end{aligned}
\end{equation}
where $M$ is the mass of the black hole and $\alpha$ is the dilation
charge. It was found that entanglement does not depend on the
coupling between the scalar field and the gravitational field and
the mass of the field. As the dilation parameter $\alpha$ increases,
entanglement is destroyed by the Hawking effect. It is interesting
to note that in the limit of $\alpha=M$, corresponding to the case
of an extreme black hole, the system has no entanglement for any
initial state, which its mutual information equals to a nonvanishing
minimum value.

\citet{wangpan2010} studied the quantum projective measurements and
generation of entangled Dirac particles in the background of a
Schwarzschild Black under the single mode approximation. They found
that the  measurements performed by Bob who locates near the event
horizon of the Schwarzschild black hole  creates entangled
particles. The particles can be detected by Alice who stays
stationary at the asymptotically flat region. In addition, the
degree of entanglement increases when the Hawking temperature
increases. \citet{dengwang2010} studied how the Hawking effect of a
black hole influence the entanglement distillability of Dirac fields
in the Schwarzschild spacetime. It was found that  entanglement
distillability of  the states are  influenced both by the Hawking
temperature of the black hole and energy of the fields. Although the
parameter of the generic entangled states affects the entanglement,
it would not change the range in which the states are entangled for
the case of generic entangled states.

\citet{martin2010a} analyzed the entanglement degradation provoked
by the Hawking effect in a bipartite system near the event horizon
of a Schwarzschild black hole beyond the single mode approximation.
They determined the degree of entanglement as a function of the
frequency of the field modes, the distance of the accelerated
observer to the event horizon, and the mass of the black hole. They
found that, in the case of Rob is far off the black hole, all the
interesting phenomena occur in the vicinity and the presence of
event horizons do not effectively degrade the entanglement. They
also discussed the localization of Alice and Rob states in the
curved spacetime.

\citet{martin2010b} studied the generation of quantum entanglement
in the formation of a black hole. It was found that a field in a
dynamical gravitational collapse the vacuum of a field can evolve to
an entangled state. They quantified and discussed the origin of this
entanglement and found that for micro-black hole formation and the
final stages of evaporating black holes, it could even reach the
maximal entanglement limit. In addition, fermions are found to be
more sensitive than bosons to the quantum entanglement generation,
which is helpful in finding experimental evidence of quantum Hawking
effect in analog gravity models.

\citet{wang2010plb} studied how the Hawking radiation influence the
redistribution of the entanglement and mutual information in the
Schwarzschild spacetime. It was shown that the physically accessible
correlations degrade while the unaccessible correlations increase
under the Hawking thermal bath. This is partly because the initial
correlations prepared in an inertial frame are redistributed between
all the bipartite subsystems. In the limit case that the temperature
tends to infinity, the accessible mutual information equals to just
half of its initial value. They also studied the influence of
Hawking radiation on the redistribution of the entanglement and
mutual information of free Dirac field modes in the Schwarzschild
spacetime \cite{wang2010plb}. The results showed that the physically
accessible correlations degrade while the unaccessible correlations
increase with increasing Hawking temperature. That is, the initial
quantum entanglement prepared in inertial frame are redistributed
between all the bipartite modes due to the influence of Hawking
effect. In the limit of infinite Hawking temperature, the physically
accessible mutual information equals to just half of its initial
value. In addition, the unaccessible mutual information between mode
$A$ and $II$ equals to the mutual information between mode $A$ and
$I$.

\citet{Hosler2012} discussed quantum communication between an
observer who free falls into the black hole and an observer hovering
over the horizon of a Schwarzschild black hole. It was found that
the communication channels degrades due to the effect of the
Unruh-Hawking noise. It was showed that for bosonic quantum
communication using single-rail and dual-rail encoding, the
classical channel capacity reduces to a finite value and the quantum
coherent tends to zero by ignoring time dilation which affects all
channels equally. That is, quantum coherence is fully removed  at
infinite acceleration, whereas classical correlation still exist in
this limit.

\citet{jieci1} studied the dynamics of the discord-type quantum
correlation, the measurement-induced disturbance, and classical
correlation of Dirac fields in the background of the
Garfinkle-Horowitz-Strominger dilation black hole. They showed that
all the above mentioned correlations are destroyed as the increase
of black hole's dilation charge. Comparing to the inertial systems,
the quantum correlation measured by QD is always not symmetric with
respect to the measured subsystems, while the measurement-induced
disturbance is always symmetric. In addition, the symmetry of QD is
found to be influenced by the spacetime curvature produced by the
dilation of the black hole.

\citet{hejuan16} discussed the MIN for Dirac particles in the
Garfinkle-Horowitz-Strominger dilation spacetime. They found that as
the dilation parameter increases, the physical accessible MIN
decreases monotonically. The physical accessible correlation is
found to be nonzero when the Hawking temperature is infinite. This
is different from the case of scalar fields and owns to the
statistical differences between the Fermi-Dirac fields and the
Bose-Einstein fields. They also derived the boundary of the MIN
related to Bell-violation and found that the former is more general
than the Bell nonlocality.

The behavior of monogamy deficit and monogamy asymmetry of quantum
steering under the influence of the Hawking effect is studied in
\cite{wangjing2018}. In the curved spacetime, the monogamy of
quantum steering shows an extreme scenario: the first part of a
tripartite system cannot individually steer two other parties, while
it can steer the collectivity of them. In addition, the monogamy
deficit of Gaussian steering are generated due to the influence of
the Hawking thermal bath.

\subsubsection{In an expanding universe}
The spacetime of a homogeneous and isotropic expanding universe is
described by the Friedmann-Lema\^itre-Robertson-Walker (FLRW)
metric, which is
\begin{equation}
 ds^2=dt^2-[a(t)]^2 (dx^2),
\end{equation}
for a two-dimensional geometry. By defining the conformal time
coordinate~$\eta$, the FLRW metric equation is rewritten as
\begin{equation}\label{MconFLRW}
 ds^2=[a(\eta)]^2(d\eta^2- dx^2),
\end{equation}
where $[a(\eta)]^2=C(\eta)$ is the conformal scale factor. To solve
the KG equation in the asymptotic past $\eta\rightarrow-\infty$ and
the asymptotic future $\eta\rightarrow+\infty$ region, we choose the
following conformal scale factor
\begin{equation}
 C(\eta)=1+\epsilon \tanh(\rho\eta)\,,
\end{equation}
where $\epsilon$ and $\rho$ are parameters controlling the total
volume and rapidity of the expansion, respectively. In the
asymptotic past and future, the FLRW universe is asymptotically
flat. The asymptotic solutions of the KG equation in the past and
asymptotic future are
\begin{equation}
 \begin{aligned}
  & u^{\text{in}}_k(x,\eta)&\underset{\eta\rightarrow-\infty}{\longrightarrow}
    \frac{1}{2\sqrt{\pi \omega_{\text{in}}}}e^{i (kx-\omega_{\text{in}}\eta)},\\
  & u^{\text{out}}_k(x,\eta)&\underset{\eta\rightarrow\infty}{\longrightarrow}
    \frac{1}{2\sqrt{\pi \omega_{\text{out}}}}e^{i (kx-\omega_{\text{out}}\eta)},
 \end{aligned}
\end{equation}
where $\omega_{\text{out/in}}=\sqrt{k^2+m^2(1\pm\epsilon)}$.
Considering the properties of hypergeometric functions, the
Bogoliubov coefficient matrix of the scalar field in the FLRW
spacetime is calculated in diagonal form. After some calculations,
the Bogoliubov transform between operators is found to be
\begin{equation}
 a_{k, in}=\alpha_k^* a_{k,out}-\beta_k^* a^\dagger_{-k,out},
\end{equation}
and
\begin{equation}
 a^\dagger_{k,in}=\alpha_k a^\dagger_{k,out}-\beta_k a_{-k,out}
\end{equation}
where
\begin{equation}
\begin{aligned}
 &\alpha_k= \sqrt{\frac{\omega_{\text{out}}}{\omega_{\text{in}}}}
            \frac{\Gamma([1-(i\omega_{\text{in}}/\rho)])
            \Gamma(-i\omega_{\text{out}}/\rho)}{\Gamma([1-(i\omega_{+}/\rho)])
            \Gamma(-i\omega_{+}/\rho)}, \\
&\label{betak}\beta_k= \sqrt{\frac{\omega_{\text{out}}}{\omega_{\text{in}}}}
                       \frac{\Gamma([1-(i\omega_{\text{in}}/\rho)])
                       \Gamma(i\omega_{\text{out}}/\rho)}{\Gamma([1+(i\omega_{-}/\rho)])
                       \Gamma(i\omega_{-}/\rho)}.
 \end{aligned}
\end{equation}

To quantize Dirac fields in the FLRW spacetime \cite{Birrelldavies,
dun1,Bergs}, an appropriate choice for the conformal factor
$a(\eta)$ is \cite{dun1,Birrelldavies}
\begin{equation}
a(\eta)=1+\epsilon(1+\tanh\rho\eta).
\end{equation}

Similarly, one may obtain the solution of the Dirac fields that
behaving as in the asymptotic past and future region of the FLRW
spacetime. Then we calculate the relation between the operators in
the asymptotic future and past region and quantize the field. For
the Dirac field, the Bogoliubov transformations \cite{Birrelldavies}
between $in$ and $out$ modes are
\begin{equation}
 \psi_{in}^{(\pm)}(k)= {\cal A}_{k}^{\pm}\psi^{(\pm)}_{out}(k)
                      +{\cal B}_{k}^{\pm}\psi^{(\mp)*}_{out}(k),
\end{equation}
where the Bogoliubov coefficients ${\cal A}^{\pm}_k$ and ${\cal
B}^{\pm}_k$ that take the form
\begin{equation}
 \begin{aligned}
 & {\cal A}^{\pm}_k=
   \sqrt{\frac{\omega_{out}}{\omega_{in}}}\frac{\Gamma(1-\frac{i\omega_{in}}{\rho})
   \Gamma(-\frac{i\omega_{out}}{\rho})}{\Gamma(1-i\zeta_{(+)}^{\mp}/\rho)
   \Gamma(-i\zeta_{(+)}^{\pm}/\rho)},\\
 &{\cal B}^{\pm}_k=\sqrt{\frac{\omega_{out}}{\omega_{in}}}
   \frac{\Gamma(1-\frac{i\omega_{in}}{\rho})
   \Gamma(\frac{i\omega_{out}}{\rho})}{\Gamma(1+i\zeta_{(-)}^{\pm}/\rho)
   \Gamma(i\zeta_{(-)}^{\mp}/\rho)}.
 \end{aligned}
\end{equation}

\citet{Ball2006} studied entanglement for scalar field modes in the
two-dimensional asymptotically flat Robertson-Walker expanding
spacetime. They showed that the expanding universe generates
entanglement between modes of the scalar field, which conversely
encodes information of the underlying spacetime structure. They
calculated the entanglement in the far future, for the scalar  field
residing in the vacuum state in the distant past. They  pointed out
how the cosmological parameters of the toy Robertson-Walker
spacetime can be extracted from quantum correlations between the
field modes.

\citet{Ahn2007} considered the entanglement of two-mode squeezed
states for scalar fields in the Riemannian spacetime. The system is
prepared as a two mode squeezed state for continuous variables from
an inertial point of view. The initial system is prepared in Unruh
mode $A$ and mode  $B$ in an inertial frame with the covariance
matrix
\begin{eqnarray}\label{inAR}
\sigma^{\rm (M)}_{AB}(s)=\left(
    \begin{array}{cc}
       \mathcal{A}_i(s) & \mathcal{E}_i(s) \\
       \mathcal{E}^T_i(s) & \mathcal{B}_i(s) \\
    \end{array}
                         \right),
\end{eqnarray}
where $\mathcal{A}_i(s)=\mathcal{B}_i(s)=\cosh(2s)\iden_2$, and
$\mathcal{E}_i(s)=\sinh (2s)Z_2$. This setting allows the use of
entanglement measure for continuous variables, which can be applied
to discuss free and bound entanglement from the point of view from
noninertial observer.

\citet{Fuentes2010} found that  entanglement was generated between
modes of Dirac fields in a two-dimensional Robertson-Walker
universe. The entanglement generated by the expansion of the
universe is lower than for the bosonic case for some fixed
conditions \cite{Ball2006}. It was also found that the entanglement
for Dirac fields codifies more information about the underlying
spacetime structure than the bosonic case, thereby allowing us to
reconstruct more information about the history of the expanding
universe. This highlights the importance of the difference between
the bosonic and the fermionic statistics to account for relativistic
effects on the entanglement of field modes.

\citet{Fengj2013} investigated quantum teleportation between the
conformal observer Alice and the inertial observer Bob in de Sitter
space with both free scalar modes and cavity modes. The fidelity of
the teleportation is found to be degraded in both cases, which is
due to the Gibbons-Hawking effect associated with the cosmological
horizon of the de Sitter space. In both schemes, the cutoff at
Planck-scale causes extra modifications to the fidelity of the
teleportation comparing with the standard Bunch-Davies choice.

\citet{Moradi14} studied the spin-particles entanglement between two
modes of Dirac field in the expanding Robertson-Walke spacetime.
They calculated the Bogoliubov transformations for spin-particles
between the asymptotic flat remote past and far future regions. It
was showed that the particles-antiparticles entanglement creation
when passing from remote past to far future due to the articles
creation, while particles entanglement in the remote past degrades
into the far future. They derived analytical expressions of
logarithmic negativity both for spin particles and for spin-less
ones as function of the density of the created particles. In
addition, they highlighted the role of spin of particles for the
dynamics of entanglement in the Robertson-Walke spacetime.

\citet{Fengj2014} studied the quantum correlations and quantum
channel of both free scalar and Dirac modes in de Sitter space. They
found that the entanglement between the free field modes is degraded
due to the radiation associated with the cosmological horizon. They
constructed proper Unruh modes admitting general $\alpha$-vacua
beyond the single-mode approximation and found a convergent feature
of both the bosonic and fermionic cases. In particular, the
convergent points of fermionic entanglement are found to dependent
on the choice of $\alpha$. Moreover, an one-to-one correspondence
between the convergent points of entanglement and zero capacity of
quantum channels in the de Sitter space was proved.

\citet{wangtian2015} studied the parameter estimation for
excitations of Dirac fields in the expanding Robertson-Walker
universe. The optimal precision of the estimation was found to
depend on the dimensionless mass $\tilde{m}$ and dimensionless
momentum $\tilde{k}$ of the Dirac particles. The precision of the
estimation was obtained by choosing the probe state as an
eigenvector of the hamiltonian. This is because the largest quantum
fisher information can be obtained by performing projective
measurements implemented by the projectors onto the eigenvectors of
specific probe states.

\citet{Pierini} investigated the effects of spin on entanglement
arising in Dirac field in the Robertson-Walker universe. They
present an approach to treat the case which only requires charge
conservation, and the case which also requires angular momentum
conservation. It was found that in both situations entanglement
originated from the vacuum have the same behaviors and does not
qualitatively deviates from the spinless case. Differences only
arise for the case in which particles or antiparticles are present
in the input state.

\citet{liunana} studied the thermodynamical properties of scalar
fields in the Robertson-Walker spacetime. They treated scalar fields
in the curved spacetime as a quantum system undergoing a
non-equilibrium transformation. The out-of-equilibrium features were
studied via a formalism which was developed to derive emergent
irreversible features and fluctuation relations beyond the linear
response regime. They applied these ideas to the expanding universe
scenario, therefore the assumptions on the relation between entropy
and quantum matter is not required. They provided a fluctuation
theorem to understand particle production due to the  universe
expansion.

\subsection{Noninertial cavity modes} \label{sec:8D}
\citet{Downes2011} proposed a scheme for storing quantum
correlations in the field modes of  moving cavities in a flat
spacetime. In contrast to previous work where quantum correlations
degradation due to the Unruh-Hawking effect, they found that
entanglement in such systems is protected. They further discussed
the establishment of entanglement and found that the generation of
maximally entangled states between the cavities is in principle
possible. Like free field modes, the dynamics of the scalar field
inside the cavity is also given by the KG equation given in Eq.
\eqref{K-G Equation}. Under the Dirichlet boundary conditions,
solutions of the KG equation are given by the plane waves
\begin{equation}
 u_{n}(t,x)= \frac{1}{\sqrt{n\pi}}\sin\left(\frac{n\pi}{L}[x-x_1]\right)
             e^{-\frac{in\pi}{L}t},
\end{equation}
and the scalar field contained within the cavity walls is
\begin{equation}
 \hat{\phi}_A(t,x)=\sum_n(u_{n}(t,x)\hat{a}_n+u_{n}^*(t,x)\hat{a}_n^{\dagger}),
\end{equation}
where $\hat{a}_n^{\dagger}$ and $\hat{a}_n$ are the creation and
annihilation operators, with $[\hat{a}_n, \hat{a}^{\dagger}_
{n^{\prime}}]=\delta_{n n^{\prime}}$. The Dirichlet boundary
conditions describe the perfectly reflecting mirrors of the scalar
field which is set to vanish on the boundary. Here, Alice's cavity
is inertial and Rob's cavity is described by a uniformly
accelerating boundary condition.

The world line of Rob's cavity is described by the Rindler
coordinates $(\eta,\xi)$ given in Eq. \eqref{Rindlerc}. We assume
that Rob is stationary at spatial location $\xi=\xi_1$ for all
$\eta$, his trajectory in the Minkowski coordinates has the form
$x_1(t)=(t^2+X_1^2)^{1/2}$, where $X_1=a^{-1}e^{a\xi_1}$, and Rob's
proper acceleration is given by $\alpha=X_1^{-1}$. Rob's cavity
consists of two mirrors, one at $\xi_1$ and the other at $\xi_2$ and
stationary with respect to him.

Then, one let Alice and Rob to meet at $t=0$ with their mirrors
aligned, which fixes the position of Alice's cavity as $x_1=X_1$ and
the length of Rob's cavity at $t=0$ to be $X_2-X_1=L$. Therefore,
the length of Rob's cavity in Rindler coordinates is $L^{\prime}=
\frac{1}{a} \ln\left(1+aL\right)$ for all $t$ for fixed $a$.  The
boundary conditions $\phi[\eta,\xi_1]=\phi[\eta,\xi_2]=0$ in this
case are time-independent since the length $L^{\prime}$ is a
constant. The solutions of the KG equation takes the form
\begin{equation}
 v_{n}(\eta,\xi)= \frac{1}{\sqrt{n\pi}}\sin\left(\frac{n\pi}{L^{\prime}}\xi\right)
                  e^{-\frac{in\pi}{L^{\prime}}\eta},
\end{equation}
where $n\in \{1,2,\ldots\}$. Therefore, the scalar field inside the
cavity is
\begin{equation}
\hat{\phi}_R(\eta,\xi)= \sum_n(v_n(\eta,\xi)
\hat{b}_n+v_n^*(\eta,\xi)\hat{b}_n^{\dagger}),
\end{equation}
from Rob's perspective, where $\hat{b}_n^{\dagger}$ and $\hat{b}_n$
are creation and annihilation operators with $[\hat{b}_n,
\hat{b}^{\dagger}_{n^{\prime}}]=\delta_{n n^{\prime}}$. The vacuum
state is defined by $\hat{b}_n|0\rangle_R=0$, $\forall n$, where the
subscript $R$ indicates Rindler cavity. Assuming the cavity's
mirrors is perfectly reflecting. Then one can obtain that if Rob
prepares the cavity in a given Rindler state, it will remain in the
same state for all times \cite{Avagyan2002}.

\begin{figure}
\includegraphics[scale=0.7]{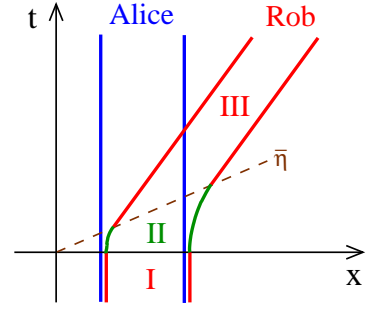}
\caption{The figure shows the trajectories of the cavities
\cite{Bruschi2012}: Alice's cavity keeps inertial,  Rob's cavity is
inertial in region~I and is again inertial in region III, Rob's
cavity is accelerated in region II. Here, $\bar\eta$ is the duration
of the acceleration.}\label{PIchi}
\end{figure}

\citet{Bruschi2012} studied whether the nonuniform motion degrades
entanglement of a relativistic quantum field that is localized both
in space and in time. The field modes in each cavity are discrete
and have the frequencies $\omega_n \coloneqq \sqrt{M^2 + \pi^2 n^2}
/ \delta$, where $M \coloneqq \mu\delta$ and the quantum number are
$n\in \{1,2,\ldots\}$. We then assume Rob undergoes accelerated
motion. The trajectories of the cavities is given in Fig.
\ref{PIchi}. They denote $U_n$ as Rob's field modes with positive
frequency $\omega_n$ before the acceleration and denote $\bar{U}_n$
as Rob's field modes after the acceleration. The two sets of modes
are related by the Bogoliubov transformation
\begin{equation}
 \bar{U}_m= \sum_n \, \bigl( \alpha_{mn} U_n + \beta_{mn} U^*_n \bigr),
\end{equation}
where the Bogoliubov coefficient matrices $\alpha$ and $\beta$ are
determined by the motion of the cavity during the acceleration
\cite{Birrelldavies}. Here, the proper acceleration at the center of
the cavity is $h/\delta$, where the parameter $h$ should satisfy
$h<2$ to ensure the acceleration at the left end of the cavity is
finite. In the region~II, the scalar field positive frequency modes
with respect to $\xi$ are a discrete set $V_n$ with $n\in
\{1,2,\ldots\}$, and their frequencies at the center of the cavity
are $\tilde\Omega_n = (\pi h n)/[2\delta \mathrm{atanh}(h/2)]$ with
respect to the proper time $\tau$.

The Bogoliubov transformation between the two sets of modes can be
computed at the junction $t=0$ \cite{Birrelldavies}. The coefficient
matrices $\alpha$ and~$\beta$, have small $h$ expansions and have
the form
\begin{equation}\label{eq:mrcoeffs}
\begin{aligned}
 & \alpha_{nn}^0 = 1-{\tfrac {1}{240}}\,{\pi }^{2}{n}^{2}{h}^{2} + O(h^4), \\[1ex]
 & \alpha_{mn}^0 = \sqrt {mn} \, {\frac { \bigl( -1+ {( -1 )}^{m-n}\bigr) }{{\pi }^{2} {( m-n )}^{3}}}h + O(h^2), \\[-1ex]
 & \beta_{mn}^0 = \sqrt {mn} \, {\frac { \bigl(1 - {( -1 )}^{m-n} \bigr)}{{\pi}^{2} {( m+n )}^{3}}}h + O(h^2),
\end{aligned}
\end{equation}
where $m\ne n$. Then one can calculate the state of the cavity modes
after the acceleration.

\citet{Friis2012} analyzed quantum entanglement and nonlocality of a
massless Dirac field confined to a cavity. The world tube of the
cavity consists of inertial and uniformly accelerated segments, and
the accelerations are assumed to be small but the travel time is
arbitrarily long. The quantum correlations between the field modes
in the inertial cavity and the accelerated cavity modes are periodic
in the durations of the individual trajectory segments. They found
that the loss of quantum correlations can be entirely avoided by
tuning the relative durations of the segments. Compared with bosonic
correlations, it is easier to calculate the quantum correlations in
the fermionic Fock space because the relevant density matrices act
in lower dimensional Hilbert space due to the fermionic statistics.
Therefore, it is possible to quantify the quantum correlations not
only in terms of the entanglement negativity but also in terms of
the CHSH inequality.

\citet{Brenna} studied the effects of different boundary conditions
and coupling forms on the response of a accelerated particle
detector in optical cavities. Specifically, they considered cavity
fields with periodic, Dirichlet, and Neumann boundary conditions.
They demonstrated that the Unruh effect does indeed occur in a
cavity, which is independent of the boundary conditions. They found
the thermalization properties of the accelerated detector: an
accelerated detector evolves to a thermal state whose temperature
increases linearly with its acceleration. In a non-perturbatively
way, it was proven  that if the switching process is smooth enough,
the detector is thermalized to the Unruh temperature, which is
independent of the type of coupling and the boundary conditions.

\citet{Bruschi2013} discussed how the accelerated motion of a
quantum system can be used to generate quantum gates. They present a
class of sample travel scenarios in which the nonuniform
relativistic motion of a cavity is used to generate two-mode quantum
gates in a quantum system with the continuous variables. They found
that the degree of entanglement between the cavity modes are
produced through resonance of the cavity which appears by repeating
periodically trajectory. In addition, they obtained analytical
expression of the generated entanglement in terms of the magnitude
and direction of the acceleration. The cavity modes are assumed to
be initially at rest and the cavity trajectories are constructed
through the Bogoliubov transformations. In the covariance matrix
formalism, the Bogoliubov transformations are represented by the
symplectic matrix $\mathcal{S}$, which has the form
\begin{eqnarray}
  s_{kk'}\label{nico}&=& \left(
  \begin{array}{cccc}
   \Re({A_{kk'}- B_{kk'}})&\Im({A_{kk'}+B_{kk'}}) \\
   -\Im({A_{kk'}- B_{kk'}})&\Re({A_{kk'}+B_{kk'}})
   \end{array}
   \right),
\end{eqnarray}
where $A_{kk'}$ and $B_{kk'}$ are the Bogoliubov coefficients
associated with the trajectory. By assuming $h=aL\ll1$, the
Bogoliubov coefficients can be expanded to the first order in $h$ as
\begin{equation}
 A_{kk'}=G_k\delta_{kk'}+A_{kk'}^{(1)},~
 B_{kk'}=B_{kk'}^{(1)},
\end{equation}
where $G_{k}=e^{i\omega_{k}T}$ are the phases of the state during
segments of free evolution, $T$ denotes the total proper time of the
segment, and the superscript $(1)$ denotes the first order in $h$.
If the cavity is prepared initially in the vacuum state, the reduced
state of the modes after an $N$-segment trajectory is found to be
\begin{equation}
 \sigma_N=(S_{kk'}^N)^{Tp} S_{kk'}^N,
\end{equation}
where
\begin{eqnarray}
S_{kk'}&=& \left( \begin{array}{cc}
s_{kk}&s_{kk'}\\
s_{k'k}&s_{k'k'}
\end{array}
\right),
\end{eqnarray}
and the transformation $S^N_{kk'}$ corresponds to two mode squeezer
that is a two mode entangling gate.

\citet{Bruschi2013b} studied the mode-mixing quantum gates and
entanglement in nonuniform accelerated cavities. It was showed that
the periodic accelerated motion of the cavity can produce entangling
quantum gates between different frequency modes. The resonant
condition in the cavities which associates with particle creation is
an important feature of the dynamical Casimir effect. It was found
that a second resonance, which has attracted less attention because
it produces negligible particles, generates a beam splitting quantum
gate. This quantum gate leads to a resonant enhancement of quantum
entanglement, which can be regarded as the most important evidence
of acceleration effects in mechanical oscillators.

\citet{Friis2013} analyzed relativistic quantum information for
quantized scalar, spinor, and photon fields in an accelerated
mechanically rigid cavity in the perturbative small acceleration
formalism. The scalar field was analyzed with Neumann and Dirichlet
boundary conditions, and the photon field was discussed under
conductor boundary conditions. The massive Dirac spinor is analyzed
with dimensions transverse to the acceleration. It was found that
for smooth accelerations, the unitarity of time evolution holds,
while for discontinuous accelerations, it fails in 4-dimensions and
higher spacetime. The experimental scenario proposed in
\cite{Bruschi2013b} for the scalar field can also apply to the
photon field.

\citet{Pozas} analyzed the harvesting of classical and quantum
correlations from vacuum for particle detectors. They demonstrated
how the spacetime dimensionality, the detectors' physical size, and
their internal energy structure would impact the detectors'
harvesting ability. They revealed several dependence on these
parameters that can optimize the harvesting of quantum entanglement
and classical correlations. Furthermore, they found that to harvest
vacuum entanglement, smooth switching is more efficient than sudden
switching, especially in the case of the detectors are spacelike
separated.

\citet{Regula2016} investigated entanglement generated between the
modes of two uniformly accelerated bosonic cavities when interacting
with a two-level system. It was found that the inertial and the
accelerated cavity become entangled by letting an atom emitting an
excitation when it passes through the cavities, but the generated
entanglement is degraded against the effects of acceleration. The
generated entanglement is affected not only by the accelerated
motions of the cavities but also by its transverse dimension which
plays the role of an effective mass. In addition, they found that
the extra spatial dimensions contribute to the mass of the field.
Therefore, if the massless bosonic field is used, the degradation
effect of entanglement should not occur.

\subsection{Unruh-DeWitt detectors} \label{sec:8E}
To model the response of an accelerated detector in a quantum field,
the Unruh-DeWitt detector model was performed \cite{UW84}. This
model consists of  a two-level non-interacting atom, which couples
to the external scalar field along its world line in a point-like
manner \cite{wald94}. The response of the Unruh-DeWitt detector
depends on its trajectory and the state of the field. For
definiteness and without loss of generality, we consider an
uniformly accelerated detector, whose world line is given by Eq.
\eqref{Rindlerc}.

Here we study two detectors, one named Alice keeps static and  the
other one named Rob moves with uniform acceleration $a$ for a time
duration $\Delta$. Alice's detector is assumed always switched off
and Rob's detector is switched on during the time duration $\Delta$.
The total Hamiltonian of the system is
\begin{equation}\label{totalh}
 H_{A\, R\, \phi} = H_A + H_R + H_{KG} + H^{R\phi}_{\rm int},
\end{equation}
where $H_{A}=\Omega A^{\dagger}A$, $H_{R}=\Omega R^{\dagger}R$, and
$\Omega$ is the energy gap of the detectors. The interaction
Hamiltonian $H^{R\phi}_{\rm int}(t)$ between Rob's detector and the
external scalar field is
\begin{equation}\label{int}
 H^{R\phi}_{\rm int}(t)= \epsilon(t)\int_{\Sigma_t} d^3 {\bf x} \sqrt{-g} \phi(x)
                         [\chi({\bf x})R +\overline{\chi}({\bf x})R^{\dagger}],
\end{equation}
where $g\equiv {\rm det} (g_{ab})$, and $g_{ab}$ is the Minkowski
metric. Moreover,
\begin{equation}
 \chi(\mathbf{x})=(\kappa\sqrt{2\pi})^{-3}
 \exp(-\mathbf{x}^{2}/2\kappa^{2}),
\end{equation}
is a Gaussian coupling function which vanishes outside a small
volume around the detector. This model describes a point-like
detector which only interacts with its neighbor fields. The total
initial state of detectors-field system has the form
\begin{equation}\label{IS}
 |\Psi_{t_0}^{AR\phi}\rangle=|\Psi_{AR}\rangle\otimes|0_{M}\rangle,
\end{equation}
where
\begin{equation}
 |\Psi_{AR}\rangle=\sin\theta |0_{A}\rangle|1_{R}\rangle+
\cos\theta|1_{A}\rangle |0_{R}\rangle,
\end{equation}
is the initial state of the detectors, and $|0_{M}\rangle$
represents that the external scalar field is in vacuum state from an
inertial perspective.

In weak coupling case, the final state $|\Psi^{R \phi}_{t =
t_0+\Delta} \rangle$ at time $t_0+\Delta$ can be calculated by
employing the first order of perturbation over the coupling constant
$\epsilon$ \cite{UW84}. After some calculations, one can find that
the final state $|\Psi^{R \phi}_{t}\rangle$ at time $t=t_0+\Delta$
is
\begin{equation}\label{primeira_ordem}
 |\Psi^{R \phi}_{t} \rangle = [I - i(\phi(f)R + \phi(f)^{\dagger}
                     R^{\dagger})] |\Psi^{R \phi}_{t_0} \rangle,
\end{equation}
where
\begin{equation}\label{phi(f1)}
 \begin{aligned}
  \phi(f) &\equiv \int d^4 x \sqrt{-g}\chi(x)f \\
          &= i [a_{RI}(\overline{u E\overline{f}})-a_{RI}^{\dagger}(u Ef)],
\end{aligned}
\end{equation}
is the operator of the external scalar field \cite{Landulfo,wald94},
and $f \equiv \epsilon(t) e^{-i\Omega t}\chi ({\bf x})$ is a compact
support complex function. In addition, $u$ is the positive frequency
part from a solution of the KG equation in the Rindler metric
\cite{Landulfo,wald94}, and $E$ is the difference between the
advanced Green function and the retarded Green functions.

\citet{Landulfo} investigated how the teleportation of a quantum
channel is affected by the Unruh effect when one of the entangled
detector is accelerated for a finite amount of proper time. They
performed a detailed analysis of how the acceleration of the
detector and the Unruh effect on the  entangled qubit system. The
mutual information and concurrence between the two detectors are
calculated and showed that the latter has a sudden death at some
fixed finite acceleration. Similarly, the teleportation fidelity
exhibits sudden death behavior via the Unruh effect. The values of
quantum entanglement and mutual information depend on the time
interval along which one of the detectors is accelerated.

\citet{Landulfo1} analyzed the dynamics of QD and classical
correlation for a pair of Unruh-DeWitt detectors when one of them is
uniformly accelerated, and showed that the discord-type quantum
correlation is completely destroyed under the influence of Unruh
thermal bath when one detector is in the limit of infinite
acceleration, while the classical correlation is nonzero for any
acceleration. In particular, unlike the quantum entanglement, the
discord-type quantum correlations exhibits sudden-change behavior at
some certain acceleration parameter. They also discussed how their
results can be interpreted when one of the detector hovers near  the
event horizon of a Schwarzschild black hole.

\citet{Ostapchuk12} discussed the dynamics of quantum entanglement
between a pair of Unruh-DeWitt detectors, one keeps inertial in the
flat spacetime, and the other non-uniformly accelerated in some
specified way. Each of the detectors coupled to the external scalar
quantum field in an indirectly way. The primary problem involving
nonuniformly accelerated detectors in an event horizon is absent and
the Unruh temperature cannot be well defined. By numerical
calculation, they demonstrated that the quantum entanglement in the
weak-coupling limit like those of an oscillator in a bath of
time-depending ``temperature" proportional to the proper
acceleration of the detector, with oscillatory modifications due to
non-adiabatic effects.

Different from the Unruh-DeWitt detector model, a localized solution
to the problem of entanglement degradation in relativistic settings
were performed by \citet{Doukas13,Dragan13}. They prepared a two
mode squeezed state between two observers, the inertial Alice and
the accelerated Rob. The initial state is
\begin{equation}
 \hat{S}_{\text{AB}} |0\rangle_\text{M} =
\exp[s(\hat{a}^\dagger\hat{b}^\dagger -\hat{a}
\hat{b})]|0\rangle_\text{M},
\end{equation}
where the annihilation operators $\hat{a}$ and $\hat{b}$ are
associated with two localized and spatially separated scalar modes
$\phi_\text{A}(x,t)$ and $\phi_\text{B}(x,t)$, respectively. From
the perspective of the accelerated observer, the covariance matrix
of the state has the form \cite{Dragan13}
\label{phi(ef)}
\begin{widetext}
\begin{equation}\label{covariance}
  \begin{aligned}
    \sigma=& \openone+2\langle\hat{n}\rangle_U
             \begin{pmatrix}
                   0&0&0&0\\
                   0&0&0&0\\
                   0&0&1&0 \\
                   0&0&0&1
             \end{pmatrix}
               +2\sinh^2 s
            \begin{pmatrix}
                  |\alpha|^2&0&0&0\\
                  0&|\alpha|^2&0&0\\
                  0&0&|\beta+\beta'^\star|^2&2\,\text{Im}(\beta\beta') \\
                  0&0&2\,\text{Im}(\beta\beta')&|\beta-\beta'^\star|^2
            \end{pmatrix} \\
         &+ \sinh 2s
            \begin{pmatrix}
                  0&0&-\text{Re}[\alpha(\beta+\beta'^\star)]&-\text{Im}[\alpha(\beta-\beta'^\star)]\\
                  0&0&-\text{Im}[\alpha(\beta+\beta'^\star)]&\text{Re}[\alpha(\beta-\beta'^\star)]\\
                  -\text{Re}[\alpha(\beta+\beta'^\star)]&-\text{Im}[\alpha(\beta+\beta'^\star)]&0&0 \\
                  -\text{Im}[\alpha(\beta-\beta'^\star)]&\text{Re}[\alpha(\beta-\beta'^\star)]&0&0
            \end{pmatrix},
     \end{aligned}
  \end{equation}
\end{widetext}
where
\begin{equation}\label{noise}
 \langle\hat{n}\rangle_U = \sum_{k}\frac{|(\psi_\text{B},w_{Ik})|^2}{e^{\frac{2\pi |k| c^2}{a}}-1},~
 w_{Ik}=\frac{1}{\sqrt{4\pi |k| c}}e^{i(k\xi-|k|c\tau)},
\end{equation}
is the average number of Unruh particles seen by an accelerated
detector in the vacuum \cite{Dragan2012}.

\citet{Dragan13} studied the amount of entanglement by using the
localized projective detection model and found that the quantum
correlations are able to extract from the initial state. It was
found that the Unruh thermal noise plays only a minor role in the
degradation process of entanglement. The dominant source of
degradation is the mismatch between the mode Rob observed in the
squeezed state and the mode which is detectable from the accelerated
frame. In addition, leakage of initial mode through the Rindler
horizon places a limit on Rob's ability to fully measure the state,
which leads to an inevitable loss of entanglement that even cannot
be corrected by changing the hardware design of the detectors.

\citet{Doukas13} investigated the quantum entanglement and discord
extractable from a two mode squeezed state as considered by two
detectors, one inertial and the other accelerated. They found that
for large accelerations, the quantum system using localized modes
produces qualitatively different properties than that of Unruh
modes. Specifically, the quantum entanglement of the given quantum
state undergoes a sudden death for some finite acceleration while
the discord asymptotes to zero in the infinite acceleration limit.

\citet{tian2014} studied the dynamics of freely falling and static
two-level detectors interacting with quantized scalar field in de
Sitter spacetime. The atomic transition rates is found to depend on
both the parameter of de Sitter spacetime and the motion of atoms.
They found that the steady states for both cases are always purely
thermal states, regardless of the initial states of the detectors.
In addition, it was found that the thermal baths will generate
entanglement between the freely falling atom and its auxiliary
partner. They also calculated the proper time for extinguishment of
the entanglement between the detectors.

\citet{Lin2015} studied quantum teleportation modeled by
Unruh-DeWitt detectors which initially coupled to a common field. An
unknown coherent state of the inertia detector is teleported to the
agent Rob with relativistic motion, using a detector pair initially
entangled and shared by these two agents. The results showed that
average fidelity of the teleportation always drops below the best
fidelity value from classical teleportation before the detector pair
becomes disentangled. The distortion of the detectors' state can
suppress the fidelity significantly even if the detectors are still
strongly entangled around the light cone. They pointed out that the
dynamics of entanglement are not directly related to the fidelity of
quantum teleportation between the detectors observed in Minkowski
frame or in quasi-Rindler frame.

\citet{Menezes} investigated the radiative processes of entangled
and accelerated atoms interacting with an electromagnetic field
prepared in the Minkowski vacuum. They discussed the structure of
the variation rate of the atomic energy for two atoms moving in
different world lines. The contributions of vacuum fluctuations and
radiation reaction were identified to the generation of entanglement
to the decay of entangled states. The situation where two static
atoms are coupled independently to two spatially separated cavities
at different temperatures is resembled by the results. In addition,
it was found that one of antisymmetric Bell state is a
decoherence-free state for equal accelerations.

\citet{wangtian2016} studied how the Unruh thermal noise influences
the quantum coherence and compared its behavior with entanglement of
the same system. They discussed the frozen condition of coherence
and find that the decoherence of detectors' quantum state is
irreversible under the effect of Unruh thermal bath without any
boundary. Comparing with entanglement which reduces to zero for a
finite acceleration, the coherence-type quantum correlation
approaches zero only in the limit of an infinite acceleration. They
found that the evolution of the detectors' state after the
interaction described by the Hamiltonian \eqref{totalh} can also be
represented by
\begin{eqnarray}
 \rho_{t}^{AR} =\sum_{\mu \nu} M^{A}_\mu \otimes M^{R}_\nu |\Psi_{AR}\rangle\langle\Psi_{AR}|
                (M^{A}_\mu \otimes M^{R}_\nu)^{\dag},
\end{eqnarray}
where $M_{\mu}^{A}$ and $M_{\mu}^{R}$ are the Kraus operators. The
Kraus operators act on Rob's state are
\begin{eqnarray}
 \begin{split}
   & M_1^{R}=\left(\begin{array}{cc}
           \sqrt{1-q}&0\\
           0&\sqrt{1-q}
           \end{array}\right), M_2^{R}=\left(\begin{array}{cc}
           0&0 \\
           v\sqrt{q}&0
           \end{array}\right), \\
   &M^{R}_3 =\left(\begin{array}{cc}
           0&v\\
           0&0
           \end{array}\right).
 \end{split}
\end{eqnarray}
Unlike $M_{\mu}^{R}$, $M_{\mu}^{A}$ is an identity matrix because
Alice's detector keeps static.

The dynamics of steering between two correlated Unruh-Dewitt
detectors when one detector interacts with external scalar field was
studied in \cite{liuwang2018}. The quantum steering is found to be
very fragile under the influence of Unruh thermal noise. In
addition, the quantum steering experience ``sudden death" for some
accelerations, which are quite different from other quantum
correlations in the same system.

\subsection{Quantum correlations and the dynamical Casimir effect} \label{sec:8F}
Like the Unruh effect, dynamical Casimir effect is an important
prediction of QFT in relativistic setting \cite{moore}. The
dynamical Casimir effect predicts that relativistic motion of
boundary conditions would generate pairs of photons from vacuo. Such
prediction has been experimentally observed in a superconducting
circuit architecture \cite{wilson}. The modulation of boundary
condition, theoretically created by a mirror at relativistic speeds,
was achieved by high-frequency modulation of the external magnetic
flux threading a superconducting quantum interferometric device
\cite{wilson}. The experimental demonstration of the dynamical
Casimir effect has triggered a renewed interest in it and has paved
the way for the analysis of the role of Casimir radiation as a
resource for quantum information tasks \cite{discordsabin}.

To understand the creation of photons from vacuum fluctuations when
the boundaries of the electromagnetic field are modulated, one
should quantize the field. In the 2011 experimental observation of
this phenomenon, the relativistic moving mirror was simulated by a
superconducting quantum interferometric device interrupting a
superconducting transmission line \cite{wilson}. The electromagnetic
field confined in the transmission line can be described by a flux
operator $\Phi(x,t)$, which obeys the KG equation. The solution of
the KG equation in the plane-waves basis is \cite{johan,johan2}:
\begin{equation}\label{flujo}
 \Phi(x,t)= \sqrt{\frac{\hbar Z_0}{4\pi}} \int\frac{d\omega}{\sqrt{\omega}}(\hat{a}
                    (\omega)e^{i(-\omega t+kx)}+\hat{b}(\omega)e^{i(-\omega t-kx)}),
\end{equation}
where $k=\omega/v$, and $v$ is the speed of light in the
transmission line, $Z_0=\sqrt{L_0/C_0}$ is the characteristic
impedance. In Eq. \eqref{flujo}, $a(\omega)$ is the annihilation
operator of photons that moves into the mirror and $b(\omega)$
denotes the annihilation operator of the photons moving away from
the mirror. For sufficiently large superconducting quantum
interferometric device plasma frequency \cite{johan}, the charging
energy is small compared with the Josephson energy $E_J(t)=
E_J[\Phi(x,t)]$. Therefore, the superconducting quantum
interferometric device can provide a boundary condition to the flux
field which is analogous to the boundary condition produced by a
relativistic moving mirror. That is,
\begin{equation} \label{boundary}
 \frac{(2\pi)^2}{\phi^2_0}\Phi(0,t)+\frac{1}{L_0}\left.\frac{\partial\Phi(x,t)}{\partial x}\right|_{x=0}=0,
\end{equation}
where $L_0$ is the characteristic inductance per unit length, and
the additional term associated with the capacitance is neglected,
$\Phi_0=h/{2e}$ is the magnetic flux quantum. Inserting Eq.
\eqref{flujo} into the boundary condition Eq. \eqref{boundary}, one
obtain
\begin{equation}
 \begin{aligned}
    \int_0^\omega \frac{d\omega}{\sqrt{\omega}}ik\frac{L_{J}(t)}{L_0}(\hat{b}(\omega)-\hat{a}(\omega))e^{-i\omega t}& \\
  = \int_0^\omega \frac{d\omega}{\sqrt{\omega}}(\hat{a}(\omega) + \hat{b}(\omega))e^{-i\omega t},&
 \end{aligned}
\end{equation}
where $L_J(t)=\frac{\phi^2_0/(2\pi)^2}{E_J(t)}$ is the tunable
Josephson inductance. Then the pair creation of photons in dynamical
Casimir effect  can be calculated using scattering theory which
describes how the time-dependent boundary condition mixes the input
and output modes. By employing the methods discussed in
\cite{johan,johan2}, we obtain the Bogoliubov transformation between
the incoming and outgoing modes, which relates the input and output
vacuum state.

\citet{Felicetti} discussed how the ultrafast modulation of the
qubit-field coupling strength between a superconducting qubit and a
single mode of a superconducting resonator mimics the motion of the
qubit at relativistic speeds. When the qubit follows an effective
oscillatory motion, they find two different regimes. The system is
found to experience unbounded photon generation or resemble the
anti-JC dynamics, which depends on the oscillation frequency.
Moreover, by combining the performed technique with the dynamical
Casimir physics, the toolbox for studying relativistic phenomena
with superconducting circuits can be enhanced.

\citet{Friis13} analyzed the effect of relativistic motion on the
fidelity of continuous variable protocol for quantum teleportation
and proposed a state-of-the-art technology experiment to test their
results. They computed the bounds for the fidelity of teleportation
when one of the observers moves with nonuniform acceleration for a
finite time, which is degraded due to the observer's motion. The
effects of time evolution can be removed by applying time dependent
local operations and the effects of acceleration on the fidelity can
be isolated in this way. In addition, the origin of the fidelity
loss of the quantum teleportation has the same physical regime for
particle generation due to motion-underlying the Unruh (or Hawking)
radiation or the dynamical Casimir effect.

\citet{Alhambra} studied the Casimir-Polder forces experienced by
atoms or molecules in optical cavities. They model the quantum
systems as qubits, and the electromagnetic field components are
modeled as scalar fields with Dirichlet or Neumann boundary
conditions. The light-matter interaction model is used to compute
the Casimir and Casimir-Polder effects. They found that the
diamagnetic term can qualitatively change the Casimir-type forces,
or in other words, it can turn a repulsive force into an attractive
force and vice versa. To be specific, when this term is present, the
atoms are attracted to plates with Dirichlet boundary conditions,
while the plate-atom forces are repulsive without this term. They
also considered the Neumann boundary condition for the atom with or
without diamagnetic coupling term in a cavity, where the forces are
found to have opposite sign to that of the Dirichlet cavity. In
addition, the microscopic-macroscopic transition was studied in this
system, and the results showed that the atoms start to affect the
Casimir force similarly to a dielectric medium for increasing number
of atoms in the cavity.

\citet{Marino} studied the thermal and nonthermal signatures of the
Unruh effect in Casimir-Polder forces. They found that the
Casimir-Polder forces between two uniformly accelerated atoms
exhibit a transition from the short distance thermal-like behavior
to a long distance nonthermal behavior. The former is predicted by
the Unruh effect and the latter is associated with the breakdown of
local inertial descriptions of the system. This effect extends the
Unruh thermal response detected by an accelerated observer to the
spatially extended system of two particles. They identified the
characteristic length scale with the acceleration of the two atoms
for this crossover. Their results were derived separating at fourth
order in perturbation theory and radiation reaction field to the
Casimir-Polder interaction between a pair of atoms separated by a
constant distance and linearly coupled to a scalar field.

\citet{steersabin} investigated the generation of the
Einstein-Podolsky-Rosen steering and Gaussian interferometric power
under the influence of the dynamical Casimir effect. They computed
the quantum steering and the interferometric power generated in the
superconducting waveguide interrupted by the superconducting quantum
interferometric device. It has been shown that, similar with
entanglement and QD \cite{discordsabin}, the value of the
experimental driving amplitude and velocity should be higher than a
critical value to overcome the initial thermal noise and to create
quantum steering. Conversely, the interferometric power is nonzero
for any experimental value of the amplitude and velocity and
increases with the increasing average number of thermal photons. In
other words, any nonzero squeezing produces interferometric power,
while a certain value of squeezing is required to generate quantum
steering.

\citet{coherencesabin} studied how the dynamical Casimir effect
influences the behavior of quantum coherence for Gaussian states in
continuous variables. They found that quantum coherence is
significantly different from zero for any value of the external pump
amplitude for the realistic experimental parameters. This means that
the Casimir radiation creates quantum coherence for any value of the
pump amplitude. In addition, quantum coherence is always greater
than QD and entanglement and exhibits a remarkable robustness again
thermal noise. They believe that quantum coherence is a more
suitable figure of merit of the quantum character for the dynamical
Casimir effect since the experimental requirements for obtaining a
dynamical Casimir effect state with finite coherence are less than
that of entanglement or QD.

\section{Conclusions} \label{sec:9}
Quantum coherence and quantum correlations are fundamental notions
of quantum theory. Although the former was defined with respect to a
single system, while the latter were for bi- and multipartite
systems, they are intimately related to, and can be transformed into
each other through an operational way. When considering their
characterization and quantification, there are many different
methods apply to different situations. They can also be quantified
in a similar manner, e.g., by using the (pseudo) geometric distance
of two states. This unified framework provides the possibility for
understanding the intrinsic connections between these two basic
notions.

We concentrated in this work the recent progresses on the above two
notions, mainly those discussed from the resource theoretic
perspective. After a short introduction, we reviewed in Secs.
\ref{sec:2} and \ref{sec:3} the various quantifiers of geometric
quantum correlations and quantum coherence, which include their
definitions and calculations, and some quantitative relations
between these measures. As these measures are defined by an optimal
(minimum or maximum) distance between the considered state and the
set of states without the quantum property one want to
characterizes, they can be categorized as the geometric
characterization of quantumness.

Building upon the above basic notions, we reviewed in Sec.
\ref{sec:4} the interpretations of the resource theory of quantum
coherence. These include the inter-conversions between coherence and
quantum correlations such as entanglement and QD established in an
operational way, their role on signifying the wave nature of quantum
particle, and the various complementarity relations. This section
also covers a review of the distillation and formation of quantum
coherence for which different free operations and communication
schemes are used.

In Sec. \ref{sec:5}, we summarized the recent investigations of
typical quantum algorithms from the perspective of quantum
coherence, which include the protocol of quantum state merging, the
Deutsch-Jozsa algorithm, the Grover search algorithm, the DQC1
algorithm, and quantum metrology such as the PD task. All these show
that the various quantitative measure of coherence can provide new
viewpoints on the origin of superiority of quantum information
processing tasks.

In Sec. \ref{sec:6}, we reviewed the recent progresses on control of
quantum correlations and quantum coherence in noisy channels. We
first showed that the various quantum correlation and quantum
coherence may be frozen for special forms of initial states, and
there is universal freezing phenomenon for the distance-based
measure of them. We also showed local and nonlocal creation of
quantum correlation and quantum coherence, the cohering power of
quantum channels, as well as the evolution equation and preservation
of quantumness.

In Sec. \ref{sec:7}, we showed some applications of quantum
coherence in the related subjects of condensed matter physics. As
explicit examples, we summarized role of quantum coherence on
studying the long-range order, VBS state, and quantum phase
transitions of the many-body systems. These reveal from one side the
potential of characterizing quantum coherence from a quantification
perspective.

Finally, we showed in Sec. \ref{sec:8} some progresses for the study
of quantum correlations and quantum coherence in relativistic
settings. These involve their behaviors for the free field modes
with and beyond single-mode approximation, for curved spacetime and
expanding universe, for noninertial cavity modes, as well as quantum
correlations for particle detectors and the dynamical Casimir
effects on these correlations. The dynamics of quantum correlations
and quantum coherence under the influence of Unruh temperature,
Hawking temperature, expansion rate of the universe, accelerated
motion of cavities and detectors, and boundary conditions of the
field, have been reviewed. The advantage and disadvantage of free
modes and local modes for the implement of quantum information
processing tasks in noninertial frames and curved spacetime are also
discussed.

Despite the main progresses summarized above which are of broad
interest, there are still many challenging problems need to be
solved in the future. We think, some of the valuable research
directions may be of the following.

The characterization and quantification of quantum coherence under
extended family of free operations. Up to now, most of the proposed
coherence measures were based on the axiomatic postulates of
\citet{coher}, where some of them may be extended in different
circumstances. There may exist other coherence measures which are
both mathematically rigorous and physically significant, e.g., those
under incoherence preserving operations, translationally invariant
operations, SIO, and GIO. Moreover, if the postulates (C1-C3) of
\citet{coher} are somewhat released, other measures of quantum
coherence that are physically relevant may exist. The physical
meanings of those coherence measures are worth exploring. We believe
that the searching process for various coherence quantifications
will deepen our understanding of quantum theory, and new findings
can also be expected.

The intrinsic connections between quantum coherence and quantum
correlations may be another topic needs to be further considered.
Although for relative entropy of coherence, there are some
progresses being achieved in the past two years along this line, for
most of the other measures the interconversion between coherence and
quantum correlations still remain to be exploited. In particular,
while the role of quantum correlations (entanglement, discord,
\emph{etc.}) in explicit quantum communication and computation tasks
have been proved, the role of quantum coherence seems not to be so
convinced. The investigation of their relations and their
interconversion can thus provide interpretations of quantum
coherence from a practical perspective. Moreover, the quantum
coherence measures are questioned as they are basis dependent, the
establishments of their connections with the basis-independent
quantum correlation measures are therefore also significant from a
theoretic point of view.

When considering a real physical system, the detrimental effects of
environments are unavoidable. Though the related decoherence process
have been analyzed extensively via decay of various quantum
correlation measures, the quantum coherence measures defined in a
rigorous framework provide in a real sense the tool for a
quantitative analysis of the decoherence process. Moreover, the
robustness of quantum correlations and quantum coherence against the
detrimental effects of environment are also different. In general,
the former are more fragile under environment coupling than the
latter. But if the two can be converted into each other efficiently,
one can store the quantum correlation of a bi- or multipartite
system by converting it to coherence of a single system, and then
convert it back into quantum correlations when being used.

As a resource theory of quantum coherence, what is really the
resource aspect of them are in fact seldom considered. As far as we
know, the coherence measures have already been related to quantum
protocols such as state merging and quantum state discrimination,
but we think these are by no means the only two roles they will be
played in explicit quantum tasks. Further research, some of the
ideas can be borrowed from the study of quantum correlations, may
help to reveal their potential role as a physical resource.
Moreover, the applications of these coherence measures in other
subjects of physics, e.g., whether they can serve as useful order
parameters for exploiting novel properties of many-body systems, may
be nontrivial topics of future research.

Just as those exciting findings of this field in the past few years,
the solve of the (no limited to) above problems, in our opinion,
will continue to impact the development of basic fundamental quantum
theory and the implementation of various new quantum technologies
which strongly depend on quantum correlations and quantum coherence.

\section*{Acknowledgments}
This work was supported by NSFC (Grants No. 11675129, 91536108,
11675052, and 11774406), National Key R \& D Program of China (Grant
Nos. 2016YFA0302104, 2016YFA0300600), New Star Project of Science
and Technology of Shaanxi Province (Grant No. 2016KJXX-27), CAS
(Grant No. XDB), Hunan Provincial Natural Science Foundation of
China (2018JJ1016), and New Star Team of XUPT, .

\newcommand{\PRL}{Phys. Rev. Lett. }
\newcommand{\RMP}{Rev. Mod. Phys. }
\newcommand{\PRA}{Phys. Rev. A }
\newcommand{\PRB}{Phys. Rev. B }
\newcommand{\PRD}{Phys. Rev. D }
\newcommand{\PRE}{Phys. Rev. E }
\newcommand{\PRX}{Phys. Rev. X }
\newcommand{\NJP}{New J. Phys. }
\newcommand{\JPA}{J. Phys. A }
\newcommand{\JPB}{J. Phys. B }
\newcommand{\PLA}{Phys. Lett. A }
\newcommand{\PLB}{Phys. Lett. B }
\newcommand{\NP}{Nat. Phys. }
\newcommand{\NC}{Nat. Commun. }
\newcommand{\SR}{Sci. Rep. }
\newcommand{\QIP}{Quantum Inf. Process. }
\newcommand{\QIC}{Quantum Inf. Comput. }
\newcommand{\AoP}{Ann. Phys. }
\newcommand{\PR}{Phys. Rep. }
\newcommand{\EPL}{Europhys. Lett. }
\newcommand{\IJTP}{Int. J. Theor. Phys. }
\newcommand{\IJMPB}{Int. J. Mod. Phys. B }
\newcommand{\IJQI}{Int. J. Quantum Inf. }
\newcommand{\CMP}{Commun. Math. Phys. }
\newcommand{\CTP}{Commun. Theor. Phys. }
\newcommand{\JMO}{J. Mod. Opt. }

\nocite{*}
\bibliographystyle{apsrmp}

\end{document}